\documentclass[10pt,a4paper,oneside]{article}                 %
\usepackage{amsmath}                                       %
\usepackage{amssymb}                                       %
\usepackage{graphicx}                                      %
\usepackage{fancyhdr}                                      %
\usepackage[T1]{fontenc}                                   %
\usepackage{txfonts}                                       %
\usepackage[latin1]{inputenc}                              %
\usepackage{pifont}                                        %
\usepackage{url,hyperref}                                           %
\usepackage{lastpage}
\usepackage{fancyhdr}
\usepackage[comma,sort&compress]{natbib}                   %
\usepackage{siunitx}
\usepackage{csquotes}

\renewcommand{\headrule}{\centerline{eLISA: Astrophysics and cosmology in the millihertz regime}}%             %

\begin{document}

\thispagestyle{empty}

\vfill
\begin{center}
{\Large \bf Doing science with eLISA:} \\[1cm]
{\Large  \bf Astrophysics and cosmology in the millihertz regime} \\[3cm]

{\large Pau Amaro-Seoane$^{1,\,13}$,
Sofiane Aoudia$^{1}$,
Stanislav Babak$^{1}$,
Pierre Bin\'etruy$^{2}$,
Emanuele Berti$^{3,\,4}$,
Alejandro Boh{\'e}$^{5}$,
Chiara Caprini$^{6}$,
Monica Colpi$^{7}$,
Neil J. Cornish$^{8}$,
Karsten Danzmann$^{1}$,
Jean-Fran\c cois Dufaux$^{2}$,
Jonathan Gair$^{9}$,
Oliver Jennrich$^{10}$,
Philippe Jetzer$^{11}$,
Antoine Klein$^{11,\,8}$,
Ryan N. Lang$^{12}$,
Alberto Lobo$^{13}$,
Tyson Littenberg$^{14,\,15}$,
Sean T. McWilliams$^{16}$,
Gijs Nelemans$^{17,\,18,\,19}$,
Antoine Petiteau$^{2,\,1}$,
Edward K. Porter$^{2}$,
Bernard F. Schutz$^{1}$,
Alberto Sesana$^{1}$,
Robin Stebbins$^{20}$,
Tim Sumner$^{21}$,
Michele Vallisneri$^{22}$,
Stefano Vitale$^{23}$,
Marta Volonteri$^{24,\,25}$, and
Henry Ward$^{26}$
}

\vfill

{\footnotesize
$^1$Max Planck Institut f\"ur Gravitationsphysik
(Albert-Einstein-Institut), Germany\\
$^2$APC, Univ. Paris Diderot, CNRS/IN2P3, CEA/Irfu, Obs. 
    de Paris, Sorbonne Paris Cit{\'e}, France\\
$^{3}$Department of Physics and Astronomy, The University of
      Mississippi, University, MS 38677, USA\\
$^{4}$Division of Physics, Mathematics, and Astronomy, California 
      Institute of Technology, Pasadena CA 91125, USA\\
$^5$UPMC-CNRS, UMR7095, Institut d'Astrophysique de Paris, F-75014, 
    Paris, France\\
$^6$Institut de Physique Th{\'e}orique, CEA, IPhT, CNRS, 
    URA 2306, F-91191Gif/Yvette Cedex, France\\
$^7$University of Milano Bicocca, Milano, I-20100, Italy\\
$^8$Department of Physics, Montana State 
   University, Bozeman, MT 59717, USA\\
$^9$Institute of Astronomy, University of Cambridge, Madingley Road, 
    Cambridge, CB3 0HA, UK\\
$^{10}$ESA, Keplerlaan 1, 2200 AG Noordwijk, The Netherlands\\
$^{11}$Institute of Theoretical Physics University of Z{\"u}rich,
       Winterthurerstr. 190, 8057 Z{\"u}rich Switzerland\\
$^{12}$Washington University in St. Louis, One Brookings Drive, St. Louis, MO
       63130, USA\\
$^{13}$Institut de Ci{\`e}ncies de l'Espai (CSIC-IEEC), Campus UAB,
Torre C-5, parells, $2^{\rm na}$ planta, ES-08193, Bellaterra,
Barcelona, Spain\\
$^{14}$Maryland Center for Fundamental Physics, Department of Physics, University of Maryland, 
       College Park, MD 20742\\       
$^{15}$Gravitational Astrophysics Laboratory, 
       NASA Goddard Spaceflight Center, 8800 
       Greenbelt Rd., Greenbelt, MD 20771, USA\\
$^{16}$Department of Physics, Princeton University, Princeton, NJ 08544, USA\\
$^{17}$Department of Astrophysics, Radboud University Nijmegen, The Netherlands\\
$^{18}$Institute for Astronomy, KU Leuven, Celestijnenlaan 200D, 3001 Leuven,
Belgium\\
$^{19}$Nikhef, Science Park 105, 1098 XG Amsterdam, The Netherlands\\
$^{20}$NASA Liason, NASA GSFC, USA\\
$^{21}$Imperial College, UK\\
$^{22}$Jet Propulsion Laboratory, California Inst. of Technology, 
       Pasadena, CA 91109, USA\\
$^{23}$University of Trento, Department of Physics, 
       I38050 Povo, Trento, Italy\\
$^{24}$Institut d'Astrophysique de Paris, 98bis Boulevard Arago, 75014 Paris, France\\
$^{25}$Astronomy Department, University of Michigan, Ann Arbor, MI 48109, USA\\ 
$^{26}$Institute for Gravitational Research, Department of Physics \& Astronomy
       Kelvin Building, University of Glasgow, Glasgow
}
\vfill

\end{center}

\newpage

\tableofcontents

\newpage

\section*{Abstract}

This document introduces the exciting and fundamentally new science and
astronomy that the European New Gravitational Wave Observatory (NGO) mission
(derived from the previous LISA proposal) will deliver. The mission (which we
will refer to by its informal name ``eLISA'') will survey for the first time
the low-frequency gravitational wave band (about $0.1$~mHz to 1~Hz), with
sufficient sensitivity to detect interesting individual astrophysical sources
out to $z=15$. The measurements described here will address the basic
scientific goals that have been captured in ESA's ``New Gravitational Wave
Observatory Science Requirements Document''; they are presented here so that
the wider scientific community can have access to them. The eLISA mission will
discover and study a variety of cosmic events and systems with high
sensitivity: coalescences of massive black holes binaries, brought together by
galaxy mergers; mergers of earlier, less-massive black holes during the epoch
of hierarchical galaxy and black-hole growth; stellar-mass black holes and
compact stars in orbits just skimming the horizons of massive black holes in
galactic nuclei of the present era; extremely compact white dwarf binaries in
our Galaxy, a rich source of information about binary evolution and about
future Type Ia supernovae; and possibly most interesting of all, the uncertain
and unpredicted sources, for example relics of inflation and of the
symmetry-breaking epoch directly after the Big Bang.  eLISA's measurements will
allow detailed studies of these signals with high signal-to-noise ratio,
addressing most of the key scientific questions raised by ESA's Cosmic Vision
programme in the areas of astrophysics and cosmology.  They will also provide
stringent tests of general relativity in the strong-field dynamical regime,
which cannot be probed in any other way.  This document not only describes the
science but also gives an overview on the mission design and orbits.  LISA's
heritage in the eLISA design will be clear to those familiar with the previous
proposal, as will its incorporation of key elements of hardware from the LISA
Pathfinder mission, scheduled for launch by ESA in 2014. But eLISA is
fundamentally a new mission, one that will pioneer the completely new science
of low-frequency gravitational wave astronomy.

\newpage

\section{Introduction}

Our view of the Universe has changed dramatically over the past
century. Less than a hundred years ago our own Galaxy, the Milky Way,
was believed to be our own island-Universe. The discovery of hundreds
of billion galaxies like our own, and of billion luminous sources such
as Quasi Stellar Objects (QSOs), changed our perception of the cosmic
landscape. New astronomical objects were discovered with the advent of
radio and X-ray Astronomy. Relativistic binaries composed of compact stars (such as
white dwarfs or neutron stars) and stellar-mass black holes are among
these sources of electromagnetic radiation. According to the accretion
paradigm, supermassive black holes at galactic centres are the
simplest explanation for the power emitted by distant, luminous QSOs,
but a conclusive test of this hypothesis is still lacking.

The remarkable discovery of the recession of galaxies and of the
fossil microwave background radiation, almost contemporary to the
discovery of X-ray sources, has further led to the emergence of a
cosmological paradigm, the Big Bang, that has revolutionized our
description of the Universe. We now know that our Universe had a
beginning and that its luminous components (in particular, galaxies
and QSOs) evolve jointly and in concordance with the evolution of the
underlying dark matter permeating the Universe.

According to General Relativity, black holes and compact binaries are
expected to be powerful sources of {\em gravitational waves}.  Rather
than ``seeing'' electromagnetic radiation, as all of astronomy has done
until present, eLISA will ``hear'' the vibrations of the fabric of
spacetime itself, emitted coherently by macroscopic bodies.  Studying
these signals will convey rich new information about the behaviour,
the structure and the history of the Universe, and it will clarify
several issues in fundamental physics.

Gravitational waves travel undisturbed through spacetime, and when
observed they offer a new and uniquely powerful way to probe the {\it
  very distant Universe}, from the extremely early Big Bang to the
early epoch of galaxy and black hole seed formation. This may allow us
to address deep questions. What powered the Big Bang? How did galaxies
and their black holes form and evolve? What is the structure of
spacetime around the massive objects we believe to be black holes?
What is the nature of the mysterious dark matter and dark energy
accelerating the expansion of the Universe?

eLISA is a space-based mission designed to measure gravitational
radiation over a broad band at frequencies ranging between $f\sim 0.1$
mHz and $f\sim 1$ Hz. In this frequency band the Universe is richly
populated by strong sources of gravitational waves.  For binary
systems the characteristic gravitational-wave frequency $f$ is twice
the Keplerian orbital frequency, which in turn is proportional to
$(M/a^3)^{1/2}$, where $M$ is the total mass of the binary and $a$ its
semi-major axis.  In the eLISA frequency band, gravitational waves are
produced by close binaries of stellar-mass objects with orbital
periods of a few to several minutes. Massive black hole binaries with
$M\sim 10^4\,M_{\odot}-10^7\,M_{\odot}$ and mass ratio $0.01\lesssim
q\lesssim 1$ on the verge of coalescing have orbital frequencies
sweeping to higher and higher values, until the binary separation $a$
becomes as small as the scale of the event horizon $GM/c^2$. Finally,
eLISA could observe binaries comprising a massive black hole and a
stellar-mass compact object (e.g., a stellar-mass black hole) skimming
the horizon of the larger black hole before being captured: these
systems are commonly referred to as extreme mass ratio inspirals
(EMRIs).  Furthermore, a stochastic background in the eLISA frequency
band can be generated by less conventional sources, such as phase
transitions in the very early Universe and/or cosmic strings.

This note is a comprehensive survey of the eLISA science case. We
consider all the relevant astrophysical and cosmological gravitational
wave sources and explore eLISA detection performances in terms of
sensitivity, SNR distributions, and parameter estimation.

In Section \ref{sec.desc-mission} we will briefly describe the {\it
  mission concept} and the basic design of the instrument, introducing
the eLISA sensitivity curve that will be used throughout the study.

Our survey of the eLISA science case will start in {Section
  \ref{sec.ultra-comp-binar}} by exploring the nearest observable
sources of gravitational waves, i.e. {\it compact stellar-mass
  binaries} in the Milky Way.  eLISA will study the gravitational wave
signals from thousands of stellar-mass close binaries in the Galaxy
and will give information on the extreme endpoints of stellar
evolution. eLISA will provide distances and detailed orbital and mass
parameters for hundreds of these binaries. This is a rich trove of
information for mapping and reconstructing the history of stars in the
Galaxy, and it can reveal details of the tidal and non-gravitational
influences on the binary evolution associated with the internal
physics of the compact remnants themselves.

Then we will summarize the science objectives that are relevant for
the {\it astrophysics of black holes} {(Section
  \ref{sec.black-hole-astr})}.  Current electromagnetic observations
are probing only the tip of the black hole mass distribution in the
Universe, targeting black holes with large masses, between $10^7\,{\rm
  M_{\odot}}$ and $10^9\,{\rm M_{\odot}}$.  Conversely, eLISA will be
able to detect the gravitational waves emitted by black hole binaries
with total mass (in the source rest frame) as small as $10^4\,{\rm
  M_{\odot}}$ and up to $10^7\,{\rm M_{\odot}}$, out to a redshift as
large as $z\sim 20$.  eLISA will detect fiducial sources out to
redshift $z\sim 10$, and so it will explore almost all the
mass-redshift parameter space relevant for addressing scientific
questions on the evolution of the black hole population.  Redshifted
masses will be measured to an unprecedented accuracy, up to the 0.1 --
1 \% level, whereas absolute errors in the spin determination are
expected to be in the range 0.01 to 0.1, allowing us to reconstruct
the cosmic evolution of massive black holes. Black holes are expected
to transit into the mass interval to which eLISA is sensitive along
the course of their cosmic evolution. Thus, eLISA will map and mark
the loci where galaxies form and cluster, using black holes as clean
tracers of their assembly.

eLISA will also bring a new revolutionary perspective to the study of
{\it galactic nuclei}, as shown {in Section
  \ref{sec.emris-astr-dense}}.  Orbits of stellar objects captured by
the massive black hole at the galactic centre evolve by gravitational
radiation.  By capturing their signal, eLISA will offer the deepest
view of nearby galactic nuclei, exploring regions that are invisible
to electromagnetic techniques.  eLISA will probe the dynamics of
compact objects in the space-time of a Kerr black hole, providing
information on the space density of those objects.

In {Section \ref{sec.black-hole-physics}} we address key questions
concerning the nature of spacetime and gravity.  GR has been
extensively tested in the weak field regime, both in the solar system
and via binary pulsar observations. eLISA will provide a unique
opportunity to {\it probe GR in the strong field limit.}  eLISA will
observe the coalescence of massive black hole binaries moving at
speeds close to the speed of light, and enable us to test the dynamics
of curved spacetime when gravitational fields are strong. By observing
a large number of orbital cycles during the last few years of the
inspiral of a stellar mass object into a massive black hole, eLISA
will allow us to measure precisely the parameters of the central
object (including its quadrupole moment) in the near Universe.  Any
deviations in the orbital motion from GR predictions will leave an
imprint in the gravitational wave phase. Thus, measurements of the
mass, spin and quadrupole moment of the central object will allow us
to check the Kerr nature of the central massive object, and to {\it
  test for the first time the black hole hypothesis.}

Lastly, {as we describe in Section \ref{sec.new-physics-early}}, eLISA
will probe {\it new physics and cosmology} with gravitational waves,
and search for unforeseen sources of gravitational waves.  The eLISA
frequency band in the relativistic early Universe corresponds to
horizon scales where phase transitions or extra dimensions may have
caused catastrophic, explosive bubble growth and efficient
gravitational wave production. eLISA will be capable of detecting a
stochastic background from such events from about 100 GeV to about
1000 TeV, if gravitational waves in the eLISA band were produced with
sufficient efficiency. 

In closing the note, we will present a summary of the science
objectives of the eLISA mission. We summarize the key theoretical and
observational goals of the eLISA science case in Section
\ref{sec.sci-obs-req}. There, in a schematic bullet-point form, we
enumerate the scientific goals related to each class of gravitational
wave sources and the observational performance of eLISA in achieving
such goals. This could serve as a compact summary of eLISA science, as
well as a reference point for other space-based gravitational wave
detector proposals.

\section{Description of the mission}
\label{sec.desc-mission}

eLISA is a European-led variant of LISA that can be launched before
2022 .  The basic principle of gravitational wave detection for eLISA
is the same as for LISA: it is a laser interferometer designed to
detect the passage of a gravitational wave by measuring the
time-varying changes of optical pathlength between free-falling
masses. Many design and technological developments were migrated from
LISA, however there are some substantial differences.

\begin{figure}
\resizebox{\hsize}{!}
         {\includegraphics[scale=1,clip]{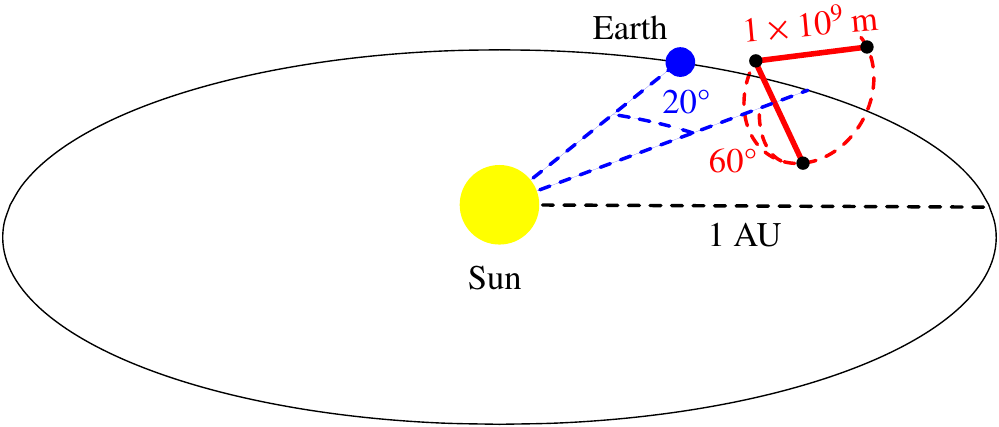}}
\caption
  {  
The eLISA orbits: The constellation is shown trailing the Earth by about
20 degrees (or \ensuremath{5 \times 10^{10} \textrm{km}}) and is inclined by 60
degrees with respect to the ecliptic. The trailing angle will vary over the
course of the mission duration from 10 degrees to 25 degrees.  The separation
between the spacecraft is \ensuremath{L = 1 \times 10^9} m.
  }
\label{fig.mission_orbits}
\end{figure}

The two measurement arms are defined by three spacecraft orbiting the Sun in a
triangular configuration (see figure~\ref{fig.mission_orbits}). A key feature
of the eLISA concept is a set of three orbits that maintain a near-equilateral
triangular formation with an armlength \ensuremath{L = 10^9 \textrm{m} },
without the need for station-keeping. Depending on the initial conditions of
the spacecraft, the formation can be kept in an almost constant distance to the
Earth or be allowed to slowly drift away to about \ensuremath{70 \times 10^{9}}
m, the outer limit for communication purposes.  A very attractive feature of
the eLISA orbits is the almost constant sun-angle of 30 degrees with respect to
the normal to the top of the spacecraft,  thereby resulting in an extremely
stable thermal environment, minimizing the thermal disturbances on the
spacecraft.

\begin{figure}
\resizebox{\hsize}{!}
         {\includegraphics[scale=1,clip]{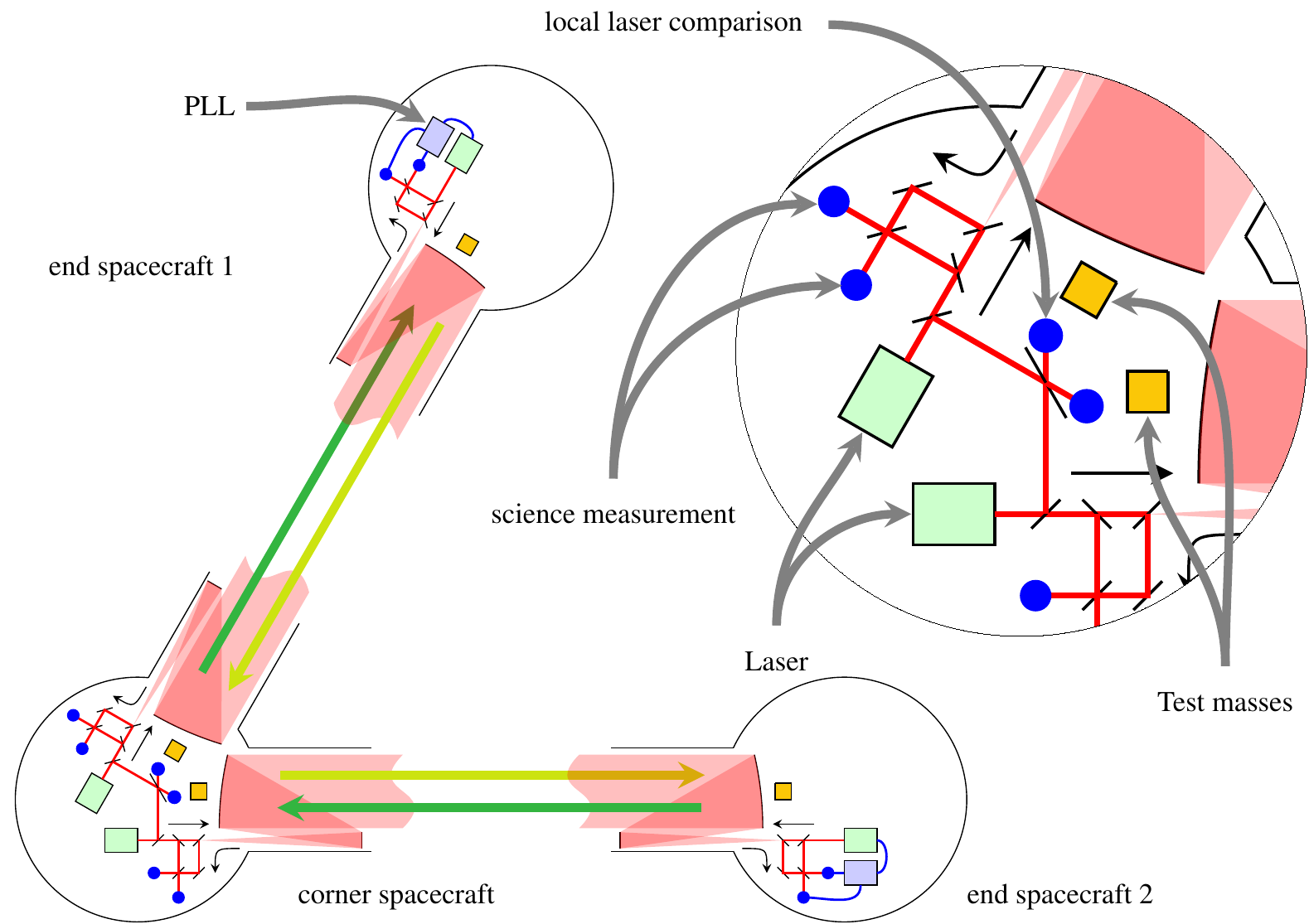}}
\caption
  {  
The constellation of the three eLISA spacecraft constitutes the science
instrument. The central spacecraft harbors two send/receive laser ranging
terminals, while the end spacecraft has one each. The laser in the end
spacecraft is phase-locked to the incoming laser light. The blue dots indicate
where interferometric measurements are taken. The sketch leaves out the test
mass interferometers for clarity.
  }
\label{fig.mission_mes_all}
\end{figure}

\begin{figure}
\resizebox{\hsize}{!}
         {\includegraphics[scale=1,clip]{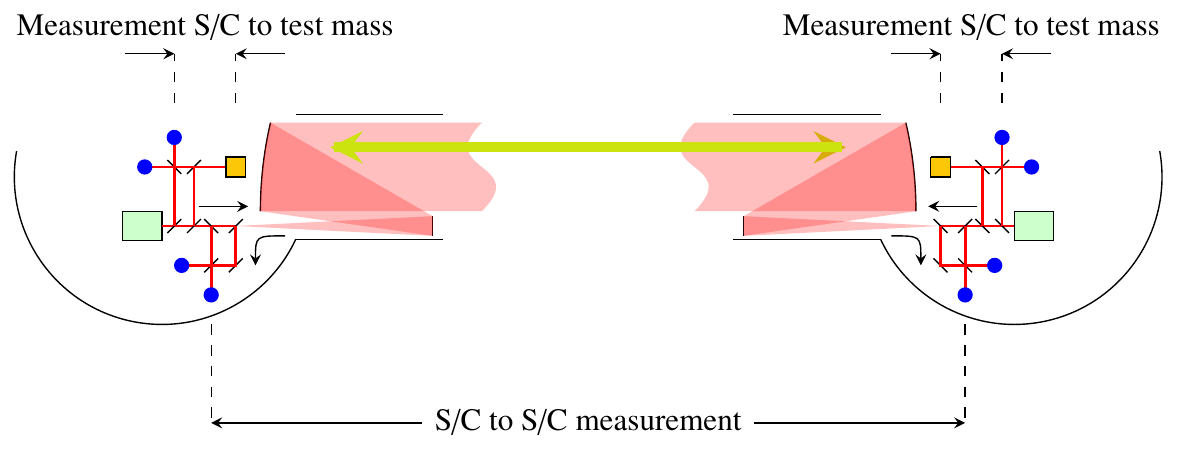}}
\caption
  {  
Partition of the eLISA measurement. Each measurement between two test masses is broken up into three different measurements: two between 
the respective test mass and the spacecraft and one between the two spacecraft (S/C). As the noise in the measurement is dominated 
by the shot noise in the S/C-S/C measurement, the noise penalty for the partitioning of the measurement is negligible. The blue (solid) dots indicate 
where the interferometric measurements are taken.
  }
\label{fig.mission_mes_mass}
\end{figure}

One of the three spacecraft serves as the ``central hub'' and defines
the apex of a ``V''. Two other, simpler spacecraft are positioned at
the ends of the V-shaped constellation. The central spacecraft houses
two free-falling ``test masses'' that define the endpoint of the two
interferometer arms. The other spacecraft contain one test mass each,
defining two more endpoints (see
figure~\ref{fig.mission_mes_all}). Each spacecraft accommodates the
interferometry equipment for measuring changes in the arm length. For
practical reasons, this measurement is broken up into three distinct
parts (see figure~\ref{fig.mission_mes_mass}): the measurement between
the spacecrafts, i.e. between the optical benches that are fixed to
each spacecraft, and the measurement between each of the test masses
and its respective optical bench.  Those measurements are recombined
in a way that allows us to reconstruct the distance between the test
masses which is insensitive to the noise in the position of the
spacecraft with respect to the test masses.

A second key feature of the eLISA concept is that the test masses are
protected from disturbances as much as possible by a careful design
and the "drag-free" operation. To establish the drag-free operation, a
housing around the test mass senses the relative position of test mass
and spacecraft, and a control system commands the spacecraft
thrusters to follow the free-falling mass. Drag-free operation reduces
the time-varying disturbances to the test masses caused by force
gradients arising in a spacecraft that is moving with respect to the
test masses. The requirements on the power spectral density of the
residual acceleration of the test mass is

\begin{equation}
S_{\rm x, acc} (f) =  2.13 \times 10^{-29} \left( 1 + { 10^{-4}{\rm Hz} \over f }  \right)
                         {\rm m}^2\,{\rm s}^{-4}\,{\rm Hz}^{-1}
\end{equation}

\noindent
or

\begin{equation}
{ S_{\rm x, acc} (f) = 1.37 \times 10^{-32} \left( 1 + { 10^{-4}{\rm Hz} \over f } \right) {{\rm Hz} \over f^4}
 {\rm m}^2\,{\rm Hz}^{-1} },
\end{equation}

\noindent
where $f$ is the frequency.

The third key feature, the distance measuring system, is a continuous interferometric laser ranging scheme, similar to that used for radar-tracking 
of spacecraft. The direct reflection of laser light, such as in a normal Michelson interferometer, is not feasible due to the large distance between 
the spacecrafts. 
Therefore, lasers at the ends of each arm operate in a "transponder" mode. A laser beam is sent out from the central spacecraft to an end spacecraft. 
The laser in the end spacecraft is then phase-locked to the incoming beam thus returning a high-power phase replica. The returned beam is received 
by the central spacecraft and its phase is in turn compared to the phase of the local laser. 
A similar scheme is employed for the second arm. In addition, the phases of the two lasers serving the two arms are compared within the central spacecraft. 
The combined set of phase measurements together with some auxiliary modulation allows to determine the relative optical path changes 
with simultaneous suppression  of the laser frequency noise  and clock noise below the secondary  (acceleration and displacement) noise. 
The displacement noise has two components: the shot noise, with a required power spectral density

\begin{equation}
 S_{\rm x,sn} (f) = 5.25 \times 10^{-23} \; {\rm m}^2\,{\rm Hz}^{-1} 
\end{equation}

\noindent
and the other (combined) measurement noise 
with a required power spectral density  

\begin{equation}
{ S_{\rm x,omn}(f) = 6.28 \times 10^{-23} \; {\rm m}^2\,{\rm Hz}^{-1}}.
\end{equation}

\begin{figure}
\resizebox{\hsize}{!}
          {\includegraphics[scale=1,clip]{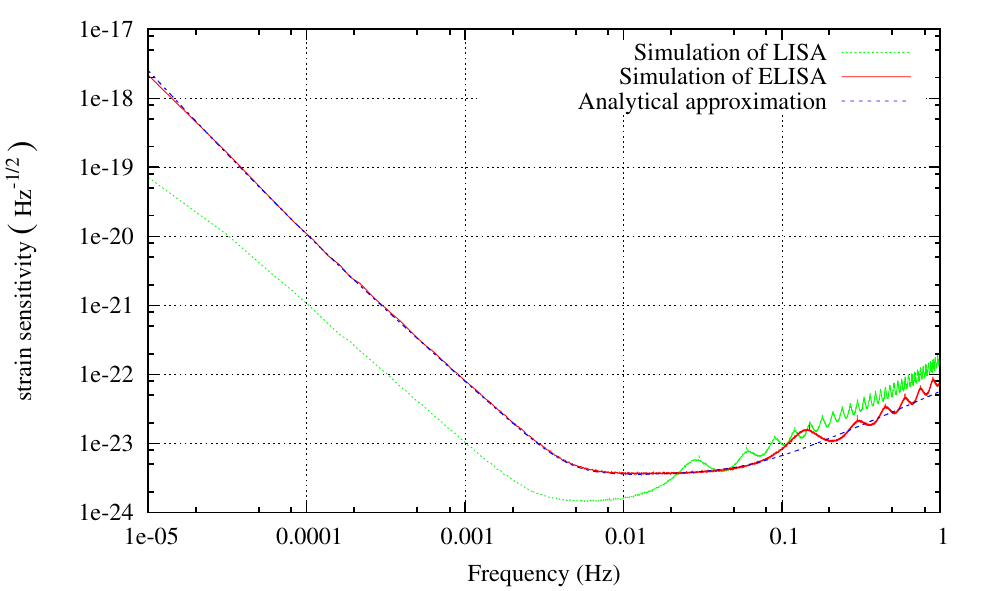}}
\caption
   {  
	Sensitivity of eLISA (averaged over all sky locations and polarisations) versus frequency: the solid red curve is 
	obtained numerically using
 the simulator LISACode 2.0 \citep{petiteau:2008PhRvD..77b3002P} and the dashed blue curve is the analytic approximation 
based on equation \ref{eq.sens}. For a reference, we also depict the sensitivity curve of LISA (dotted, green curve).
   }
\label{fig.mission_sensitivity}
\end{figure}

\noindent
According to the requirements,  eLISA achieves the strain noise amplitude spectral density (often called sensitivity) showed 
in figure~\ref{fig.mission_sensitivity} which can be analytically approximate 
as  $\tilde{h}(f) =  2 \; \widetilde{ \delta L }(f) / L =  \sqrt{S(f)} $,  where : 
\begin{align}
  \label{eq.sens}
  S(f) & = {20 \over 3} \, {  4 \, S_{\rm x,acc}(f) + S_{\rm x,sn}(f) + S_{\rm x,omn}(f)  \over L^2} \left(   1 + \left( { f \over 0.41 \left({c \over 2 L} \right) } \right)^2 \right) \,,
\end{align}
This allows to detect a strain of about \ensuremath{3.7 \times 10^{-24}} in a 2-year measurement with an SNR of 1
(displacement sensitivity of  \ensuremath{11 \times 10^{-12} \textrm{m}/ \sqrt{\rm Hz}} over a path length
 of \ensuremath{1 \times 10^{9} \textrm{m}}).  The feasible reduction of disturbances on test masses and the displacement 
sensitivities achievable by the laser ranging system yield a useful measurement frequency bandwidth  from \ensuremath{3 \times 10^{-5}} Hz 
to 1 Hz (the requirement is \ensuremath{10^{-4}} Hz to 1 Hz; the goal is \ensuremath{3 \times 10^{-5}} Hz to 1 Hz).

\section{Ultra-Compact Binaries}
\label{sec.ultra-comp-binar}

\subsection{Overview}
\label{sec.introduction}

The most numerous sources in the low-frequency gravitational wave band
are ultra-compact binary stars: double stars in which two
compact objects, such as white dwarfs and neutron stars, orbit each
other with short periods.  They have relatively weak gravitational
wave signals in comparison to massive black hole binaries, but are
numerous in the Galaxy and even the Solar neighbourhood. % The prospects

Several
thousand systems are expected to be detected individually, with their
parameters determined to high precision, while the combined signals of
the millions of compact binaries in the eLISA band will form a
foreground signal.  This is in
contrast to less than 50 ultra-compact binaries known today. The number of detections will allow for detailed study of the entire WD
binary population. In particular, the most numerous sources are
double white dwarfs, which are one of the candidate progenitors of type
Ia supernovae and related peculiar supernovae. eLISA will determine the
merger rate of these binaries. The detailed knowledge of the
ultra-compact binary population also constrains the formation of these
binaries and thus many preceding phases in binary evolution. This has
a strong bearing on our understanding of many high-energy phenomena in
the Universe, such as supernova explosions, gamma-ray bursts and X-ray
sources, as they share parts of the evolution history of the binaries
detectable by eLISA.

As many of the Galactic sources are rather close (within a few
\text{kpc}), they will be detectable at high SNR (often larger than
\ensuremath{50}), allowing detailed studies of individual binaries.
For many hundreds, the 
frequency and phase evolution can be studied, enabling the study of
the physics of tides and mass transfer in unprecedented detail.  The
extreme conditions of short orbital periods, strong gravitational
fields and high mass-transfer rates are unique in astrophysics.

The information provided by eLISA will be different from what can be
deduced by electromagnetic observations.  In particular, eLISA's
capability to determine distances and inclinations, as well as the
fact that the gravitational wave signals are unaffected by
interstellar dust, provide significant advantages over other detection
techniques.  Compared to Gaia, eLISA will observe a quite
different population.  Gravitational wave observations allow us to
determine the distances to binaries that are right in the Galactic
centre rather than to those close to the Sun.  The distance
determinations will make it possible to map the distribution of many
compact binaries in the Galaxy, providing a new method to study
Galactic structure. The inclination determinations allow the study of
binary formation by comparing the average angular momentum of the
binaries to that of the Galaxy. Electromagnetic observations and
gravitational wave observations are complementary to one another;
dedicated complementary observing programs as well as public data
releases will allow simultaneous and follow-up electromagnetic
observations of binaries identified by eLISA.

 A number of guaranteed
detectable sources are known to date from electromagnetic
observations.  Some of these can be used to verify instrument
performance by looking for a gravitational signal at twice the orbital
period and comparing the signal with expectations.  In addition, once
eLISA has detected several nearby binaries and determined their sky
position they can be observed optically, thus providing an additional
quantitative check on instrument sensitivity.

\subsection{Instrument verification}
\label{sec.verify-lisa-test}

\begin{figure}
\resizebox{\hsize}{!}
          {\includegraphics[scale=1,clip]{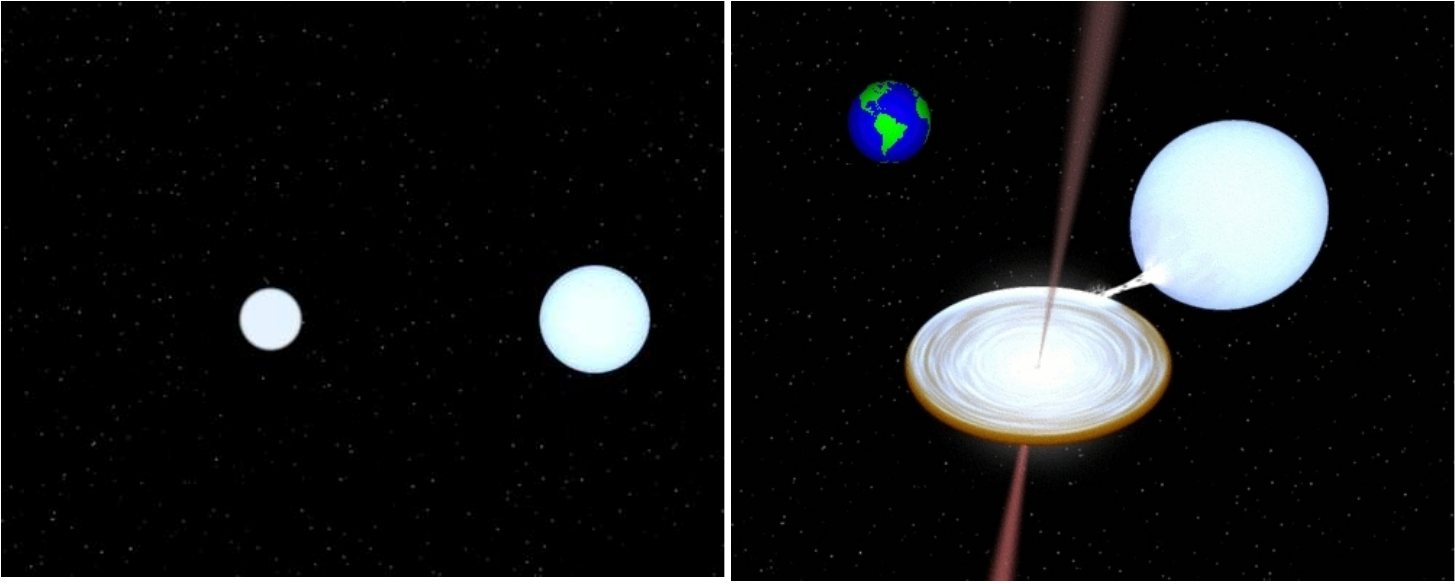}}
\caption
   {  
Artist impression of a detached double white dwarf binary (left) and
an interacting binary in which a neutron star accretes material from a white
dwarf donor. The Earth is shown to set the scale. Courtesy BinSim by Rob
Hynes.
   }
\label{fig.binsim}
\end{figure}

There are currently about 50 know ultra-compact binaries. They come in
two flavours: systems in which the two stars are well apart, called
detached binaries, and systems in which the two stars are so close
together that mass is flowing from one star to the other, called
interacting binaries (see figure \ref{fig.binsim}).

\begin{figure}
\resizebox{\hsize}{!}
          {\includegraphics[scale=1,clip]{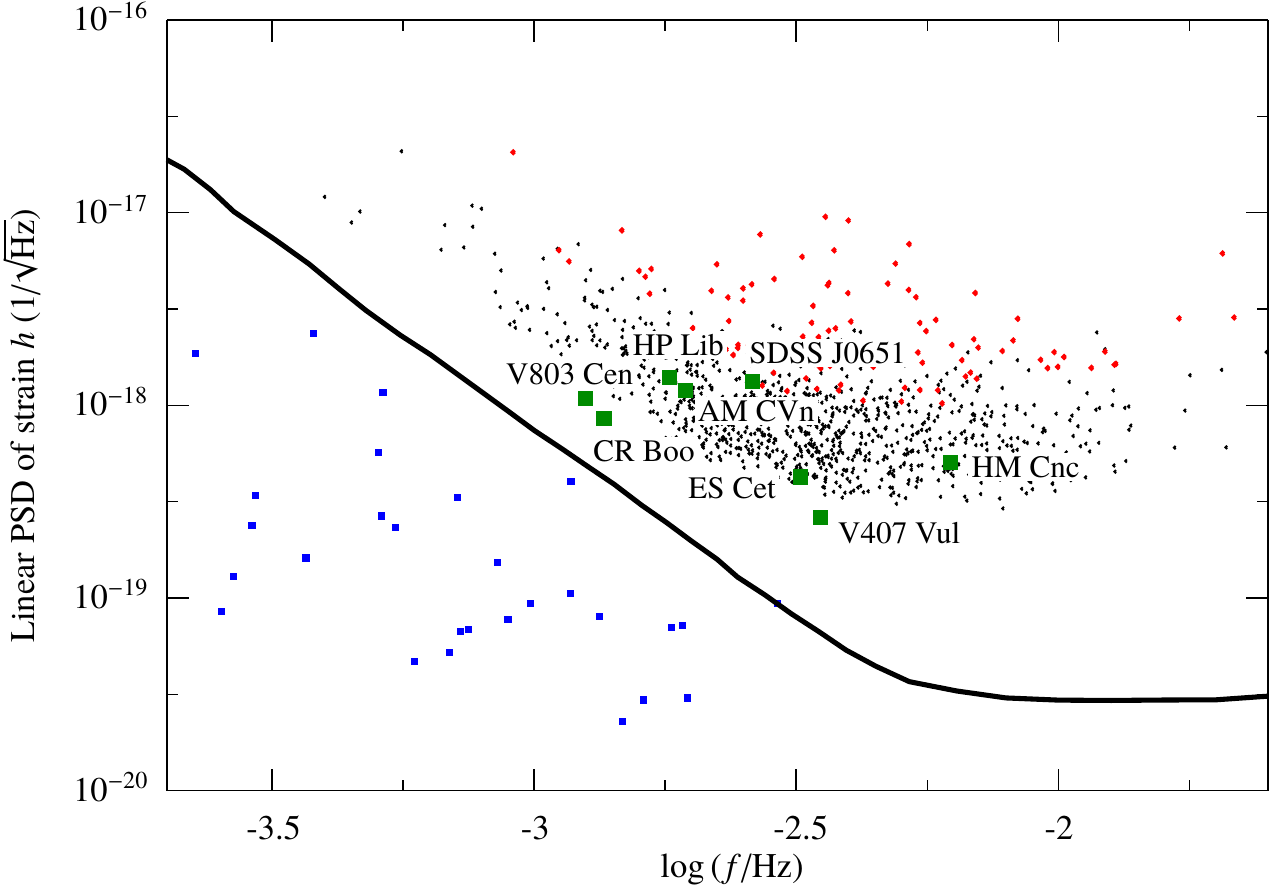}}
\caption
   {Strain amplitude spectral density versus frequency for the verification binaries and
the brightest binaries,expected from a simulated Galactic population of
ultra-compact binaries. The solid line shows the sensitivity of eLISA.
Verification binaries are indicated as green squares with their names indicated,
blue squares are other known binaries. Strongest 100 simulated binaries are
shown in red, strongest 1000 as black dots. The integration time for the binaries
is two years.  Based on
\citet{roelofs:2006:kuh,roelofs:2010:se,brown:2011ApJ...737L..23B} for the
known binaries and \citet{nelemans:2004:spc} for the simulation.  
   }
\label{fig.6-3}
\end{figure}

A subset of the known ultra-compact binaries have been
recognised as instrument verification sources, as they should be
detected in a few weeks to months and thus can be used to verify the
performance of the instrument \citep{stroeer:2006:lvb}. The most
promising verification binaries, shown as green squares in figure \ref{fig.6-3}, are the shortest-period interacting
binaries HM Cnc (RX~J0806.3+1527), V407~Vul, ES~Cet and the recently
discovered 12 minute period detached system SDSS~J0651+28 \citep{brown:2011ApJ...737L..23B}, whose lightcurve
is shown in figure \ref{fig.sdssj0651}. For a decade it has remained unclear if
the measured periods of HM~Cnc and V407~Vul were actually orbital
periods, but recent results from the Keck
telescope on HM~Cnc \citep{roelofs:2010:se} show conclusively that
this system has an orbital period of 5.4~minutes. As V407~Vul has
almost identical properties, this implies that this also really is a
binary with an orbital period of 9.5~minutes.
As the signal from the verification binaries is essentially
monochromatic with a well known frequency within the eLISA mission
time, astrophysical effects such as those discussed in section
\ref{sec.study-astr-comp} will not hamper their detection.
As more and more wide field and synoptical surveys are completed, the
number of ultra-compact binaries is gradually increasing and is
expected to continue to do so in the future. Already several new
binaries have been found in the SDSS
and the PTF
\citep{rau:2010ApJ...708..456R,levitan:2011ApJ...739...68L} while
surveys such as Pan-Starrs, the EGAPS and in the future LSST will also
find new systems.
However, most of the systems found so far have relatively long orbital
periods (longer than about 30~minutes).  Two pilot surveys in
principle capable of finding ultra-compact binaries with periods less
than 30~minutes are underway or will start soon: the RATS \citep{barclay:2011MNRAS.413.2696B}
and the OmegaWhite survey.

Interacting ultra-compact binaries with neutron star accretors are
strong X-ray sources and new discoveries are expected, both through
the continued monitoring of the sky to search for X-ray transients
with RXTE, MAXI and other satellites, as well as through
dedicated X-ray and optical surveys of the Galactic bulge that are
currently happening \citep{jonker:2011ApJS..194...18J}.
With these developments we expect that several tens of verification
sources should be available for eLISA, allowing detailed tests of the
performance of the instrument.

\subsection{eLISA as a workhorse: thousands of new binaries}
\label{sec.lisa-as-workhorse}

Ultra-compact binaries will completely dominate the number of source
detections by eLISA.  Current estimates suggest the numbers of
resolved compact binaries that will
be detected by eLISA to be in the thousands
\citep{webbink:2010:wts}. 
We
provide a visual impression in figure \ref{fig.6-3} by showing the 100 (red dots) and 1000 (black dots) strongest binaries from a
MonteCarlo realization of the galaxy compact white dwarf binary population.
The shortest period systems will be
the most numerous, the majority having periods between 5 and
10~minutes. eLISA will revolutionise our knowledge of such a
population, especially given that only two of the known fifty sources
have periods less than ten minutes. As these systems are relatively
short lived and faint, there is no hope to detect them in
significant numbers by any other means than via gravitational
radiation, as there are only several thousand expected to exist in the
whole Galaxy. Their detection will allow us to test different models
for the common-envelope phase, a significant uncertainty in our
understanding of binary evolution and many high-energy phenomena.  The
internal statistical accuracy delivered by the sheer number of
detected sources will ensure that the common-envelope phase will be
put to the most critical test expected in the midterm future. 
The same population can be used to constrain models for
type Ia supernovae and peculiar supernovae, as well as the formation
of ultra-compact binaries in globular clusters.

\subsubsection*{The outcome of the common envelope phase}
\label{sec.outc-comm-envel}

Only a minority of the
stars in the Universe are single, leaving the majority to be part of
a binary, a triple or a higher-order system.  On the order of half of
the binaries formed with sufficiently small orbital separation, so
that the stars will interact during the evolution of the components
into giants or super giants.  Especially for low-mass stars, the
majority of interactions are unstable and will lead to runaway mass
transfer.  Based on the observed short orbital periods of binaries
that have passed this stage it is argued that somehow the companion of
the giant ends up inside the giant's outer layers.  During that common
envelope phase, (dynamical) friction reduces the velocity of the companion,
leading to orbital shrinkage and transfer of angular momentum from the
orbit into the envelope of the giant.  Along with angular momentum,
orbital energy is deposited in the envelope, whose matter is then
unbound from the giant's core, leading to a very compact binary
consisting of the core of the giant and the original companion
\citep{paczynski:1976:secbs}.

Virtually all compact binaries and most of the systems giving rise to
high-energy phenomena (such as X-ray binaries, relativistic binary
pulsars and possibly gamma-ray bursts) have experienced at least one
common-envelope phase.  Given the importance of this phase in
high-energy astrophysics, our understanding of the physics and our
ability to predict the outcome of the common-envelope phase are poor.
Theoretical progress to understand the phase from first physical
principles is slow
\citep[e.g.][]{taam:2000:cee,taam:2010NewAR..54...65T} and the standard
formalism described above has been challenged by observational tests
\citep{nelemans:2005MNRAS.356..753N,demarco:2011MNRAS.411.2277D}.
Comparison of the parameters of the thousands of binaries detected by
eLISA with model predictions will provide a direct test of the
different proposed outcomes of the common-envelope phase and our
understanding of the preceding binary evolution in general.

\subsubsection{Type Ia supernovae and sub-luminous supernovae}

Type Ia supernovae have been the heralds of a new paradigm in
Cosmology: cosmic acceleration
\citep{riess:1998ApJ...504..935R,perlmutter:1999AIPC..478..129P} for
which the 2011 Nobel Prize in Physics was awarded. However, there are
different scenarios proposed for the progenitors of SN~Ia. One is the merger of two
(carbon-oxygen) white dwarfs that are brought together via
gravitational wave radiation \citep{pakmor:2010:sltIa} which is exactly
the population  eLISA will be probing. By determining the
number of systems in the Galaxy and their period distribution, the
rate at which they will merge will be measured. By comparing that to
the inferred SNIa rate for an Sbc galaxy, the viability of this
progenitor scenario will be determined. The significant efforts in the
past decade to find more supernovae and the advent of wide field
optical surveys have revealed a host of new types of supernovae
\citep{perets:2010Natur.465..322P,kasliwal:2010ApJ...723L..98K,sullivan:2011ApJ...732..118S,perets:2011ApJ...730...89P}. Some
of these have been suggested to originate in the interaction between
two white dwarfs at very short periods, again exactly the population
to which eLISA is sensitive
\citep{perets:2010Natur.465..322P,waldman:2011ApJ...738...21W}.

\subsubsection*{Formation of ultra-compact binaries in globular clusters}
\label{sec.lisa-as-workhorse-1}

Globular clusters have a strong
overabundance of bright X-ray sources per unit mass compared to the
field, probably due to dynamical interactions.  Many of these have
turned out to be so-called ultra-compact X-ray binaries, in which a
neutron star accreted material from a white dwarf companion is a very
compact orbit, exactly the type of sources that eLISA may see. However,
it is not clear if the same enhancement will operate for the much more
numerous white dwarf binaries.
The angular resolution that can be achieved with eLISA is such that
globular clusters can be resolved, so that the cluster sources can be
distinguished from the Galactic disc sources. %\warning{(FIGURE?)}.
This enables eLISA to determine the number of ultra-compact binaries in
globular clusters and thus to provide a direct test of the overabundance of
white dwarfs binaries in globular clusters. That in turn can be used to test
models for dynamical interactions in clusters.

\subsubsection*{The foreground of Galactic gravitational waves}
\label{sec.lisa-as-workhorse-2}
\begin{figure}
  \centering
  \resizebox{\hsize}{!}
	{\includegraphics[scale=1,clip]{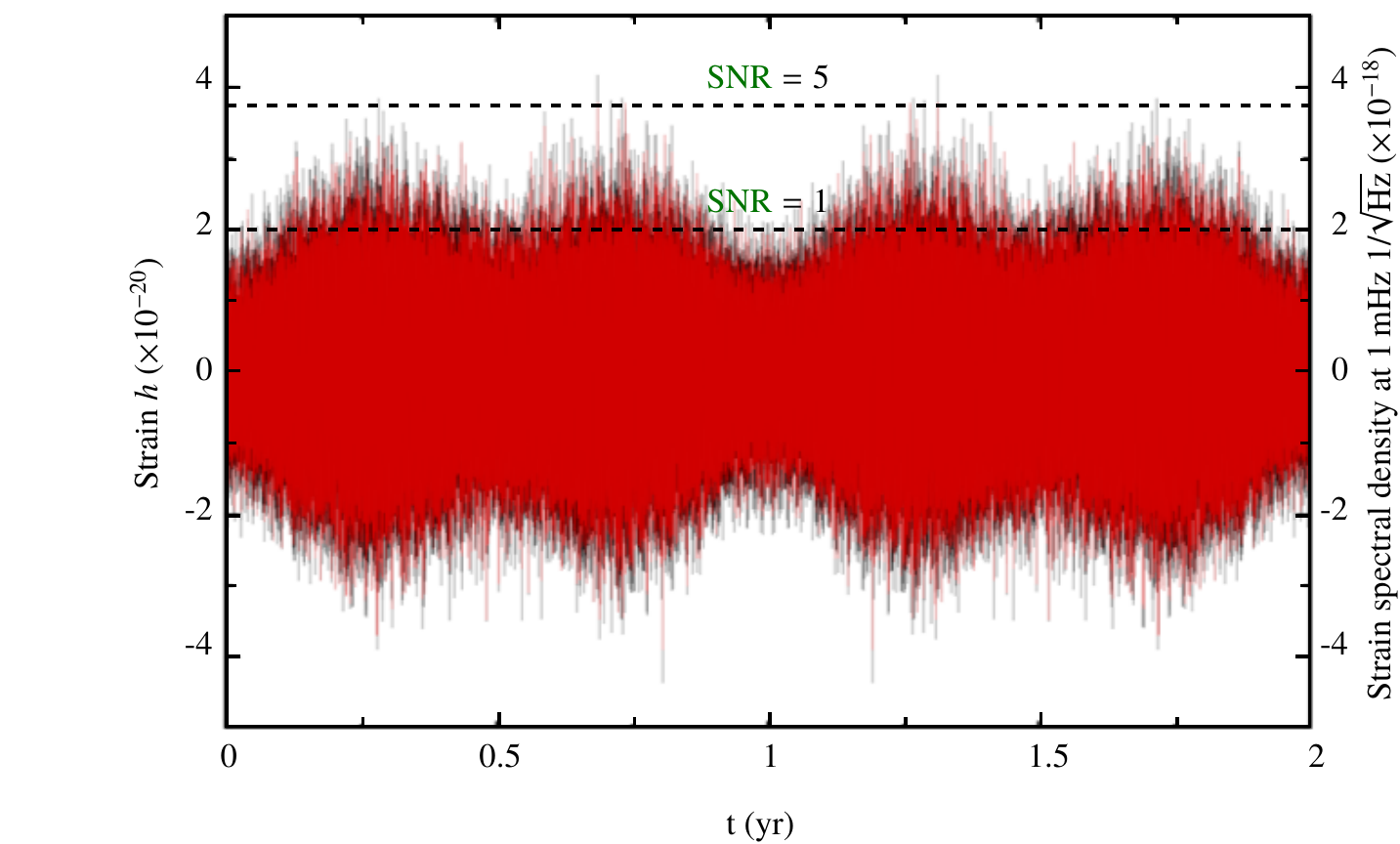} }
  \caption{ Level of the Galactic gravitational wave signal as a function of 
time. Black is the total signal, the red after removal of the resolved 
binaries. The yearly variation of the Galactic foreground is clearly seen. 
Based on the \citet{ruiter:2009:hwdb} Galactic model.}
  \label{fig.matt_fig5}
\end{figure}

 At frequencies below a few
\text{mHz} the number of sources in the Galaxy is so large
\citep[\ensuremath{6 \times 10^{7}} to \ensuremath{8 \times 10^{7}}, see e.g.][]{yu:2010:gw, ruiter:2010ApJ...717.1006R} that only a small percentage, the
brightest sources, will be \emph{individually} detected.  The vast
majority will form an unresolved \emph{foreground} signal in the
detector, which is quite different from and much stronger than any
diffuse extragalactic \emph{background} \citep{farmer:2003:gwb}.

 This foreground
is often described as an additional noise component, which is
misleading for two reasons.  The first is that there is a lot of
astrophysical information in the foreground.  The overall level of the
foreground is a measure of the total number of ultra-compact binaries,
which gives valuable information given the current uncertainty levels
in the normalisation of the population models.  The spectral shape of
the foreground also contains information about the homogeneity of the
sample, as simple models of a steady state with one type of binary
predict a very distinct shape.  In addition, the geometrical
distribution of the sources can be detected by eLISA.

Due to the concentration of sources in
the Galactic centre and the inhomogeneity of the eLISA antenna pattern,
the foreground is strongly modulated over the course of a year (see figure \ref{fig.matt_fig5}), with
time periods in which the foreground is more than a factor two lower
than during other periods \citep[][]{edlund:2005:wdw}. The
characteristics of the modulation can be used to learn about the
distribution of the sources in the Galaxy, as the different Galactic
components (thin disk, thick disk, halo) contribute differently to the
modulation and their respective amplitude can be used, for
example, to set upper limits on the halo population
\citep[e.g.][]{ruiter:2009:hwdb}.

\subsection{Studying the astrophysics of compact binaries using eLISA}
\label{sec.study-astr-comp}

Although the effect of gravitational radiation on the orbit will
dominate the evolution of the binaries detected by eLISA, additional
physical processes will cause strong deviations from the simple
point-mass approximation.  The two most important interactions that
occur are tides -- when at least one of the stars in a binary system is
not in co-rotation with the orbital motion or when the orbit is
eccentric -- and mass transfer.  Because many binaries will be easily
detected, these interactions do not hamper their discovery, but
instead will allow tests of the physics underlying these deviations.
By providing a completely complementary approach, gravitational wave
measurements are optimal to the study of short period systems, in
contrast to the current bias towards bright electromagnetic systems
and events.

\subsubsection*{Physics of tidal interaction}
\label{sec.study-astr-comp-1}

\begin{figure}
  \centering
  \resizebox{\hsize}{!}
	{\includegraphics[scale=1,clip]{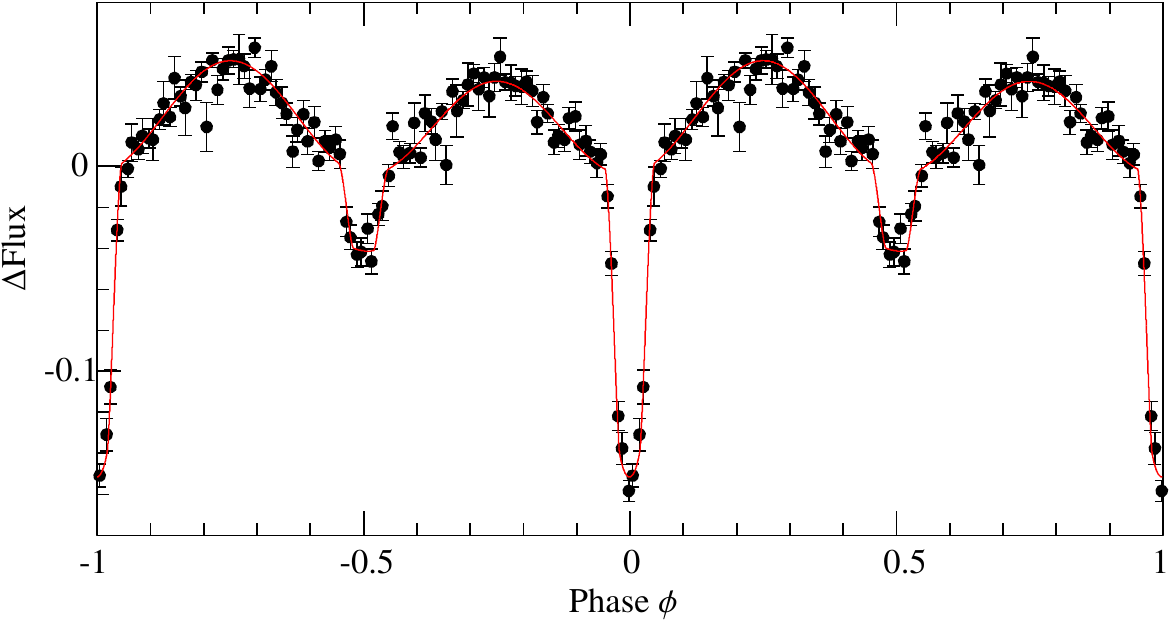} }
  \caption{Lightcurve of SDSS J0651+28, folded on the 12 minute orbital period. 
Except for the two eclipses at phase $\phi=0$ and $\phi=0.5$, the sinusoidal variation due 
to the tidal distortion of the primary white dwarf. From \citet{brown:2011ApJ...737L..23B}}
\label{fig.sdssj0651}
\end{figure}

 eLISA measurements of
individual short-period binaries will give a wealth of information on
the physics of tides and the stability of the mass transfer.  For
detached systems with little or no interaction, the frequency
evolution is well understood as that of two point masses. The strain
amplitude $h$, the frequency $f$ and its derivatives then are
connected by
\begin{align}
  \label{eq.5}
  h &\propto {\cal M}^{5/3} f^{2/3} D^{-1}\,,\\
  \label{eq.6}
  \dot f &\propto {\cal M}^{5/3} f^{11/3}\,,\\  \label{eq.7}
  \ddot f &= \frac{11}{3} \frac{\dot f}{f}\,,
\end{align}
where ${\cal M} = \left(m_1
m_2\right)^{3/5}/\left(m_1+m_2\right)^{1/5}$ is the chirp mass, $m_1, m_2$ are the masses of the binary
constituents and $D$ is the distance.  Thus the measurement of $h$,
$f$, $\dot f$ provides chirp mass and distance; the additional
measurement of $\ddot f$ gives a direct test of the dominance of
gravitational wave radiation in the frequency evolution.  Tidal
interaction between white dwarfs in detached systems before the onset
of mass transfer will give rise to
distinct deviations of the frequency evolution as compared to systems
with no or little tidal interaction.  The strength of the tidal
interaction is virtually unknown, with estimates ranging over many
orders of magnitude \citep{marsh:2004:mtwd}, although the high
temperature of the white dwarf in the recently discovered 12 min
double white dwarf may suggest efficient tidal heating
\citep{piro:2011ApJ...740L..53P}.  Knowledge of the strength of the
tides is not only important for understanding the physics of tides in
general and of white dwarf interiors, it also has important consequences
for the tidal heating (and possibly
optical observability) of eLISA sources and for the stability of mass
transfer between white dwarfs
\citep{racine:2007:ndts,willems:2010ApJ...713..239W,marsh:2011CQGra..28i4019M,fuller:2011arXiv1108.4910F}.

In globular clusters, dynamical interactions may produce eccentric
double white dwarf systems, which can be used to constrain white dwarf
properties and masses \citep{valsecchi:2011arXiv1105.4837V}.

\subsubsection*{Physics of mass-transfer stability}
\label{sec.study-astr-comp-2}

 Detached ultra-compact
binaries will evolve to shorter and shorter periods due to the angular
momentum loss through gravitational wave radiation.  At sufficiently
short orbital periods (a few minutes) one of the stars becomes larger
than its Roche lobe -- the equipotential
surface that crosses the minimum of the potential between the two
stars -- and material leaks out of the potential well of one
star upon the other star.  Depending on the difference between the
change of the radius of this star and the Roche lobe upon mass
transfer, there may be positive or negative feedback, leading to
either limited, stable mass transfer, or a runaway mass-transfer
instability.

For double white dwarfs and white dwarf-neutron star binaries the
stability of the ensuing mass transfer has important consequences, for
the number of detectable sources, as well as for a number of open
astrophysical questions.  The stable systems will form interacting
binaries, AM~CVn systems or
ultra-compact X-ray binaries, that
can be detected through their gravitational wave emissions.  eLISA will detect a
number of detached double white dwarfs and AM~CVn
systems that are so close to the onset of mass transfer that the
stability of the mass transfer can be tested directly by comparing
their numbers. In addition, eLISA will detect several ultra-compact
X-ray binaries at the very early stages of mass transfer, providing a
test of the mass transfer stability in these systems as well \citep{marsh:2011CQGra..28i4019M}.

For AM~CVn systems, a major uncertainty in the
mass-transfer stability is again the tidal interaction between the two white dwarfs.  Most
likely the mass transfer will proceed via the direct impact
configuration: due to the proximity of the two stars, the mass
transfer stream lands directly on the surface of the accreting white
dwarf, rather than wrapping around the accreting stars and interacting
with itself to form a flat accretion disk in the plane of the orbit
\citep{webbink:1984:dwd,marsh:2002MNRAS.331L...7M}.  The stability of
the mass transfer depends critically on the tidal interaction between
the two white dwarfs \citep{marsh:2004:mtwd}: In the absence of any
tidal interaction, there will be additional angular momentum loss from
the orbit due to the transfer of angular momentum from the orbit to
the accreting star which will consequently spin up.  This is different
from cases where the accretion is via a disc for which most of the
angular momentum generally is stored in the disc and eventually via
very efficient tidal interaction put back into the orbit.  Efficient
tidal coupling between the accreting star and the companion has the
ability to return the angular momentum back to the orbit
\citep[see][]{racine:2007:ndts,dsouza:2006:dmt}, thus reducing the
magnitude of the spin-up.
The difference between efficient and inefficient tidal coupling is
rather dramatic: the fraction of double white dwarfs estimated to
survive the onset of mass transfer can drop from about \ensuremath{20}\,\text{\%} to
\ensuremath{0.2}\,\text{\%} \citep{nelemans:2001:ps2} depending on assumptions about
the tidal coupling.  This difference is easily measurable with
eLISA. Short-term variations in the secular evolution of the systems
experiencing mass transfer will change the frequency evolution, but
are likely to be rare and will not prevent the detection of these
systems \citep{stroeer:2009:stv}.

\begin{figure}
  \centering
  \resizebox{\hsize}{!}
	{\includegraphics[scale=1,clip]{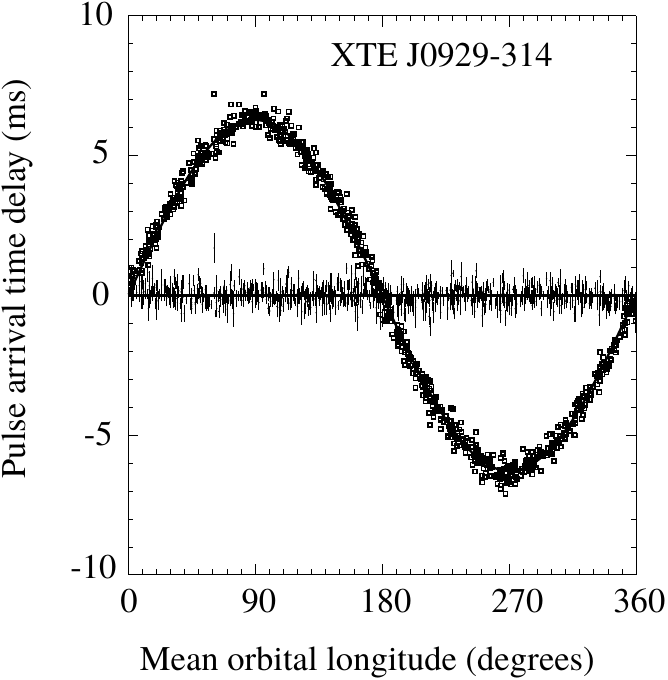} }
  \caption{Imprint of the 40 min orbital period on the arrival times
    of the X-ray pulsations in the ultra-compact X-ray binary XTE
    J0929-314. From \citet{galloway:2002:hlamsp}.}
  \label{fig.xray}
\end{figure}

For
ultra-compact X-ray binaries (see e.g. figure \ref{fig.xray}), the stability issue is completely
different.  At the onset, the mass transfer is orders of magnitude
above the Eddington limit
for a neutron star (the mass transfer rate at which the potential
energy liberated in the accretion can couple to the infalling gas to
blow it away).  For normal stars and white dwarfs, this would likely
lead to a complete merger of the system, but the enormous amount of
energy liberated when matter is falling into the very deep potential
well of a neutron star allows matter to be dumped on it at rates up to
a thousand times the Eddington limit if the white dwarf has a low mass
\citep[see][]{yungelson:2002:ucxb}.  This allows the formation of
ultra-compact X-ray binaries from white dwarf-neutron star pairs.
eLISA will unambiguously test this prediction by detecting several tens
of ultra-compact X-ray binaries with periods between 5 and 20~minutes.

\subsubsection*{Double white dwarf mergers}
\label{sec.study-astr-comp-3}

The 80\% to 99.8\%
of the double white dwarfs that experience run-away mass transfer and
merger give rise to quite spectacular phenomena. Mergers of double white dwarfs have been proposed as
progenitors of single subdwarf O and B stars, R Corona Borealis stars and maybe all massive white dwarfs
\citep[e.g.][]{webbink:1984:dwd}.  In addition, the merger of a
sufficiently massive double white dwarf can be a trigger for type Ia
supernova events \citep[see][]{pakmor:2010:sltIa}.  Alternatively, if
the merger does not lead to an explosion, a (rapidly spinning) 
 neutron star will be formed.  This is one
possible way to form isolated millisecond radio
pulsars as well as magnetars, which have been proposed
as sites for short gamma-ray bursts
\citep[e.g.][]{levan:2006:grb}.

Although it is not expected that eLISA will witness the actual
merger of a double white dwarf as the event rate in our Galaxy is too low, it will certainly detect the
shortest-period binaries known, expected at a period of about two
minutes, and give an extremely good estimate of their merger rate. In
addition, if the actual merger takes many orbits as recently found in
simulations \citep{dan:2011ApJ...737...89D}, eLISA may observe
them directly.

By measuring (chirp) masses and coalescence times, eLISA will directly
determine the merger rate for double white dwarfs with different
masses, which can then be compared with the rates and population of
their possible descendants determined by other means \citep{stroeer:2011arXiv1109.4978S}.

\subsubsection*{Neutron star and black hole binaries}
\label{sec.study-astr-comp-4}

The current observational and theoretical estimates of the formation
rate of neutron star binaries are highly uncertain and predict several
tens of
neutron star binaries to be detected by eLISA
\citep[e.g.][]{nelemans:2001:ps2,belczynski:2010:dco}. 
The number of ultra-compact stellar-mass black hole binaries in the
Galaxy is even more uncertain \citep[e.g.][]{belczynski:2002:bco};
furthermore, these binaries are likely to be detectable only through
their gravitational wave emission as they are electromagnetically quiet.

eLISA will thus constrain the formation rate estimates and the numbers of
neutron star binaries and ultra-compact stellar mass black hole
binaries. As these systems can be seen throughout the Galaxy, the samples for all
these populations will be \emph{complete at the shortest
  periods}. Thus the sample will be independent of selection effects,
such as those present in radio pulsar surveys and X-ray surveys, that
pick up only transient X-ray sources.
In addition, by the time eLISA will fly,  Advanced
LIGO and Virgo will likely have detected a number
of double neutron star mergers from far away galaxies, so these
measurements together will test our ability to extrapolate our
population models from our own galaxy to the rest of the Universe.

A special situation might arise for the case of millisecond X-ray pulsars, 
in ultra-compact X-ray binaries.  In the last decade, observations of
X-ray pulsations from many 
ultra-compact X-ray binaries have enabled
astrophysicists to determine the rotation rate of the neutron star in
the binary using the NASA mission RXTE \citep{wijnands:2010HiA....15..121W}.  
As had been expected on theoretical grounds, neutron stars are
spinning rapidly (several hundred times per second) due to the angular
momentum gained from infalling matter.  The measurements give credence
to the idea that these rapidly spinning neutron stars observed as
millisecond radio pulsars are descendants of accreting neutron stars
in binary systems \citep[e.g.][]{bhattacharya:1991:febmps}.
However, the exact role of ultra-compact binaries in the formation of
these pulsars has yet to be established.  The distribution of spin
periods discovered in X-ray binaries suggests additional neutron star
angular momentum loss on top of the plasma physics interaction between
the accretion and magnetic field of the spinning neutron stars
\citep{chkrabarty:2003:npms} which could be due to strong
gravitational wave emission (\cite{bildsten:1998:grr}; but
  see \cite{watts:2008:gw} and \cite{patruno:2011arXiv1109.0536P}).
In that case, ultra-compact X-ray binaries might be the only sources
that could be studied simultaneously with eLISA and ground based
detectors, with eLISA detecting the orbital period and the ground based
detector detecting the neutron star spin period.

\subsection{Studies of galactic structure with eLISA}
\label{sec.new-studies-galactic}

\begin{figure}
  \centering
  \resizebox{\hsize}{!}
	{\includegraphics[scale=1,clip]{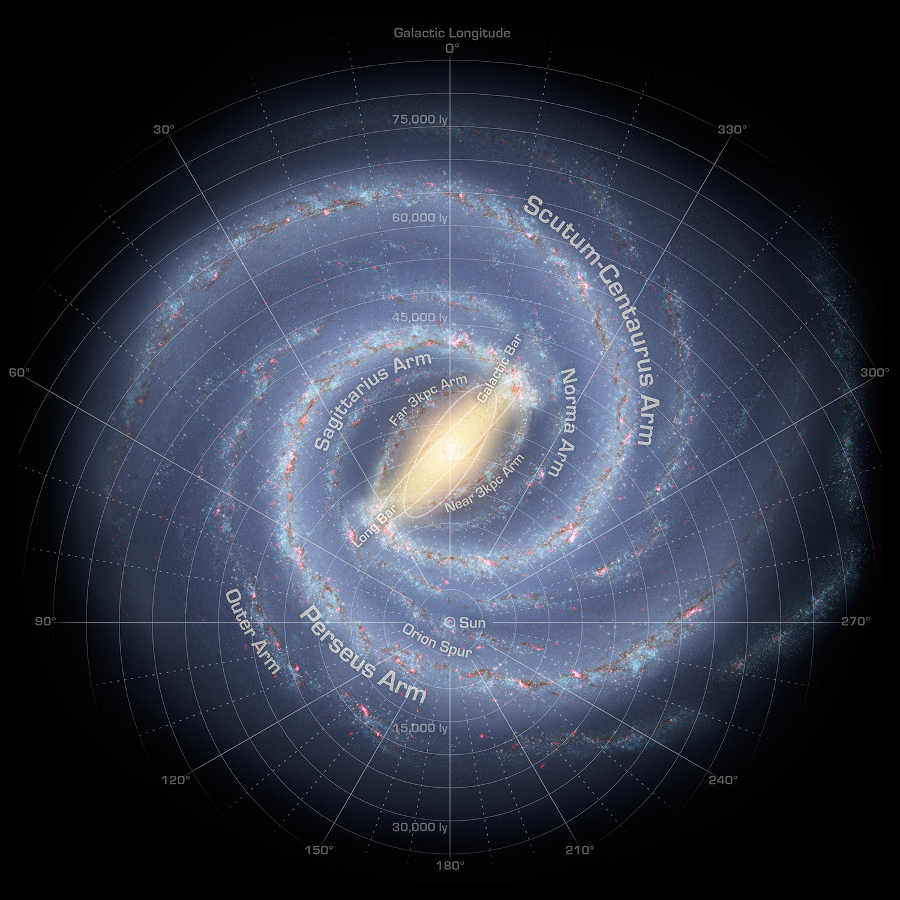} }
  \caption{Spitzer GLIMPSE model of the Milky Way, showing bulge, bar and spiral 
arms. The resolved binaries are expected to trace the old stellar populations 
of the Milky Way.  Courtesy NASA/JPL-Caltech/R. Hurt (SSC/Caltech)}
  \label{fig.glimpse}
\end{figure}
One of the major capabilities of
eLISA is that it will determine distances for hundreds of compact
binaries by measuring their $\dot f$ (see equation \ref{eq.6}).  The ability of
eLISA to determine distances depends on the mission lifetime, as larger
life times lead to more accurate $\dot f$ measurements.
The directional dependence of the Galactic foreground as well as the
directional accuracy for the resolved systems allow a statistical
assessment of the contributions of the different Galactic components,
such as the Galactic bulge (with its bar), the
thin and thick disc and the Galactic halo (for a realistic MW model, see in figure \ref{fig.glimpse}).

The Galactic center is one of the most interesting areas of the
Galaxy, with a central massive black hole surrounded by a dense
assembly of stars with intriguing properties.  Dynamical effects, in
particular mass segregation, will lead to many interactions close to
the central black hole so that wide binaries will become tighter or
will be disrupted \citep[for a review see][]{alexander:2005:spmbh}.
This likely leads to an increase in the number of ultra-compact
binaries as well as the possibility of EMRIs (see
\ref{sec.extreme-mass-ratio-1}).  eLISA will put much more stringent
constraints on these populations than current observations \citep[see
e.g.][]{roelofs:2007:sdss}, which are limited by the electromagnetic
faintness of the sources, or theoretical predictions, which are
limited by our current understanding of the processes leading to
compact binary formation. Distance determinations to the many
ultra-compact binaries around the Galactic centre will allow for an
independent distance determination.

 The level and shape of the
double white dwarf foreground as well as the distribution of resolved
sources will provide information on the \emph{scale height} of the
ultra-compact binary population \citep{benacquista:2006:cds} in the
disc of the Galaxy.  

 The distribution of sources in the
\emph{Galactic halo} will be significantly different from the other
Galactic components.  In principle the halo population is expected to
be much smaller than the rest of the Galaxy
\citep{ruiter:2009:hwdb,yu:2010:gw}, but it might be enhanced as the
formation and evolution of binaries in the halo may have been quite
different.  Such old and metal-poor population can be studied locally
only in globular clusters, where the
formation and evolution of binaries is generally completely altered by
dynamical effects.  Two of the known AMCVn systems may
belong to the halo.  They have very low metal abundances and have
anomalous velocities.  If true this implies that a large number of
AMCVn stars are in the halo, maybe as many as in the
rest of the Galaxy.

The eLISA directional sensitivity
will immediately pick up any strong halo population if it exists.

Finally, for many of the resolved sources the eLISA measurements will
also provide an accurate estimate of their orbital inclination. For the first
time, this will give hints on the dynamics of the formation of
binaries from interstellar clouds, because the angular momentum
vectors of the binaries is related (in a statistical way) to the
overall angular momentum of the Galaxy.

\section{Astrophysical Black Holes}
\label{sec.black-hole-astr}

\subsection{Overview}
Astrophysical black holes appear to come in nature into two flavours:
the ``stellar mass'' black holes of
\ensuremath{3}\,\text{\ensuremath{\mathrm{M}_\odot}} to approximately
\ensuremath{100}\,\text{\ensuremath{\mathrm{M}_\odot}} resulting from
the core collapse of very massive stars, and the ``supermassive''
black holes of
\ensuremath{10^{6}}\,\text{\ensuremath{\mathrm{M}_\odot}} --
\ensuremath{10^{9}}\,\text{\ensuremath{\mathrm{M}_\odot}} that,
according to the accretion paradigm, power the luminous QSO.  The
former light up the X-ray sky, albeit only in our neighbourhood, as
stellar mass black holes fade below detection limits outside our
local group.  The latter are detected as active nuclei, over the whole 
cosmic time accessible to our current telescopes.  Electromagnetic evidence 
of black holes in the mass range 
\ensuremath{10^2}\,\text{\ensuremath{\mathrm{M}_\odot}}--
\ensuremath{10^{6}}\,\text{\ensuremath{\mathrm{M}_\odot}} is less
common, due to the intrinsic difficulty of detecting such faint
sources in external galaxies.  However, it is in this mass interval,
elusive to electromagnetic observations, that the history of
supermassive black hole growth is imprinted.

Supermassive black holes inhabit bright galaxies, and are ubiquitous
in our low-redshift Universe.  The discovery of close correlations
between the mass of the supermassive black hole with key properties of
the host has led to the notion that black holes form and evolve in
symbiosis with their galaxy host.  In agreement with the current
paradigm of hierarchical formation of galactic structures and with
limits imposed by the cosmic X-ray background light, astrophysical
black holes are believed to emerge from a population of \emph{seed}
black holes with masses in the range
\ensuremath{100}\,\text{\ensuremath{\mathrm{M}_\odot}} --
\ensuremath{10^{5}}\,\text{\ensuremath{\mathrm{M}_\odot}}, customarily
called \emph{intermediate mass black holes}.  The mass and spin of
these black holes change sizably in these interactions as they evolve
over cosmic time through intermittent phases of copious accretion and
merging with other black holes in galactic halos.  In a galactic
merger, the black holes that inhabit the two colliding galaxies spiral
in under the action of dynamical friction, and \emph{pair} on
sub-galactic scales forming a Keplerian binary: \emph{binary black
  holes} thus appear as the inescapable outcome of galaxy assembly.
When two massive black holes coalesce, they become one of the loudest
sources of gravitational waves in the Universe.

eLISA is expected to target
coalescing binaries of \ensuremath{10^{5}}\,\text{\ensuremath{\mathrm{M}_\odot}} -- \ensuremath{10^{7}}\,\text{\ensuremath{\mathrm{M}_\odot}} during the epoch of
widespread cosmic star formation and up to $z\sim 20$, and to capture the signal
of a coalescing binary of \ensuremath{10^{4}}\,\text{\ensuremath{\mathrm{M}_\odot}} -- \ensuremath{10^{5}}\,\text{\ensuremath{\mathrm{M}_\odot}} beyond the era of the
earliest known QSO ($z\sim 7$). Gravitational waveforms carry information on the
spins of the black holes that eLISA will measure with exquisite
precision, providing a diagnostic of the mechanism of black hole
growth.  The detection of coalescing black holes not only will shed
light into the phases of black hole growth and QSO evolution, but will
pierce deep into the hierarchical process of galaxy formation.

\subsection{Black holes in the realm of the observations}
\label{sec.black-holes-realm}

\begin{figure}
  \centering
  \resizebox{\hsize}{!}
	{\includegraphics[scale=1,clip]{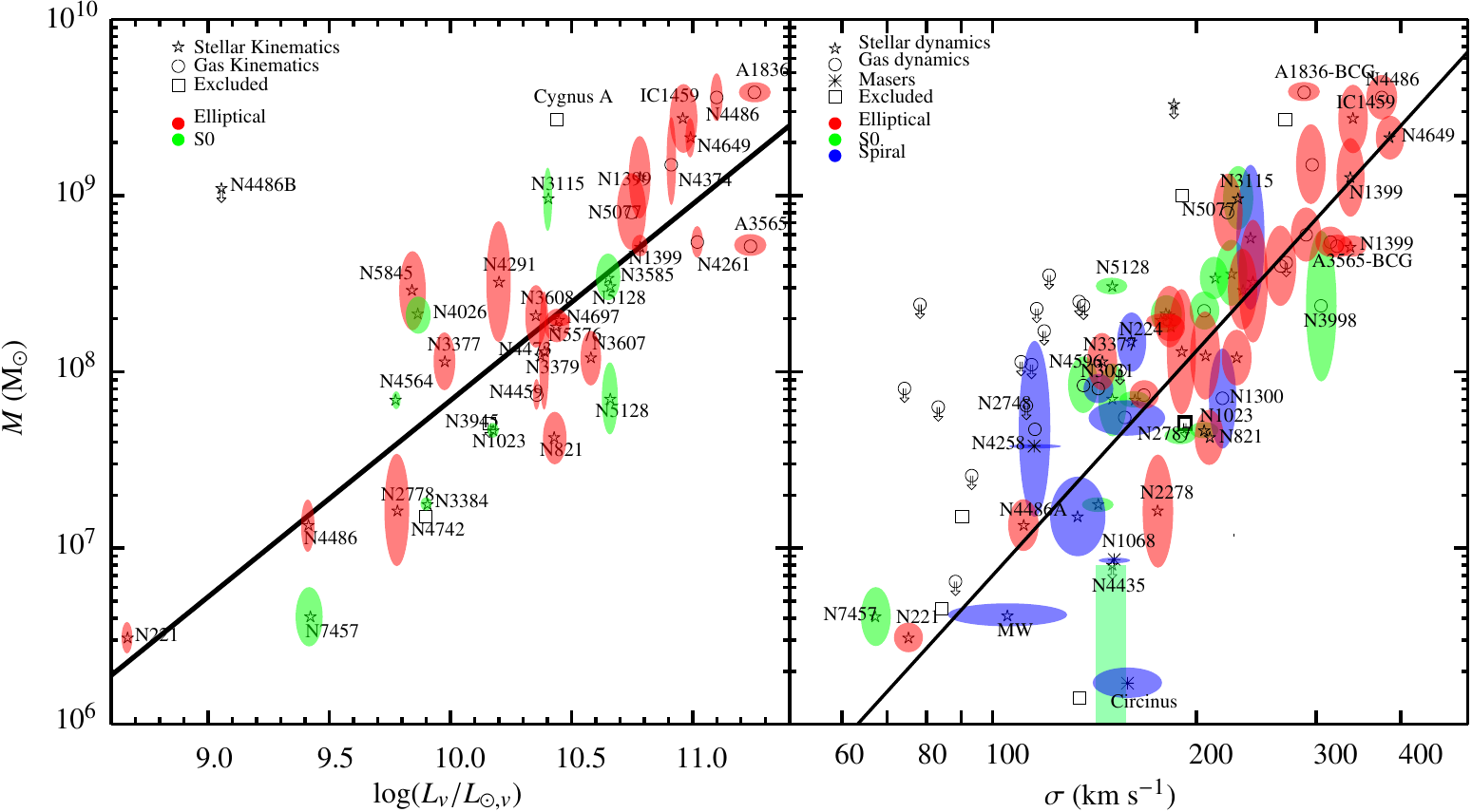} }
  \caption{The correlation between the black hole mass
    $\ensuremath{\mathit{M}_\bullet}$ and the luminosity of the host
    galaxy's stellar bulge (left), and host galaxy's bulge velocity
    dispersion $\sigma$ (right) for all detections in galaxies near
    enough for current instruments to resolve the region in which the
    black hole mass dominates the dynamics \citep[adapted
      from][]{gultekin:2009:ms}}
\label{fig.bh_1}
\end{figure}

\subsubsection*{Dormant and active supermassive black holes}
\label{sec.dorm-active-superm}

QSOs are active nuclei so luminous that they often outshine their galaxy host.
They are sources of electromagnetic energy, with radiation emitted across the
spectrum, almost equally, from X-rays to the far-infrared, and in a fraction of
cases, from $\gamma-$rays to radio waves.  Their variability on short
timescales revealed that the emitting region is compact, only a few light hours
across.

There is now scientific consensus that the electromagnetic power from QSO and
from the less luminous AGN results from \emph{accretion} onto a supermassive
black hole of \ensuremath{10^{6}}\,\text{\ensuremath{\mathrm{M}_\odot}} --
\ensuremath{10^{9}}\,\text{\ensuremath{\mathrm{M}_\odot}}
\citep{Salpeter:1964,zeldovich:1964:rgw,krolik:1999:agn}.  Escaping energy in
the form of radiation, high velocity plasma outflows, and ultra relativistic
jets can be generated with high efficiency ($\varepsilon \sim
\ensuremath{10}\,\text{\%}$, higher than nuclear reactions) just outside the
event horizon, through viscous stresses on parcels of gas orbiting in the
gravitational potential of the black hole.  The accretion paradigm has thus
been, and still is, at the heart of the hypothesis of black holes as being
``real'' sources in our cosmic landscape. eLISA will offer the new perspective
of revealing these black holes as powerful sources of gravitational waves,
probing the smallest volumes of the large scale Universe.

Massive black holes are
tiny objects compared to their host galaxies. The event horizon of a Kerr black hole of mass
$\ensuremath{\mathit{M}_\bullet}$ scales as $R_\text{horizon}\sim
G\ensuremath{\mathit{M}_\bullet}/c^2$, and it is far smaller than the
optical radius of the galaxy host: $R_\text{horizon}\sim
\ensuremath{10^{-11}}\,\text{R\ensuremath{_\text{gal}}}$. The distance
out to which a black affects the kinematic of stars and gas (the
gravitational influence radius), $R_\text{grav}\sim
G\ensuremath{\mathit{M}_\bullet}/\sigma^2$, is also small compared to
the optical radius of the host, $R_\text{grav}\sim
\ensuremath{10^{-4}}\,\text{R\ensuremath{_\text{gal}}}$ (where
$\sigma$ is the velocity dispersion of the stars of the galactic
bulge).

For a long time, QSO
and more generally the less luminous {AGN} phenomena were understood
as caused by a process exclusively confined to the nuclear region of
the host. This picture of disjoint black hole and galaxy evolution
changed with the advent of the HST \citep{ferrarese:2005:smbh}.

Observations of almost all
bright galaxy spheroids in the near universe reveal that the
velocities of stars and gas start to rise in a Keplerian fashion at their centres,
highlighting the presence of a dark point-mass which dominates the central gravitational
potential. The same observations provide the mass of this dark object, 
hypothesised to be a quiescent black hole. The
proximity of these galaxies to Earth allowed for a full optical
characterisation of the host, and this ultimately led to the discovery of tight correlations -- depicted in figure \ref{fig.bh_1}, from
\cite{gultekin:2009:ms} -- between the black hole mass
$\ensuremath{\mathit{M}_\bullet}$ and the optical luminosity and velocity dispersion $\sigma$
of the stars measured far from the black hole
\citep{gultekin:2009:ms,ferrarese:2000:smbh,gebhardt:2000:bhm,tremaine:2002:slope,graham:2011MNRAS.412.2211G}.
The relations state that galaxy spheroids with higher stellar velocity
dispersions, i.e. with deeper gravitational potential wells and higher stellar masses and luminosities, host heavier central black holes
with little dispersion in the correlation. Thus more massive galaxies grow
more massive black holes: 
the black hole sees the galaxy that it inhabits, and the
galaxy sees the black hole at its centre despite its
small influence radius \citep{magorrian:1998:dmdo,marconi:2003ApJ...589L..21M,haering:2004:bhmbr}.

 Consensus is rising that the
$\ensuremath{\mathit{M}_\bullet} - \sigma$ relation of figure
\ref{fig.bh_1} is fossil evidence of a \emph{co-evolution of black
  holes and galaxies}. The relation may have been established along
the course of galactic mergers and in episodes of self-regulated accretion 
\citep{mihos:1996ApJ...464..641M,dimatteo:2005:eiq,hopkins:2006:umm,hopkins:2008ApJS..175..356H,hopkins:2009MNRAS.398..303H,croton:2006MNRAS.365...11C,somerville:2008MNRAS.391..481S,johannson:2009ApJ...690..802J,lamastra:2010MNRAS.405...29L}.
However, the origin of the $\ensuremath{\mathit{M}_\bullet}-\sigma$
relation
\citep{silk:1998A&A...331L...1S,king:2003ApJ...596L..27K,wyithe:2003ApJ...595..614W,ciotti:2010ApJ...717..708C},
and its evolution at look-back times is still unclear
\citep{robertson:2006ApJ...641...90R,peng:2007ApJ...671.1098P,treu:2007ApJ...667..117T,woo:2008ApJ...681..925W}. The
similarity between the evolution, over cosmic time, of the luminosity
density of QSOs and the global star formation rate
\citep{boyle:1998MNRAS.293L..49B,kauffmann:2000MNRAS.311..576K} points
to the presence of a symbiotic growth, which is still under study
\citep{schawinski:2010ApJ...714L.108S,schawinski:2011ApJ...727L..31S}.

The census of black holes, from the study of the kinematics of stars and gas in nearby galaxies, 
has further led to the estimate of the black hole local mass
density: $\ensuremath{\rho_\bullet} \sim \ensuremath{2 \times 10^{5}}\,\text{\ensuremath{\mathrm{M}_\odot}\ Mpc\ensuremath{^{-3}}} - \ensuremath{5 \times 10^{5}}\,\text{\ensuremath{\mathrm{M}_\odot}\ Mpc\ensuremath{^{-3}}}$
\citep{Aller:2002,marconi:2004:lsmbh,lauer:2007:mnbh,tundo:2007:bhmf}. 
Whether this mass density traces the initial conditions, i.e. the mass
since birth, obtained at most by rearranging individual masses via
coalescences, or the mass acquired via major episodes of accretion in
active AGN phases can only be inferred using additional information:
that resulting from the AGN demographics and from studies of the X-ray
cosmic background.

Two arguments provide information about how much of the black hole growth
occurred through accretion of gas, in phases when the black hole is active as
AGN. The first is the existence of a limiting luminosity for an accreting black
hole, corresponding to when the radiation pressure force equals gravity. Above
this limit material that would be responsible for the emission can not fall onto
the black hole, as it is pushed away. This limit is the Eddington luminosity
$L_{\rm E}=4\pi G\ensuremath{\mathit{M}_\bullet} m_{\rm p} c/\sigma_T\sim
\ensuremath{10^{46}}\,\text{erg s\ensuremath{^{-1}}}
(\ensuremath{\mathit{M}_\bullet}/(\ensuremath{10^{8}}\,\text{\ensuremath{\mathrm{M}_\odot}})$
($\sigma_T$ and $m_{\rm p}$ are the Thomson cross section and proton mass). The
AGN luminosity $L$ is normally a fraction $f_{\rm E}\lesssim 1$ of the
Eddington luminosity, since as soon as $L$ approaches $L_{\rm E}$ the radiation
pressure force against gravity self-regulates the accretion flow to $L\sim
L_{\rm E}$, providing also a lower bound on $M$.
The second argument is that ``light is mass'', i.e.\ that any light
output from accretion (at a luminosity level $L=\varepsilon {\dot
  M}c^2$) increases the black hole's mass at a rate
$\text{d}\ensuremath{\mathit{M}_\bullet}/\text{d}t=(1-\varepsilon)
{\dot M}$, where ${\dot M}$ is the rest-mass accreted per unit time
and $\varepsilon$ the accretion efficiency, i.e.\ how much of the
accreted mass is converted into radiation. Accordingly, the black
hole's mass increases exponentially in relation to the self-regulated
flow, with an $e$-folding time $\tau_{\rm BH}\approx \ensuremath{4.7
  \times 10^{8}}\, \varepsilon [f_{\rm E}(1-\varepsilon )]^{-1}
\text{yr}$.  For $\varepsilon\approx \ensuremath{0.1} $ -- typical of
radiatively efficient accretion onto a non-rapidly rotating black hole
\citep{shapiro:1979:gc} -- and $f_{\rm E}\approx \ensuremath{0.1}$, this
timescale is short (about \ensuremath{3}\,\text{\%}) compared to the
age of the Universe, indicating that black holes can enhance their
mass via accretion by orders of magnitude.

Active black holes in galaxies are known to contribute to the rise of a cosmic
X-ray background resulting mostly from unresolved and obscured AGN of mass
\ensuremath{10^{8}}\,\text{\ensuremath{\mathrm{M}_\odot}} --
\ensuremath{10^{9}}\,\text{\ensuremath{\mathrm{M}_\odot}}, in the redshift
interval $0.5<z<3$ \citep{merloni:2004:ahg}.  As energy from accretion is
equivalent to mass, the X-ray light present in the background mirrors the
increment experienced by the black holes over cosmic history due to accretion.
This mass-density increment is found to be
$\Delta\ensuremath{\rho_\bullet}\approx \ensuremath{3.5 \times 10^{5}}\,
(\varepsilon/0.1)^{-1}\,\text{\ensuremath{\mathrm{M}_\odot}
Mpc\ensuremath{^{-3}}}$ \citep{marconi:2004:lsmbh,yu:2002:oc,soltan:1982:moq}.

As the contribution to the local (zero redshift) black hole mass density
$\ensuremath{\rho_\bullet}$ results from black holes of comparable mass
\ensuremath{10^{8}}\,\text{\ensuremath{\mathrm{M}_\odot}} --
\ensuremath{10^{9}}\,\text{\ensuremath{\mathrm{M}_\odot}}, the close match
between the two independent measures, $\ensuremath{\rho_\bullet}$ and
$\Delta\ensuremath{\rho_\bullet}$, indicates that radiatively efficient
accretion ($\varepsilon \approx 0.1$) played a large part in the building of
supermassive black holes in galaxies, from redshift $z\sim3$ to now.  It
further indicates that information residing in the initial mass distribution of
the, albeit unknown, black hole seed population is erased during events of
copious accretion, along the course of cosmic evolution.

\subsection*{Massive black holes in the cosmological framework}
\label{sec.massive-black-holes}

\begin{figure}
\resizebox{\hsize}{!}
          {\includegraphics[scale=1,clip]{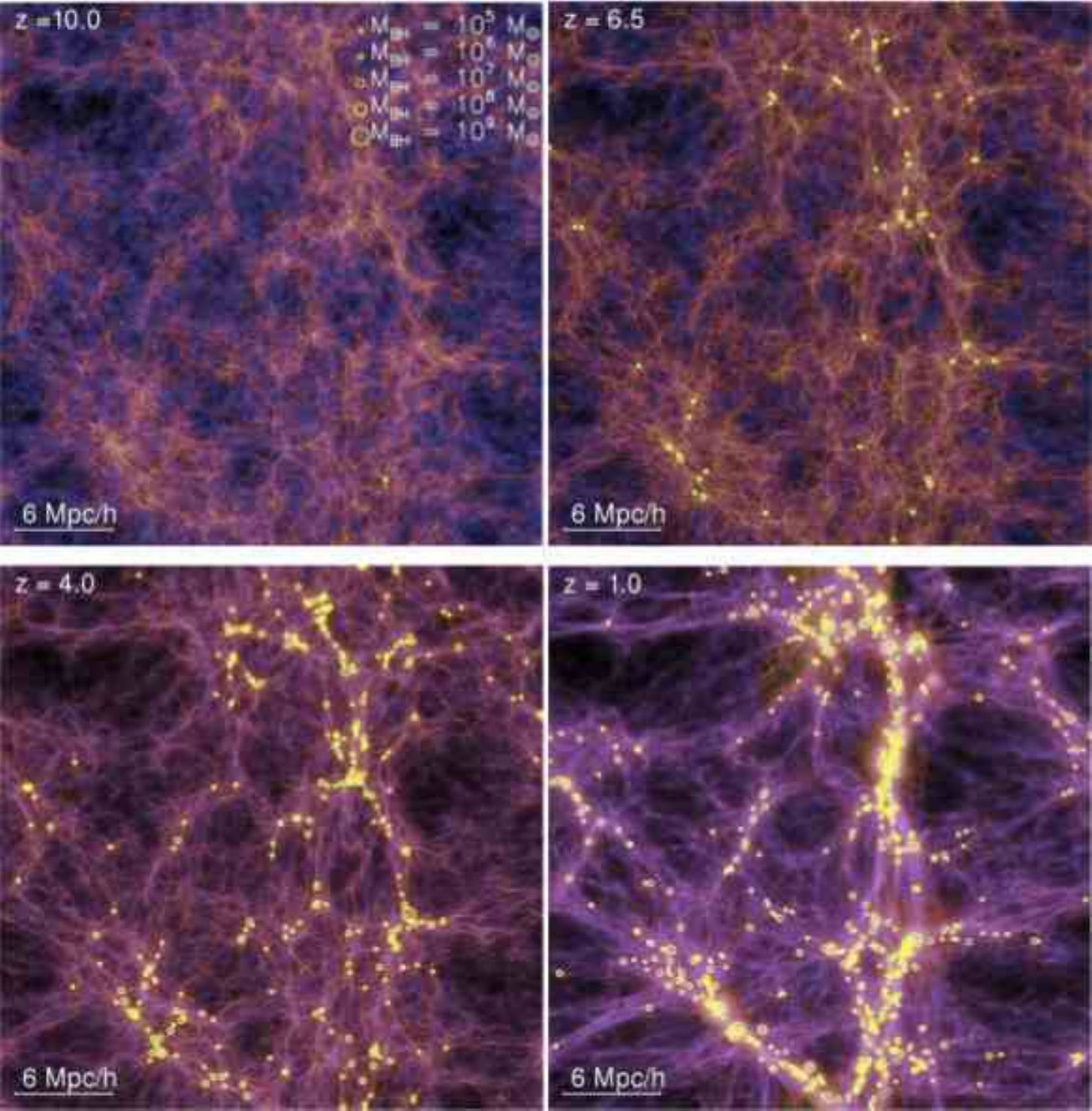}}
\caption
   {A state-of-art hydrodynamical simulation by
     \citet{dimatteo:2008:dcs} visualising the cosmic evolution of the
     baryonic density field and of their embedded black holes, in the
     $\Lambda$CDM cosmology. Each panel shows the same region of space
     (\ensuremath{33.75}\,\text{$h^{-1}$ Mpc} on a side) at different
     redshift, as labelled. The circles mark the positions of the
     black holes, with a size that encodes the mass, as indicated in
     the top left panel (numerical force resolution limits the lowest
     black hole mass to
     \ensuremath{10^{5}}\,\text{\ensuremath{\mathrm{M}_\odot}}). The
     projected baryonic density field is colour-coded with brightness
     proportional to the logarithm of the gas surface density. The
     images show that the black holes emerge in halos starting at high
     redshift (as early as $z \sim 10$) and later grow by accretion
     driven by gas inflows that accompany the hierarchical build-up of
     ever larger halos through merging. As the simulation evolves the
     number of black holes rapidly increases and larger halos host
     increasingly larger black holes. No black holes as massive as
     \ensuremath{10^{9}}\,\text{\ensuremath{\mathrm{M}_\odot}} are
     present in the simulated box because they are extremely rare.  }
\label{fig.bhweb}
\end{figure}

\begin{figure}
  \centering
  \resizebox{\hsize}{!}
	{\includegraphics[scale=1,clip]{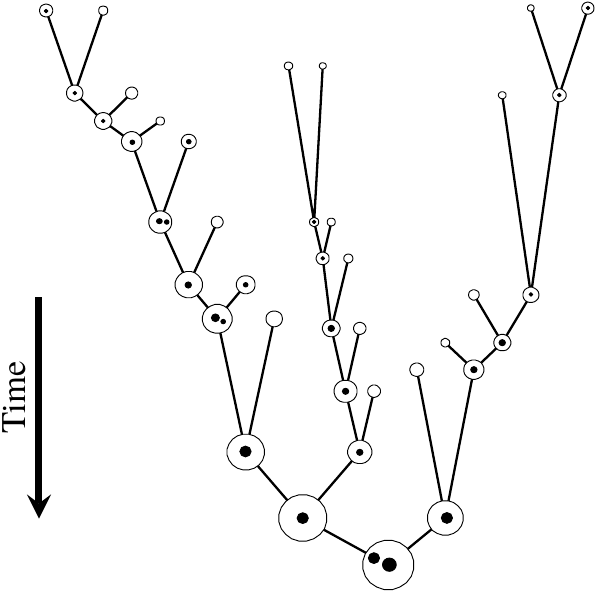} }
  \caption{A cartoon of the merger-tree history for the assembly of a
    galaxy and its central black holes. Time increases along the
    arrow.  Here the final galaxy is assembled through the merger of
    twenty smaller galaxies housing three seed black holes, and four
    coalescences of binary black holes.}
  \label{fig.bh_3}
\end{figure}

\noindent
These key findings hint in favour of the existence, at any redshift,
of an underlying population of black holes of a smaller variety, with
masses of \ensuremath{10^{4}}\,\text{\ensuremath{\mathrm{M}_\odot}} --
\ensuremath{10^{7}}\,\text{\ensuremath{\mathrm{M}_\odot}}, that grew in
mass along cosmic history inside their galaxies, through episodes of
merging and accretion. The evolution of black holes mimics closely
that of their host galaxies within the currently favoured cosmological
paradigm: a Universe dominated by cold dark matter (CDM).

Observations show that the mass content of the Universe is dominated
by CDM, with baryons contributing only at a \ensuremath{10}\,\text{\%}
level to the CDM, and that the spectrum of primordial density
perturbations contains more power at lower masses
\citep{mo:2010gfe..book.....M}. Thus, at the earliest epoch, the
Universe was dominated by small density perturbations. Regions with
higher density grow in time, to the point where they decouple from the
Hubble flow and collapse and virialise, forming self gravitating
halos. The first objects that collapse under their own self-gravity
are small halos that grow bigger through mergers with other halos and
accretion of surrounding matter. This is a bottom up path, and the
process is known as hierarchical clustering. As halos cluster and
merge to build larger ones, baryons follow the CDM halo potential well
and, similarly, black holes form and evolve in the same bottom-up
fashion
\citep{white:1978MNRAS.183..341W,haiman:1998ApJ...503..505H,haehnelt:1998MNRAS.300..817H,wyithe:2002ApJ...581..886W,volonteri:2003:amh}.

State-of-the-art hydrodynamical cosmological simulations
\citep{dimatteo:2008:dcs} illustrate (figure \ref{fig.bhweb}) where and when
the massive black holes form and how they are connected with the evolving
background baryonic density field.  As illustrated in figure \ref{fig.bhweb}
and as inferred in statistical models based on the extended Press-Schechter
(EPS) formalism, most of the black holes transit into the mass interval for
which eLISA is sensitive during their cosmic evolution
\citep{volonteri:2003:amh}.  Figure \ref{fig.bh_3} sketches and simplifies
conceptually the complex net terminating with the formation of a bright galaxy
at zero redshift, highlighting the sites where black holes form, cluster within
halos, pair with other black holes, and eventually coalesce.

\subsection*{Black holes in the sensitivity window of eLISA}
\label{sec.black-holes-sens}

\noindent
 Is there any observational
evidence of black holes of this variety in the Universe that may be
observed by eLISA? The Milky Way hosts in its bulge a black hole of
$(4\pm 0.06\pm0.35)\times
\ensuremath{10^{6}}\,\text{\ensuremath{\mathrm{M}_\odot}}$
\citep{ghez:2005:sogc,gillessen:2009:mso}, providing an example of a
black hole that does not fall into the population that
can be traced by luminous QSO.  Black holes in the mass range
\ensuremath{10^{5}}\,\text{\ensuremath{\mathrm{M}_\odot}} --
\ensuremath{10^{7}}\,\text{\ensuremath{\mathrm{M}_\odot}} are now
increasingly found in low mass spiral galaxies and dwarfs with and
without a bulge
\citep{greene:2004ApJ...610..722G,barth:2004ApJ...607...90B,greene:2008ApJ...688..159G,jiang:2011ApJ...737L..45J,kuo:2011ApJ...727...20K,xiao:2011ApJ...739...28X,jiang:2011arXiv1107.4105J} and evidence exists that some of these low mass black holes of $M <
\ensuremath{10^{5}}\,\text{\ensuremath{\mathrm{M}_\odot}}$ cohabit nuclear star clusters
\citep{ferrarese:2006:smbh,wehner:2006ApJ...644L..17W,barth:2009ApJ...690.1031B,seth:2008ApJ...678..116S,graham:2009MNRAS.397.2148G,bekki:2010ApJ...714L.313B}.

Dwarf galaxies in the galactic field are believed to undergo a quieter
merger and accretion history than their brighter analogues.  They may
represent the closest example of low mass halos from
which galaxy assembly took off.  Late type dwarfs are thus the
preferred site for the search of \emph{pristine} black holes
\citep{volonteri:2009MNRAS.400.1911V}.    NGC~4359, a close-by bulgeless, disky dwarf houses
in its centre a black hole of only \ensuremath{3.6 \times
  10^{5}}\,\text{\ensuremath{\mathrm{M}_\odot}}
\citep{peterson:2005ApJ...632..799P}.  This key discovery shows that
nature provides a channel to black hole formation also in potential
wells much shallower than that of the massive spheroids.

These middleweight mass black holes are numerous at high redshifts \citep{dimatteo:2008:dcs}, but 
are invisible with today instrumentation, given their low intrinsic luminosity and far-out distance. 
Furthermore, they become invisible to electromagnetic
observations near $z\gtrsim 11$ as close to this redshift the
intergalactic medium becomes opaque to their light, due to intervening absorption of
the neutral hydrogen \citep{fan:2006AJ....132..117F,miralda-escude:1998ApJ...501...15M}.  
ULAS~J1120+0641 holds the record of being the further
distant known QSO, at redshift $z={7.085 \pm 0.003}$, 
and hosts a bright, very massive black hole of $\sim 2\times 10^9\,M_{\odot}$ \citep{mortlock:2011Natur.474..616M}.
Its light was emitted before the end of the
reionisation, i.e. before the theoretically predicted transition of
the interstellar medium from an electrically neutral to an ionised
state \citep{fan:2006AJ....132..117F}.

\subsection{Galaxy mergers and black hole coalescence}
\label{sec.merg-galax-coal-1}

\begin{figure}
  \centering

  \resizebox{\hsize}{!}
	{\includegraphics[scale=1,clip]{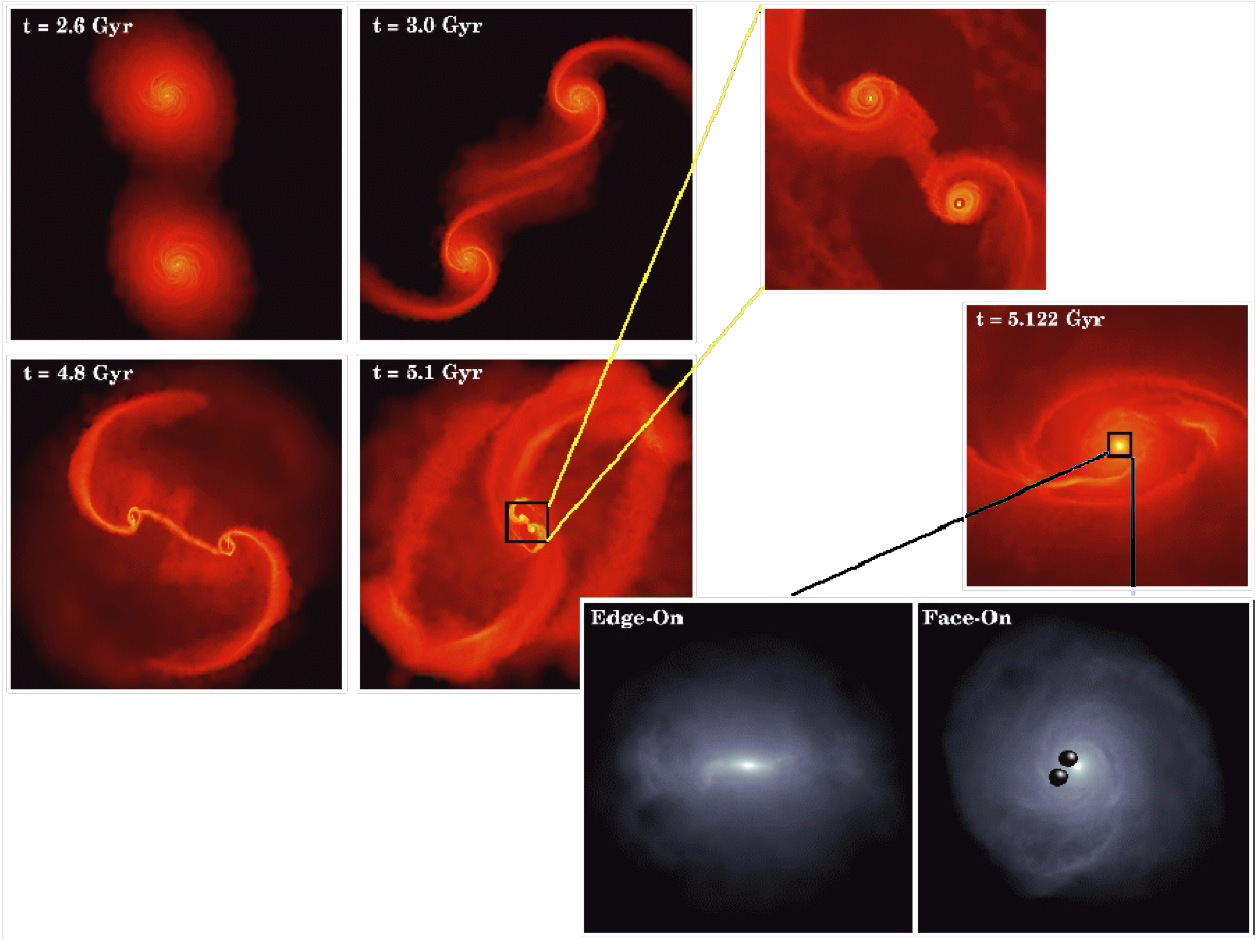} }
  \caption{ The different stages of the merger between two identical
    Milky-Way-like gas-rich disc galaxies
    \citep[from][]{mayer:2007Sci...316.1874M}. The panels show the
    density maps of the gas component in logarithmic scale, with
    brighter colours for higher densities. The four panels to the left
    show the large-scale evolution at different times. The boxes are
    \ensuremath{120}\,\text{kpc} on a side (top) and
    \ensuremath{60}\,\text{kpc} on a side (bottom). During the
    interaction tidal forces tear the galactic disks apart, generating
    spectacular tidal tails and plumes. The panels to the right show a
    zoom in of the very last stage of the merger, about 100 million
    years before the two cores have fully coalesced (upper panel), and
    2 million years after the merger (middle panel), when a massive,
    rotating nuclear gaseous disc embedded in a series of large-scale
    ring-like structures has formed. The boxes are now
    \ensuremath{8}\,\text{kpc} on a side. The two bottom panels, with
    a grey colour scale, show the detail of the inner
    \ensuremath{160}\,\text{pc} of the middle panel; a massive nuclear
    disc, shown edge-on (left) and face-on (right), forms in the
    aftermath of the merger (of
    $\ensuremath{10^{9}}\,\text{\ensuremath{\mathrm{M}_\odot}}$).  The
    two black holes continue to sink inside the disc and form a
    Keplerian binary; they are shown in the face-on image.}
  \label{fig.bh_6}
\end{figure}

\noindent
A grand collision between two galaxies of comparable mass (called
major merger) is not destructive event, but rather a transformation, 
as the two galaxies, after merging, form a new galaxy
with a new morphology.  Individual stars do not collide during the
merger, as they are tiny compared to the distances between them. The
two galaxies pass through each other and complex, time-varying
gravitational interactions redistribute the energy of each star in
such a way that a new bound galaxy forms.  Gas clouds instead collide
along the course of the merger: new stars form, and streams of gas
flow in the nuclear region of the newly forming galaxy.  The massive
black holes in the grand collision behave like stars. A key question
for the eLISA science case is: \emph{do black holes coalesce as their
  galaxies merge?}

The fate of black holes in merging galaxies can only be traced using
numerical simulations at the limits of current numerical
resolution. Not only isolated black holes are tiny, but also binary
black holes are. They form a tight binary system within a galaxy when
the mass in stars enclosed in the binary orbit becomes negligible
compared to the total mass of the binary $M$, and their Keplerian
velocity exceeds the velocity of the stars, $\sigma$. This occurs when
their relative separation $a_{\rm B}$ decays below about
$GM/\sigma^2$, i.e.\ when $a_{\rm B}\lesssim 10^{-4}-10^{-5}\,R_{\rm
  gal}$.  Binary black holes on the verge of coalescing within less
than a Hubble time are even smaller, as they touch when their
separation is of the size of the event horizon.  The timescale for
coalescence by gravitational waves only is a sensitive function of the
binary separation, scaling as $a^4$
\citep{peters:1964PhRv..136.1224P}.  Therefore, gravitational waves
guide the inspiral only when $a$ is less than a critical value $a_{\rm
  GW}\sim 0.003\, a_{\rm B}
(M/\ensuremath{10^{6}}\,\text{\ensuremath{\mathrm{M}_\odot}})^{1/4}$
(using scaling relations) that is of \ensuremath{0.01}\,\text{pc} --
\ensuremath{0.001}\,\text{pc} for a circular binary in the eLISA mass
interval. Typical orbital periods at $a_{\rm GW}$ are of a few years
to tens of years, and the hole's relative velocities are as high as
\ensuremath{3000}\,\text{km/s} -- \ensuremath{5000}\,\text{km/s}.

Black holes have to travel a distance from \ensuremath{0.1}\,\text{kpc} -- \ensuremath{10}\,\text{kpc} down
to \ensuremath{0.01}\,\text{pc} -- \ensuremath{0.001}\,\text{pc}, before entering the gravitational wave
inspiral regime, in a galaxy. Given the huge dynamical range,
different physical mechanisms are guiding their sinking
\citep{colpi:2011:asl.2010.1210}.  We can distinguish four phases for the dynamics
of black holes on their way to and after merging:  
\begin{enumerate}
\item the {\em pairing phase}, when the black holes pair on galactic
  scales following the dynamics of the galaxies they inhabit until
  they form a Keplerian binary (on pc scales);
\item the {\em binary phase}, when the Keplerian binary continues to
  harden at the centre of the galaxy remnant;
\item the {\em gravitational wave phase}, when black hole inspiral is dominated by
  loss of energy and angular momentum by gravitational waves; and finally
\item the {\em recoiling phase}, when the now single black hole either
  oscillates or escapes the galaxy following gravitational recoil.
\end{enumerate}

In major merger of galaxies the black holes pair under the action of
dynamical friction against the dark matter background, that brakes the disc/bulge which they inhabit
\citep{begelman:1980Natur.287..307B,chandrasekhar:1943:dfI,ostriker:1999ApJ...513..252O,colpi:1999ApJ...525..720C}.
Pairing occurs on the typical timescale of a galactic merger of a few
billion years.  A few million years after the new galaxy has formed, a
Keplerian binary forms on the scale of \ensuremath{1}\,\text{pc} -- \ensuremath{10}\,\text{pc}, under the
action of dynamical friction by stars and gas
\citep{mayer:2007Sci...316.1874M}.  Figure \ref{fig.bh_6} shows the
evolution of the two gas-discs in merging galaxies similar to the
Milky Way.  The galaxies host a central black hole, and the black
holes end forming a Keplerian binary embedded in a massive nuclear
disc \citep{mayer:2007Sci...316.1874M}.  The subsequent hardening of
the binary orbit (phase II) is controlled by the inflow of stars from
larger radii, and by the gas rotating in a circum-binary disc
\citep{merritt:2005:mbh,colpi:2011:asl.2010.1210,escala:2004:rgm,dotti:2006:lis,dotti:2007:smbh,dotti:2009MNRAS.396.1640D,callegari:2009ApJ...696L..89C}.
In gas rich environments, and for black holes of mass smaller than
about \ensuremath{10^{7}}\,\text{\ensuremath{\mathrm{M}_\odot}}, gas-dynamical torques on the binary suffice to
drive the system down to the gravitational wave inspiral domain
\citep{cuadra:2009MNRAS.393.1423C,hayasaki:2008ApJ...682.1134H,hayasaki:2009ApJ...691L...5H,macfayden:2008ApJ...672...83M,armitage:2005:esb,ivanov:1999MNRAS.307...79I,gould:2000ApJ...532L..29G}
if the gas does not fragment in stars
\citep{lodato:2009MNRAS.398.1392L}.  

Stars are ubiquitous, and in stellar bulges the black holes lose
orbital energy and angular momentum by ejecting stars that scatter
individually off the black holes \citep{quinlan:1996:dembhb,
  milosavljevic:2001ApJ...563...34M,makino:2004ApJ...602...93M,merritt:2004:clc,berczik:2005ApJ...633..680B,merritt:2005:mbh,merritt:2006ApJ...648..976M,merritt:2007ApJ...671...53M,sesana:2006:mbhb,sesana:2007ApJ...660..546S,sesana:2008ApJ...686..432S,perets:2008:mpem}.
These stars approach the binary from nearly radial orbit, and shrink
the binary down to the gravitational wave phase, if they are present in sufficient
number to carry away the energy for the binary to decay down to
$a_{\rm GW}$.  These stars, ejected with high velocities, are lost by
the galaxy, and the timescale of sinking of the binary depends on the
rate at which new stars are supplied from far-out distances.
Self-consistent high resolution direct N-body simulations
\citep{khan:2011ApJ...732...89K,preto:2011ApJ...732L..26P,
  berczik:2006:embsbh} indicate that the stellar
potential of the remnant galaxy retains, in response to the anisotropy
of the merger, a sufficiently high degree of rotation and triaxiality
to guarantee a large reservoir of stars on centrophilic orbits that
can interact with the black holes down to the transit from the binary
phase to the gravitational wave phase.  This seems to be a universal process.  When
coalescence occurs, the merger remnant ``recoils''because of the
anisotropic emission of gravitational waves 
\citep{baker:2008ApJ...682L..29B}, moving away from the 
gravitational centre of the galaxy.  The kicked
black hole, may return after a few
oscillations down to the nuclear regions of the host galaxy, or escape
the galaxy depending on the magnitude of the kick
\citep{gualandris:2008ApJ...678..780G,blecha:2008MNRAS.390.1311B,blecha:2011MNRAS.412.2154B,guedes:2011ApJ...729..125G}.

\subsection{Dual, binary and recoiling  AGN in the cosmic landscape}
\label{sec.dual-binary-reco}

\begin{figure}
\resizebox{\hsize}{!}
          {\includegraphics[scale=1,clip]{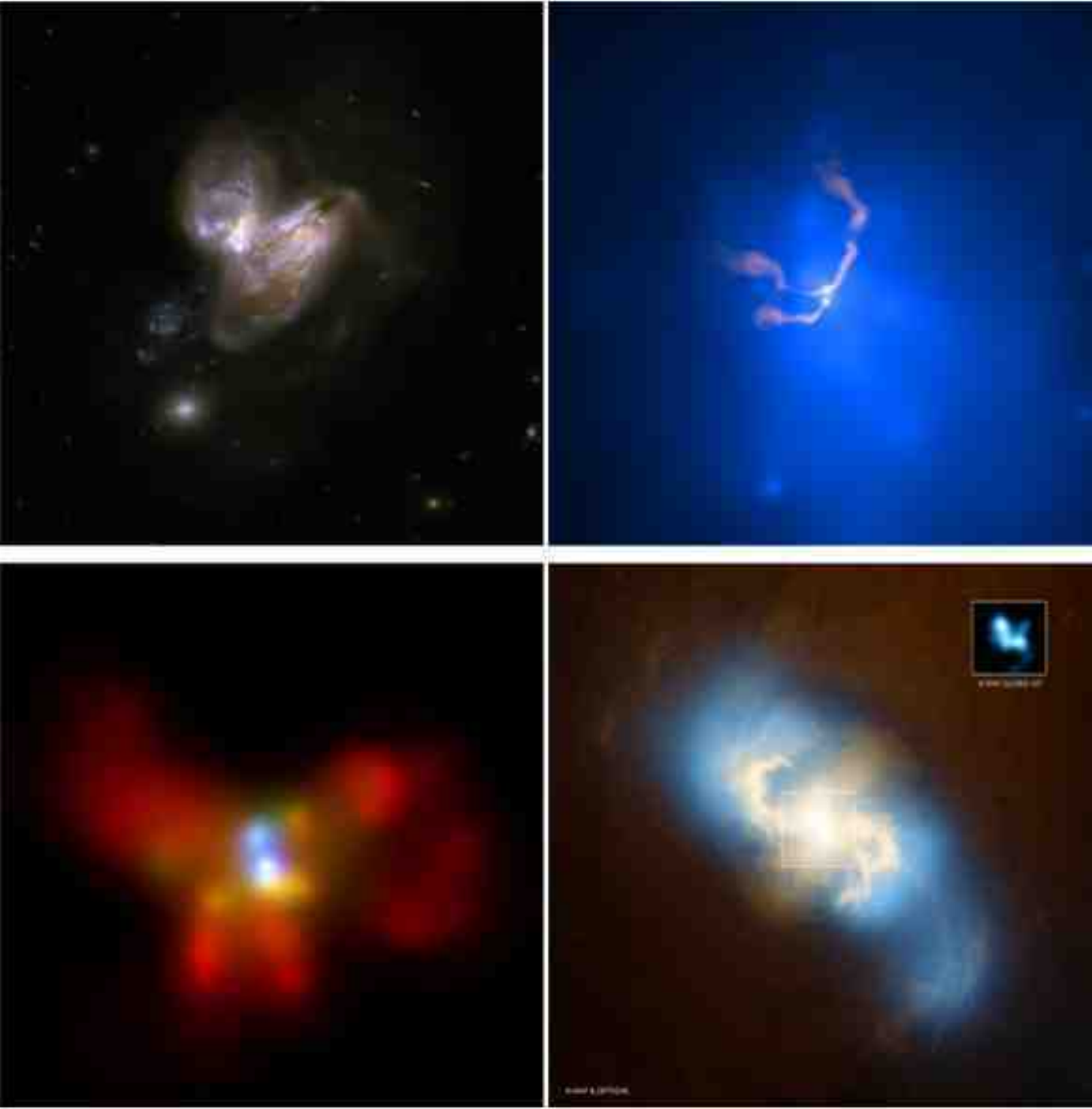}}
\caption
   {  
Active black holes in colliding galaxies. Arp~299 (upper left panel) is the
interacting system resulting from the collision of two gas-rich spirals, and
hosts a dual AGN, i.e.\ two black holes ``active'' during the pairing phase.
The accreting black holes are visible in the X-rays and are located at the
optical centres of the two galaxies, at a separation of
\ensuremath{4.6}\,\text{kpc}\citep{ballo:2004ApJ...600..634B}. X-ray view of
NGC6240 (lower left panel) an ultra luminous infrared galaxy considered to be a
merger in a well advanced phase \citep{komossa:2003ApJ...582L..15K}. X-ray
observations with the Chandra Observatory let to the discovery of two strong
hard X-ray unresolved sources embedded in the diluted soft X-ray emission (red)
of a starburst. The dual AGN are at a separation of
\ensuremath{700}\,\text{pc}. Composite X-ray (blue)/radio (pink) image of the
galaxy cluster Abell~400 (upper right panel) showing radio jets immersed in a
vast cloud of multimillion degree X-ray emitting gas that pervades the cluster.
The jets emanate from the vicinity of two supermassive black holes (a dual
radio-loud AGN) housed in two elliptical galaxies in the very early stage of
merging. Composite optical and X-ray image of NGC~3393 (lower right panel), a
spiral galaxy with no evident signs of interaction. In its nucleus, two active
black holes have been discovered at a separation of only
\ensuremath{150}\,\text{pc} \citep{fabbiano:2011Natur.477..431F}. The closeness
of the black holes embedded in the bulge, provide a hitherto missing
observational point to the study of galaxy-black hole evolution: the phase when
the black holes are close to forming a Keplerian binary. The regular spiral
morphology and predominantly old circum-nuclear stellar population of this
galaxy, indicates that a merger of a dwarf with a large spiral led to the
formation of the binary
\citep{callegari:2009ApJ...696L..89C,callegari:2011ApJ...729...85C}.
   }
\label{fig.bh_4}
\end{figure}

\noindent

Surprisingly, the closest example of an imminent merger is in our Local Group.  
Andromeda (M31) along with a handful of lesser galaxies does not follow 
Hubble's law of cosmic expansion: it is falling toward us at a speed of about
\ensuremath{120}\,\text{km/s}.  M31 is a member
of a group of galaxies, including the Milky Way, that form a
gravitationally bound system, the Local Group.  M31 and the Milky Way 
each house a massive black hole \citep{vandermarel:1994MNRAS.268..521V} and 
are on a collision course, with a merger possibly before the Sun expands into a
red giant ($\sim 4$ billion years) \citep{cox:2008MNRAS.386..461C}.  
Observations are now
revealing the presence  of many colliding galaxies in the Universe, and
in a number of cases two active black holes are visible through their
X-ray or radio emission.

The existence of
binary AGN, i.e.\ of two active black holes bound in a Keplerian
fashion, is still debatable at the observational level, as they are
rare objects \citep{volonteri:2009:subpc}.  Two cases deserve
attention.  The first case is 0402+379, a radio source in an
elliptical galaxy showing two compact flat-spectrum radio nuclei, only
\ensuremath{7}\,\text{pc} apart \citep{rodriguez:2006:csb,rodirguez:2009:smbh}.  The
second case is OJ~287, a source displaying a periodic variability of
12 years
\citep{valtonen:2006ApJ...643L...9V,valtonen:2008Natur.452..851V,valtonen:2010ApJ...709..725V,valtonen:2011ApJ...729...33V}
that has been interpreted as being a Keplerian binary with evidence of
orbital decay by emission of gravitational waves.  A number of
sub-parsec binary black hole candidates have been proposed
\citep{tsalmantza:2011ApJ...738...20T,eracleous:2011arXiv1106.2952E}
based on the recognition that gas clouds orbiting one/two black
hole(s) can leave an imprint in the optical spectra of the AGN
\citep{montuori:2011MNRAS.412...26M,shen:2010ApJ...725..249S,bogdanovic:2008ApJS..174..455B,
  boroson:2009:subpcbh,shields:0004-637X-707-2-936,decarli:2010ApJ...720L..93D,barrows:2011NewA...16..122B}. Follow-up
observations will be necessary to assess their true nature.

Recoiling AGN, i.e.\ recoiling black holes observed in an active phase
\citep{loeb:2007PhRvL..99d1103L,devecchi:2009MNRAS.394..633D,oleary:2009MNRAS.395..781O,merritt:2009ApJ...699.1690M},
have been searched recently, and there has been a claim of a discovery
\citep{komossa:2008ApJ...678L..81K}, even though alternative
interpretations are also viable \citep{dotti:2009MNRAS.398L..73D,
  bogdanoovic:2009ApJ...697..288B}.  Two spatially off-set AGN have
been found in deep surveys with kinematic properties that are
consistent with being two recoiling black holes
\citep{jonker:2010MNRAS.407..645J,civano:2010ApJ...717..209C}.

The most remarkable, albeit indirect, evidence of coalescence events
is found in bright elliptical galaxies that are believed to be product
of mergers.  Bright elliptical galaxies show light deficits (cores) in
their surface brightness profiles, i.e.\ lack of stars in their nuclei
\citep{kormendy:2009ApJ...691L.142K}, and this missing light
correlates with the mass of the central black hole
\citep{merritt:2006ApJ...648..976M,merritt:2007ApJ...671...53M}. Thus,
cores are evidence of binary black holes scouring out
the nuclear stars via three-body scattering
\citep{merritt:2007ApJ...671...53M} or even via post-merger relaxation
following a kick
\citep{boylankolchin:2004ApJ...613L..37B,gualandris:2008ApJ...678..780G,guedes:2009ApJ...702..890G}.
Lastly, mergers change the black hole's spin directions due to
conservation of angular momentum. Reorientation of the black hole spin
following coalescence is now believed to be at the heart of X-shaped
radio galaxies, where an old jet coexists with a new jet of different
orientation
\citep{merritt:2002Sci...297.1310M,liu:2004MNRAS.347.1357L}. This
would be again a sign of a fully accomplished coalescence event.

\subsection{Seed black holes}
\label{sec.seed-black-holes}

Models of hierarchical structure formation predict that galaxy sized
dark matter halos start to become common at redshift $z\sim 10-20$
\citep{mo:2010gfe..book.....M}. This is the beginning of the nonlinear
phase of density fluctuations in the Universe, and hence also the
epoch of baryonic collapse leading to star and galaxy formation.
Different populations of seed black holes have been proposed in the
range \ensuremath{100}\,\text{\ensuremath{\mathrm{M}_\odot}} --
\ensuremath{10^{6}}\,\text{\ensuremath{\mathrm{M}_\odot}}
\citep{volonteri:2010A&ARv..18..279V}.  

\emph{Small} mass seeds 
(\ensuremath{100}\,\text{\ensuremath{\mathrm{M}_\odot}} --
\ensuremath{1000}\,\text{\ensuremath{\mathrm{M}_\odot}}) may result from
the core collapse of the first generation of massive stars (Pop~III)
that form from unstable metal-free gas clouds, at $z\sim 20$ and in
halos of \ensuremath{10^{6}}\,\text{\ensuremath{\mathrm{M}_\odot}}
\citep{haiman:1996ApJ...464..523H,
  tegmark:1997ApJ...474....1T,Abel:2002,omukai:2001ApJ...561L..55O,bromm:2002:ffs,ripamonti:2002MNRAS.334..401R,omukai:2003ApJ...589..677O}. Pop~III stars
as massive as \ensuremath{260}\,\text{\ensuremath{\mathrm{M}_\odot}}
or larger collapse into a black hole of similar mass after only about
\ensuremath{2}\,\text{Myr}
\citep{heger:2003ApJ...591..288H,madau:2001:mbhIII}.
The formation of Pop~III stars remains a poorly understood process
\citep{zinnecker:2007ARA&A..45..481Z} and the maximum mass reached by individual
stars is unknown \citep[see, e.g.,][]{clark:2011ApJ...727..110C}.

\emph{Large} seeds form in heavier halos (of
\ensuremath{10^{8}}\,\text{\ensuremath{\mathrm{M}_\odot}}) from the
collapse of unstable gaseous discs of
\ensuremath{10^{4}}\,\text{\ensuremath{\mathrm{M}_\odot}} --
\ensuremath{10^{6}}\,\text{\ensuremath{\mathrm{M}_\odot}}.  This
route, ending with the formation of a very massive quasi-star, assumes
that fragmentation is suppressed possibly by turbulence and by an
intense ultraviolet background light, in an environment of low
metallicity
\citep{haehnelt:1993MNRAS.263..168H,loeb:1994:cbg,bromm:2003:ffs,koushiappas:2004:mbhs,begelman:2006:fsm,lodato:2006:smbh,dotan:2011MNRAS.tmp.1661D,dijkstra:2008MNRAS.391.1961D}.
Collapsing clouds can have significant angular momentum
\citep{bullock:2001ApJ...555..240B}, and thus additional momentum
transport is required for the self-gravitating disc to form a
supermassive star
\citep{shlosman:1989Natur.338...45S,begelman:2009ApJ...702L...5B}.
The very massive quasi-star of
about \ensuremath{10^{5}}\,\text{\ensuremath{\mathrm{M}_\odot}},
burns hydrogen and helium in its core and 
once formed, the star collapses into a black hole when metallicity is
below some threshold, as the alternative is its entire explosion 
\citep{montero:2011arXiv1108.3090M,shibata:2002ApJ...572L..39S}. Recently, other formation 
routes have been explored for the large seeds
(\ensuremath{10^{3}}\,\text{\ensuremath{\mathrm{M}_\odot}} --
\ensuremath{10^{4}}\,\text{\ensuremath{\mathrm{M}_\odot}}).  They
comprise the formation of a massive star via stellar runaway collisions
in young dense star clusters
\citep{portegieszwart:2002ApJ...576..899P,guerkan:2004ApJ...604..632G,devecchi:2009ApJ...694..302D},
and the relativistic collapse of stellar mass black holes in nuclear
star clusters invaded by a large inflow of gas
\citep{davies:2011ApJ...740L..42D}. Lastly, much heavier seeds resulting
from direct collapse of nuclear gas in a gas-rich galactic merger have
been proposed as the origin of black holes 
\citep{mayer:2010Natur.466.1082M}.

The subsequent step is to follow the evolution of the black hole seeds 
according to the growth of the halos they inhabit, and the mode of accretion 
\citep{volonteri:2010:qce,devecchi:2009ApJ...694..302D,volonteri:2008MNRAS.383.1079V}. This
is an inherently model dependent process. Observations of nearby
dormant black holes and of AGN at higher redshift help in constraining
their evolution, but theoretical models still have some unconstrained
degrees of freedom.

\subsection{ Evolving massive black hole spins via coalescence and accretion events}
\label{sec.design-evol-mass}

\noindent
Astrophysical black holes are fully described by the mass $\ensuremath{\mathit{M}_\bullet}$ and
angular momentum ${\bf J}$, referred to as spin. The modulus $J$ of ${\bf J}$ is usually
specified in terms of the dimensionless spin parameter $\ensuremath{a_\bullet}$
defined so that $J=\ensuremath{a_\bullet} (G\ensuremath{\mathit{M}_\bullet}^2/c).$ For a specified mass $\ensuremath{\mathit{M}_\bullet}$, a
black hole described by GR cannot have $\ensuremath{a_\bullet} >1$,
without showing a naked singularity (and this is forbidden by the
Cosmic Censorship conjecture).

Both coalescences and accretion change $\ensuremath{\mathit{M}_\bullet}$, $J$ (or $\ensuremath{a_\bullet}$) and
the orientation of ${\bf J}$ in a significant manner \citep{berti:2008:cbhse,volonteri:2007:bhs}.

\subsubsection*{Spins in black hole coalescences}
\label{sec.spins-black-hole}

With the advent of numerical relativity, it became possible to
accurately determine the evolution of the initial spins of the black
holes to the final spin of the remnant black hole in a merger event
\citep{pretorius:2005:ebs,baker:2006:gwe,campanelli:2006:aeo,rezzolla:2008PhRvD..78d4002R,centrella:2010ARNPS..60...75C}.

Numerical relativity simulations for equal mass, non spinning black
holes find a spin $a_{\bullet}= \num{0.68646} \pm \num{0.00004}$
\citep{scheel:2009PhRvD..79b4003S} for the merged black hole, 
resulting from the angular momentum of the orbit.  Extrapolation of
black hole coalescences with large initial spins (larger than
approximately \ensuremath{0.9}) exactly aligned with the orbital angular
momentum find a final $a_{\bullet}= \num{0.951} \pm \num{0.004}$
\citep{marronetti:2008PhRvD..77f4010M}.  
When mergers occur with retro- and pro-grade orbits equally
distributed, as it is expected in the case of astrophysical black
holes \citep{hughes:2003:bhm}, the average spin of the merger
remnant is about \ensuremath{0.7}, close to the expectation for non
spinning holes \citep{berti:2008:cbhse}.

For almost any configuration of spins and mass ratio, the emission
pattern of the gravitational wave is anisotropic, leading to a
\emph{gravitational recoil}
\citep{campanelli:2007PhRvL..98w1102C,gonzalez:2007PhRvL..98w1101G,lousto:2011PhRvD..83b4003L}.

Numerical studies show that initially nonspinning black holes or
binaries with spins aligned with the orbital angular momentum are
recoiling with a velocity below about
\ensuremath{200}\,\text{km/s}. By contrast, the recoil is dramatically
larger, up to approximately \ensuremath{5000}\,\text{km/s}, for
binaries of comparable mass and black holes with large spins in
peculiar non-aligned configurations
\citep{lousto:2011arXiv1108.2009L}.  Thus, unexpectedly, spins
(regulated by coalescence and accretion) affect the retention fraction
of black holes in galactic halos, and this has consequences on the
overall evolution of black holes in galaxies
\citep{volonteri:2010MNRAS.404.2143V,schnittman:2007ApJ...667L.133S,schnittman:2007ApJ...662L..63S,kesden:2010ApJ...715.1006K}.

\subsubsection*{Spins and black hole accretion}
\label{sec.spin-accretion}

The evolution of mass and spin of astrophysical black holes are strongly
correlated, also when considering accretion.  Spins determine directly
the radiative efficiency $\varepsilon (a_\bullet)$, and so the rate at
which mass is increasing. In radiatively efficient accretion discs
\citep{shakura:1973A&A....24..337S} $\varepsilon$ varies from
\ensuremath{0.057} (for $a_\bullet=0$) to \ensuremath{0.151} (for $a_\bullet=0.9$)
and \ensuremath{0.43} (for $a_\bullet=1$).  Accretion on the other hand
determines black hole spins since matter carries with it angular
momentum (the angular momentum at the innermost stable circular orbit
of a Kerr black hole).  A non-rotating black hole is spun-up to
$a_\bullet=1$ after increasing its mass by a factor $\sqrt 6$, for
prograde accretion \citep{bardeen:1970Natur.226...64B}. Conversely, a
maximally rotating black hole is spun-down by retrograde accretion to
$a_\bullet\sim 0$, after growing by a factor $\sqrt {3/2}$.
 
Accretion imposes limits on the black hole spin.  Gas accretion from a
geometrically thin disc limits the black-hole spin to
$a^{\text{acc}}_{\bullet}=\num{0.9980} \pm \num{0.0002}$, as photons emitted
by the disc and with angular momentum anti-parallel to the black hole spin
are preferentially captured, having a larger cross section, limiting
its rotation \citep{thorne:1974:dabh}.  The inclusion of a jet, as studied in
magneto-electrodynamic simulations, reduces this limit to $a^{\rm
  jet}_\bullet\simeq 0.93$ \citep{gammie:2004:bhse}, and changes in the
accretion geometry produce a similar effect \citep{popham:1998ApJ...504..419P}.

How black holes are fed from the large scale down to the hole's
influence radius ($R_{\rm grav}$) is presently unknown, and the spin
is sensitive to the way gas is accreted with time \citep{volonteri:2007:bhs}.
Two limiting modes
of accretion can occur.  \emph{Coherent accretion} refers to accretion
from a geometrically thin disc, lasting longer than a few black hole
mass growth $e$-folding times.  During coherent accretion the black 
hole can more than double
its mass, bringing its spin up to the limit imposed by basic physics,
either $a_\bullet^{\rm acc}$ or $a_\bullet^{\rm jet}$.  By contrast,
\emph{chaotic accretion} refers to a succession of accretion episodes
that are incoherent, i.e. randomly oriented. The black hole can then
spin-up or down, and spin-down occurs when counter-rotating material is
accreted, i.e.\ when the angular momentum ${\bf L}$ of the disc is
strongly misaligned with respect to $\bf{J}$ (i.e.\
$\bf{J}\cdot\bf{L}<0$).  If accretion proceeds via
short-lived, uncorrelated episodes with co-rotating and
counter-rotating material equally probable, spins tend to be small
\citep{king:2006MNRAS.373L..90K,king:2007MNRAS.377L..25K,moderski:1998MNRAS.301..142M}:
counter-rotating material spins the black hole down
more than co-rotating material spins it up, as the innermost stable
orbit of a counter-rotating test particle is located at a larger
radius than that of a co-rotating particle, and accordingly carries
a larger orbital angular momentum.

The direction of the black hole spin is also an important element in
the study of black holes. In a viscous accretion disc that is
misaligned with the spin of the black hole, Lense-Thirring precession
of the orbital plane of fluid elements warps the disc, forcing the gas
close to the black hole to align (either parallel or anti-parallel)
with the spin of the black hole. Warping is a rapid process that
causes alignment of the disc out to
\ensuremath{100}\,\text{\ensuremath{R_{\rm horizon}}} --
\ensuremath{10^{3}}\,\text{\ensuremath{R_{\rm horizon}}}, depending on
$a_\bullet$ \citep[]{bardeen:1975ApJ...195L..65B}.  Following
conservation of total angular momentum, the black hole responds to the
warping through precession and alignment, due to dissipation in the
disc \citep{scheuer:1996MNRAS.282..291S,perego:2009MNRAS.399.2249P}
evolving into a configuration of minimum energy where the black hole
and disc are aligned \citep{king:2005MNRAS.363...49K}. This process is
short (\ensuremath{10^{5}}\,\text{yrs}) compared to the typical
accretion time scale, allowing astrophysical black holes to evolve into
a quasi-aligned spin-orbit configuration prior to coalescence
\citep{dotti:2010MNRAS.402..682D}. % This has led to the recognition

According to these theoretical findings, masses and spins evolve
dramatically following coalescence and accretion events.  The spin offers the
best diagnostics on whether the black holes prior to coalescence have
experienced either coherent or chaotic accretion episodes. Both, mass and spin,
are directly encoded into the gravitational waves emitted during the the merger
process.  eLISA will measure the masses and spins of the black holes prior to
coalescence, offering unprecedented details on how black hole binaries have
been evolving along of the course of the galactic merger and along cosmic
history.

\subsection{Cosmological massive black hole merger rate}
\label{sec.cosm-mass-black}

\noindent
As eLISA creates a new exploratory window on the evolution of black
holes, covering a mass and redshift range that is out of the reach of
current (and planned) instruments, its expected detection rate is
observationally unconstrained. Today we can probe dormant black holes
down to masses of about
\ensuremath{10^{5}}\,\text{\ensuremath{\mathrm{M}_\odot}}
\citep{magorrian:1998:dmdo,xiao:2011ApJ...739...28X} in the local
Universe only, and their massive (i.e.\ heavier than
\ensuremath{10^{8}}\,\text{\ensuremath{\mathrm{M}_\odot}}) active
counterparts out to redshift $\gtrsim6$
\citep{fan:2006:sq_4,fan:2004:sq_3,fan:2003:sq_2,fan:2001:sq_1}. Any
estimate of the eLISA detection rate necessarily has to rely on
extrapolations based on theoretical models matching the properties of
the observable black hole population.

Observationally, the black hole merger rate can be inferred only at relatively 
low redshift, by counting the fraction of close pairs in deep galaxy surveys. 
Given a galaxy density per co-moving megaparsec cube $n_G$, a fraction 
of close pairs $\phi$, and a characteristic merger timescale ${\cal T}_M$, 
the merger rate density \emph{of galaxies} (number of mergers per year
per co-moving megaparsec cube), is given by
${\dot{n}}_M=\phi n_G/(2{\cal T}_M)$. 

Estimates of ${\dot n}_M$ have been produced by several groups in the
last decade
\citep{patton:2002:dcgp,lin:2004:deep2,Bell:2006:geg,depropris:2007:mgc,lin:2008:rew,deravel:2009:vimos,xu:2011:cosmos},
using deep spectroscopic galaxy surveys like COMBO, COSMOS
and DEEP2. Surveys are obviously flux limited, and usually an
absolute magnitude cutoff (which translate into a stellar mass lower
limit) is applied to obtain a complete sample of galaxy pairs across a
range of redshifts. The galaxy merger rate is therefore fairly well
constrained only at redshift $z\lesssim1$ for galaxies with stellar
mass larger than approximately \ensuremath{10^{10}}\,\text{\ensuremath{\mathrm{M}_\odot}}. From compilation of
all the measurement \citep{xu:2011:cosmos}, typical average massive
galaxy merger rates $\dot n_M$ at $z<1$ lie in the range
$\ensuremath{5 \times 10^{-4}}<\dot n_M<\ensuremath{2 \times 10^{-3}} h^3_{100} \text{Mpc\ensuremath{^{-3}} Gyr\ensuremath{^{-1}}}$. By
applying the black hole-host relations \citep{gultekin:2009:ms}, the
galaxy stellar mass cutoff is converted into a lower limit to the
hosted black hole mass.  Assuming a black hole occupation fraction of
one (appropriate for massive galaxies) and integrating over the
appropriate co-moving cosmological volume, this translates into an
\emph{observational} estimate of the \emph{massive black hole merger
  rate} for $z<1$ and $M>\text{few}\times\ensuremath{10^{6}}\,\text{\ensuremath{\mathrm{M}_\odot}}$.

These estimates can be compared to the rate predicted by Monte Carlo
merger trees \citep{volonteri:2003:amh} based on the EPS formalism
\citep{press:1974:fgcg,lacey:1993:mrhm,sheth:1999:lsb}, which are
used to reconstruct the black hole assembly, and thus to infer
eLISA detection rates, in the $\Lambda$CDM cosmology.
The evolutionary path (outlined in the previous sections) can be
traced back to very high redshift ($z>20$) with high resolution via
numerical EPS Monte Carlo realisations of the merger hierarchy. \citet{sesana:2008:sgwb} carried a detailed comparison of the
merger rate predicted by such models in the $z<1$ and
$M>\text{few}\times\ensuremath{10^{6}}\,\text{\ensuremath{\mathrm{M}_\odot}}$ range with those inferred by galaxy
pair counting, finding a generally broad consistency within a factor
of \ensuremath{2}.

 On the theoretical
side, massive black hole merger rates can be computed from
semi-analytic galaxy formation models coupled to massive N-body
simulations tracing the cosmological evolution of dark matter halos
\citep{delucia:2006:fheg,bertone:2007:rgm,guo:2011:fds}, such as the
Millennium Run \citep{springel:2005:sefc}. Such models are generally
bound to the limiting resolution of the underlying N-body simulations,
and are therefore complete only for galaxy masses larger than
approximately
\ensuremath{10^{10}}\,\text{\ensuremath{\mathrm{M}_\odot}}.  In a
companion study, \cite{sesana:2009:gwrbb} also showed that the
merging black hole mass functions predicted by EPS based merger trees
is in excellent agreement with those extracted by semi-analytic galaxy
formation model in the mass range
$M>\ensuremath{10^{7}}\,\text{\ensuremath{\mathrm{M}_\odot}}$, where
semi-analytic models can be considered complete.

Merger rates obtained by EPS merger trees are therefore 
firmly anchored to low redshift observations and to theoretical 
galaxy formation models. 
Nevertheless, the lack of observations in the mass range of interest
for eLISA leaves significant room for modelling, and theoretical
astrophysicists have developed a large variety of massive black holes
formation scenarios that are compatible with observational constraints
\citep{volonteri:2003:amh,koushiappas:2004:mbhs,begelman:2006:fsm,lodato:2006:smbh}.
The predicted coalescence rate in the eLISA window depends on the
peculiar details of the models, ranging from a handful up to few
hundred events per year
\citep{haehnelt:1994:lfg,wyithe:2003:lfg,sesana:2004:lfg,enoki:2004:gws,sesana:2005:gws,rhook:2005:rer,koushiappas:2006:tms,sesana:2007:imprint}.
A recent compilation, encompassing a wide variety of assembly history
models, can be found in \citep{sesana:2011:rmbh}.

\subsection{Massive black hole binaries as gravitational waves sources: what can eLISA discover?}
\label{sec.massive-black-hole}

\noindent
In the eLISA window of detectability massive black hole binary
coalescence is a three-step process comprising the \emph{inspiral,
  merger, and ring-down} \citep{flanagan:1998:mgwa}. The inspiral
stage is a relatively slow, adiabatic process in which the black holes
spiral together on nearly circular orbits.  The black holes have a
separation wide enough so that they can be treated analytically as
point particles through the Post Newtonian (PN) expansion
of their binding energy and radiated flux \citep{blanchet:2006:grp}.
The inspiral is followed by the dynamical coalescence, in which the
black holes plunge and merge together, forming a highly distorted,
perturbed remnant.  At the end of the inspiral, the black hole
velocities approach $v/c \sim 1/3$. At this stage the PN approximation
breaks down, and the system can only be described by a numerical
solution of the Einstein equations.  The distorted remnant settles
into a stationary Kerr black hole as it \emph{rings down}, by emitting
gravitational radiation. This latter stage can be, again, modelled
analytically in terms of black hole perturbation theory.  At the end
of the ring-down the final black hole is left in a quiescent state,
with no residual structure besides its Kerr spacetime geometry.

In recent years there has been a major effort in constructing
accurate waveforms inclusive of the inspiral merger and ring-down
phases \citep{pretorius:2005:ebs,campanelli:2006:aeo,baker:2006:gwe}.
Even a few orbital cycles of the full waveform are computationally
very demanding. ``Complete'' waveforms can be designed by stitching
together analytical PN waveforms for the early inspiral with a
(semi)phenomenologically described merger and ring-down phase
\citep{santamaria:2010PhRvD..82f4016S,pan:2011arXiv1106.1021P,damour:2011PhRvD..83b4006D},
calibrated against available numerical data.  In the following
estimations we will mostly employ phenomenological waveforms
constructed in frequency domain, as described in
\citet{santamaria:2010PhRvD..82f4016S}.  Self-consistent waveforms of
this type (the so called PhenomC waveforms) are available for
non-spinning binaries and for binaries with aligned spins. In the case
of binaries with misaligned spins, we use ``hybrid'' waveforms
obtained by stitching precessing PN waveforms for the inspiral with
PhenomC waveforms for the merger/ring-down. This stitching is
performed by projecting the orbital angular momentum and individual
spins onto the angular momentum of the distorted black hole after
merger.  Given a waveform model, a first measure of the eLISA
performance is the SNR of a binary merger with parameters in the
relevant astrophysical range.

\subsubsection*{Detector performance}
\label{sec.detect-perf}

\begin{figure}
  \centering
  \resizebox{\hsize}{!}
	{\includegraphics[scale=1,clip]{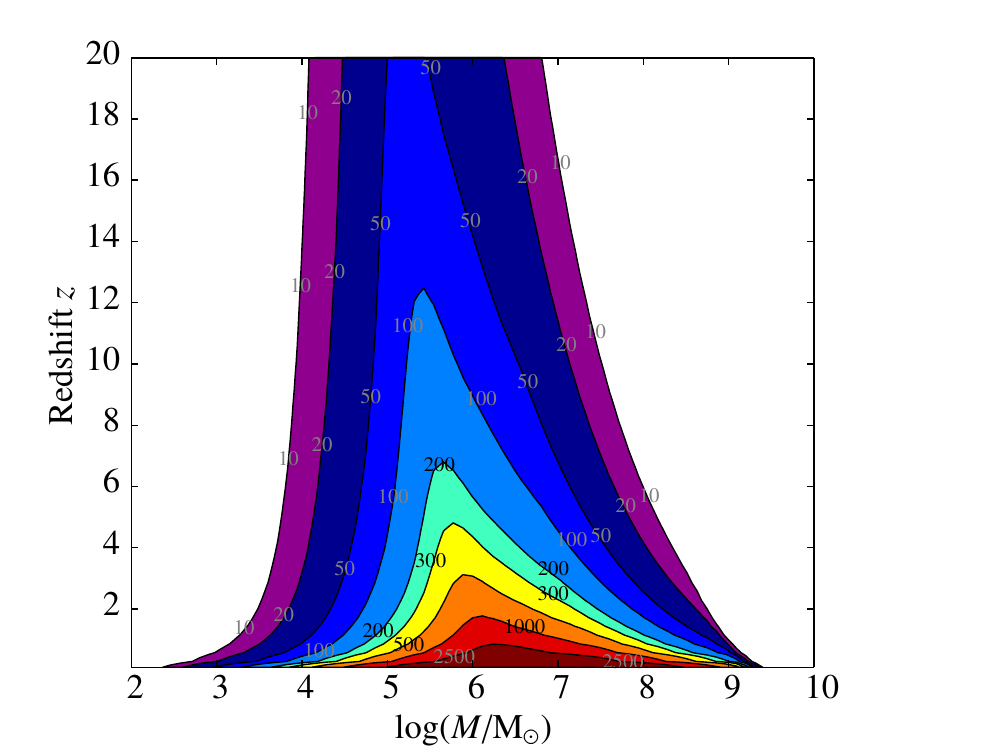} }
  \caption{Constant-contour levels of the sky and polarisation
    angle-averaged signal-to-noise ratio (SNR) for equal mass
    non-spinning binaries as a function of their total mass $M$ and
    cosmological redshift $z$. The total mass $M$ is measured in the
    rest frame of the source. The SNR is computed using PhenomC
    waveforms \citep{santamaria:2010PhRvD..82f4016S}, which are
    inclusive of the three phases of black hole coalescence (in
    jargon: inspiral, merger, and ring-down, as described in the
    text).}
  \label{fig.SNRmz}
\end{figure}

\begin{figure}
\resizebox{\hsize}{!}
          {\includegraphics[scale=1,clip]{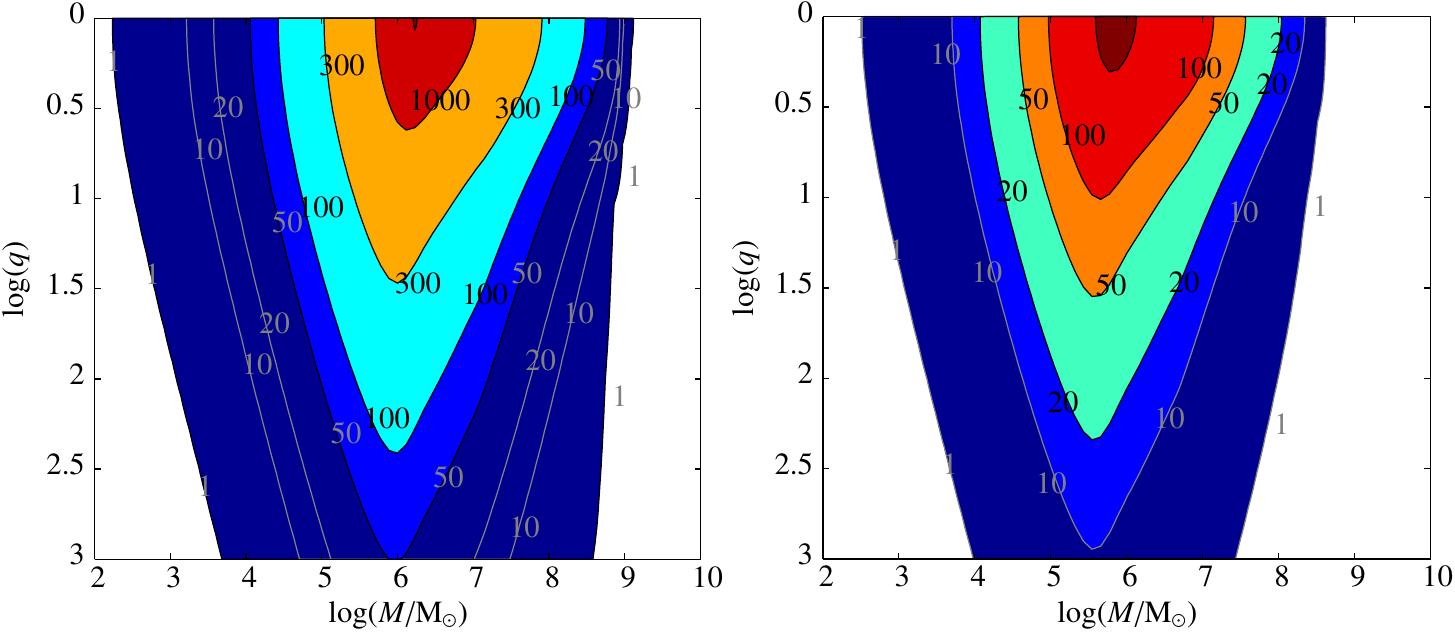}}
\caption
   {The figure shows constant-contour levels of the sky and
     polarisation angle averaged signal-to-noise ratio (SNR) for
     non-spinning binaries, at cosmological redshift $z = 1$ (left
     panel) and $z = 4$ (right panel) in the $M$--$q$ plane. Here
     $M$ is the total mass of the binary in the source rest frame, and
     $q$ is the mass ratio. The SNR is computed from the full non
     spinning PhenomC waveform inclusive of inspiral, merger and
     ring-down, as in figure \ref{fig.SNRmz}.}
\label{fig.SNRmqz14}
\end{figure}

Figure \ref{fig.SNRmz} shows eLISA SNRs for equal mass, non-spinning
coalescing binaries. Here we use PhenomC waveforms and we compute the
SNR as a function of the rest-frame total binary mass $M$ and of the
redshift $z$, averaging over all possible source sky locations and
wave polarisation, for two-year observations.  The plot highlights the
exquisite capabilities of the instrument in covering almost all the
mass-redshift parameter space relevant to massive black hole
astrophysics.  It is of importance to emphasise that current
electromagnetic observations are probing only the tip of the massive
black hole distribution in the Universe. Our current knowledge of
massive black holes is bound to instrument flux limits, probing only
the mass range
\ensuremath{10^{7}}\,\text{\ensuremath{\mathrm{M}_\odot}} --
\ensuremath{10^{9}}\,\text{\ensuremath{\mathrm{M}_\odot}} at
$0\lesssim z\lesssim 7$. Conversely, eLISA will be able to detect the
gravitational waves emitted by sources with total mass (in the source
rest frame) as small as
\ensuremath{10^{4}}\,\text{\ensuremath{\mathrm{M}_\odot}} at
cosmological distances inaccessible to any other astrophysical probe.
A binary with total mass in the interval
\ensuremath{10^{4}}\,\text{\ensuremath{\mathrm{M}_\odot}} --
\ensuremath{10^{7}}\,\text{\ensuremath{\mathrm{M}_\odot}} can be
detected out to a redshift as remote as $z\sim 20$ with a
$\text{SNR}\ge 10$.  By contrast, a binary as massive as a few
\ensuremath{10^{8}}\,\text{\ensuremath{\mathrm{M}_\odot}} can be
detected with high SNR in our local Universe ($z\lesssim 1$).
Binaries with total mass between
\ensuremath{10^{5}}\,\text{\ensuremath{\mathrm{M}_\odot}} --
\ensuremath{10^{7}}\,\text{\ensuremath{\mathrm{M}_\odot}} can be
detected with a $\text{SNR}\gtrsim100$, between $0\lesssim z\lesssim
5.$ These intervals in mass and redshift can be considered as optimal
for a deep and extensive census of the black hole population in the
Universe.

Figure \ref{fig.SNRmqz14} shows constant-contour levels of the SNR
expected from binaries with different mass ratios $q$ (defined as
$q=m_2/m_1$, where $m_2$ is the mass of the less massive black hole in
the binary) located at redshift $z=1$ and $z=4$.  The plots show the
SNR reduction that occurs with decreasing $q$, as unequal mass
binaries have lower strain amplitudes than equal mass binaries.  They
also show how SNR decreases with increasing redshift, and thus with
increasing luminosity distance.  Notice however that even at $z=4$,
binaries in the mass range
\ensuremath{10^{5}}\,\text{\ensuremath{\mathrm{M}_\odot}} --
\ensuremath{10^{7}}\,\text{\ensuremath{\mathrm{M}_\odot}} with mass
ratio $q\lesssim\ensuremath{10^{-1}}$ can be detected with
$\text{SNR}>20$.

\subsubsection*{Parameter estimation}
\label{sec.parameter-estimation}

\begin{figure}
  \centering
  \resizebox{\hsize}{!}
	{\includegraphics[scale=1,clip]{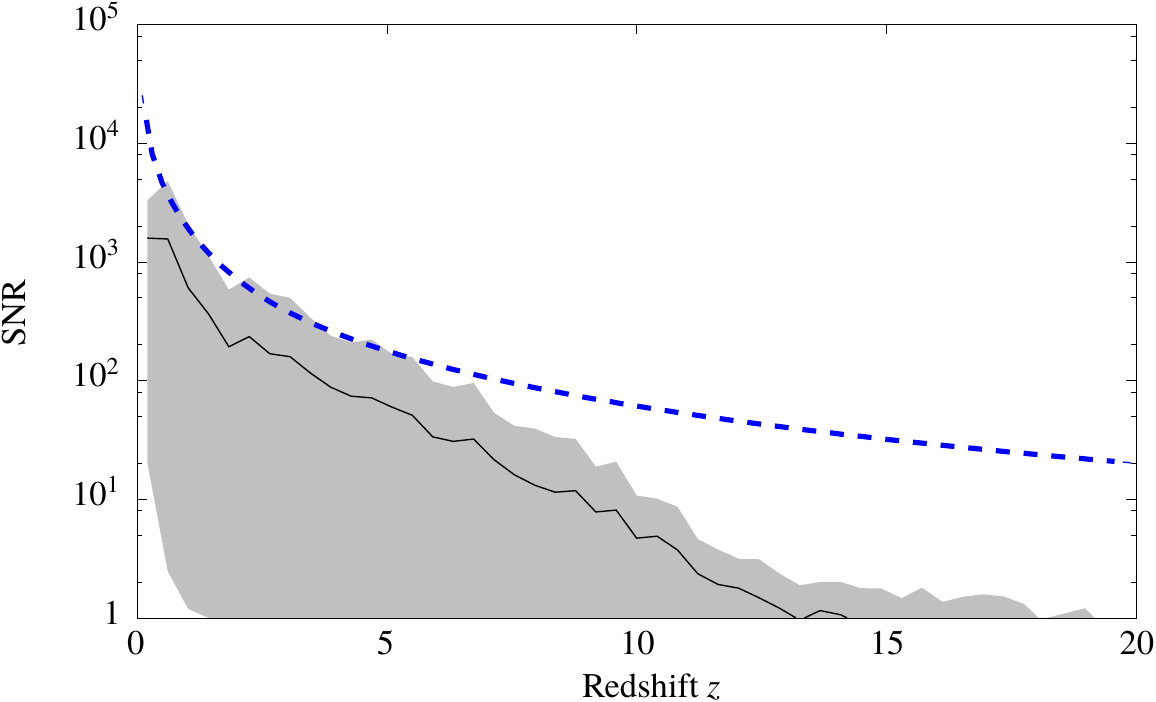} }
  \caption{SNR distribution as a function of cosmological redshift,
    computed using the inspiral, merger and ring-down waveform PhenomC
    for spinning binaries \citep{santamaria:2010PhRvD..82f4016S}.  The
    solid line corresponds to the mean value, and the grey area to the
    distribution corresponding to 10th and 90th percentile of SNR
    distribution. These results are based on a catalogue of 15360
    sources obtained combining 25 realisations of each of the four
    fiducial massive black hole evolution models. For reference, the 
    dashed-blue line indicates the sky-averaged SNR, for one year of 
    integration, computed for an equal mass coalescing binary of 
    \ensuremath{10^{6}}\,\text{\ensuremath{\mathrm{M}_\odot}}.
    }
  \label{fig.SNRzshadow}
\end{figure}

\begin{figure}
  \centering
  \resizebox{\hsize}{!}
	{\includegraphics[scale=1,clip]{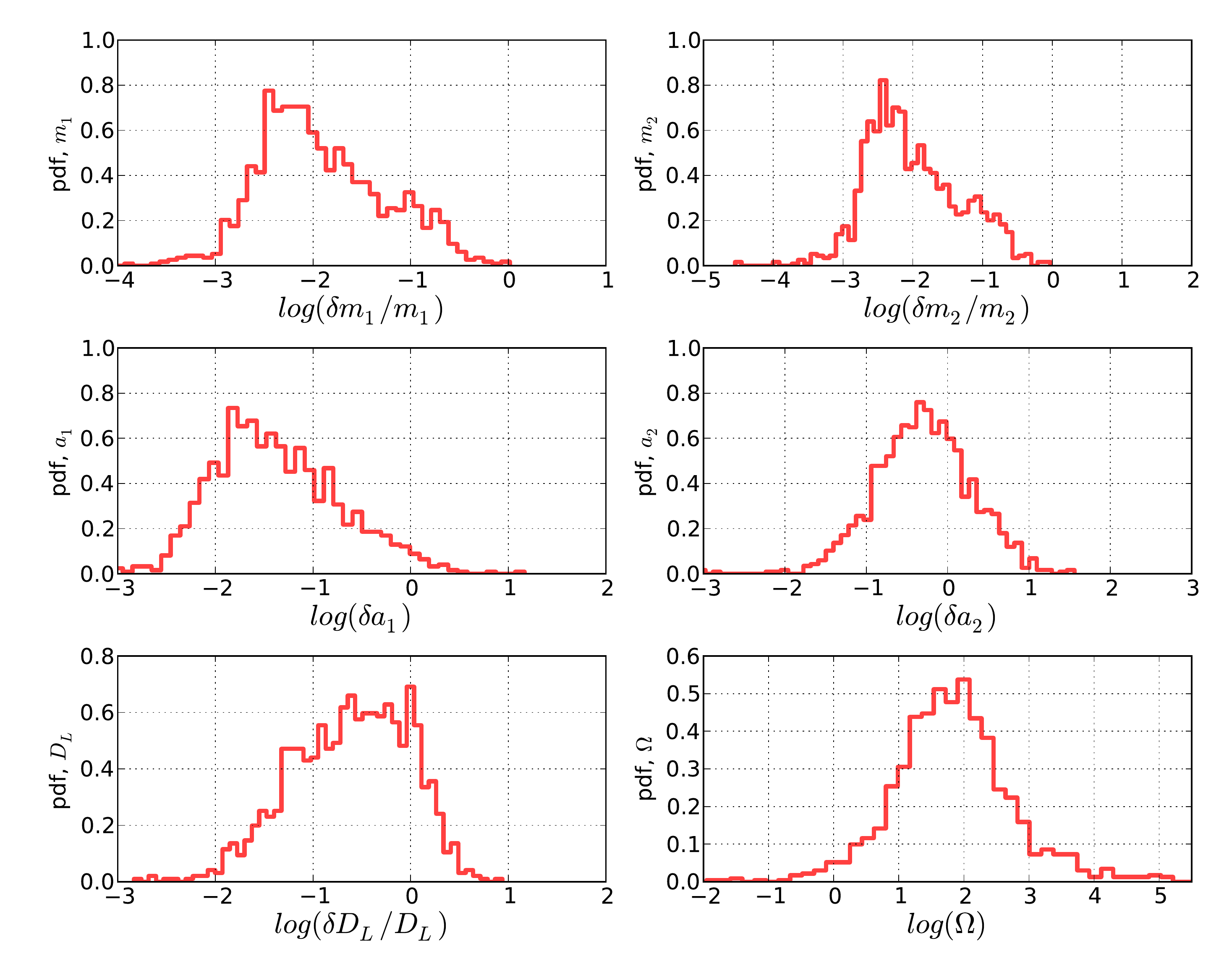} }
  \caption{Parameter estimation accuracy evaluated on a source
    catalogue, obtained combining ten Monte Carlo realisations of the
    coalescing massive black hole binary population, predicted by the
    four SE-SC-LE-LC models. The top panels show the distributions of
    the fractional errors in the estimation of the redshifted masses
    of the primary (left) and secondary (right) black hole. The middle
    panels show the absolute error distributions on the measurement of
    the primary (left) and secondary (right) black hole spin, while
    the bottom panels show the $D_L$ fractional error distribution on
    the luminosity distance $D_L$, and the sky location accuracy
    $\Delta\Omega$ (in \text{deg\ensuremath{^2}}).  Errors are evaluated considering
    full waveforms.}
  \label{fig.errors}
\end{figure}

\noindent Figures \ref{fig.SNRmz} and \ref{fig.SNRmqz14} describe the
detectability of single events, and for these individual events, it is
possible to extract information on the physical parameters of the
source.  Waveforms carry information on the redshifted mass (the mass
measured at the detector is $(1+z)$ times the mass at the source
location) and on the spin of the individual black holes prior to
coalescence. The measure of the mass and spin is of importance in
astrophysics.  Except for the Galactic centre
\citep{ghez:2005:sogc,gillessen:2009:mso}, the mass of astrophysical
black holes is estimated with uncertainties ranging from
\ensuremath{15}\,\text{\%} to a factor of about \ensuremath{2},
depending on the technique used and the type of source.  As far as
spin is concerned, its measure is only indirect, and it is derived
through modelling of the spectrum, or of the shape of emission lines,
mainly by fitting the skewed relativistic K$\alpha$ iron line. There
are few notable examples, but uncertainties are still large. By
contrast, spins leave a distinctive peculiar imprint in the
waveform.

In section \ref{sec.seed-black-holes} and \ref{sec.design-evol-mass}
we explored different routes to seed black hole formation and to their
subsequent assembly and growth through mergers and accretion
episodes. Different physically motivated assumptions lead to different
black hole evolution scenarios, and, as we highlighted above, the lack
of observational constraints allowed theoretical astrophysicists to
develop a large variety of massive black holes formation scenarios.
To assess the astrophysical impact of eLISA, we simulate observations
assuming a fiducial set of four cosmological black hole evolution
scenarios: SE refers to a model where the seeds have small (S) mass
about \ensuremath{100}\,\ensuremath{\mathrm{M}_\odot} (from Pop III
stars only) and accretion is coherent, i.e.\ resulting from extended
(E) accretion episodes: SC refers to a model where seeds are small but
accretion is chaotic (C), i.e., resulting from uncorrelated episodes;
and finally, LE and LC refer to models where the seed population is
heavy (L stands for large seeds of
\ensuremath{10^{5}}\,\ensuremath{\mathrm{M}_\odot}) and accretion is
extended and chaotic, respectively. The models are almost the same
used in previous studies by the LISA Parameter Estimation Task Force
\citep{arun:2009:petf}. The only difference is that in the extended
accretion model, spins are not assumed to be perfectly aligned to the
binary orbital angular momentum. The angles of misalignment relative
to the orbit are drawn randomly in the range 0 to 20 degree,
consistent with the finding of recent hydrodynamical simulations of
binaries forming in wet mergers
\citep{dotti:2010MNRAS.402..682D}. These models encompass a broad
range of plausible massive black hole evolution scenarios, and we use
them as a testbed for eLISA capabilities in a fiducial astrophysical
context.  Each massive black hole binary, coalescing at redshift $z$,
is characterised by the (rest frame) total mass $M=m_1+m_2$ (with
$m_1$ and $m_2$ the mass of the primary and secondary black hole),
mass ratio $q=m_2/m_1$, spin vectors $\bf{J}_1$ and $\bf{J}_2$
; spin magnitudes are denoted by $a_1$ and $a_2$. The orientations of
the spins are drawn as described above for the extended (E) accretion
models, and completely random for the chaotic (C) accretion models.
Here we generate several Monte Carlo realisations of each model and we
sum up all the generated sources in a single ``average'' catalogue (we
will consider models separately in the next section). Catalogues are
generated by selecting $M$, $q$, $z$, $a_1$, $a_2$ according to the
distribution predicted by the individual models, and by randomising
other source parameters (sky location, polarisation, inclination,
initial phase, coalescence time) according to the appropriate
distribution.

Figure \ref{fig.SNRzshadow} shows the average source SNR as a function
of the source redshift.  According to the simulated models, eLISA will
detect sources with $\text{SNR}\gtrsim 10$ out to $z\lesssim 10$.
Note that the astrophysical capabilities of eLISA are not limited by
the detector design, but by the population of astrophysical
sources. If there were a coalescing black hole binary of
\ensuremath{10^{4}}\,\ensuremath{\mathrm{M}_\odot} --
\ensuremath{10^{6}}\,\ensuremath{\mathrm{M}_\odot} out to redshift
$z\sim 20$, eLISA would reveal such a source.  Our models, and
accordingly our SNR distribution, do not have such an event.

Figure \ref{fig.errors} shows \emph{error distributions} in the source
parameter estimation, for all the events in the combined catalogue. We
used a hybrid approach of joining inspiral with PhenomC waveforms, as
described above, to evaluate uncertainties based on the Fisher matrix
approximation.

The figure illustrates the importance and power of including the full
waveform modelling, comprising all the stages of the binary merger, in
order to reach a high level of precision measurements. It is found
that individual black hole redshifted masses can be measured with
unprecedented precision, i.e.\ with an error of
\ensuremath{0.1}\,\text{\%} -- \ensuremath{1}\,\text{\%}, on both
components. No other astrophysical tool has the capability of reaching
a comparable accuracy.  As far as spins are concerned, the analysis
shows that the spin of the primary massive black hole can be measured
with an exquisite accuracy, to a \ensuremath{0.01} -- \ensuremath{0.1}
absolute uncertainty. This precision in the measure mirrors the fact
that the primary black hole leaves a bigger imprint in the
waveform. The measurement is more problematic for $a_2$ that can be
either determined to an accuracy of \ensuremath{0.1}, or remain
completely undetermined, depending on the source mass ratio and spin
amplitude. We emphasise that the spin measure is a neat, direct
measurement, that does not involve complex, often degenerate,
multi-parametric fits of high energy emission processes.

The source luminosity distance error $D_L$ has a wide spread, usually
ranging from $50\%$ to only few percent. Note that this is a direct
measurement of the luminosity distance to the source, which, again,
cannot be directly obtained (for cosmological objects) at any
comparable accuracy level by any other astrophysical means.  eLISA
is a full sky monitor, and the localisation of the source in the sky
is also encoded in the waveform pattern. Sky location accuracy is
typically estimated in the range \ensuremath{10} -- \ensuremath{1000} square degrees.

\subsection{Reconstructing the massive black hole cosmic history 
through eLISA observations}
\label{sec.astr-with-grav}

\noindent
eLISA will be an \emph{observatory}. The goal is not only to
detect sources, but also to extract valuable astrophysical information
from the observations.  While measurements for
individual systems are interesting and potentially very useful for
making strong-field tests of GR (see section \ref{sec.insp-merg-ringd}), it
is the properties of the set of massive black hole binary mergers that
are observed which will carry the most information for astrophysics.
Gravitational wave observations of multiple binary mergers may be used
together to learn about their formation and evolution through cosmic
history.

As any observatory, eLISA will observe a set of signals.  After signal
extraction and data analysis, these observations will provide a
catalogue of coalescing binaries, with measurements of several
properties of the sources (masses, mass ratio, spins, distances, etc)
and estimated errors.  The interesting questions to ask are the
following: \emph{can we discriminate among different massive black
  hole formation and evolution scenarios on the basis of gravitational
  wave observations alone?} Given a set of observed binary
coalescences, what information can be extracted about the underlying
population?  For example, will gravitational wave observations alone
tell us something about the mass spectrum of the seed black holes at
high redshift, that are inaccessible to conventional electromagnetic
observations, or about the poorly understood physics of accretion?
These questions were extensively tackled in \citep{sesana:2011:rmbh} in
the context of LISA. % We

\subsubsection*{Selection among a discrete set of models}
\label{sec.select-among-discr}

First we consider a discrete set of models. As argued above, in the 
general picture of massive black hole cosmic evolution, the population 
is shaped by the \emph{seeding process} and the \emph{accretion history}. 
The four models we study here are the SE, SC, LE, and LC models introduced
in the previous section \citep{arun:2009:petf}.
As a first step, we test here if eLISA observations will provide 
enough information to enable us to discriminate between those models, 
assuming that the Universe is well described by one of them.
 
Each model predicts a \emph{theoretical} distribution of coalescing
massive black hole binaries.  A given dataset $D$ of observed events
can be compared to a given model $A$ by computing the likelihood
$p(D|A)$ that the observed dataset $D$ is a realisation of model
$A$. When testing a dataset $D$ against a pair of models $A$ and $B$,
we assign probability $p_A=p(D|A)/(p(D|A)+p(D|B))$ to model $A$, and
probability $p_B=1-p_A$ to model $B$.  The probabilities $p_A$ and
$p_B$ are a measure of the relative confidence we have in model $A$
and $B$, given an observation $D$.  Once eLISA data is available, each
model comparison will yield this single number, $p_A$, which is our
confidence that model $A$ is correct. Since the eLISA data set is not
currently available, we can only work out how likely it is that we will
achieve a certain confidence with future eLISA observations.

We therefore generate \ensuremath{1000} independent realisations of
the population of coalescing massive black hole binaries in the
Universe predicted by each of the four models.  We then simulate
gravitational wave observations by producing datasets $D$ of observed
events (including measurement errors), which we statistically compare
to the theoretical models.  We consider only sources that are observed
with SNR larger than eight in the detector. We set a confidence
threshold of $0.95$, and we count what fraction of the
\ensuremath{1000} realisations of model $A$ yield a confidence
$p_A>0.95$ when compared to an alternative model $B$. We repeat this
procedure for every pair of models. For simplicity, in modelling gravitational wave
observations, we focus on circular, non-spinning binaries; therefore,
each coalescing black hole binary in the population is characterised
by only three intrinsic parameters -- redshift $z$, mass $M$, and mass
ratio $q$ -- and we compare the \emph{theoretical} trivariate
distribution in these parameters predicted by the models to the
observed values in the dataset $D$.  In terms of gravitational
waveform modelling, our analysis can therefore be considered extremely
conservative.

Results are shown in the left-hand panel of table \ref{tab.tab1}, for
a one year observation.  
The vast majority of the pair comparisons yield a
\ensuremath{95}\,\text{\%} confidence in the true model for almost all
the realisations --- we can perfectly discriminate among different
models. Similarly, we can always rule out the alternative (false)
model at a \ensuremath{95}\,\text{\%} confidence level. Noticeable
exceptions are the comparisons of models LE to LC and SE to SC, i.e.,
among models differing by accretion mode only. This is because the
accretion mode (efficient versus chaotic) particularly affects the
spin distribution of the coalescing systems, which was not considered
here. To extend this work, we added to our analysis the distribution
of the merger remnant spins $S_r$, and compared the \emph{theoretical}
distribution predicted by the models to the observed values (including
determination errors once again).  The spin of the remnant can be
reasonably determined in about \ensuremath{30}\,\text{\%} of the cases
only; nevertheless, by adding this information, we are able to almost
perfectly discriminate between the LE and LC and the SE and SC models,
as shown in the right hand panel of table \ref{tab.tab1}.

\begin{table}

\centering
\caption{Summary of all possible comparisons of the pure models. 
  Results are for one year of observation with eLISA. We take a fixed confidence  
  level of $p=0.95$. The numbers
  in the upper-right half of each table show the fraction of realisations in
  which the \emph{row} model will be chosen at more than this confidence level
  when the \emph{row} model is true. The
  numbers in the lower-left half of each table show the fraction of
  realisations in which the \emph{row} model \emph{cannot be ruled out} at that
  confidence level when the \emph{column} model is true. In the left table we
  consider the trivariate $M$, $q$, and $z$ distribution of observed events;
  in the right table we also include the observed distribution of remnant spins, $S_r$.
}
\begin{tabular}{@{}lccccm{2em}lcccc@{}}
\hline
&\multicolumn{4}{c}{Without spins}&&
&\multicolumn{4}{c}{With spins}\\

& SE & SC & LE & LC  & & & SE & SC & LE & LC\\
\hline
\hline(l)

SE & $\times$ & 0.48     & 0.99     & 0.99     && SE & $\times$ & 0.96     & 0.99     & 0.99     \\
SC & 0.53     & $\times$ & 1.00     & 1.00     && SC & 0.13     & $\times$ & 1.00     & 1.00     \\
LE & 0.01     & 0.01     & $\times$ & 0.79     && LE & 0.01     & 0.01     & $\times$ & 0.97     \\
LC & 0.02     & 0.02     & 0.22     & $\times$ && LC & 0.02     & 0.02     & 0.06     & $\times$ \\

\hline
\end{tabular}
\label{tab.tab1}
\end{table}

\subsubsection*{Constrains on parametric models}
\label{sec.constr-param-models}
\begin{figure}
  \centering
  \resizebox{\hsize}{!}
	{\includegraphics[scale=1,clip]{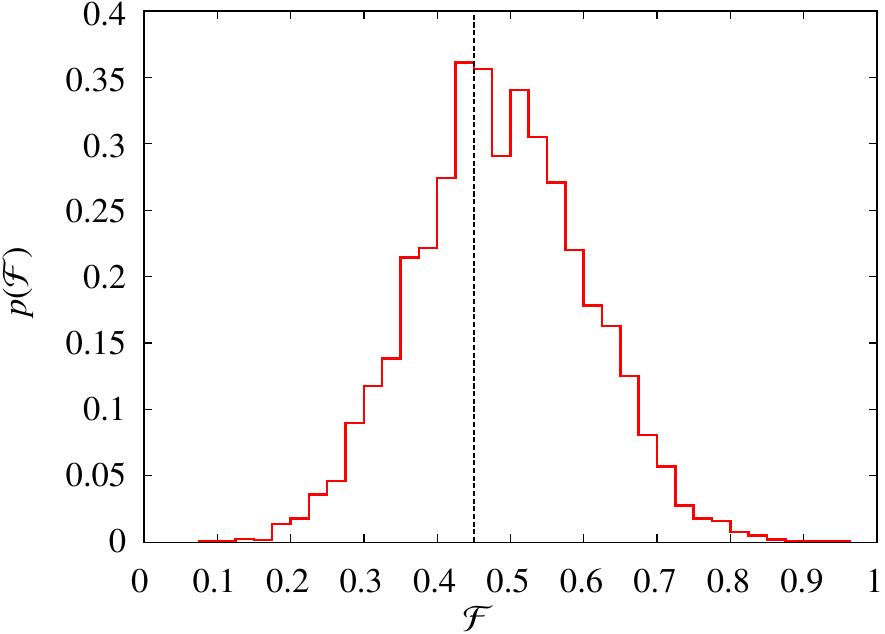} }
  \caption{Likelihood distribution of the mixing fraction ${\cal F}$,
    for a particular realisation of the model ${\cal F}{\mathrm
      [SE]}+(1-{\cal F}){\mathrm [LE]}$. The true mixing parameter,
    marked by a dashed vertical line, was ${\cal F}=0.45$.}
  \label{fig.bh_7}
\end{figure}

In the preceding section we demonstrated the potential of eLISA to
discriminate among a discrete set of ``pure'' models given a
priori.  However, the true massive black hole population in the
Universe will probably result from a mixing of known physical
processes, or even from a completely unexplored physical mechanism. A
meaningful way to study this problem is to construct parametric models
that depend on a set of key physical parameters, $\lambda_i$,
describing, for instance, the seed mass function and redshift
distribution, the accretion efficiency etc. and to investigate the
potential of eLISA to constrain these parameters. Such a
parametric family of models is not available at the moment, but we can
carry out a similar exercise by mixing two of our pure models, $A$ and
$B$, to produce a model in which the number of events of a particular
type is given by ${\cal F}$[A]$+(1-{\cal F})[B]$, where $[A]$ is the
number of events of that type predicted by model $A$, $[B]$ is the
corresponding number predicted by model $B$ and ${\cal F}$ is the
``mixing fraction''. In this case we generate datasets $D$ from a
mixed model with a certain unknown ${\cal F}$, and we estimate the
${\cal F}$ parameter by computing the likelihood that the data $D$ is
drawn from a mixed distribution, as a function of ${\cal F}$. A
specific example is shown in figure \ref{fig.bh_7}. Here the underlying model is
${\cal F}$[SE]$+(1-{\cal F})$[LE], with ${\cal F}=0.45$. eLISA
observations will allow us to pin-down the correct value of the mixing
parameter with an uncertainty of $\sim 0.1$. More complex examples of
multi-model mixing can be found in \citep{sesana:2011:rmbh}. Although highly
idealised, this exercise demonstrate the potential of eLISA
observations to constrain the physics and astrophysics of massive black hole along
their entire cosmic history, in a mass and redshift range inaccessible
to conventional electromagnetic observations.

\section{Extreme mass ratio inspirals and astrophysics of dense stellar systems}
\label{sec.emris-astr-dense}

\subsection{The Galactic Centre: a unique laboratory}
\label{sec.galact-centr-uniq}

The discovery, in the local Universe, of dark, massive objects lurking
at the centres of nearly all bright galaxies is one of the key
findings of modern-day astronomy, the most spectacular being the case
of the dark object in our own Galaxy
\citep{schoedel:2003:sdca,ghez:2003:fmsl,eisenhauer:2005:sinfoni,ghez:2005:sogc,ghez:2008:mdp,gillessen:2009:mso}. %
The nucleus of the Milky Way is one hundred times closer to Earth than
the nearest large external galaxy Andromeda, and one hundred thousand
times closer than the nearest QSO. Due to its proximity, it is the
only nucleus in the Universe that can be studied and imaged in great
detail.  The central few parsecs of the Milky Way house gas cloud
complexes in both neutral and hot phases, a dense luminous star
cluster, and a faint radio source SgrA$^*$ of extreme compactness
(\ensuremath{3} -- \ensuremath{10} light minutes across).
Observations, using diffraction-limited imaging and spectroscopy in
the near-infrared, have been able to probe the densest region of the
star cluster and measure the stellar dynamics of more than two hundred
stars within a few light days of the dynamic centre. The latter is
coincident, to within \ensuremath{0.1}\,\text{arcsec}, with the
compact radio source SgrA$^*$. The stellar velocities increase toward
SgrA$^*$ with a Kepler law, implying the presence of a $(4\pm
0.06\pm0.35)\times
\ensuremath{10^{6}}\,\text{\ensuremath{\mathrm{M}_\odot}}$ central
dark mass \citep{gillessen:2009:mso}. This technique has also led to
the discovery of nearly thirty young stars that orbit the innermost
region: the so called S0 (or S) stars.  These young stars are seen to
move on Keplerian orbits, with S02 (or S2) the showcase star orbiting
the putative black hole on a highly eccentric (\ensuremath{0.88})
orbit with a period of \ensuremath{15.9}\,\text{years}. The periapsis
of this orbit requires a lower limit on the density of the dark
point-like mass concentration of more than
\ensuremath{10^{13}}\,\text{\ensuremath{\mathrm{M}_\odot}
  pc\ensuremath{^{-3}}} \citep{maoz:1998:dc}. Additionally, a lower
limit of more than
\ensuremath{10^{18}}\,\text{\ensuremath{\mathrm{M}_\odot}
  pc\ensuremath{^{-3}}} can be inferred from the compactness of the
radio source \citep{genzel:2010RvMP...82.3121G}.  These limits provide
compelling evidence that the dark point-mass at SgrA$^*$ is a black
hole.  A cluster of dark stars of this mass and density (e.g. composed
of neutron stars, stellar black holes or sub-stellar entities such as
brown dwarfs, planets and rocks) can not remain in stable equilibrium
for longer than \ensuremath{10^{7}}\,\text{years}
\citep{maoz:1998:dc,genzel:2000MNRAS.317..348G,genzel:2006Natur.442..786G},
and the only remaining, albeit improbable, hypothesis is a
concentration of heavy bosons (a boson star,
\citep{colpi:1986PhRvL..57.2485C}) or of hyperlight black holes
\citep[$\ensuremath{\mathit{M}_\bullet}<\ensuremath{0.005}\,\text{\ensuremath{\mathrm{M}_\odot}}$,]{maoz:1998:dc}.
Overall, the measurements at the Galactic Centre are consistent with
a system composed of a massive black hole and an extended
close-to-isotropic star cluster, with the young S0 (or S) stars the
only population showing a collective rotation pattern in their orbits
\citep{genzel:2010RvMP...82.3121G}.

\subsection{Extreme Mass Ratio Inspirals in galactic nuclei}
\label{sec.extreme-mass-ratio-1}

Can we probe the nearest environs of a massive black hole other than
the Galactic Centre?  Massive black holes are surrounded by a variety
of stellar populations, and among them are compact stars (stellar
black holes, neutron stars and white dwarfs).  White dwarfs, neutron
stars, and stellar-mass black holes all share the property that they
reach the last stable orbit around the central massive black hole
before they are tidally disrupted.  
A compact star can either plunge directly toward the event horizon of
the massive black hole, or gradually spiral in and fall into the hole,
emitting gravitational waves.  The latter process is the one of
primary interest for eLISA.  Gravitational waves produced by inspirals
of stellar-mass compact objects into massive black holes are
observable by eLISA.  The mass of the compact object is typically of
the order of a few solar masses, while the mass of the central black
holes detectable by eLISA is from
\ensuremath{10^{4}}\,\ensuremath{\mathrm{M}_\odot} to
\ensuremath{10^{7}}\,\ensuremath{\mathrm{M}_\odot}. Because the mass
ratio for these binaries is typically around \ensuremath{10^{5}},
these sources are commonly referred to as EMRI.

The extreme mass ratio ensures that the inspiralling object
essentially acts as a test particle in the background space-time of
the central massive black hole. EMRI detections thus provide the
best means to probe the environment of an astrophysical black hole and
its stellar surroundings.  White dwarfs, neutron stars, and
stellar-mass black holes can all in principle lead to observable
EMRI signals. However, stellar-mass black holes, being more
massive, are expected to dominate the observed rate for eLISA, for
two reasons: mass segregation tends to concentrate the heavier compact
stars nearer the massive black hole, and black hole inspirals have higher signal-to-noise,
and so can be seen within a much larger volume.

Three different mechanisms for the production of EMRI have been
explored in the literature. The oldest and best-understood mechanism
is the diffusion of stars in angular-momentum space, due to two-body
scattering.  Compact stars in the inner \ensuremath{0.01}\,\text{pc} will sometimes
diffuse onto very high eccentricity orbits, such that gravitational
radiation will then shrink the orbit's semi-major axis and eventually
drive the compact star into the massive black hole.  Important
physical effects setting the overall rate for this mechanism are mass
segregation, which concentrates the more massive stellar-mass black
holes ($\simeq \ensuremath{10}\,\ensuremath{\mathrm{M}_\odot}$) close to the central black hole, and
resonant relaxation, which increases the rate of orbit diffusion in
phase-space \citep{hopman:2006:ems}: the orbits of stars close to a
massive black hole are nearly Keplerian ellipses, and these orbits
exert long term torques on each other, which can modify the angular
momentum distribution of the stars and enhance the rate of EMRI
formation \citep{guerkan:2007:rr}.  However, subtle relativistic
effects can reduce the estimated rates from relaxation processes
\citep{merritt:2011PhRvD..84d4024M}.  In addition to the two-body
scattering mechanism, other proposed channels for EMRI are tidal
disruption of binaries that pass close to the central black hole
\citep{miller:2005:bes}, and creation of massive stars (and their
rapid evolution into black holes) in the accretion discs surrounding
the central massive black hole \citep{levin:2007:ssmbh}.  Tidal break
up of incoming stellar binaries may already have been seen in the
Milky Way following the remarkable discovery of a number of so-called
hypervelocity stars observed escaping from our Galaxy
\citep[e.g.,][]{brown:2009:asd}. They are believed to be the outcome of
an ejection following the break-up of two bound stars by the tidal
field of SgrA$^*$. All these mechanisms give specific predictions on
the eccentricity and inclination of EMRI events that can be
extracted from the gravitational wave signal \citep{miller:2005:bes}.

When the orbiting object is close enough (within a few horizon radii
from the large black hole) gravitational radiation dominates energy
losses from the system, and the semimajor axis of the orbit shrinks.
Radiation is emitted over hundreds of thousand of orbits as the object
inspirals to the point where it is swallowed by the central massive
black hole.  Over short periods of time, the emitted radiation can be
thought of as a snapshot that contains detailed information about the
physical parameters of the binary.  The detection of the emitted
gravitational wave signal will give us very detailed information about
the orbit, the mass, and spin of the massive black hole as well as the mass
of the test object \citep{hopman:2009:emri,gair:2010:emri,preto:2010:ApJ...708L..42P}.

The measurement of even \emph{a few} EMRI will give astrophysicists a
totally new and different way of probing dense stellar systems
determining the mechanisms that shape stellar dynamics in the galactic
nuclei and will allow us to recover information on the emitting system
with a precision which is not only unprecedented in the history of
astrophysics, but beyond that of any other technique
\citep{amaro-soane:2007:tr,porter:2009GWN.....1....4P,babak:2010:mldc}.

\subsection{A probe of galactic dynamics}
\label{sec.probe-galact-dynam-1}

The centre-most part of the stellar spheroid, i.e.\ the \emph{galactic
  nucleus}, constitutes an extreme environment in terms of stellar
dynamics. With stellar densities higher than
\ensuremath{10^{6}}\,\text{\ensuremath{\mathrm{M}_\odot}
  pc\ensuremath{^{-3}}} and relative velocities exceeding
\ensuremath{100}\,\text{km/s}, collisional processes (i.e.\, collective
gravitational encounters among stars) are important in shaping the
density profiles of stars.  The mutual influence between the massive
black hole and the stellar system occurs thanks to various mechanisms.
Some are global, like the capture of stars via collisional relaxation,
or accretion of gas lost by stars through stellar evolution, or
adiabatic adaptation of stellar orbits to the increasing mass of the
black hole. Others involve a very close interaction, like the tidal
disruption of a star or the formation of an EMRI.

The distribution of stars around a massive black hole is a classical
problem in stellar dynamics
\citep{bahcall:1976ApJ...209..214B,bahcall:1977ApJ...216..883B}, and
of importance for EMRI is the distribution of stellar black holes.
Objects more massive than the average star, such as stellar black
holes, tend to segregate at the centre of the stellar distribution in
the attempt to reach, through long-distance gravitational encounters,
equipartition of kinetic energy.  A dense, strongly mass-segregated
cusp of stellar black hole is expected to form near a massive black
hole, and such a cusp plays a critical role in the generation of EMRI.
The problem of the presence of a dark cusp has been addressed, for the
Galactic Centre, by different authors, from a semi-analytical and
numerical standpoint
\citep{sigurdsson:1997:csm,miralda-escoude:2000:bhgc,freitag:2006JPhCS..54..252F,freitag:2006:srg,hopman:2006:rrn}.
A population of stellar black holes can leave an imprint on the
dynamics of the S0 (or S) stars at the Galactic Centre, inducing a
Newtonian retrograde precession on their orbits
\citep{mouawad:2005:wcgc}.  Current data are not sufficient to provide
evidence of such deviations from Keplerian orbits, so that the
existence of a population of stellar black holes is yet to be
confirmed \citep{gillessen:2009:mso,merritt:2009:eog}.

\subsection{A probe of the masses of stellar and massive black holes}
\label{sec.probe-masses-stellar}

It is very difficult to measure the mass of black holes, both of the
massive and stellar variety.  In the case of massive black holes,
methods based on following the innermost kinematics are difficult for
low-mass black holes in the range
\ensuremath{10^{5}}\,\ensuremath{\mathrm{M}_\odot} --
\ensuremath{10^{7}}\,\ensuremath{\mathrm{M}_\odot}.  These black holes
have low intrinsic luminosities even when they are active, making
detection hard. Performing dynamical measurements at these masses
through stellar kinematics requires extremely high spatial
resolution. Nowadays with adaptive optics we could optimistically hope
to get a handful of measurements through stellar kinematics about
\ensuremath{5}\,\text{kpc} away, although future
\ensuremath{20}\,\text{m} -- \ensuremath{30}\,\text{m} telescopes can
reach out to the Virgo cluster (\ensuremath{16.5}\,\text{Mpc}).
Exquisite gas-dynamical measurements are possible for only a handful
of active black holes using water megamaser spots in a Keplerian
circumnuclear disk \citep{kuo:2011ApJ...727...20K}. Still, the black
hole in the centre of our own galaxy lies in this range, and placing
constraints on the mass function of low-mass black holes has key
astrophysical implications. Observations show that the masses of black
holes correlate with the mass, luminosity and the stellar velocity
dispersion of the host \citep{gultekin:2009:ms}.  These correlations
imply that black holes evolve along with their hosts throughout cosmic
time. One unanswered question is whether this symbiosis extends down
to the lowest galaxy and black hole masses due to changes in the
accretion properties \citep{mathur:2005ApJ...633..688M}, dynamical
effects \citep{volonteri:2007:bhs}, or cosmic bias
\citep{volonteri:2009MNRAS.400.1911V}.  eLISA will discover the
population of massive black holes in galaxies smaller than the Milky
Way, that are difficult to access using other observational
techniques, and provide insights on the co-evolution of black holes
and their hosts.

Difficulties, albeit of a different nature, exist in measuring the
masses and mass distribution of stellar black holes.  Stellar black
holes are observed as accreting X-ray sources in binaries.  According
to stellar evolution, black holes result from the core collapse of
very massive stars, and their mass is predicted to be in excess of the
maximum mass of a neutron star, which is still not fully
constrained. Depending on the state of nuclear matter, this limit
varies from about \ensuremath{1.6}\,\ensuremath{\mathrm{M}_\odot} to
about \ensuremath{3}\,\ensuremath{\mathrm{M}_\odot}
\citep{shapiro:1986bhwd.book.....S}.  The maximum mass of a stellar
black hole is not constrained theoretically, and is known to depend
sensitively on the metallicity of the progenitor star.  The masses of
stellar black holes are inferred using Kepler's third law, or through
spectral analysis of the emission from the hole's accretion
disc. These techniques can be used only for black holes in a binary
system.  Current measurements of the black hole mass indicate a range
for stellar black holes from about
\ensuremath{5}\,\ensuremath{\mathrm{M}_\odot} up to
\ensuremath{20}\,\ensuremath{\mathrm{M}_\odot}, but uncertainties in
the estimate can be as large as a factor of two
\citep{orosz:2003IAUS..212..365O}.  In addition, stellar
black holes in interacting binaries are a very small and probably
strongly biased fraction of the total stellar black hole
population. They are formed from stars that have lost their hydrogen
mantle due to mass transfer (and thus formed in a different way than
the vast majority of stellar black holes.  eLISA will measure the mass of the
stellar black holes again with unprecedented precision providing
invaluable insight on the process of star formation in the dense
nuclei of galaxies, where conditions appear extreme.

\subsection{Detecting extreme mass ratio inspirals with eLISA}
\label{sec.detecting-emris-with}

\begin{figure}
\resizebox{\hsize}{!}
          {\includegraphics[scale=1,clip]{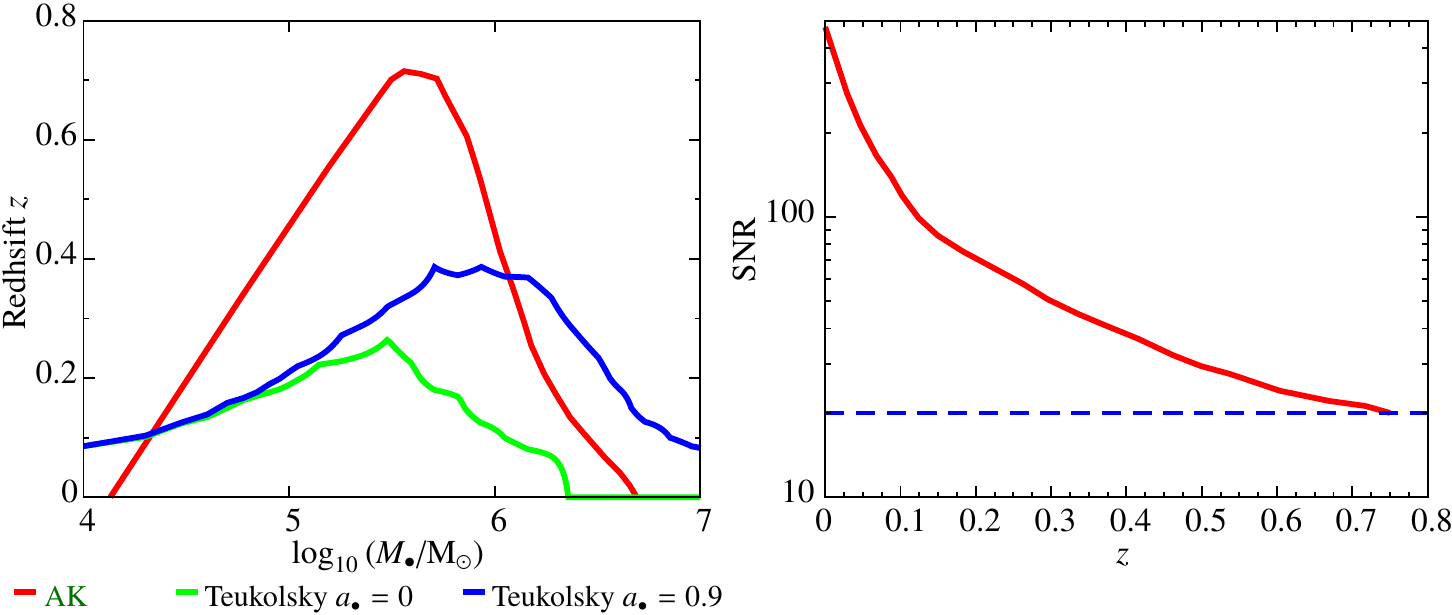}}
\caption
   { \emph{Left}: The maximum detectable redshift $z$ (or horizon) for
     EMRIs assuming a two year mission lifetime, and a
     $\text{SNR}=20$. The red curve is computed using the analytic
     waveform model and has a maximum redshift of $z\sim 0.7$ for
     black hole masses at the source rest-frame of \ensuremath{4
       \times 10^{5}}\,\ensuremath{\mathrm{M}_\odot}. The green and
     blue curves are sky-averaged horizons, computed using the
     Teukolsky waveform model, for two different values of the spin of
     the central black holes, $\ensuremath{a_\bullet} = 0$ (green) and
     $\ensuremath{a_\bullet} = 0.9$ (blue).\newline \emph{Right}: The
     distribution of maximum EMRI SNRs versus redshift for eLISA. The
     dashed horizontal line denotes the SNR threshold of 20.  }
\label{fig.emrihorizon}
\end{figure}

\begin{figure}
  \centering
  \resizebox{\hsize}{!}
	{\includegraphics[scale=1,clip]{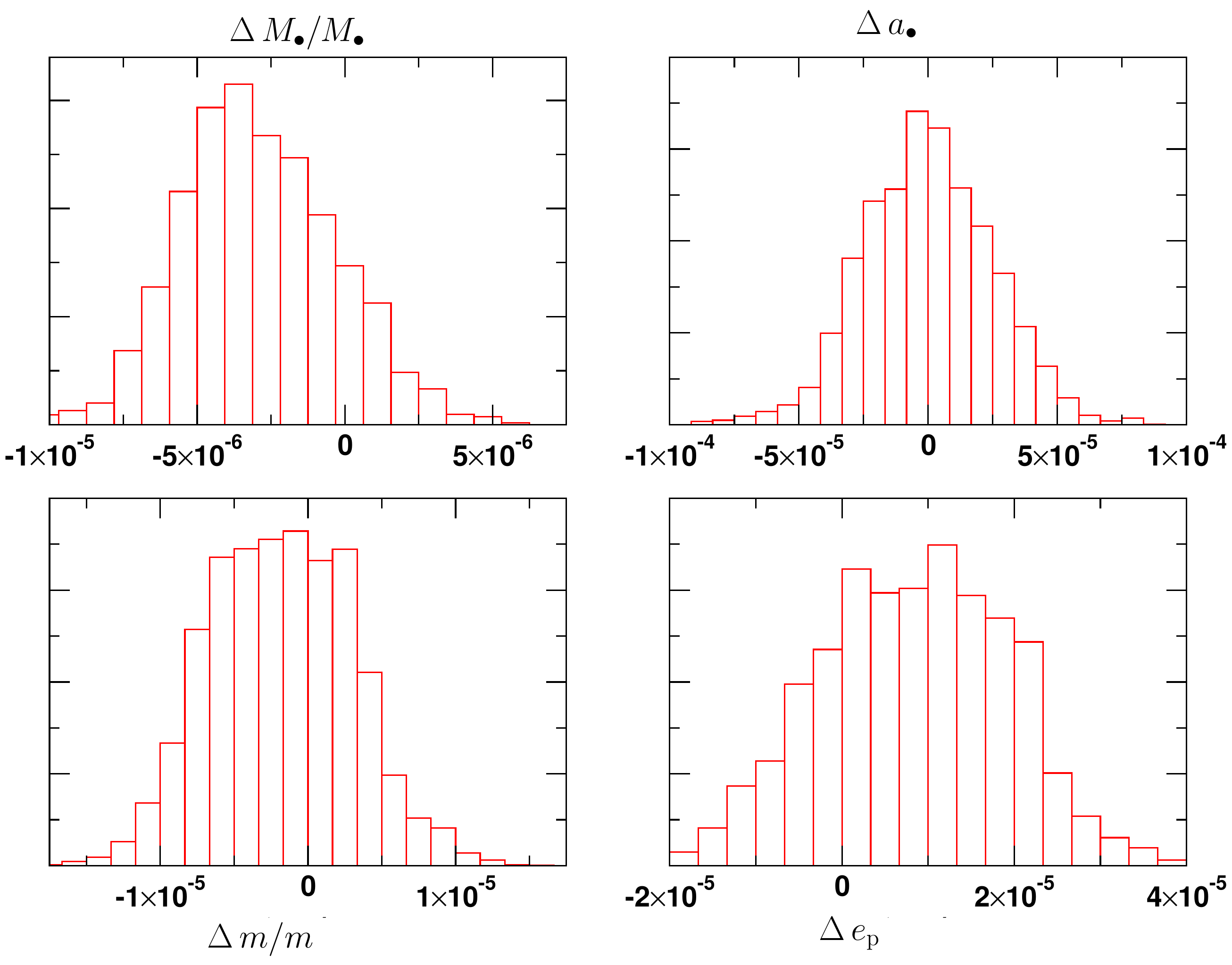} }
  \caption{The distribution of errors from a Markov Chain Monte Carlo
    analysis for a source at $z=0.55$ with an SNR of 25. The plot
    shows the error distributions for the central black holes mass
    $\ensuremath{\mathit{M}_\bullet}$ and spin
    $\ensuremath{a_\bullet}$, the mass of the compact object $m$ and
    the eccentricity at plunge $e_p$.}
  \label{fig.emrimcmc}
\end{figure}

EMRI are compact stars moving on relativistic orbits around a massive
black hole.  As the compact object spends most of its time in the
strong field regime, its orbit is very complex and difficult to model.
While not fully realistic, a set of phenomenological waveforms have
been developed \citep{barack:2004:lcs}, the AK waveforms, which fully
capture the complexity of the model.  These waveforms are defined by a
14 dimensional parameter set, of which the most physically relevant
are the masses of the central black hole and of the compact object,
$\ensuremath{\mathit{M}_\bullet}$ and $m$ respectively, the spin of
the massive black hole $\ensuremath{a_\bullet}$, the eccentricity of
the orbit at plunge, $e_p$, the sky position of the source with
respect to the detector, and the luminosity distance to the source,
$D_L$.  In addition to these approximate models, more accurate EMRI
waveform models have been computed using black hole perturbation
theory, in which the inspiraling object is regarded as a small
perturbation to the background spacetime of the large black hole. The
perturbation theory framework was first outlined in
\citep{teukolsky:1973:rbh} and gave rise to the Teukolsky
equation. However, the solution of this equation is computationally
expensive, and results have only recently been obtained for a
selection of generic orbits \citep{drasco:2006:gws}. Nonetheless,
results have been fully tabulated for certain restricted types of
orbit. For the calculations described here we will use data for
circular-equatorial orbits
\citep{finn:2000:gwc,gair:2009CQGra..26i4034G}. We can use both models
to compute the maximum detectable redshift, or the horizon for EMRI
detection, as a function of mass.

To calculate the detection limit of EMRI for eLISA using the AK
waveforms, we must perform a Monte Carlo simulation over the waveform
parameters.  We explore the mass range
$\ensuremath{10^{4}}\,\text{\ensuremath{\mathrm{M}_\odot}} \lesssim M
\lesssim \ensuremath{5 \times
  10^{6}}\,\text{\ensuremath{\mathrm{M}_\odot}}$.  As not much is
known about the distribution of spins or eccentricities for EMRI, we
consider uniform distributions for the spins in the range $-0.95 \leq
a \leq 0.95$, and for eccentricities at plunge in the interval $0.05
\leq e_p \leq 0.4$.  We fix the mass of the inspiraling body to
\ensuremath{10}\,\text{\ensuremath{\mathrm{M}_\odot}} to represent the
inspiral of a stellar black hole, as these are expected to dominate
the event rate \citep{gair:2004:ere}. The detection horizon for
neutron star and white dwarf inspirals is significantly less than for
black holes.  The final assumption required is to set a threshold of
detection.  While a SNR threshold of \ensuremath{30} was thought to be
justified in the past, advances in search algorithms have recently
demonstrated that EMRI with SNR about \ensuremath{20} is sufficient
for detection
\citep{cornish:2011CQGra..28i4016C,gair:2008:cmh,babak:2010:mldc},
allowing us to assume an SNR threshold of \ensuremath{20} in this
analysis. Assuming a mission lifetime $T$ of two years, and plunge times between
$\ensuremath{0}\,\text{yr} \leq t_p \leq \ensuremath{5}\,\text{yr}$, a
large scale Monte Carlo simulation was run over all 14 parameters.  In
figure \ref{fig.emrihorizon} (left) we plot the maximum detectable
redshift $z$ (also referred to as horizon) as a function of intrinsic
mass of the massive black hole.
Systems with intrinsic mass in the range from
$\ensuremath{10^{4}}\,\text{\ensuremath{\mathrm{M}_\odot}} \leq M \leq
\ensuremath{5 \times 10^{6}}\,\text{\ensuremath{\mathrm{M}_\odot}}$
are detectable in the local Universe at redshift of $z \lesssim 0.1$,
while systems in the range from
$\ensuremath{10^{5}}\,\text{\ensuremath{\mathrm{M}_\odot}} \leq M \leq
\ensuremath{10^{6}}\,\text{\ensuremath{\mathrm{M}_\odot}}$ should be
detectable by eLISA to $z\sim 0.7$, corresponding to a co-moving
volume of about \ensuremath{70}\,\text{Gpc\ensuremath{^{3}}}.

Figure \ref{fig.emrihorizon} (left) also shows the maximum detectable
redshift $z$ as a function of the mass of the central massive black
hole, computed for circular-equatorial inspirals using the Teukolsky
equation for the same masses of the inspiralling compact object and
massive black holes. 
This curve shows the sky-averaged horizon, i.e., the maximum redshift
at which the SNR averaged over inclinations and orientations of the
EMRI system reaches the threshold value of $20$. Tabulated Teukolsky
results are only available for selected values of the spin of the
central black hole, so we show the horizon assuming all the central
black holes have either spin $\ensuremath{a_\bullet}=0$ or
$\ensuremath{a_\bullet}=0.9$. 
The Teukolsky horizon appears significantly lower than the AK horizon,
but this is a result of the sky-averaging approximation -- the sky
averaged SNR is expected to be a factor of about $2.5$ lower than the
SNR of an ``optimally oriented'' binary. The AK horizon was computed
using a Monte Carlo simulation over orientations and sky locations for
the source and will therefore approach the value for an
optimally-oriented binary. 
The difference between the sky-averaged Teukolsky horizon and the AK
horizon is therefore consistent with the expected level of
difference. The maximum horizon for the Teukolsky curves is at a
similar value for the mass of the central black hole as the AK results
-- somewhat lower for $\ensuremath{a_\bullet}=0$ and higher for
$\ensuremath{a_\bullet}=0.9$, as we would expect since inspirals into
more rapidly spinning black holes emit radiation at higher
frequencies, which shifts the peak sensitivity to higher masses. For
the same reason, we see that the eLISA horizon is at a higher redshift
for more rapidly spinning central black holes.

In figure \ref{fig.emrihorizon} (right) we plot the distribution of
maximum SNR as a function of redshift for the Monte Carlo simulation
performed using the AK waveforms. A nearby EMRI will be detectable
with SNR of many tens, with SNRs of 30 being available out to $z =
0.5$. EMRI can be detected with an SNR of 20 up to $z\simeq 0.7$, up
to a volume of about \ensuremath{70}\,\text{Gpc\ensuremath{^{3}}},
encompassing the last 6 billion years of the Universe.

EMRIs are the most complex sources to model and to search for.
However, if they can be detected, this complexity will allow us to
estimate the parameters of the system with great accuracy
\citep{cornish:2011CQGra..28i4016C,gair:2008:cmh,babak:2010:mldc}.
For an EMRI detected with a certain SNR, the parameter estimation
accuracy does not strongly depend on the detector configuration, since
any detected EMRI will be observed for many waveform cycles. For this
reason the parameter estimation accuracy achievable with eLISA is
essentially the same as reported in the published LISA literature
\citep{barack:2004:lcs,huerta:2009PhRvD..79h4021H}. For any EMRI
observed with SNR above the detection threshold of \ensuremath{20}, we
expect to measure the mass $\ensuremath{\mathit{M}_\bullet}$ and spin
$\ensuremath{a_\bullet}$ of the central massive black hole with a
precision to better than a part in \ensuremath{10^{4}}.  This is
illustrated in Figure~\ref{fig.emrimcmc}, that shows the results from
a Markov Chain Monte Carlo analysis \citep{cornish:2006:mes} of a
source at $z=0.55$ with $\text{SNR}=25$.  The plots show the
distribution of errors for a particular source that would be recovered
by analysing the data from the detector.  Results are shown for the
mass
$\ensuremath{\mathit{M}_\bullet}/\text{\ensuremath{\mathrm{M}_\odot}}$
and spin $\ensuremath{a_\bullet}$ of central black hole, the mass
$m/\ensuremath{\mathit{M}_\bullet}$ of the stellar black hole, and the
eccentricity at plunge $e_p$.  Our analysis also shows that the
luminosity distance $D_L$ to the source is determined with an accuracy
of less than \ensuremath{1}\,\text{\%} and the source sky location can
be determined to around \ensuremath{0.2} square degrees.  While the
SNR is quite low for this source, the accuracy in the estimation of
parameters is very good.

\subsection{Estimating the event rates of extreme mass ratio inspirals for eLISA}
\label{sec.estim-event-rates}

We can use the horizon distances described in the preceding section to
compute the likely number of EMRI events that eLISA will detect, if we
make further assumptions about the EMRI occurring in the
Universe. This depends on the black hole population and on the rate at
which EMRI occur around massive black holes with particular
properties. The latter is poorly known, and we will use results from
\citeauthor{hopman:2009:emri} (\citeyear{hopman:2009:emri}) and
\citeauthor{amaro-seoane:2011CQGra..28i4017A}
(\citeyear{amaro-seoane:2011CQGra..28i4017A}) for the rate of
inspirals involving black holes.  The rate $\Gamma_\bullet$ is found
to scale with the central black hole mass,
$\ensuremath{\mathit{M}_\bullet}$, as $ \Gamma_\bullet \sim
\ensuremath{400}\,\text{Gyr\ensuremath{^{-1}}} \left(
\ensuremath{\mathit{M}_\bullet}/\ensuremath{3 \times
  10^{6}}\,\text{\ensuremath{\mathrm{M}_\odot}} \right)^{-0.19}$.
We do not consider neutron star and white dwarf inspirals in these
rate estimates as the expected number of detections with eLISA is less
than one in both cases, due to the considerably reduced horizon
distance for these events. We therefore fix the mass of the
inspiraling body at
\ensuremath{10}\,\text{\ensuremath{\mathrm{M}_\odot}}, as in the
previous section.

To model the black hole population, we take the mass function of black
holes to be in the intrinsic mass range
$\ensuremath{10^{4}}\,\ensuremath{\mathrm{M}_\odot} \lesssim
\ensuremath{\mathit{M}_\bullet} \lesssim \ensuremath{5 \times
  10^{6}}\,\ensuremath{\mathrm{M}_\odot}$.  Using the assumption that
there is no evolution in the black hole mass function, we sampled
sources from a uniform distribution in co-moving volume.  These
assumptions are consistent with the mass function derived from the
observed galaxy luminosity function using the
$\ensuremath{\mathit{M}_\bullet}-\sigma$ relation, and excluding Sc-Sd
galaxies
\citep{Aller:2002,gair:2004:ere,gair:2009CQGra..26i4034G}. For the
results using the AK waveform model, we choose the spin of the central
object uniformly in the range $0 \leq a \leq 0.95$, the eccentricity
of the orbit at plunge uniformly in the range $0.05\leq e_p \leq 0.4$
and all angles to be uniform or uniform in cosine as appropriate. For
the Teukolsky based results we do not need to specify the angles, as we
use a sky and orientation averaged sensitivity, and we do not specify
the eccentricity or inclination as the orbits are all circular and
equatorial (although we assume equal numbers of prograde and
retrograde inspirals). As before, the Teukolsky results are available
for fixed values of the spin only, so we estimate the event rate
assuming that all the black holes have spin \ensuremath{0},
\ensuremath{0.5} or \ensuremath{0.9}.

It is important also to correctly randomise over the plunge time of
the EMRI.  For the AK calculation, we choose the plunge time uniformly
in $\ensuremath{0}\,\text{yr} \leq t_p \leq
\ensuremath{5}\,\text{yr}$, with time measured relative to the start
of the eLISA observation and assuming an eLISA lifetime of 2
years. Although sufficiently nearby events with plunge times greater
than 5 years in principle could be detected, it was found that such
events contribute less than one event to the total event rate. For the
Teukolsky calculation, we evaluated the observable lifetime for every
event, which is the amount of time during the inspiral that eLISA
could start to observe that will allow sufficient SNR to be
accumulated over the mission lifetime to allow a detection
\citep{gair:2009CQGra..26i4034G}.

In table \ref{tab.EMRIratetab} we give the results of this calculation
for different waveform models and black hole spins. The predicted
number of events depends on the assumptions about the waveform model
and the spin of the black holes, but it is in the range of
\ensuremath{25} -- \ensuremath{50} events in two years.  The number of
events predicted for the AK model is higher because the presence of
eccentricity in the system tends to increase the amount of energy
radiated in the eLISA band. The analytic kludge estimates include
randomisation over the black hole spin, orbital eccentricity and
inclination, so the true detection rate is likely to be closer to this
number, although this depends on the unknown astrophysical
distribution of EMRI parameters.  Even with as few as 10 events,
\citeauthor{gair:2010:emri} (\citeyear{gair:2010:emri}) show that the
slope of the mass function of massive black holes in the mass range
\ensuremath{10^{4}}\,\ensuremath{\mathrm{M}_\odot} --
\ensuremath{10^{6}}\,\ensuremath{\mathrm{M}_\odot} can be determined
to a precision of about \ensuremath{0.3}, which is the current level
of observational uncertainty.

\begin{table}
  \centering
  \caption{Estimated number of EMRI events detectable by
    eLISA. 
    The first three columns shows the results computed using the
    Teukolsky waveform model, assuming all black holes have fixed spin
    of $0$, $0.5$ or $0.9$. The last column shows results computed
    using the analytic kludge waveform model.}
\label{tab.EMRIratetab}

\begin{tabular}{@{}*{5}{c}}
\hline
&\multicolumn{3}{c}{Teukolsky}&\\
\multicolumn{1}{c}{Waveform model}&
\multicolumn{3}{c}{with black hole spin}&
\multicolumn{1}{c}{Analytic Kludge}\\
&$\ensuremath{a_\bullet} = 0$&$\ensuremath{a_\bullet} = 0.5$&$\ensuremath{a_\bullet} = 0.9$&\\
\hline
Number of events&30&35&55&50\\
\hline
\end{tabular}

\end{table}

\subsection{Black hole coalescence events in star clusters}
\label{sec.low-mass-massive}

In closing this section on astrophysical black holes, we explore
briefly the possibility that an instrument like eLISA will detect
coalescences between low-mass massive black holes (called also
intermediate-mass black holes) in the mass range
\ensuremath{10^{2}}\,\ensuremath{\mathrm{M}_\odot} --
\ensuremath{10^{4}}\,\ensuremath{\mathrm{M}_\odot}. These coalescence
events do \emph{not} result from the assembly of dark matter halos,
but rather they are \emph{local} coalescence events occurring in star
clusters under extreme (and largely unexplored) astrophysical
conditions.  Given the tiny radius of gravitational influence (about
\ensuremath{0.01}\,\text{pc}) of such light black holes on the
surrounding dense stellar environment, their detection is extremely
difficult, and their existence has never been confirmed, though
evidence has been claimed in a number of globular clusters
\citep[see][ and references
  therein]{miller:2004:imbh,miller:2009CQGra..26i4031M}.

An intermediate-mass black hole may form in a young cluster if the
most massive stars sink to the cluster's centre due to mass
segregation before they evolve and explode. There, they start to
physically collide. The most massive star gains more and more mass and
forms a runaway star that may collapse to form an intermediate-mass black
hole
\citep{portegieszwart:2000ApJ...528L..17P,guerkan:2004ApJ...604..632G,portegieszwart:2004Natur.428..724P,freitag:2006JPhCS..54..252F,freitag:2006MNRAS.368..121F}.
Intermediate-mass black holes can be observed by eLISA via the
inspiral of compact objects such as stellar mass black holes
\citep{konstatantinidis:2011arXiv1108.5175K}, or when they form a
binary. The formation of an intermediate-mass black hole binary can
occur via star-cluster star-cluster collisions like those found in the
Antenn{\ae} galaxy
\citep{amaro-seoane:2006ApJ...653L..53A,amaro-seoane:2010:mbhb}, or
via formation in situ \citep{guerkan:2006ApJ...640L..39G}.
   
eLISA will observe intermediate-mass black hole binaries with
$\text{SNR}>10$ out to a few Gpc
\citep{santamaria:2010PhRvD..82f4016S}, and it will detect
stellar-mass black holes plunging into an intermediate-mass black hole
in a massive star cluster in the local Universe
\citep{konstatantinidis:2011arXiv1108.5175K}. Event rates are hard to
predict due to large uncertainties in the dynamical formation of
intermediate mass black holes in star clusters, but we may observe as
many as a few events per year
\citep{amaro-seoane:2006ApJ...653L..53A}. The detection of even a
single event would have great importance for astrophysics, probing
the existence of black holes in this unexplored mass range and
shedding light on the complex dynamics of dense stellar clusters.

\section{Confronting General Relativity with
 Precision Measurements \\ of Strong Gravity}
\label{sec.black-hole-physics}

\subsection{Setting the stage}
\label{sec.setting-stage}

\noindent
GR is a theory of gravity in which gravitational fields are manifested
as curvature of spacetime.  GR has no adjustable parameters other than
Newton's gravitational constant, and it makes solid, specific
predictions.  Any test can therefore potentially be fatal to its
viability, and any failure of GR can point the way to new
physics. Confronting GR with experimental measurements, particularly
in the strong gravitational field regime, is therefore an essential
enterprise. In fact, despite its great successes, we know that GR
cannot be the final word on gravity, since it is a classical theory
that necessarily breaks down at the Planck scale.  As yet there is no
complete, quantum theory of gravity, and gravitation is not unified
with the other fundamental forces. Under such a premise, several
stress tests of GR have been proposed, each of them potentially fatal
to the theory, however all of them involve low energies and
length-scales much larger than the Planck scale.

Although so far GR has passed all the tests to which it has been
subjected \citep{will:2006:gre}, most of these tests were set in the
 weak-field regime,
in which the parameter $\epsilon = v^2/c^2 \sim GM/(Rc^2)$ is much
smaller than one.  Here $v$ is the typical velocity of the orbiting
bodies, $M$ their total mass, and $R$ their typical separation.

For the tests of GR that have been carried out in our Solar System, expected second-order GR
corrections to the Newtonian dynamics are of the order $\epsilon \sim
10^{-6}-10^{-8}$ , and so to date it has been sufficient to expand GR
equations to the first Post-Newtonian (PN) order.  Solar System tests
are completely consistent with GR to this order of approximation.

Binary pulsars, which are essentially very stable and accurate clocks
with typical orbital velocities $v/c \sim 10^{-3}$ ($\epsilon\sim
10^{-6}$), are excellent laboratories for precision tests of GR
\citep{lorimer:2008:bmp}.  Current observations of several binary
pulsars are perfectly consistent with the GR predictions, with orbits
again calculated to the first PN order. Observations of the first
binary pulsar to be discovered, PSR~1913+16,
also provided the first astrophysical evidence for gravitational
radiation, a 2.5-PN-order effect.  Loss of energy due to
gravitational-wave emission (radiation reaction) causes the binary
orbit to shrink slowly; its measured period derivative $\dot P$ agrees
with GR predictions to within \ensuremath{0.2}\,\text{\%}, consistent
with measurement error bars \citep{weisberg:2005:rbp}. Another
 double pulsar system, SR J0737-3039 A and B,
allows additional tests of GR that were not available prior to its
discovery \citep{kramer:2006:tgr}.  In that system, the orbital period
derivative is consistent with GR at the \ensuremath{0.3}\,\text{\%}
level, and the Shapiro delay agrees to within
\ensuremath{0.05}\,\text{\%} with the predictions of GR
\citep{kramer:2009:tr}. 

However, the gravitational fields responsible for the orbital motion
in known binary pulsars are not much stronger than those in the Solar
System: the semimajor axis of the orbit of
PSR~1913+16 is about
\ensuremath{1.4}\,\text{\ensuremath{\mathrm{R}_\odot}}. Such weak
fields limit the ability of binary pulsars to probe nonlinear GR
\emph{dynamics}. They do provide important tests of strong-field
\emph{static} gravity, as the redshift at the surface of a neutron
star is of order \ensuremath{0.2}.

\begin{figure}
  \centering
  \resizebox{\hsize}{!}
	%{\includegraphics[scale=1,clip]{strainM1e5z15-WDWD_SRDcopy_manual_self_standing} }
        {\includegraphics[scale=1,clip]{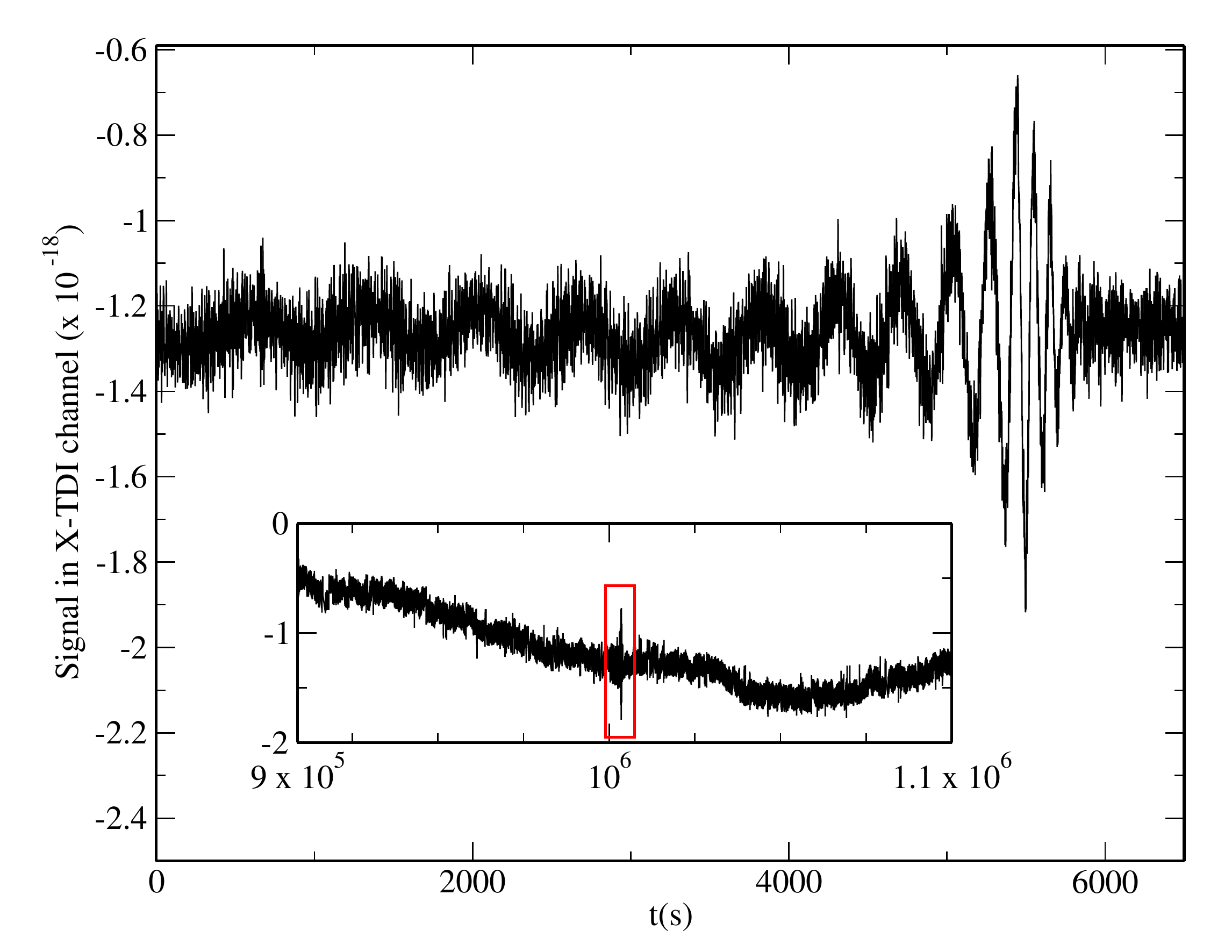} }
  \caption{Gravitational wave signal for the final few orbits, plunge,
    merger and ringdown of a coalescing binary. The total mass of
    system $M(1+z) = \ensuremath{2 \times
      10^{6}}\,\text{\ensuremath{\mathrm{M}_\odot}}$, mass ratio
    $m_1/m_2 = 2$, spin magnitudes $a_1 = 0.6\; a_2 = 0.55$,
    misalignment between spins and orbital angular momentum few
    degrees, the distance to the source $z=5$.  The inset shows the
    signal on a larger data span.}
  \label{fig.fig_4-1}
\end{figure}

eLISA observations of coalescing massive black hole binaries, or of
stellar-mass compact objects spiralling into massive black holes, will
allow us to confront GR with precision measurements of physical
regimes and phenomena that are not accessible through Solar System or
binary pulsar measurements.  The merger of comparable-mass black hole
binaries produces an enormously powerful burst of gravitational
radiation, which eLISA will be able to measure with amplitude SNR as
high as a few hundred, even at cosmological distances.

In the months prior to merger, eLISA will detect the gravitational
waves emitted during the binary inspiral; from that inspiral waveform,
the masses and spins of the two black holes can be determined to high
accuracy.  Given these physical parameters, numerical relativity will
predict very accurately the shape of the merger waveform, and this can
be compared directly with observations, providing an ideal test of
pure GR in a highly dynamical, strong-field regime.

Stellar-mass compact objects spiralling into massive black holes will
provide a qualitatively different test, but an equally exquisite one.
The compact object travels on a near-geodesic of the spacetime of the
massive black hole. As it spirals in, its emitted radiation
effectively maps out the spacetime surrounding the massive black
hole. Because the inspiralling body is so small compared to the
central black hole, the inspiral time is long and eLISA will typically
be able to observe of order \ensuremath{10^{5}} cycles of inspiral waveform, all of
which are emitted as the compact object spirals from 10 horizon radii
down to a few horizon radii.  Encoded in these waves is an extremely
high precision map of the spacetime metric just outside the central
black hole.  Better opportunities than these for confronting GR with
actual strong-field observations could hardly be hoped for.  

 The LIGO and
Virgo detectors should come online
around 2015, and their sensitivity is large enough that they should
routinely observe stellar mass black hole coalescences, where the binary
components are of roughly comparable mass.  However, even the
brightest black hole mergers that LIGO and Virgo should observe will
still have an amplitude SNR about 10 to 100 times smaller than the
brightest massive black hole coalescences that eLISA will observe.  The
precision with which eLISA can measure the merger and ringdown waveforms
will correspondingly be better by the same factor when compared to
ground-based detectors.  The situation is similar for 
the EMRI described in the previous section: while
ground-based detectors may detect binaries with mass ratios of about
\ensuremath{10^{-2}} (e.g., a neutron star spiralling into a \ensuremath{100}\,\ensuremath{\mathrm{M}_\odot}
black hole), in observations lasting approximately \ensuremath{10^{2}} -- \ensuremath{10^{3}}
cycles, the precision with which the spacetime can be mapped in such
cases is at least two orders of magnitude worse than what is
achievable with eLISA's EMRI sources.  Thus eLISA will test our
understanding of gravity in the most extreme conditions of strong and
dynamical fields, and with a precision that is two orders of magnitude
better than that achievable from the ground.

GR has been extraordinarily fruitful in correctly predicting new
physical effects, including gravitational lensing, the gravitational
redshift, black holes and gravitational waves.  GR also provided
the overall framework for modern cosmology, including the expansion of
the Universe. However, our current understanding of the nonlinear,
strong gravity regime of GR is quite limited. Exploring
gravitational fields in the dynamical, strong-field regime could
reveal new objects that are unexpected, but perfectly consistent with
GR, or even show violations of GR. 

The best opportunity for making 
such discoveries is with an instrument of high sensitivity. Ground-based 
detectors like LIGO and Virgo will almost certainly always have to detect 
signals by extracting them from deep in the instrumental noise, and they 
will therefore depend on prior predictions of waveforms. eLISA, on the other 
hand, will have enough sensitivity that many signals will show themselves
well above noise; unexpected signals are much easier to recognize with 
such an instrument.

\subsection{Testing strong-field gravity: The inspiral, merger, and ringdown of massive black hole binaries}
\label{sec.insp-merg-ringd}

\nopagebreak

\noindent

eLISA's strongest sources are expected to be coalescing black hole
binaries where the components have roughly comparable masses, $0.1 <
m_2/m_1 < 1$.  Their signal at coalescence will be visible by eye in
the data stream, standing out well above the noise, as illustrated in
figure \ref{fig.fig_4-1}.

As discussed in section \ref{sec.massive-black-hole}, black hole
binary coalescence can be schematically decomposed into three stages
(inspiral, merger, and ringdown), all of which will be observable by
eLISA for a typical source.  The inspiral stage is a relatively slow,
adiabatic process, well described by the analytic PN approximation.
The inspiral is followed by the dynamical merger of the two black
holes, that form a single, highly distorted black hole remnant.  Close
to merger, the black hole velocities approach $v/c \sim 1/3$ and the
PN approximation breaks down, so the waveform must be computed by
solving the full Einstein equations via advanced numerical techniques.
The distorted remnant black hole settles down into a stationary
rotating solution of Einstein's equations (a Kerr black hole) by
emitting gravitational radiation. This is the so called ``ringdown''
phase, where the gravitational wave signal is a superposition of
damped exponentials QNM, and therefore similar to the sound of a
ringing bell.

While numerical relativity is required to understand the gravitational
radiation emitted during merger, the post-merger evolution -- i.e.,
the black hole ``quasinormal ringing'' -- can be modelled using black
hole perturbation theory. The final outcome of the ringdown is the
Kerr geometry, with a stationary spacetime metric that is determined
uniquely by its mass and spin, as required by the black hole%
 ``no-hair'' theorem.

For equal-mass black hole binaries with total mass $M$ in the range
$\ensuremath{2 \times 10^{5}}\,\ensuremath{\mathrm{M}_\odot} < M(1+ z)
< \ensuremath{2 \times 10^{6}}\,\ensuremath{\mathrm{M}_\odot}$, where
$z$ is the cosmological redshift of the source, the inspiral SNR and
post-inspiral (merger plus ring-down) SNR are within an order of
magnitude of each other.  From a typical eLISA observation of the
inspiral part of the signal, it will be possible to determine the
physical parameters of the binary to extremely high accuracy.  Using
these parameters, numerical relativity can predict very precisely the
merger and ringdown waves.  Measurements of the individual masses and
spins will allow us to predict the mass and the spin of the remnant
black hole \citep{rezzolla:2008ApJ...674L..29R}, which can be directly
tested against the corresponding parameters extracted from the
ringdown.  The merger and ringdown waveforms will typically have an
SNR of \ensuremath{10^{2}} -- \ensuremath{10^{3}} for binary black
holes with total mass
$\ensuremath{10^{5}}\,\ensuremath{\mathrm{M}_\odot} < M(1+z) <
\ensuremath{6 \times 10^{8}}\,\ensuremath{\mathrm{M}_\odot}$ at $z=1$,
so an extremely clean comparison will be possible between the observed
waveforms and the predictions of GR.

\subsubsection*{The inspiral stage: comparing inspiral rate with predictions of General Relativity}

\noindent
With orbital velocities $v/c$ typically in the range \ensuremath{0.05}
-- \ensuremath{0.3}, most of the inspiral stage can be well described
using high-order PN expansions of the Einstein equations.  The
inspiral waveform is a chirp: a sinusoid that increases in frequency
and amplitude as the black holes spiral together. Depending on the
source parameters, eLISA will be able to observe the final stages of
the inspiral, for up to one year in some favourable cases.  To give a
practical reference, when the gravitational-wave frequency sweeps past
\ensuremath{0.3}\,\text{mHz}, the time remaining until merger is
approximately
\begin{align}
t &= \ensuremath{106.8}\,\text{days} \left(\frac{0.25}{\eta}\right) 
\left(  \frac{M(1+z)}{\ensuremath{2 \times 10^{5}}\,\text{\ensuremath{\mathrm{M}_\odot}}} \right)^{-5/3} 
\left( \frac{f}{\ensuremath{0.3}\,\text{mHz}}\right)^{-8/3}
\end{align}
where, as above, $M = m_1 + m_2$ is the total mass of the binary and
$\eta = m_1m_2/M^2$ is the symmetric mass
ratio.  eLISA will observe the last \ensuremath{10^{2}} --
\ensuremath{10^{4}} gravitational wave inspiral cycles, depending on the total mass
and distance of the source.  
Since the inspiral signal is quite well understood theoretically,
matched filtering can be used to recognise these inspirals up to a
year before the final merger, at a time when the total SNR is still
small.  Moreover, as the total SNR in the inspiral is quite large in
many cases, and such signals are long lived, matched filtering based
on the inspiral waveform alone can determine the system parameters to
very high accuracy.  Both masses can be determined to within a
fractional error of about \ensuremath{10^{-2}} --
\ensuremath{10^{-1}}, and the spin of the primary black hole can be
measured to an accuracy of \ensuremath{10}\,\text{\%} or better.

The nonlinear structure of GR (and possible deviations from GR) could
be encoded in a phenomenological way by considering hypothetical
modifications of the gravitational wave amplitude and phasing, as
proposed by different authors
\citep{arun:2006:pns,yunes:2009PhRvD..80l2003Y}.  The relatively large
strength of the inspiral gravitational wave signal will allow a
sensitive test of GR by comparing the rate of the observed inspiral
(phase evolution) to predictions of the PN approximation to GR
\citep{mishra:2010PhRvD..82f4010M,cornish:2011PhRvD..84f2003C,huwyler:2011arXiv1108.1826H,li:2011arXiv1110.0530L}.

\subsubsection*{The merger stage: spectacular bursts}

\noindent
The inspiral is followed by a dynamical merger that produces a burst
of gravitational waves.  This is a brief event, comprising a few
cycles lasting about $\ensuremath{5 \times 10^{3}}\,\text{s} \left( M/
\ensuremath{10^{6}}\,\text{\ensuremath{\mathrm{M}_\odot}}\right)
\left(0.25/\eta\right)$, yet very energetic: during the merger the
gravitational wave luminosity is
$L_\text{GW}\sim\ensuremath{10^{23}}\,\text{\ensuremath{\mathrm{L}_\odot}}$,
emitting more power than all the stars in the observable Universe.
The final merger of massive binaries occurs in the very strong-field,
highly nonlinear and highly dynamical regime of GR, and is the
strongest gravitational wave source that eLISA is expected to see.
eLISA will be able to see the merger of two
\ensuremath{10^{4}}\,\text{\ensuremath{\mathrm{M}_\odot}} black hole
beyond redshift $z=20$, and for mergers of two
\ensuremath{10^{6}}\,\text{\ensuremath{\mathrm{M}_\odot}} black hole
at $z = 1$ the SNR will be about \ensuremath{2000}. 
As mentioned above, eLISA observations of the inspiral yield a good
measurement of the masses and spins of the black holes.  With these in
hand, numerical relativity will make a very specific prediction for
the merger and ringdown radiation from the system.  Comparison with
the waveform that eLISA actually observes will allow us to confront
the predictions of GR with an ultra-high precision measurement in the
fully nonlinear and dynamical regime of strong gravity for the first
time.

\subsubsection*{The ringdown stage: black hole spectroscopy}

\noindent
Although numerical relativity waveforms from colliding holes naturally
include the ringdown waves, these waves are also well understood
analytically.  GR predicts, as a consequence of the ``no-hair''
theorem, that every excited black hole emits gravitational waves until
it reaches a time-independent state characterised entirely by its mass
and spin.  These ringdown waves consist of a set of superposed black
hole QNM waves with exponentially damped sinusoidal time dependence,
plus a far weaker ``tail''.  The modes are strongly damped as their
energy is radiated away to infinity, so the final ringdown stage is
brief, lasting only a few cycles.

The QNM of Kerr black hole can be computed using perturbation theory:
the spacetime metric is written as the Kerr metric plus a small
perturbation, and Einstein's equations are expanded to first-order in
that perturbation.  The solutions can be decomposed into a sum of
damped exponentials with complex eigenfrequencies
\citep{chandrasekhar:1975:qnm} that can be computed to essentially
arbitrary accuracy \citep{leaver:1985RSPSA.402..285L}.

While there are infinitely many modes (corresponding to the angular
order and overtone number of the perturbation from the stationary
state), the lowest-order modes are the most readily excited and the
least strongly damped, so in practice only a few modes are likely to
be observed.

The frequencies and damping times of these ringdown QNM
\citep[tabulated in][]{berti:2009CQGra..26p3001B} are completely
determined by the mass and the spin of the remnant black hole.

A data analysis strategy based on multi-mode searches will be
necessary for an accurate estimation of the mass and spin of the final
black hole
\citep{berti:2006:gws,berti:2007PhRvD..76j4044B}. Furthermore, if we
can measure at least two different QNM in a ringdown signal, the
ringdown radiation itself will provide a strong-field test of the
hypothesis that the central massive objects in galactic nuclei are
indeed Kerr black holes. The reason is that a two-mode signal contains
four parameters (the frequencies and damping times of each mode),
which must all be consistent with the \emph{same} mass and spin values
\citep{dreyer:2004:bhs}.  Just like we can identify chemical elements
via  spectroscopic
measurements, we can uniquely identify a black hole (determine its
mass and spin) from the spectrum of its ringdown radiation.

If GR is correct but the observed radiation is emitted
from a different source (exotic proposals include boson stars and
gravastars, among others), the spectrum would most certainly be
inconsistent with the QNM spectrum of Kerr black holes in GR
\citep{yoshida:1994PhRvD..50.6235Y,berti:2006:boson,chirenti:2007CQGra..24.4191C,pani:2009PhRvD..80l4047P}.  The
same should occur if GR does not correctly describe gravity in the
extremes of strong fields and dynamical spacetimes. The fact that
black hole oscillations should produce different radiation spectra in
different theories of gravity is true in general
\citep{barausse:2008xv}, and the spectrum was studied in 
some specific extensions of GR, such as
Einstein-dilaton-Gauss-Bonnet gravity \citep{pani:2009PhRvD..79h4031P}.  
The possibility of testing the no-hair theorem with QNM depends on
the accuracy with which frequencies and damping times can be measured,
which in turn depends on the SNR of the ringdown signal. As shown in
\citep{berti:2006:lom,berti:2007PhRvD..76j4044B}, SNR larger than \ensuremath{50} should be sufficient to identify the presence of a second mode and use it for
tests of the no-hair theorem. This is only marginally achievable with
advanced Earth-based detectors, but SNR of this order should be the
norm for the black hole mergers detectable by eLISA. Furthermore, recent
work showed that multi-mode ringdown waveforms could encode
information on parameters of the binary \emph{before merger}, such as
the binary's mass ratio \citep{kamaretsos:2011arXiv1107.0854K}, and this would 
provide further consistency checks on the strong-field dynamics of general
relativity.

\subsection{Extreme mass ratio inspirals: precision probes of Kerr spacetime}
\label{sec.extreme-mass-ratio}

\noindent

\begin{figure}
  \centering
  \resizebox{\hsize}{!}
	%{\includegraphics[scale=1,clip]{strainM1e5z15-WDWD_SRDcopy_manual_self_standing} }
        {\includegraphics[scale=1,clip]{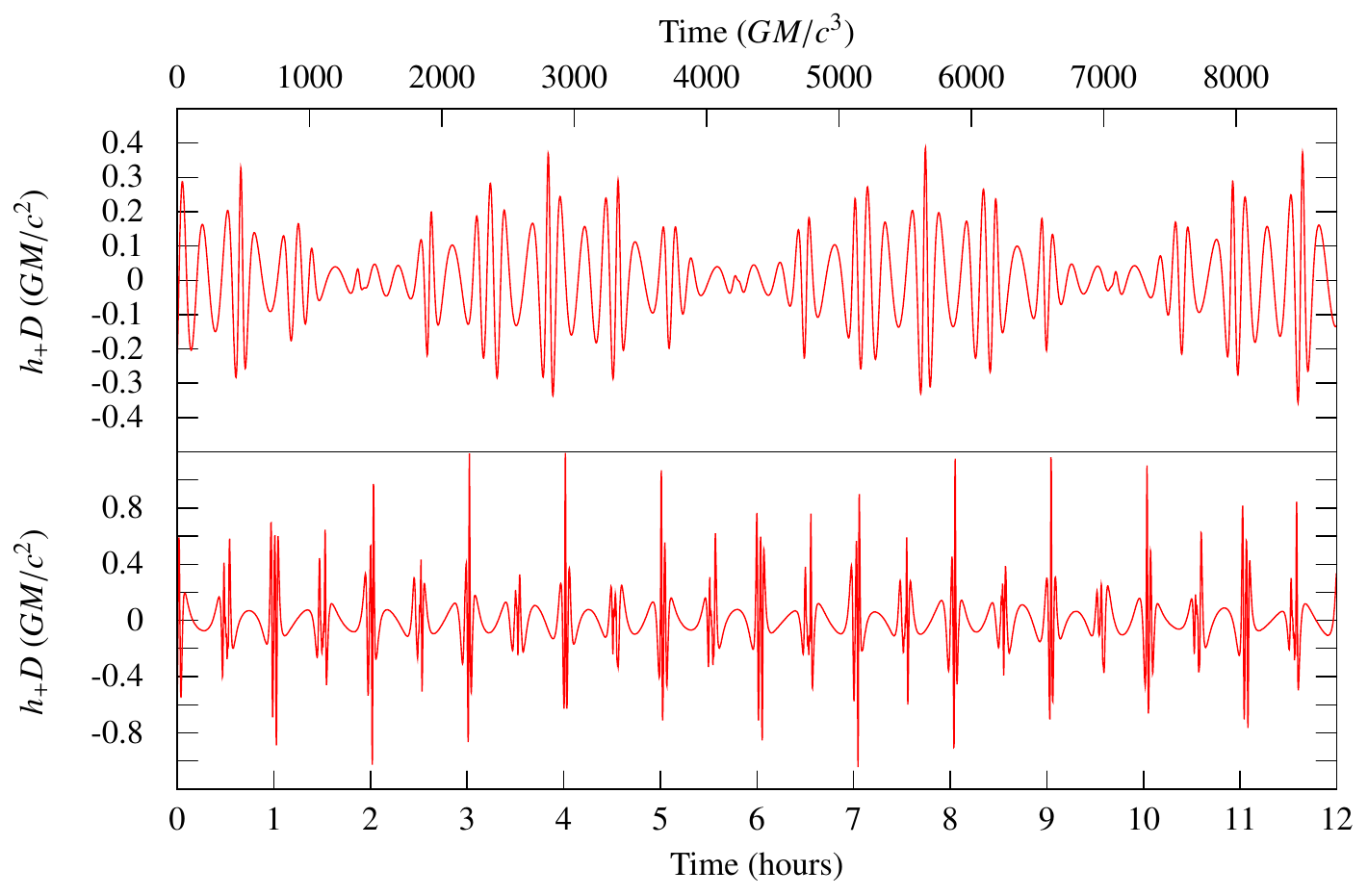} }
  \caption{Segments of generic EMRI waveforms \citep{drasco:2006:gws}. These are the plus-polarised waves
produced by a test mass orbiting a $10^6 M_{\odot}$ black hole that is spinning at 90 \% of the maximal rate allowed by
general relativity, a distance $D$ from the observer. Top panel: Slightly eccentric and inclined retrograde orbit
modestly far from the horizon. Bottom panel: Highly eccentric and inclined prograde orbit much closer to the
horizon. The amplitude modulation visible in the top panel is mostly due to Lense-Thirring precession of the
orbital plane. The bottom panel's more eccentric orbit produces sharp spikes at each pericentre passage.}
  \label{fig.fig_4-7}
\end{figure}

EMRI are expected to be very clean astrophysical systems, except
perhaps in the few percent of galaxies containing accreting massive
black holes, where interactions with the accretion disk could possibly
affect the EMRI dynamics.  Over timescales of the order of a day, the
orbits of the smaller body are essentially geodesics in the spacetime
of the massive black hole.  On longer timescales, the loss of energy and angular
momentum due to gravitational-wave emission causes the smaller body to
spiral in; i.e., the 
geodesic's ``constants'' of motion change
slowly over time. Over a typical eLISA observation time (years), 
EMRI orbits are highly relativistic (radius smaller than 10 Schwarzschild 
radii) and display extreme forms of 
periastron and
orbital plane precession due to the dragging of
inertial frames by the massive black hole's spin. 
Figure \ref{fig.fig_4-7}
shows two sample waveforms, corresponding to short stretches of time.

Given the large amount of gravitational wave cycles collected in a
typical EMRI observation (about \ensuremath{10^{5}}), a fit of the
observed gravitational waves to theoretically calculated templates
will be very sensitive to small changes in the physical
parameters of the system.  As mentioned above, this sensitivity makes the search
computationally challenging, but it allows an extremely accurate
determination of the source parameters, once an EMRI signal is
identified.  Assuming that GR is correct and the central massive
object is a  black hole, eLISA should be able
to determine the mass and spin of the massive black hole to fractional accuracy
of about \ensuremath{10^{-4}} -- \ensuremath{10^{-3}} for gravitational wave signals with an SNR of 20
\citep{barack:2004:lcs}.

This level of
precision suggests that we can use EMRI as a highly precise
observational test of the ``Kerr-ness'' of the central massive
object.  That is, if we do \emph{not} assume that the larger object is
a black hole, we can use gravitational waves from an EMRI to map the
spacetime of that object.  The spacetime outside a stationary
axisymmetric object is fully determined by its mass moments $M_l$ and
current multipole moments $S_l$.  Since these moments fully
characterise the spacetime, the orbits of the smaller object and the
gravitational waves it emits are determined by the
multipolar structure of the spacetime.  By observing these
gravitational waves with eLISA we can therefore precisely characterise the spacetime of the
central object. Extracting the moments from  the EMRI waves is
analogous to geodesy, in
which the distribution of mass in the Earth  is determined by studying
the orbits of satellites.  Black hole geodesy, also known as
holiodesy, is very powerful because %
Kerr black holes have a very special %
multipolar structure.  A Kerr
black hole with mass $\ensuremath{\mathit{M}_\bullet}$ and spin parameter $\ensuremath{a_\bullet}$ (in units with $G=c=1$)
has multipole moments given by

\begin{align}
  M_l + \mathrm{i} S_l &= (\mathrm{i}\ensuremath{a_\bullet} )^l \ensuremath{\mathit{M}_\bullet}^{l+1} 
\end{align}

Thus, $M_0 = \ensuremath{\mathit{M}_\bullet}$, $S_1 = \ensuremath{a_\bullet}\ensuremath{\mathit{M}_\bullet}^2$, and $M_2 = -\ensuremath{a_\bullet}^2M^3$ , and
similarly for all other multipole moments; they are all completely
determined by the first two moments, the black hole mass and spin.
This is nothing more than the black
hole ``no-hair'' theorem: the properties of a black hole are
entirely determined by its mass and spin.

For inspiraling trajectories that are slightly eccentric and slightly
non-equatorial, in principle all the multipole moments are redundantly
encoded in the emitted gravitational waves \citep{ryan:1995:gwi},
through the time-evolution of the three fundamental
frequencies of the orbit: the fundamental frequencies associated with the $r$,
$\theta$, and $\phi$ motions \citep{drasco:2004:rbh}, or,
equivalently, the radial frequency and the two precession frequencies.

The mass quadrupole moment $M_2$ of a Kerr black hole can be measured
to within $\Delta M_2 \approx
\ensuremath{10^{-2}}\,\text{\ensuremath{\mathit{M}_\bullet}\ensuremath{^{3}}}
-
\ensuremath{10^{-4}}\,\text{\ensuremath{\mathit{M}_\bullet}\ensuremath{^{3}}}$
for signals with an SNR of 30 \citep{barack:2004:lcs}, At the same time
$\Delta
\ensuremath{\mathit{M}_\bullet}/\ensuremath{\mathit{M}_\bullet}$ and
$\Delta S_1/\ensuremath{\mathit{M}_\bullet}^2$ will be estimated to an
accuracy of \ensuremath{10^{-4}} -- \ensuremath{10^{-3}}.

Any inconsistency with the Kerr multipole structure could signal a
failure of GR, the discovery of a new type of compact object, or a
surprisingly strong perturbation from some other material or
object. For a review of the different hypotheses regarding the nature
of the central object see
\citep{sopuerta:2010:emri,babak:2011CQGra..28k4001B}.

Other tests of the Kerr nature of the central massive object have also
been proposed.  EMRI signals can be used to distinguish definitively
between a central massive black hole and a boson star
\citep{kesden:2005:gws}.  In the black hole case the gravitational wave signal ``shuts
off'' shortly after the inspiraling body reaches the last stable orbit
(and then plunges through the event horizon), while for a massive
boson star, the signal does not fade, and its frequency derivative
changes sign, as the body enters the boson star and spirals toward its
centre. Similarly, if the central object's horizon is replaced by some
kind of membrane (this is the case for the so-called gravastars) the
orbital radiation produced by the orbiting body could resonantly
excite the QNM of the gravastar, with characteristic signatures in the
gravitational wave energy spectrum that would be detectable by eLISA
\citep{pani:2009PhRvD..80l4047P}.

Other studies within GR considered axisymmetric solutions of the
Einstein field equations for which the multipole moments can differ
from the Kerr metric, such as the Manko-Novikov solution.  These
studies revealed ergodic orbital motion in some parts of the parameter
space \citep{gair:2008:pbh} as a result of the loss of the third integral of
motion.  A similar study suggested that the inspiralling body could
experience an extended resonance in the orbital evolution when the ratio of intrinsic
frequencies of the system is a rational number
\citep{lukes-gerakopoulos:2010PhRvD..81l4005L}.  If detected, these features would be
a robust signature of a deviation from the Kerr metric.

These and similar studies of ``bumpy'' Kerr black holes -- spacetime
metrics with a multipolar stucture that deviates from the Kerr
spacetime by some ``tunable'' amount \citep{ryan:1995:gwi,collins:2004:tfm,glampedakis:2006:msl,
hughes:2006:abs,vigeland:2011PhRvD..83j4027V,vigeland:2010PhRvD..81b4030V} -- focussed on
understanding whether the best fit to eLISA data is consistent with
the Kerr solution \emph{within general relativity}. However, an even
more exciting prospect is that modifications in EMRI waveforms might
arise because the true theory of gravity is in fact different from GR.
For example, black holes in dynamical Chern-Simons theory (a
parity-violating, quantum-gravity inspired extension of GR) deviate
from Kerr black holes in the fourth multipole moment $\ell=4$. This
affects geodesic motion, and therefore the phasing of the gravitational wave signal
\citep{sopuerta:2009PhRvD..80f4006S,pani:2011PhRvD..83j4048P}. 

Gravitational wave observations of black hole-black hole binaries
cannot discriminate between GR and
scalar-tensor theories of gravity.  The reason is that
black holes do not support scalar fields; i.e., they have no scalar
hair. However, eLISA could place interesting bounds on scalar-tensor
theories using observations of neutron stars spiralling into massive
black holes \citep{berti:2005:esb,yagi:2010PhRvD..81f4008Y}. These limits will be
competitive with -- but probably not much more stringent than -- Solar
System and binary pulsar measurements \citep{esposito-farese:2004:stg}.

Finally, eLISA observations of
compact binaries could provide interesting bounds on Randall-Sundrum
inspired braneworld models \citep{mcwilliams:2010PhRvL.104n1601M,yagi:2011PhRvD..83h4036Y}. A
general framework to describe deviations from GR in different
alternative theories and their imprint on the gravitational wave signal from EMRI can
be found in \citep{gair:2011PhRvD..84f4016G}.

Most high-energy modifications to GR predict the existence of light
scalar fields (axions). If such scalar fields exist, as pointed out
long ago by Detweiler and others
\citep{detweiler:1980PhRvD..22.2323D}, rotating black holes could
undergo a superradiant ``black hole bomb'' instability for some values
of their spin parameter. Depending on the mass of axions,
string-theory motivated ``string axiverse'' scenarios predict that
stable black holes cannot exist in certain regions of the mass/angular
momentum plane \citep{arvanitaki:2011PhRvD..83d4026A}.  Furthermore,
this superradiant instability could produce a surprising result: close
to the resonances corresponding to a superradiant instability the EMRI
would stop, and the orbiting body would float around the central black
hole.  These ``floating orbits'' (for which the net gravitational
energy loss at infinity is entirely provided by the black hole's
rotational energy) are potentially observable by eLISA, and they could
provide a smoking gun of high-energy deviations from general
relativity \citep{Cardoso:2011xi,Yunes:2011aaarXiv1112.3351}.

In conclusion we remark that, if GR must be modified, the ``true'' theory of
gravity should lead to similar deviations in all observed EMRI. For this
reason, statistical studies of EMRI to test GR would alleviate possible
disturbances that may cause deviations in individual systems, such as
interactions with an accretion disk
\citep{barausse:2007PhRvD..75f4026B,barausse:2008PhRvD..77j4027B,kocsis:2011PhRvD..84b4032K},
perturbations due to a second nearby black hole
\citep{yunes:2011PhRvD..83d4030Y} or by a near-by star, which could allow us to
investigate different models of how stars distribute around a massive black
hole \citep{2012ApJ...744L..20A}.

\subsection{Intermediate mass ratio binaries}

\noindent
A loud gravitational wave source for eLISA would be the IMRI of binaries comprising a middleweight (or
equivalently intermediate-mass) 
black hole, with mass in the range of a few times \ensuremath{10^{2}}\,\ensuremath{\mathrm{M}_\odot} to a few times \ensuremath{10^{4}}\,\ensuremath{\mathrm{M}_\odot}, along with either a
massive black hole (\ensuremath{10^{6}}\,\ensuremath{\mathrm{M}_\odot}) or a solar-mass black hole.  Currently there is no
fully convincing evidence for the existence of intermediate-mass black  holes, primarily due to
the enormous observational difficulties of resolving the central
region of dwarf galaxies and/or globular clusters, the two most likely
places where they might reside. eLISA is one of the most promising
observatories for discovering these middleweight black holes.

The strength of the gravitational wave signal from an IMRI lies between that of massive black
hole binaries and EMRI, and the signal itself carries features of both limiting
types, including a relatively fast frequency evolution and comparable
contribution of several harmonics to the total strength of the signal.
According to the proposed eLISA sensitivity, IMRI could be seen up to redshift
$z\sim 4$.  There are good reasons to expect that IMRI orbits may have
measurable eccentricity
\citep{2006ApJ...653L..53A,2006AIPC..873..250A,2009ApJ...692L..50A,2010MNRAS.401.2268A,amaro-seoane:2010ApJ...722.1197A,sesana:2010:scm}.
It may also be possible in some cases to observe the gravitational spin-spin
coupling between the two black holes (equivalent to the Lense-Thirring effect).
The precision in the measurements of the source parameters will lie between
that of EMRI and comparable-mass binaries.

\subsection{The mass of the graviton}
\label{sec.mass-graviton}

\noindent
In GR, gravitational waves travel with the speed of light and the
graviton is hence massless. Alternative theories with a massive
graviton predict an additional frequency-dependent phase shift of the
observed waveform. The dominant effect can be expressed at 1-PN order,
and would change the PN coefficient $\psi_2$ in the stationary-phase
approximation to the Fourier transform of the waveform as follows:

\begin{align}
\psi_2 &\rightarrow \psi_2  - \frac{128\pi^2}{3} 
\frac{G\eta^{3/5} M}{c^2}\frac{D}{\lambda_g^2 (1+z)},
\end{align}

\noindent
where $\eta$ is again the symmetric mass
ratio.  This term alters the time of arrival of waves of different
frequencies, causing a dispersion, and a corresponding modulation in
the phase of the signal that depends on the Compton wavelength $\lambda_g$ and the distance $D$ to
the binary.  Hence, by tracking the phase of the inspiral waves, eLISA
should set bounds in the range $\lambda_g \in [\ensuremath{2 \times
    10^{16}}\,\text{km}, \ensuremath{10^{18}}\,\text{km}]$ on the
graviton Compton wavelength \citep{berti:2011arXiv1107.3528B,huwyler:2011arXiv1108.1826H},
improving current Solar System bound on the graviton mass, $m_g < \ensuremath{4 \times
  10^{-22}}\,\text{eV}$ ($\lambda_g > \ensuremath{3 \times
  10^{12}}\,\text{m}$) by several orders of magnitude.

Statistical observations of an ensemble of black hole coalescence
events could be used to yield stringent constraints on other theories
whose deviations from GR are parametrized by a set of global
parameters: examples considered so far in the literature include
theories with an evolving gravitational constant
\citep{yunes:2010PhRvD..81f4018Y}, massive Brans-Dicke theories
\citep{Alsing:2011erarXiv1112.4903} and Lorentz-violating
modifications of GR \citep{mirshekari:2011arXiv1110.2720M}.

\section{Cosmology}
\label{sec.new-physics-early}

\subsection{New physics and the early Universe}
\begin{figure}
  \centering
  \resizebox{\hsize}{!}
	{\includegraphics[scale=1,clip]{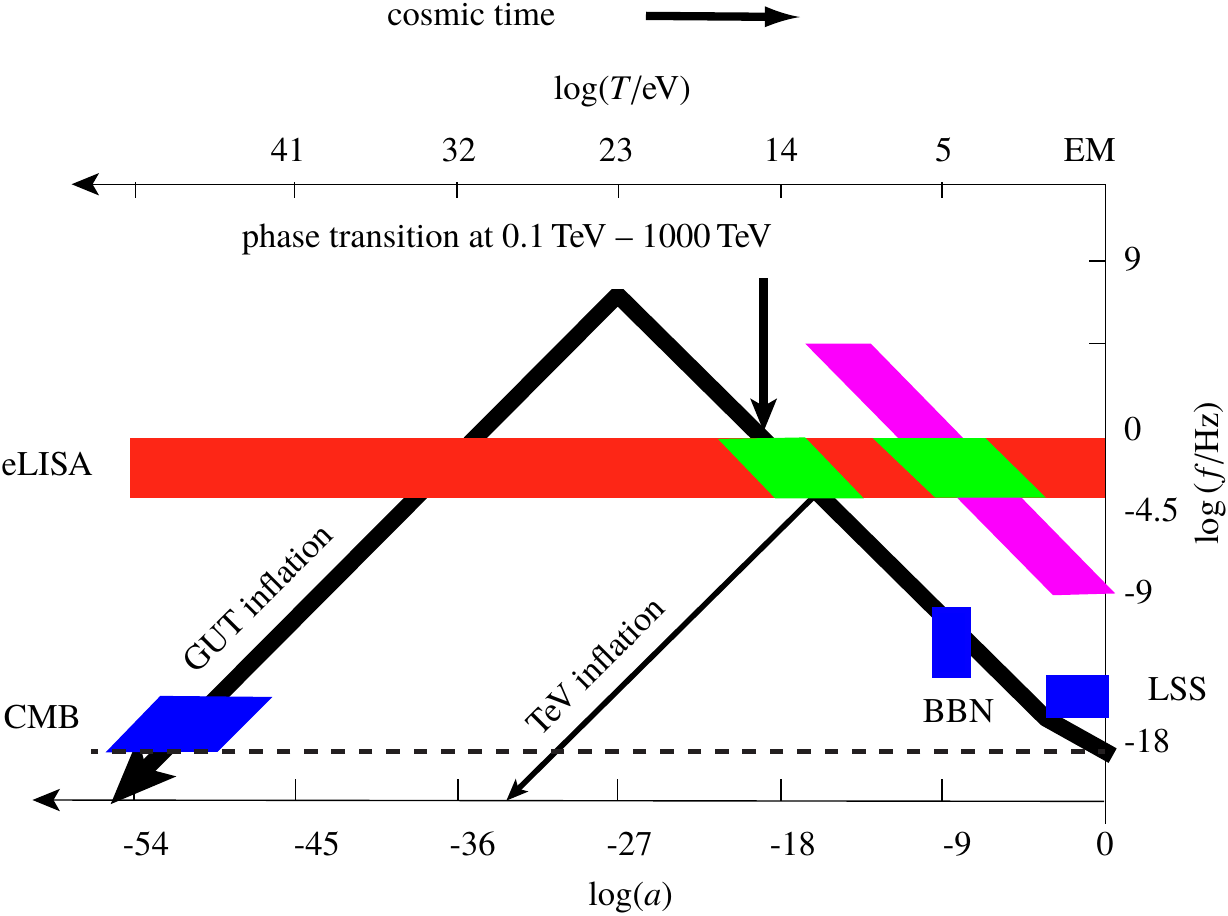} }
  \caption{%
    The observed (redshifted) frequency of wave-generating phenomena
    is shown as a function of cosmic scale factor $a$, with the
    present epoch at the right. The redshifted Hubble rate (horizon
    scale) is shown in black for a standard Grand Unified Theory (GUT)
    and a lower temperature Terascale (TeV) inflationary cosmology.
    Blue regions are accessible to electromagnetic (EM) observations:
    the Universe since recombination (right box) and cosmic microwave
    background (CMB) fluctuations (left box).  The red bar shows the
    range of cosmic history accessible through eLISA  from processes
    within the horizon up to about \ensuremath{1000}\,\text{TeV}.}
  \label{fig.fig_7-1}
\end{figure}

\noindent
Gravitational waves penetrate all of cosmic history, which allows eLISA to
explore scales, epochs, and new physical effects not accessible in
any other way (see figure \ref{fig.fig_7-1}).  Indeed a detectable
gravitational wave background in the eLISA band is predicted by a
number of new physical ideas for early cosmological evolution
\citep{maggiore:2000:gwe,hogan:2006:gws}.  Two important mechanisms
for generating stochastic backgrounds are
phase transitions in the
early Universe and
cosmic strings.

Gravitational waves produced after the Big Bang form a fossile radiation:
expansion prevents them from reaching thermal equilibrium with the other 
components because of the weakness of the gravitational interaction. 
Important information on the first instants of the Universe is 
thus imprinted in these relics and can be decoded. The mechanical effect of 
expansion is simply to redshift the corresponding frequency. Assuming that 
the wavelength is set by the 
apparent horizon size $c/H_* = c a/\dot a$
at the time of production (when the temperature of the Universe is $T_*$), the 
redshifted  frequency is
\begin{align}
  \label{eq.8}
  f_0 &= \dot a(t) \approx \ensuremath{10^{-4}}\,\text{Hz} \sqrt{ H_*(t) \times
    \frac{\ensuremath{1}\,\text{mm}}{c}} \approx \ensuremath{10^{-4}}\,\text{Hz}
  \left(\frac{k_B T_*}{\ensuremath{1}\,\text{TeV}}\right)
\end{align}

Thus, the eLISA frequency band of about \ensuremath{0.1}\,\text{mHz}
to \ensuremath{100}\,\text{mHz} 
today corresponds to the horizon at
and beyond the Terascale frontier of fundamental
physics.  This allows eLISA to probe bulk motions at times about
\ensuremath{3 \times 10^{-18}} -- \ensuremath{3 \times
  10^{-10}} seconds 
after the Big Bang, a period not
directly accessible with any other technique.  Taking a typical broad
spectrum into account, eLISA has the sensitivity to detect
cosmological backgrounds caused by new physics active in the range of
energy from \ensuremath{0.1}\,\text{TeV} to
\ensuremath{1000}\,\text{TeV}, if more than a modest fraction
$\Omega_{\text{GW}}$ of about \ensuremath{10^{-5}} 
of the energy
density is converted to gravitational radiation at the time of production.

Various sources of gravitational wave background of cosmological origin are
presented in detail in \cite{2012arXiv1201.0983B}. Here we will only briefly
summarize the main mechanisms leading to the potentially observable
backgrounds.

A standard example of new physics is a first-order phase transition resulting 
in bubble nucleation and growth, and subsequent bubble collisions and turbulence.
Phase transitions also often lead to the formation of one-dimensional
topological defects known as cosmic strings. Among possible topological
defects, cosmic strings are unique from a cosmological point of view because,
whereas their energy density should grow with the expansion, they interact
and form loops which decay into gravitational waves. Thus cosmic
strings tend to form networks with a typical scaling behaviour, losing
energy mainly through gravitational radiation with a very broad
and uniquely identifiable spectrum.  Besides topological defects, 
cosmic strings could also find their origin among the fundamental objects of
string theory, the theory that is aiming at providing a unified framework for 
all particles and forces of nature. Indeed, although fundamental strings were devised as
submicroscopic objects, it has been progressively realized
\citep{copeland:2004JHEP...06..013C} that some of these strings could
be stretched to astronomical size by the cosmic expansion. eLISA
will be our most sensitive probe for these objects by several orders of
magnitude and so offers the possibility of detecting direct evidence
of fundamental strings.

In order to distinguish backgrounds of gravitational waves from those
waves emitted by point sources, it is essential to make use of the
successive positions of eLISA around the Sun, and thus to wait a
sufficient amount of time (of the order of a few months). It is more
difficult to disentangle an isotropic cosmological (or astrophysical)
background from an instrumental one, all the more because the eLISA ``Mother-Daughter''
configuration, providing only two measurement arms, does not allow to
use Sagnac calibration \citep{hogan:2001:esg}. Luckily, in the case of phase transitions as well as
cosmic strings, the spectral dependence of the signal is well
predicted and may allow to distinguish cosmological backgrounds as
long as they lie above the eLISA sensitivity curve.

\subsubsection*{First-order cosmological phase transitions: Bulk motion from 
bubble nucleation, cavitation, collisions, turbulence
}
\label{sec.new-physics-early-1}

Abundant evidence suggests that the physical vacuum was not always in
its current state, but once had a significantly higher free energy.
This idea is fundamental and general: it underlies symmetry breaking
in theories such as the Standard Model and its %
supersymmetric extensions, and cosmological models including almost
all versions of inflation.  Common to all these
schemes is the feature that a cold, nearly uniform free energy
contained in the original %
(\emph{false}) vacuum is liberated in a phase transition to a final
(\emph{true}) vacuum, and eventually converted into thermal energy
of radiation and hot plasma.

In many theories beyond the Standard Model, the conversion between vacuum 
states corresponds to a
first-order phase transition.  In
an expanding Universe this leads to a cataclysmic process.  After
supercooling below the critical temperature $T_*$ for the transition, a
thermal or quantum jump across an energy barrier leads to the
formation of bubbles of the new phase.  The bubbles rapidly expand and collide.
The internal energy
is thus converted to organised flows of mass-energy, whose bulk
kinetic energy eventually dissipates via turbulence and finally
thermalises.  The initial bubble collision and subsequent turbulent
cascade lead to relativistic flows and acceleration of matter that
radiate gravitational waves on a scale not far below the horizon scale
\citep{witten:1984:csp,hogan:1986:grc,Kamionkowski:1993fg,Huber:2008hg,Caprini:2009yp}.

The gravitational wave energy density $\ensuremath{\Omega_{\text{GW}}}$
typically depends  on two parameters: $H_*/\beta$ is the duration of the
transition in Hubble units and $\alpha$ is the fraction of energy density
available in the source (false vacuum, relativistic motion). Typically
$\ensuremath{\Omega_{\text{GW}}} \sim \Omega_{\text{{rad}}} \left(
H_*/\beta\right)^2 (\kappa\, \alpha)^2 / (1 + \alpha)^2$, where
$\Omega_{\text{rad}}$ is the the fraction of radiation energy today, and
$\kappa$ the fraction of vacuum energy which is converted into bulk kinetic
energy during the phase transition. Strong first order phase transitions are
obtained for $\alpha\gg 1$ but, in the context of specific models, increasing
$\alpha$ may increase $\beta$ as well.

\subsubsection*{Dynamics of warped sub-millimetre extra dimensions}
\label{sec.new-physics-early-2}

Superstring theory provides examples of strong first order phase transitions 
in the Terascale region. It requires,  for mathematical
consistency, several  extra dimensions.  The sizes of these 
dimensions, their shapes, and how they are stabilised are yet to be determined.
If they exist, gravity can penetrate into
them, so they must be small or warped -- with a size below the sub-millimetre 
scale limit set by
direct laboratory tests of the gravitational inverse-square law. The
scales probed by %
Standard Model particles and fields are much smaller than this, but
fields other than gravity might be confined to a 3-dimensional
subspace or  (mem)brane plunged in the higher dimensional space.

Since the the Hubble length at the Terascale is about a millimetre, 
the current threshold where possible new effects of extra dimensions might appear 
happens to be about the same for experimantal gravity in the laboratory as for the 
cosmological regime accessible to eLISA.
It is even possible that new properties of gravity on this scale are
related to cosmic dark energy, whose energy density is about
$(\ensuremath{0.1}\,\text{mm})^{-4}$ in particle physics units.

\begin{figure}
  \centering

 \resizebox{\hsize}{!}
	{\includegraphics[scale=1,clip]{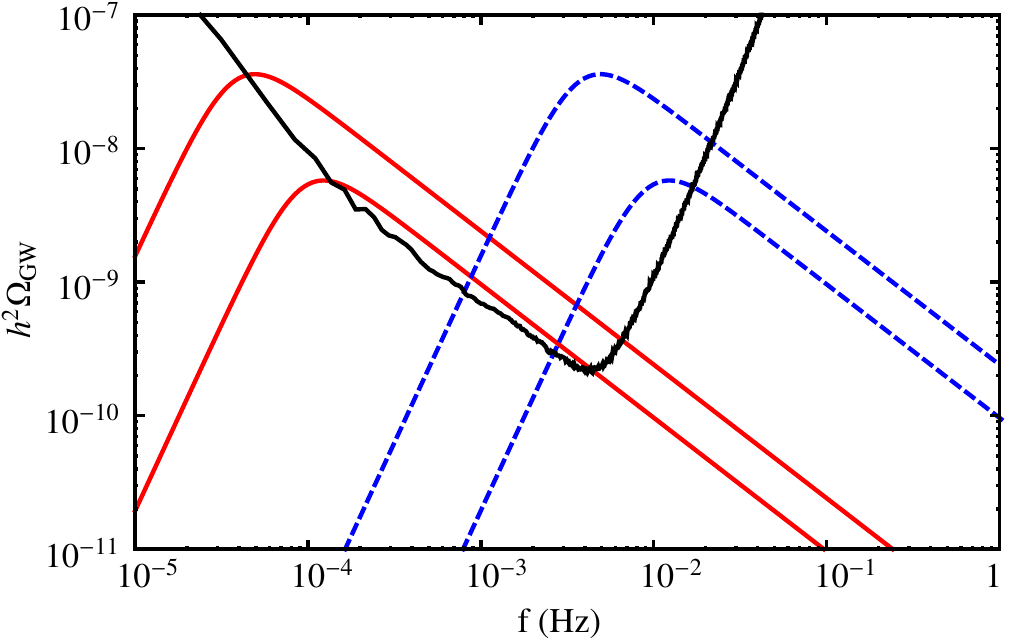} }
 \caption{%
   Predictions for the holographic phase transition
   \citep{konstandin:2010PhRvD..82h3513K} corresponding to the model
   of Randall and Sundrum with a TeV brane stabilized. In black, the
   sensitivity curve of eLISA expressed in terms of the
   gravitational wave background density $\ensuremath{\Omega_{\text{GW}}}$. In red, signals
   corresponding to a phase transition temperature of \ensuremath{10^{2}} GeV.  
   In dashed blue, a transition temperature of \ensuremath{10^{4}} GeV. 
   From top to bottom, curves correspond to $\beta/H_*= 6$ and $\beta/H_*=15$.
  \label{fig.fig_7-2}
  }
\end{figure}

 The dynamics
associated with the stabilisation of extra dimensions at a certain
size or warp radius might introduce a source of free internal energy
released coherently on a \emph{mesoscopic}, i.e.\ sub-millimetre to
nanometre scale, leading to a detectable background
\citep{hogan:2000:gwm,randall:2006:gww}.  If the extra dimensions are
much smaller than the Hubble length when the stabilisation occurs, the
behaviour of the extra dimensions is nearly equivalent to scalar field
behaviour as viewed in conventional 3-dimensional space, with effects
similar to the phase transitions discussed above (see 
figure \ref{fig.fig_7-2}).

\subsubsection*{Backgrounds, bursts, and harmonic notes from cosmic strings}
\label{sec.backgr-bursts-from}

As we have seen above, models of physics and
cosmology based on string theory, as well as their field-theory
counterparts, often predict the cosmological formation of cosmic
superstrings \citep{copeland:2004JHEP...06..013C}  that form after inflation and are stretched to
enormous length by the cosmic expansion.  In equivalent field-theory
language, cosmic strings arise from certain types of phase
transitions, and stable relics of the high-energy phase persist as
topological defects: in the form of one-dimensional strings that
resemble flux tubes or trapped vortex lines.

The primordial network of strings produces  isolated, oscillating loops
that ultimately radiate almost all of their energy into gravitational
waves.  Their gravitational radiation is mainly governed by a single
dimensionless parameter $G\mu/c^4$ reflecting the fundamental physics
of the strings, where $G$ is Newton's constant and $\mu$ is the energy per 
unit length, or tension.
This parameter is known to be very small, as current
limits on gravitational wave backgrounds already indicate that if
cosmic strings exist, they must be so light that they would have few
observable effects apart from their gravitational radiation. 

\begin{figure}
  \centering
 \resizebox{\hsize}{!}
	{\includegraphics[scale=1,clip]{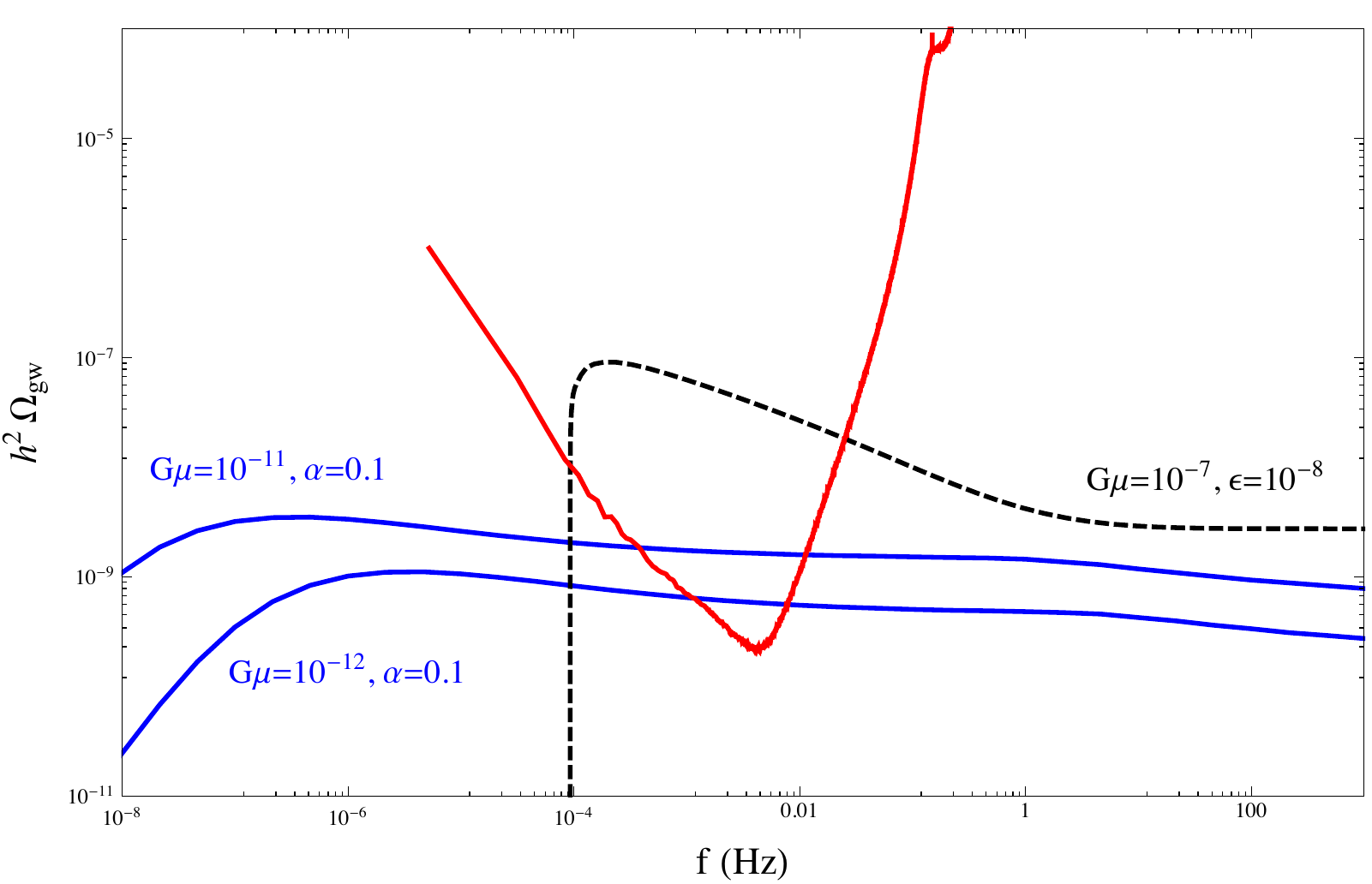} }
  \caption{%
Typical string background expected for eLISA (whose sensitivity curve is in red)
in the  large loop scenario in blue ($\alpha$ determines the loop size as a 
fraction of the horizon size) and in the small loop scenario in dashed
($\epsilon \equiv \alpha /(50 G\mu)$). See \cite{2012arXiv1201.0983B} for more details. 
}
  \label{fig.fig_7-3}
\end{figure}

Figure \ref{fig.fig_7-3} compares eLISA sensitivity (in red) with predicted 
stochastic background spectra in two distinct scenarios: large loops in blue
(where newly formed loops are about $\alpha = 0.1$ times the horizon size) for
two values of $G\mu/c^4$ spanning a range of scenarios motivated by brane 
world inflation, and small loops in dashed (with $\alpha = 50 \epsilon G \mu$)
for one value of $G\mu/c^4$. We note that the spectrum from cosmic strings is 
distinguishably different from
that of phase transitions or any other predicted source: it has nearly
constant energy per logarithmic frequency interval over many decades
at high frequencies, and falls off after a peak at low frequencies, since 
large string loops are rare and radiate slowly. In the small loop scenario, 
the peak frequency shifts to lower values when increasing $\epsilon$, whereas 
the amplitude decreases with $G\mu/c^4$. This allows an interesting interplay 
between measurements at eLISA, ground interferometers and millisecond pulsar 
arrays:
depending on the parameters, one may have detection of the string background
at one, two or three of these different types of detectors. In the large loop 
scenario,  eLISA 
sensitivity in terms
of $G\mu/c^4$ is several orders of magnitude deeper than even the
best possible future sensitivity from pulsar timing.

If the strings are not too much lighter than $G\mu/c^4 \sim
\ensuremath{10^{-10}}$, occasional distinctive bursts might be seen from loops,
produced by a sharply bent bit of string moving at nearly the speed of
light \citep{damour:2005:grc,siemens:2006:gwb}.  These rare events,
known as kinks or cusps, are recognisable, if they are intense enough
to stand out above the background, from their universal waveform which
derives just from the geometry of the string. Cusps are localized in
time whereas kinks are propagating along the strings. In the case of
fundamental strings, the presence of junctions between strings leads
to a proliferation of kinks
\citep{binetruy:2010PhRvD..82l6007B,bohe:2011PhRvD..84f5016B}.

Although individual burst events, if detected, give the
clearest signature of a string source, the first detectable sign of a
superstring loop population is likely their integrated stochastic
background as shown in figure \ref{fig.fig_7-3}.

\subsubsection*{Terascale inflationary reheating}
\label{sec.new-physics-early-3}

Inflation represents an extraordinarily coherent
behaviour of an energetic scalar field that is nearly uniform across
the observable Universe.  After inflation, the internal potential
energy of this field is converted into a thermal mix of relativistic
particles, in a process known as reheating.  The reheating temperature
might be as cool as \ensuremath{1}\,\text{TeV}, especially in some brane-world models
where the Planck scale is itself not far above the
Terascale.

There is no reason to assume a quiet, orderly reheating process: the
decay of the inflaton energy may be violently
unstable.  In many scenarios, the conversion begins with
macroscopically coherent but inhomogeneous motions that eventually
cascade to microscopic scales.  Quantum coherent processes such as
preheating transform the energy into coherent classical motions that can 
generate backgrounds on the order of \ensuremath{10^{-3}} or more of the total 
energy density
\citep{khlebnikov:1997:rgw,easther:2006:sgw,GarciaBellido:2007dg,Dufaux:2007pt,Dufaux:2008dn}.  
The characteristic frequency of the background can fall in the eLISA band if the final reheating occurred 
at \ensuremath{0.1}\,\text{TeV} to \ensuremath{1000}\,\text{TeV}.

\subsubsection*{Exotic inflationary quantum vacuum fluctuations}
\label{sec.new-physics-early-4}

The amplification of %
quantum vacuum fluctuations during inflation leads to a background of
primordial %
gravitational waves.  An
optimistic estimate of this background in the case of conventional
inflation limits these to less than about \ensuremath{10^{-10}} of the CMB
energy density, far below eLISA's sensitivity; in many inflation models
it is much less \citep{chongchitnan:2006:pdd}.  However, some
unconventional versions of inflation, particularly pre-Big-Bang or
bouncing brane scenarios, predict possibly detectable backgrounds in
the eLISA band \citep[see
e.g.][]{brustein:1995:rgw,buonanno:2003:tasi,buonanno:1997:srg}.
Although some key parameters remain unknown, which limits the
predictive power of these models, they are significantly constrained
by gravitational wave backgrounds.  If such a background is detected,
its spectrum also contains information about the Universe at the time
perturbations re-enter the horizon (the second horizon
intersection in figure \ref{fig.fig_7-1}).

\subsection{Cosmological measurements with eLISA}
\label{sec.cosm-meas-with}

As discussed in section \ref{sec.black-hole-astr} we can probe the assembly of
cosmic structures through observations of black hole binaries up to high
redshifts. In addition to that, gravitational wave sources could serve as
standard sirens for cosmography \citep{holz:2005:ugw}, because chirping binary
systems allow direct measurements of the luminosity distance to the source. The
principle is elegant and simple \citep{schutz:1986:hgw}: the chirping time
$\tau$ of an inspiral/merger event, together with its orbital frequency
$\omega$ and strain $h$, gives the absolute luminosity distance to the source,
$D_L \sim c/(\omega^2 \tau h)$, with a numerical factor depending on details of
the system that are precisely determined by the measured waveform.  However,
eLISA cannot independently determine the redshift of a source, since in
gravitational wave astronomy, the measured source frequency and chirp time are
always combined with cosmic redshift $\omega = \omega_\text{source} /(1+z),
\tau = (1+z)\tau_\text{source}$, i.e., the redshift is degenerate with the
source intrinsic parameters. An independent measurement of redshift is
therefore needed. This may be accomplished by getting the optical redshift to
the host galaxy, for instance by identifying an electromagnetic radiation
counterpart to the event.

In the last decade, several mechanisms producing electromagnetic counterparts
to black hole binary coalescences have been proposed
\citep[e.g.,][]{armitage:2002ApJ...567L...9A,milosavljevic:2005:amb,phinney:2009astro2010S.235P};
an exhaustive review can be found in \citep{schnittman:2011CQGra..28i4021S}.
While there are still uncertainties in the nature and strength of such
counterparts, we might expect some of them to be observable at least in the
local Universe (say, $z\le 1$). Our parameter estimation simulations show that,
at low redshift, we could expect to localize at least
\ensuremath{50}\,\text{\%} of the inspiralling black holes to better than
\ensuremath{400} square degrees and about \ensuremath{11}\,\text{\%} to better
than \ensuremath{10} square degrees.  Merger and ringdown (if observed) should
further improve those numbers. As a practical example, wide area surveys like
LSST \citep{lsst:2009arXiv0912.0201L} in optical or the VAST project using the
Australian Square Kilometer Array Pathfinder
\citep{johnston:2007PASA...24..174J} in radio will have the capability of
covering such large area in the sky to high depth several times per day during
and right after the merger event, looking for distinctive transients. Any
identified counterpart will provide precise measurements of the source redshift
and sky location.  We can use this information to perform directional search
(fixing the sky location of the gravitational wave source) in the eLISA data and the resulting
uncertainty in the luminosity distance drops to less than
\ensuremath{1}\,\text{\%} for \ensuremath{60}\,\text{\%}
(\ensuremath{5}\,\text{\%} for \ensuremath{87}\,\text{\%}) of the sources.
Those numbers are comparable with (or even lower than) the weak lensing error
at these low redshifts \citep{wang:2002ApJ...572L..15W}.  Ultra-precise
measurements of the redshift and the luminosity distance will allow us to
cross--check the SNIa measurements
\citep{riess:1998ApJ...504..935R,perlmutter:1999AIPC..478..129P}, and because
of the very different systematics from the usual cosmological distance ladder
estimates, will be a strong check on hidden systematic errors in these
measurements. This will improve the estimation of cosmological parameters, such
as $H_0$ and $w$.

Without electromagnetic identification  of the host, we can check statistical 
consistency between all the possible hosts detected within the measurement 
error box, to infer cosmological parameters as suggested in 
\citep{petiteau:2011ApJ...732...82P}. To realize this scheme one needs a rather good source 
sky location and distance determination, which is possible with  eLISA 
only at low redshifts ($z<2$). In the local Universe,
the same technique applied to EMRI will allow precision measurement of 
$H_0$ \citep{macleod:2008:phc} at a level of a few percent.

\section{Conclusions: science and observational requirements}
\label{sec.sci-obs-req}

In this document we have presented the science that eLISA will be able to do,
which ranges from ultra-compact binaries to cosmology and tests of GR.

In particular, we note that eight of the known ultra-compact binaries
will be detected by eLISA as verification binaries. Upcoming
wide-field and synoptical surveys will most likely discover more
verification binaries before eLISA's launch. eLISA will detect about
3,000 double white dwarf binaries individually.  Most have orbital
periods between 5 and 10 minutes and have experienced at least one
common-envelope phase, so they will provide critical tests of physical
models of the common-envelope phase.  These sources are exactly the
population which has been proposed as progenitors of normal as well as
peculiar (type Ia) supernovae. eLISA will tell us if the formation of
all ultra-compact binaries is enhanced in globular clusters by
dynamical interactions.  The millions of ultra-compact binaries that
will not be individually detected by eLISA will form a detectable
foreground from which the global properties of the whole population
can be determined. The binaries detected by eLISA will improve our
knowledge of tidal interactions in white dwarfs, mass-transfer
stability and white dwarf mergers. eLISA will unravel the Galactic
population of short-period neutron star and black hole binaries, and
thus determine their local merger rate.  eLISA will measure the sky
position and distance of several hundred binaries, constraining the
mass distribution in the Galaxy and providing an independent distance
estimate to the Galactic centre.  The level and shape of the Galactic
foreground will constrain the relative contributions of thin disc,
thick disc and halo populations and their properties. For several
hundred sources the orbital inclination will be determined to better
than 10 degrees, allowing to test if binaries are statistically
aligned with the Galactic disc.

One of the most promising science goals of the mission are
supermassive black holes, which appear to be a key component of
galaxies. They are ubiquitous in near bright galaxies and share a
common evolution.  The intense accretion phase that supermassive black
holes experience when shining as QSOs and AGN erases information on how
and when the black holes formed. eLISA will unravel precisely this
information. Very massive black holes are expected to transit into the
mass interval to which eLISA is sensitive along the course of their
cosmic evolution.  eLISA will then map and mark the loci where
galaxies form and cluster, using black holes as clean tracers of their
assembly by capturing gravitational waves emitted during their
coalescence, that travelled undisturbed from the sites where they
originated. On the other hand, middleweight black holes of
\ensuremath{10^{5}}\,\text{\ensuremath{\mathrm{M}_\odot}} are observed
in the near universe, but our knowledge of these systems is rather
incomplete.  eLISA will investigate a mass interval that is not
accessible to current electromagnetic techniques, and this is
fundamental to understand the origin and growth of supermassive black
holes.  Due to the transparency of the universe to gravitational waves
at any redshift, eLISA will explore black holes of
\ensuremath{10^{5}}\,\text{\ensuremath{\mathrm{M}_\odot}} --
\ensuremath{10^{7}}\,\text{\ensuremath{\mathrm{M}_\odot}} out to a
redshift $z\lesssim 20$, tracing the growth of the black hole
population.

eLISa will also shed light on the path of black holes to coalescence
in a galaxy merger. This is a complex process, as various physical
mechanisms involving the interaction of the black holes with stars and
gas need to be at play and work effectively, acting on different
scales (from kpc down to $10^{-3}$ pc). Only at the smallest scales
gravitational waves are the dominant dissipative process driving the
binary to coalescence. eLISA will trace the last phase of this
evolution. Dual AGN, i.e. active black holes observed during their
pairing phase, offer the view of what we may call the galactic
precursors of black hole binary coalescences. They are now discovered
in increasing numbers, in large surveys. By contrast, evidence of
binary and recoiling AGN is poor, as the true nature of a number of
candidates is not yet fully established.  eLISA only will offer the
unique view of an imminent binary merger by capturing its loud
gravitational wave signal.

There exist major uncertainties in the physical mechanism(s) conducive to the
gravitational collapse of a star (or perhaps of a very massive quasi-stars)
leading to the formation of the first black holes in galaxies.
The mass of seed black holes ranges from a few hundred to a
few thousand solar masses.  Seed black holes later grow, following different
evolutions according to their different formation path and clustering inside
dark matter halos, and eLISA aims at disentangling different routes of
evolution. eLISA will considerably reduce uncertainties on the nature of
the seed population, as the number of observed mergers and the inferred masses
will allow to decide among the different models or, in the case of concurrent
models, determine their prevalence.

According to the theoretical findings we have presented, massive black
hole masses and spins evolve through coalescence and accretion events.
Black hole spins offers the best opportunity to determine whether
accretion episodes prior to coalescence are coherent or chaotic.
Masses and spins are directly encoded into the gravitational waves
emitted during the the merger process.  eLISA will measure the masses
and spins of the black holes prior to coalescence, offering
unprecedented details on how black hole binaries have been evolving
via mergers and accretion along cosmic history.
At present, coalescence rates, as a function of redshift and in
different mass bins, can only be inferred theoretically, using
statistical models for the hierarchical build-up of cosmic structures.
These models, firmly anchored to low redshift observations, indicate
that the expected detection rates for eLISA range between few and few
hundred per year.

Current electromagnetic observations are probing only the tip of the
massive black hole distribution in the universe, targeting black holes
with large masses, between $10^7\,M_{\odot}-10^9\,M_{\odot}$.
Conversely, eLISA will be able to detect the gravitational waves
emitted by black hole binaries with total mass (in the source rest
frame) as small as $10^4\, {\rm M_{\odot}}$ and up to $10^7\,{\rm M_{\odot}}$, out
to a redshift as remote as $z\sim 20$.  eLISA will detect fiducial
sources out to redshift $z\lesssim10$ with SNR $\gtrsim 10$ and so it
will explore almost all the mass-redshift parameter space relevant for
addressing scientific questions on the {evolution of the black hole
  population}.  Redshifted masses will be measured to an unprecedented
accuracy, up to the {0.1}--{1}{\%} level, whereas absolute errors in
the spin determination are expected to be in the range {0.01}--{0.1},
allowing us to reconstruct the cosmic evolution of massive black
holes. eLISA observations hence have the potential of constraining the
astrophysics of massive black holes along their entire cosmic history,
in a mass and redshift range inaccessible to conventional
electromagnetic observations.

On smaller scales, eLISA will also bring a new revolutionary
perspective, in this case relative to the study of {galactic
  nuclei}. eLISA will offer the deepest view of galactic nuclei,
exploring regions to which we are blind using current electromagnetic
techniques and probing the dynamics of stars in the space-time of a
Kerr black hole, by capturing the gravitational waves emitted by
stellar black holes orbiting the massive black hole.  EMRI detections
will allow us to infer properties of the stellar environment around a
massive black hole, so that our understanding of stellar dynamics in
galactic nuclei will be greatly improved.  Detection of EMRIs from
black holes in the eLISA mass range, that includes black holes similar
to the Milky Way's, will enable us to probe the population of central
black holes in an interval of masses where electromagnetic
observations are challenging.  eLISA's EMRIs can be detected up to
$z\simeq 0.5-0.7$ allowing to explore a volume of several tens of
Gpc$^3$ and discover massive black holes in dwarf galaxies that are
still elusive to electromagnetic observations.  eLISA may also
measure the mass of stellar-mass black holes. This will provide
invaluable information on the mass spectrum of stellar black holes,
and on the processes giving rise to compact stars.  eLISA will detect
EMRI events out to redshift $z\sim 0.7$, in normal galaxies with high
SNR, and in the mass interval, $10^4\,{{\rm M_\odot}} \lesssim M\lesssim
5\times 10^6\,{\rm M_\odot}$.  eLISA will measure the mass and spin of
the large, massive black hole with a precision to better than a part
in $10^4$.  This will enable us to characterise the {population of
  massive black holes in nuclei} in an interval of masses where
electromagnetic observations are poor, incomplete or even missing,
providing information also on their spins.  eLISA will also measure
with equivalent precision the mass of the stellar black hole in the
EMRI event, and also the orbital eccentricity at plunge. These
observations will provide insight on the way stars and their remnants
are forming and evolving in the extreme environment of a galactic
nucleus. The estimated detection rates based on the best available
models of the black hole population and the {EMRI} rate per galaxy,
are about $50$ events with a two year eLISA mission, with a factor of
$\lesssim 2$ uncertainty from the waveform modelling and lack of
knowledge about the likely system parameters (larger uncertainties 
are of astrophysical nature). Even with a handful of events, EMRIs
will be a powerful astrophysical probe of the formation and evolution
of massive and stellar black holes. We also note that the detection
with eLISA of even a single coalescence event involving two
intermediate mass black holes in colliding star clusters, present in
the very local universe, would be a major discovery, and it would have
a strong impact in the field of stellar dynamics and stellar evolution
in star forming regions.

General Relativity has been extensively tested in the weak field
regime both in the solar system and by using binary pulsars. eLISA
will provide a unique opportunity of confronting GR in the highly
dynamical strong field regime of massive black holes.  eLISA will be
capable of detecting inspiral and/or merger plus ring-down parts of
the gravitational wave signal from coalescing massive black holes
binaries of comparable mass. For the nearby events ($z\sim 1$) the
last several hours of the gravitational wave signal will be clearly
seen in the data, allowing direct comparison with the waveforms
predicted by GR.  The inspiral phase could be observed by eLISA up to
a year before the final merger with relatively large SNR. Comparison
of the observed inspiral rate with the predictions of GR will provide
a valuable test of the theory in the regime of strong, dynamical
gravitational fields.

The merger of two black holes could be observed by eLISA throughout
the Universe if it falls into the detector band. The observation of
the merger could be confronted directly with the predictions of GR
and, if the inspiral is also observed, could be used for a consistency
check between the two parts of the gravitational wave
signal. According to GR the merger leads to a single ringing Kerr
black hole characterised by its mass and spin. Detecting two or more
quasinormal modes (the individual damped exponential components of the
so-called ringdown radiation) will allow us to check whether the final
object indeed is described only by two parameters in accord with the
no-hair theorem of GR. eLISA will give us a unique opportunity to
observe middleweight mass black holes in the local Universe. If
observed, these systems would provide an additional testbed for GR.

eLISA will be capable of setting an upper limit on the mass of
graviton that is at least four orders of magnitude better than the
current limit based on observations in the Solar System. The discovery
of coalescing binary black holes, signposts of (pre-)galactic mergers,
will test, albeit indirectly, the hypothesis which is at the heart of
the current paradigm of galaxy formation, i.e. their assembly in a
bottom-up fashion. Furthermore coalescing binary black holes can be
regarded as standard sirens, and they may allow a direct measurement
of the luminosity distance to the source.  If coalescence is
accompanied by an electromagnetic signal that permits the measurement
of the optical redshift of the source eLISA will improve upon the
estimation of cosmological parameters, such as the Hubble constant and
the dark-energy parameter $w$.  eLISA will have unique capabilities in
detecting signatures from (or setting meaningful constraints on) a
wide range of cosmological phenomena and fundamental physics.
Gravitational radiation backgrounds are predicted in cosmological
models that include first order phase transitions, late-ending
inflation, and dynamically active mesoscopic extra dimensions. eLISA
will provide the most sensitive direct probes of such phenomena near
TeV energies.

We state now the eLISA science requirements (SR) which summarize the
research needed to fulfill the eLISA objectives.  For each science
requirement, one or more observational requirements (OR) are
defined. The observational requirements are stated in terms of
observable quantities necessary to meet the science requirements, and
in terms of the precision with which such quantities must be
measured.

\begin{itemize}
\item \emph{\textbf{ Galactic binaries }}
\begin{itemize}
\item \emph{\textbf{SR 1.1} : Elucidate the formation and evolution of Galactic stellar-mass compact binaries and thus constrain the outcome of the common envelope phase and the progenitors of (type Ia) supernovae.}
\begin{itemize}
\item \textbf{OR 1.1.1} : eLISA shall have the capability to detect at least 1000 binaries at SNR > 10 with orbital periods shorter than approximately six hours and determine their period. eLISA shall maintain this detection capability for at least one year.
\item \textbf{OR 1.1.2} : eLISA shall detect all neutron star and black hole binaries in the Milky Way with periods shorter than 35 minutes if they exist.
\item  \textbf{OR 1.1.3} : eLISA shall have the capability to measure the level of the unresolved Galactic foreground. eLISA shall maintain this detection capability for at least one year.
\end{itemize}
\item \emph{\textbf{SR 1.2} : Determine the spatial distribution of stellar mass binaries in the Milky Way.}
\begin{itemize}
\item \textbf{OR 1.2.1} : eLISA shall have the capability to determine the position of at least 500 sources with better than ten square degree angular resolution and the frequency derivative to a fractional uncertainty of 10 \%. 
\item \textbf{OR 1.2.2} : eLISA shall measure the inclination of at least 500 binaries to better than 10 degrees.
\end{itemize}
\item \emph{\textbf{SR 1.3} : Improve our understanding of white dwarfs, their masses, and their interactions in binaries, and enable combined gravitational and electromagnetic observations.}
\begin{itemize}
\item  \textbf{OR 1.3} : eLISA shall have the capability to measure the frequency derivative of all detected binary systems with gravitational wave frequencies above 10 mHz to better than 10 \%.
\end{itemize}
\end{itemize}
\item \emph{\textbf{ Massive black hole binaries }}
\begin{itemize}
\item \emph{\textbf{SR 2.1} : Trace the formation, growth and merger history of massive black hole with masses \ensuremath{10^5 \mathrm{M}_\odot - 10^7 \mathrm{M}_\odot } during the epoch of growth of QSO and widespread star formation ($0 < z < 5$) through their coalescence in galactic halos.}
\begin{itemize}
\item  \textbf{OR 2.1.1} : eLISA shall have the capability to detect the mergers of similar masses massive black hole (mass ratio \ensuremath{m_2/m_1 > 0.1}) with total mass in the range \ensuremath{10^5 \, \mathrm{M}_\odot <  m_{1} + m_{2} < 10^7  \, \mathrm{M}_\odot } up to redshift $z = 20$. The SNR of those sources with redshift $z < 5$ should be sufficient to enable determination of the massive black hole  masses (relative errors smaller than 1 \%) and the spin of the largest massive black hole  (error smaller than 0.1) and an estimation of the luminosity distance (relative error smaller than 50 \%).
\item  \textbf{OR 2.1.1} : eLISA shall have the capability to detect the mergers of massive black hole with total mass in the range \ensuremath{10^5 \mathrm{M}_\odot <  m_{1} + m_{2} < 10^7 \mathrm{M}_\odot } and mass ratio \ensuremath{m_{2}/m_{1}} about 0.01 up to redshift $z = 8$. The SNR of those sources with redshift $z < 5$ shall be sufficient to enable determination of the massive black hole masses (relative errors smaller than a few percents).
\end{itemize}
\item \emph{\textbf{SR 2.2} : Capture the signal of coalescing massive black hole binaries with masses \ensuremath{ 2 \times 10^{4} \, \mathrm{M}_\odot -   10^{5} \, \mathrm{M}_\odot  } in the range of \ensuremath{ \, 5 < z <10} when the universe is less than 1 Gyr old.}
\begin{itemize}
\item  \textbf{OR 2.2.1} :  eLISA shall have the capability to detect the mergers of comparable mass massive black hole (mass ratio \ensuremath{m_2/m_1 > 0.1}) with total mass in the range \ensuremath{ 2 \times 10^4 \, \mathrm{M}_\odot <  m_{1} + m_{2} < 10^5  \, \mathrm{M}_\odot } beyond redhift \ensuremath{z = 5} and up to \ensuremath{ z = 15} for equal mass systems with sufficient SNR to enable determination of the massive black hole masses (relative errors smaller than 1 \%) and the spin of the largest massive black hole (error smaller than 0.1) and an estimation of the luminosity distance (relative error smaller than 50 \%).
\item  \textbf{OR 2.2.2} : eLISA shall have the capability to detect some of the mergers of massive black hole with total mass in the range \ensuremath{ 2 \times 10^4 \, \mathrm{M}_\odot <  m_{1} + m_{2} < 10^5  \, \mathrm{M}_\odot } and mass ratio \ensuremath{ 0.01 < m_{1} / m_{2} < 0.1} beyond redshift \ensuremath{ z = 5 } with sufficient SNR to enable determination of the massive black hole masses with relative errors smaller than a few percent.
\end{itemize}
\end{itemize}
\item \emph{\textbf{  Extreme (and intermediate) mass ratio inspiral }}
\begin{itemize}
\item \emph{\textbf{SR 3.1} :  Characterise the immediate environment of massive black hole in  \ensuremath{ z < 0.7 } galactic nuclei from EMRI capture signals. }
\begin{itemize}
\item  \textbf{OR 3.1} : eLISA shall have the capability to detect gravitational waves emitted during the last two years of inspiral for a stellar-mass compact object (\ensuremath{ m_{2} \sim 5 \, \mathrm{M}_{\odot}  - 20 \,  \mathrm{M}_{\odot} }) orbiting a massive black hole (\ensuremath{ m_{1} \sim 10^{5} \, \mathrm{M}_{\odot}  - 10^{6} \,  \mathrm{M}_{\odot} }) up to  \ensuremath{ z = 0.7 } with an  SNR > 20 . The detection of those sources shall be sufficient to determine the mass of the massive black hole with an relative error smaller than 0.1 \%, the spin of the massive black hole with an error smaller than  \ensuremath{ 10^{-3} }, and the mass of the compact object with a relative error smaller than 0.1 \%, as well as the orbital eccentricity before the plunge with an error smaller than  \ensuremath{10^{-3} }.
\end{itemize}
\item \emph{\textbf{SR 3.2}  : Discovery of intermediate-mass black holes from their captures by massive black hole.  }
\begin{itemize}
\item  \textbf{OR 3.2} : eLISA shall have the capability to detect gravitational waves emitted by a  \ensuremath{10^2 \, \mathrm{M}_\odot - 10^4 \mathrm{M}_\odot }   intermediate-mass black hole spiralling into an massive black hole with mass  \ensuremath{ 3 \times 10^5 \mathrm{M}_\odot - 10^7 \mathrm{M}_\odot }  out to \ensuremath{ z  \sim 2 - 4 } (for a mass ratio around \ensuremath{10^{-2}} to \ensuremath{10^{-3}}).
\end{itemize}
\end{itemize}
\item \emph{\textbf{ Confronting General Relativity with Precision Measurements of Strong Gravity }}
\begin{itemize}
\item \emph{\textbf{SR 4.1} :  Detect gravitational waves directly and measure their properties precisely. }
\begin{itemize}
\item  \textbf{OR 4.1.1} : eLISA shall have capability to detect and study three or more optically observable verification binaries between 1 mHz and 10 mHz with SNR > 10 in two years of mission lifetime.
\item  \textbf{OR 4.1.2} :  eLISA shall be capable of observing the gravitational waves from at least 50 \% of all $z \sim 2$ coalescing binary systems consisting of compact objects with masses between \ensuremath{ 10^{5} \, \mathrm{M}_{\odot} } and \ensuremath{ 10^{6} \, \mathrm{M}_{\odot} } and mass ratios between 1~:~1 and 1~:~3. eLISA shall detect these systems with SNR $\ge$ 5 in each of five equal logarithmic frequency bands between 0.1 mHz (or the lowest observed frequency) and the highest inspiral frequency.
\end{itemize}
\item \emph{\textbf{SR 4.2}  Test whether the central massive objects in galactic nuclei are consistent with the Kerr black holes of General Relativity. }
\begin{itemize}
\item  \textbf{OR 4.2} :  eLISA shall have the capability to detect gravitational waves emitted during the last year of inspiral for a \ensuremath{ 10 \, \mathrm{M}_{\odot} }  black hole orbiting a  \ensuremath{ 10^{5} \, \mathrm{M}_{\odot} } and \ensuremath{ 10^{6} \, \mathrm{M}_{\odot} } black hole up to \ensuremath{ z = 0.7} with SNR $>$ 20. eLISA shall have a science mission duration with adequate observation time for extreme mass-ratio inspirals (EMRIs) to sweep over a range of \ensuremath{r/M} to map space-time.
\end{itemize}
\item \emph{\textbf{SR 4.3}  : Perform precision tests of dynamical strong-field gravity. }
\begin{itemize}
\item  \textbf{OR 4.3.1} : eLISA shall have the capability to observe the inspiral radiation from massive black hole with masses between  \ensuremath{ 10^{5} \, \mathrm{M}_{\odot}  -  10^{6} \, \mathrm{M}_{\odot} } and mass ratio \ensuremath{m_2/m_1 > 1/3} to \ensuremath{z \le 5 } with an average SNR > 30, measuring the mass to better than 1 \% and spin parameters to better than 0.1. The SNR should be sufficient to check consistency of the inspiral waveform with the predictions of the General Theory of Relativity.
\item  \textbf{OR 4.3.2} : eLISA shall have the capability to observe the merger and ring-down radiation from massive black hole with masses between  \ensuremath{ 10^{5} \, \mathrm{M}_{\odot}  -  10^{6} \, \mathrm{M}_{\odot} }  and mass ratio \ensuremath{m_2/m_1 > 1/3} to  \ensuremath{z \le 8} with an average SNR > 60, measuring the mass to better than 1 \% and spin parameters to better than 0.3. The SNR should be sufficient to check consistency with the predictions of the General Theory of Relativity based on inspiral measurements.
\end{itemize}
\end{itemize}
\item \emph{\textbf{ Cosmology } }
\begin{itemize}
\item \emph{\textbf{SR 5.1}  : Measure the spectrum of cosmological backgrounds, or set upper limits on them in the \ensuremath{10^{-4} \textrm{Hz} -10^{-1} \textrm{Hz}} band. }
\begin{itemize}
\item  \textbf{OR 5.1} : eLISA shall be capable of setting an upper limit on the spectrum of a stochastic gravitational wave background in the  \ensuremath{10^{-4} \textrm{Hz} -10^{-1} \textrm{Hz}} band.
\end{itemize}
\item \emph{\textbf{SR 5.2} : Search for gravitational wave bursts from cosmic string cusps and kinks.  }
\begin{itemize}
\item  \textbf{OR 5.2} : eLISA shall be capable of detecting gravitational wave bursts from cusps or kinks, or of setting cosmologically interesting 
constraints on cosmic (super-)strings. 
\end{itemize}
\end{itemize}
\item \emph{\textbf{ Discovery }}
\begin{itemize}
\item \emph{\textbf{SR 6.1} : Search for unforeseen sources of gravitational waves  }
\begin{itemize}
\item  \textbf{OR 6.1} : eLISA shall be sensitive over discovery space for unforeseen effects (e.g. even at frequencies where we cannot predict likely signals from known classes of astrophysical sources). eLISA shall allow for reliable separation of real strain signals from instrumental and environmental artifacts.
\end{itemize}
\end{itemize}
\end{itemize}

\section*{Acknowledgements}

The following scientists contributed significantly through work regarding the
previous study (in alphabetical order): Berangere Argence (APC), Stuart Aston
(Birmingham), Gerard Auger (APC), John Baker (GSFC), Simon Barke (AEI), Matthew
Benacquista (UTB), Iouri Bykov (AEI), Martin Caldwell (RAL), Jordan Camp
(GSFC), John Conklin (Stanford), Dan deBra (Stanford), Luciano Di Fiore
(Naples), Christian Diekmann (AEI), Juan Jose Esteban Delgado (AEI), Roland
Fleddermann (AEI), Antonio Garcia (AEI), Catia Grimani (Urbino) Felipe Guzman
(AEI/GSFC), Hubert Halloin (APC), Tupper Hyde (GSFC), Ian Harris (JPL), Gerhard
Heinzel (AEI), Martin Hewitson (AEI) Steven Hochman (U Florida), Daniel
Hollington (Imperial College), Nick Jedrich (GSFC), Mac Keiser (Stanford),
Christian Killow (Glasgow), William Klipstein (JPL), Joachim Kullmann (AEI),
Jeffrey Livas (GSFC), Achmed Mansoor (GSFC), Kirk McKenzie (JPL), Stephen
Merkowitz (GSFC), Shawn Mitryk (U Florida), Anneke Monsky (AEI), Guido
M{\"u}ller (U Florida), Miquel Nofrarias (AEI), Kenji Numata (GSFC), Frank Ohme
(AEI), Eric Plagnol (APC), Moshe Pniel (JPL), Scott Pollack (University of
Washington), Alix Preston (UF), Volker Quetschke (UTB), Emma Robinson (AEI),
Dave Robertson (Glasgow), Albrecht R{\"u}diger (AEI), Josep Sanjuan (IEEC
Barcelona),  B. Sathyaprakash (U Cardiff), Daniel Shaddock (ANU), Diana Shaul
(Imperial College), Ben Sheard (AEI), Robert Spero (JPL), Frank Steier (AEI),
Ke-Xun Sun (Stanford), Dylan Sweeney (U Florida), David Tanner (UF), Michael
Troebs (AEI), Glenn de Vine (JPL), Vinzenz Wand (AEI), Gudrun Wanner (AEI),
Brent Ware (JPL), Peter Wass (U Trento/Imperial College), Bill Weber (U
Trento), Yinan Yu (U Florida), Alberto Vecchio (U Birmingham), and Andreas
Zoellner (Stanford).

We acknowledge the rest of the members of the former joint science team for
LISA, whose work constituted a crucial starting point for our article, even if
they are not included in the authorship list of this paper: Thomas A Prince
(JPL), Peter Bender (JILA), Sasha Buchman (Stanford University), Joan Centrella
(NASA GFSC), Massimo Cerdonio (INFN Padua), Mike Cruise (University of
Birmingham), Curt J Cutler (JPL), Lee Sam Finn (PennState University), Jens
Gundlach (University of Washington), Craig Hogan (Fermilab), Jim Hough
(University of Glasgow), Scott A. Hughes (MIT), Piero Madau (UC Santa Cruz),
Yannick Mellier (IAP), Sterl Phinney (Caltech), Douglas O. Richstone
(University of Michigan), Kip Thorne (Caltech), and Jean-Yves Vinet
(Observatoire de C{\^o}te d'Azur).

It is a pleasure for P.A.S to thank Ant{\'o}n for the very interesting discussions
about black holes and astrophysics.  This work has been supported by the
Transregio 7 ``Gravitational Wave Astronomy'' financed by the Deutsche
Forschungsgemeinschaft DFG (German Research Foundation). E.Berti was supported
by NSF Grant PHY-0900735 and by NSF CAREER Grant PHY-1055103.  A.  Klein was
supported by the Swiss National Science Foundation. T. B. Littenberg was
supported by NASA Grant 08-ATFP08-0126. R. N.  Lang was supported by an
appointment to the NASA Postdoctoral Program at the Goddard Space Flight
Center, administered by Oak Ridge Associated Universities through a contract
with NASA.  M. Vallisneri performed this work at the Jet Propulsion Laboratory,
California Institute of Technology, under contract with the National
Aeronautics and Space Administration.

\label{lastpage}

\newpage


\begin{thebibliography}{}

\bibitem[{Abel} et~al., 2002]{Abel:2002}
{Abel}, T., {Bryan}, G.~L., and {Norman}, M.~L. (2002).
\newblock {The Formation of the First Star in the Universe}.
\newblock {\em Science}, 295:93--98.

\bibitem[{Alexander}, 2005]{alexander:2005:spmbh}
{Alexander}, T. (2005).
\newblock {Stellar processes near the massive black hole in the Galactic center
  [review article]}.
\newblock {\em Phys.~Rep.}, 419:65--142.

\bibitem[{Aller} and {Richstone}, 2002]{Aller:2002}
{Aller}, M.~C. and {Richstone}, D. (2002).
\newblock {The Cosmic Density of Massive Black Holes from Galaxy Velocity
  Dispersions}.
\newblock {\em AJ}, 124:3035--3041.

\bibitem[Alsing et~al., 2011]{Alsing:2011erarXiv1112.4903}
Alsing, J., Berti, E., Will, C., and Zaglauer, H. (2011).
\newblock {Gravitational radiation from compact binary systems in the massive
  Brans-Dicke theory of gravity}.
\newblock {\em ArXiv e-prints}.

\bibitem[{Amaro-Seoane}, 2006]{2006AIPC..873..250A}
{Amaro-Seoane}, P. (2006).
\newblock {Gravitational waves from coalescing massive black holes in young
  dense clusters}.
\newblock In {S.~M.~Merkovitz \& J.~C.~Livas}, editor, {\em Laser
  Interferometer Space Antenna: 6th International LISA Symposium}, volume 873
  of {\em American Institute of Physics Conference Series}, pages 250--256.

\bibitem[{Amaro-Seoane} et~al., 2012]{2012ApJ...744L..20A}
{Amaro-Seoane}, P., {Brem}, P., {Cuadra}, J., and {Armitage}, P.~J. (2012).
\newblock {The Butterfly Effect in the Extreme-mass Ratio Inspiral Problem}.
\newblock {\em ApJ Letts}, 744:L20.

\bibitem[{Amaro-Seoane} et~al., 2010a]{amaro-seoane:2010:mbhb}
{Amaro-Seoane}, P., {Eichhorn}, C., {Porter}, E.~K., and {Spurzem}, R. (2010a).
\newblock {Binaries of massive black holes in rotating clusters: dynamics,
  gravitational waves, detection and the role of eccentricity}.
\newblock {\em MNRAS}, 401:2268--2284.

\bibitem[{Amaro-Seoane} et~al., 2010b]{2010MNRAS.401.2268A}
{Amaro-Seoane}, P., {Eichhorn}, C., {Porter}, E.~K., and {Spurzem}, R. (2010b).
\newblock {Binaries of massive black holes in rotating clusters: dynamics,
  gravitational waves, detection and the role of eccentricity}.
\newblock {\em MNRAS}, 401:2268--2284.

\bibitem[{Amaro-Seoane} and {Freitag}, 2006a]{amaro-seoane:2006ApJ...653L..53A}
{Amaro-Seoane}, P. and {Freitag}, M. (2006a).
\newblock {Intermediate-Mass Black Holes in Colliding Clusters: Implications
  for Lower Frequency Gravitational-Wave Astronomy}.
\newblock {\em ApJ}, 653:L53--L56.

\bibitem[{Amaro-Seoane} and {Freitag}, 2006b]{2006ApJ...653L..53A}
{Amaro-Seoane}, P. and {Freitag}, M. (2006b).
\newblock {Intermediate-Mass Black Holes in Colliding Clusters: Implications
  for Lower Frequency Gravitational-Wave Astronomy}.
\newblock {\em ApJ Letts}, 653:L53--L56.

\bibitem[{Amaro-Seoane} et~al., 2007]{amaro-soane:2007:tr}
{Amaro-Seoane}, P., {Gair}, J.~R., {Freitag}, M., {Miller}, M.~C., {Mandel},
  I., {Cutler}, C.~J., and {Babak}, S. (2007).
\newblock {TOPICAL REVIEW: Intermediate and extreme mass-ratio inspirals ---
  astrophysics, science applications and detection using LISA}.
\newblock {\em Class.~Quantum~Grav.}, 24:113--+.

\bibitem[{Amaro-Seoane} et~al., 2009]{2009ApJ...692L..50A}
{Amaro-Seoane}, P., {Miller}, M.~C., and {Freitag}, M. (2009).
\newblock {Gravitational Waves from Eccentric Intermediate-Mass Black Hole
  Binaries}.
\newblock {\em ApJ Letts}, 692:L50--L53.

\bibitem[{Amaro-Seoane} and {Preto}, 2011]{amaro-seoane:2011CQGra..28i4017A}
{Amaro-Seoane}, P. and {Preto}, M. (2011).
\newblock {The impact of realistic models of mass segregation on the event rate
  of extreme-mass ratio inspirals and cusp re-growth}.
\newblock {\em Class.~Quantum~Grav.}, 28(9):094017--+.

\bibitem[{Amaro-Seoane} and {Santamar{\'{\i}}a},
  2010]{amaro-seoane:2010ApJ...722.1197A}
{Amaro-Seoane}, P. and {Santamar{\'{\i}}a}, L. (2010).
\newblock {Detection of IMBHs with Ground-based Gravitational Wave
  Observatories: A Biography of a Binary of Black Holes, from Birth to Death}.
\newblock {\em ApJ}, 722:1197--1206.

\bibitem[{Armitage} and {Natarajan}, 2002]{armitage:2002ApJ...567L...9A}
{Armitage}, P.~J. and {Natarajan}, P. (2002).
\newblock {Accretion during the Merger of Supermassive Black Holes}.
\newblock {\em ApJ}, 567:L9--L12.

\bibitem[{Armitage} and {Natarajan}, 2005]{armitage:2005:esb}
{Armitage}, P.~J. and {Natarajan}, P. (2005).
\newblock {Eccentricity of Supermassive Black Hole Binaries Coalescing from
  Gas-rich Mergers}.
\newblock {\em ApJ}, 634:921--927.

\bibitem[{Arun} et~al., 2009]{arun:2009:petf}
{Arun}, K.~G., {Babak}, S., {Berti}, E., {Cornish}, N., {Cutler}, C., {Gair},
  J., {Hughes}, S.~A., {Iyer}, B.~R., {Lang}, R.~N., {Mandel}, I., {Porter},
  E.~K., {Sathyaprakash}, B.~S., {Sinha}, S., {Sintes}, A.~M., {Trias}, M.,
  {Van Den Broeck}, C., and {Volonteri}, M. (2009).
\newblock {Massive black-hole binary inspirals: results from the LISA parameter
  estimation taskforce}.
\newblock {\em Class.~Quantum~Grav.}, 26(9):094027--+.

\bibitem[{Arun} et~al., 2006]{arun:2006:pns}
{Arun}, K.~G., {Iyer}, B.~R., {Qusailah}, M.~S.~S., and {Sathyaprakash}, B.~S.
  (2006).
\newblock {Probing the nonlinear structure of general relativity with black
  hole binaries}.
\newblock {\em Phys.~Rev.~D}, 74(2):024006--+.

\bibitem[{Arvanitaki} and {Dubovsky}, 2011]{arvanitaki:2011PhRvD..83d4026A}
{Arvanitaki}, A. and {Dubovsky}, S. (2011).
\newblock {Exploring the string axiverse with precision black hole physics}.
\newblock {\em Phys.~Rev.~D}, 83(4):044026--+.

\bibitem[{Babak} et~al., 2010]{babak:2010:mldc}
{Babak}, S., {Baker}, J.~G., {Benacquista}, M.~J., {Cornish}, N.~J., {Larson},
  S.~L., {Mandel}, I., {McWilliams}, S.~T., {Petiteau}, A., {Porter}, E.~K.,
  {Robinson}, E.~L., {Vallisneri}, M., {Vecchio}, A., {Data Challenge Task
  Force}, t.~M.~L., {Adams}, M., {Arnaud}, K.~A., {B{\l}aut}, A., {Bridges},
  M., {Cohen}, M., {Cutler}, C., {Feroz}, F., {Gair}, J.~R., {Graff}, P.,
  {Hobson}, M., {Shapiro Key}, J., {Kr{\'o}lak}, A., {Lasenby}, A., {Prix}, R.,
  {Shang}, Y., {Trias}, M., {Veitch}, J., {Whelan}, J.~T., and {participants},
  t.~C.~. (2010).
\newblock {The Mock LISA Data Challenges: from challenge 3 to challenge 4}.
\newblock {\em Class.~Quantum~Grav.}, 27(8):084009--+.

\bibitem[{Babak} et~al., 2011]{babak:2011CQGra..28k4001B}
{Babak}, S., {Gair}, J.~R., {Petiteau}, A., and {Sesana}, A. (2011).
\newblock {Fundamental physics and cosmology with LISA}.
\newblock {\em Class.~Quantum~Grav.}, 28(11):114001--+.

\bibitem[{Bahcall} and {Wolf}, 1976]{bahcall:1976ApJ...209..214B}
{Bahcall}, J.~N. and {Wolf}, R.~A. (1976).
\newblock {Star distribution around a massive black hole in a globular
  cluster}.
\newblock {\em ApJ}, 209:214--232.

\bibitem[{Bahcall} and {Wolf}, 1977]{bahcall:1977ApJ...216..883B}
{Bahcall}, J.~N. and {Wolf}, R.~A. (1977).
\newblock {The star distribution around a massive black hole in a globular
  cluster. II Unequal star masses}.
\newblock {\em ApJ}, 216:883--907.

\bibitem[{Baker} et~al., 2008]{baker:2008ApJ...682L..29B}
{Baker}, J.~G., {Boggs}, W.~D., {Centrella}, J., {Kelly}, B.~J., {McWilliams},
  S.~T., {Miller}, M.~C., and {van Meter}, J.~R. (2008).
\newblock {Modeling Kicks from the Merger of Generic Black Hole Binaries}.
\newblock {\em ApJ}, 682:L29--L32.

\bibitem[{Baker} et~al., 2006]{baker:2006:gwe}
{Baker}, J.~G., {Centrella}, J., {Choi}, D.-I., {Koppitz}, M., and {van Meter},
  J. (2006).
\newblock {Gravitational-Wave Extraction from an Inspiraling Configuration of
  Merging Black Holes}.
\newblock {\em Phys.~Rev.~Lett.}, 96(11):111102--+.

\bibitem[{Ballo} et~al., 2004]{ballo:2004ApJ...600..634B}
{Ballo}, L., {Braito}, V., {Della Ceca}, R., {Maraschi}, L., {Tavecchio}, F.,
  and {Dadina}, M. (2004).
\newblock {Arp 299: A Second Merging System with Two Active Nuclei?}
\newblock {\em ApJ}, 600:634--639.

\bibitem[{Barack} and {Cutler}, 2004]{barack:2004:lcs}
{Barack}, L. and {Cutler}, C. (2004).
\newblock {LISA capture sources: Approximate waveforms, signal-to-noise ratios,
  and parameter estimation accuracy}.
\newblock {\em Phys.~Rev.~D}, 69:082005--+.

\bibitem[{Barausse} and {Rezzolla}, 2008]{barausse:2008PhRvD..77j4027B}
{Barausse}, E. and {Rezzolla}, L. (2008).
\newblock {Influence of the hydrodynamic drag from an accretion torus on
  extreme mass-ratio inspirals}.
\newblock {\em Phys.~Rev.~D}, 77(10):104027--+.

\bibitem[{Barausse} et~al., 2007]{barausse:2007PhRvD..75f4026B}
{Barausse}, E., {Rezzolla}, L., {Petroff}, D., and {Ansorg}, M. (2007).
\newblock {Gravitational waves from extreme mass ratio inspirals in nonpure
  Kerr spacetimes}.
\newblock {\em Phys.~Rev.~D}, 75(6):064026--+.

\bibitem[Barausse and Sotiriou, 2008]{barausse:2008xv}
Barausse, E. and Sotiriou, T.~P. (2008).
\newblock {Perturbed Kerr Black Holes can probe deviations from General
  Relativity}.
\newblock {\em Phys.Rev.Lett.}, 101:099001.

\bibitem[{Barclay} et~al., 2011]{barclay:2011MNRAS.413.2696B}
{Barclay}, T., {Ramsay}, G., {Hakala}, P., {Napiwotzki}, R., {Nelemans}, G.,
  {Potter}, S., and {Todd}, I. (2011).
\newblock {Stellar variability on time-scales of minutes: results from the
  first 5 yr of the Rapid Temporal Survey}.
\newblock {\em MNRAS}, 413:2696--2708.

\bibitem[{Bardeen}, 1970]{bardeen:1970Natur.226...64B}
{Bardeen}, J.~M. (1970).
\newblock {Kerr Metric Black Holes}.
\newblock {\em Nature}, 226:64--65.

\bibitem[{Bardeen} and {Petterson}, 1975]{bardeen:1975ApJ...195L..65B}
{Bardeen}, J.~M. and {Petterson}, J.~A. (1975).
\newblock {The Lense-Thirring Effect and Accretion Disks around Kerr Black
  Holes}.
\newblock {\em ApJ}, 195:L65+.

\bibitem[{Barrows} et~al., 2011]{barrows:2011NewA...16..122B}
{Barrows}, R.~S., {Lacy}, C.~H.~S., {Kennefick}, D., {Kennefick}, J., and
  {Seigar}, M.~S. (2011).
\newblock {Unusual double-peaked emission in the SDSS quasar J093201.60 +
  031858.7}.
\newblock {\em NewA}, 16:122--127.

\bibitem[{Barth} et~al., 2004]{barth:2004ApJ...607...90B}
{Barth}, A.~J., {Ho}, L.~C., {Rutledge}, R.~E., and {Sargent}, W.~L.~W. (2004).
\newblock {POX 52: A Dwarf Seyfert 1 Galaxy with an Intermediate-Mass Black
  Hole}.
\newblock {\em ApJ}, 607:90--102.

\bibitem[{Barth} et~al., 2009]{barth:2009ApJ...690.1031B}
{Barth}, A.~J., {Strigari}, L.~E., {Bentz}, M.~C., {Greene}, J.~E., and {Ho},
  L.~C. (2009).
\newblock {Dynamical Constraints on the Masses of the Nuclear Star Cluster and
  Black Hole in the Late-Type Spiral Galaxy NGC 3621}.
\newblock {\em ApJ}, 690:1031--1044.

\bibitem[{Begelman} et~al., 1980]{begelman:1980Natur.287..307B}
{Begelman}, M.~C., {Blandford}, R.~D., and {Rees}, M.~J. (1980).
\newblock {Massive black hole binaries in active galactic nuclei}.
\newblock {\em Nature}, 287:307--309.

\bibitem[{Begelman} and {Shlosman}, 2009]{begelman:2009ApJ...702L...5B}
{Begelman}, M.~C. and {Shlosman}, I. (2009).
\newblock {Angular Momentum Transfer and Lack of Fragmentation in
  Self-Gravitating Accretion Flows}.
\newblock {\em ApJ}, 702:L5--L8.

\bibitem[{Begelman} et~al., 2006]{begelman:2006:fsm}
{Begelman}, M.~C., {Volonteri}, M., and {Rees}, M.~J. (2006).
\newblock {Formation of supermassive black holes by direct collapse in
  pre-galactic haloes}.
\newblock {\em MNRAS}, 370:289--298.

\bibitem[{Bekki} and {Graham}, 2010]{bekki:2010ApJ...714L.313B}
{Bekki}, K. and {Graham}, A.~W. (2010).
\newblock {On the Transition from Nuclear-cluster- to Black-hole-dominated
  Galaxy Cores}.
\newblock {\em ApJ}, 714:L313--L317.

\bibitem[{Belczynski} et~al., 2010]{belczynski:2010:dco}
{Belczynski}, K., {Benacquista}, M., and {Bulik}, T. (2010).
\newblock {Double Compact Objects as Low-frequency Gravitational Wave Sources}.
\newblock {\em ApJ}, 725:816--823.

\bibitem[{Belczynski} et~al., 2002]{belczynski:2002:bco}
{Belczynski}, K., {Kalogera}, V., and {Bulik}, T. (2002).
\newblock {A Comprehensive Study of Binary Compact Objects as Gravitational
  Wave Sources: Evolutionary Channels, Rates, and Physical Properties}.
\newblock {\em ApJ}, 572:407--431.

\bibitem[{Bell} et~al., 2006]{Bell:2006:geg}
{Bell}, E.~F., {Phleps}, S., {Somerville}, R.~S., {Wolf}, C., {Borch}, A., and
  {Meisenheimer}, K. (2006).
\newblock {The Merger Rate of Massive Galaxies}.
\newblock {\em ApJ}, 652:270--276.

\bibitem[{Benacquista} and {Holley-Bockelmann}, 2006]{benacquista:2006:cds}
{Benacquista}, M. and {Holley-Bockelmann}, K. (2006).
\newblock {Consequences of Disk Scale Height on LISA Confusion Noise from Close
  White Dwarf Binaries}.
\newblock {\em ApJ}, 645:589--596.

\bibitem[{Berczik} et~al., 2005]{berczik:2005ApJ...633..680B}
{Berczik}, P., {Merritt}, D., and {Spurzem}, R. (2005).
\newblock {Long-Term Evolution of Massive Black Hole Binaries. II. Binary
  Evolution in Low-Density Galaxies}.
\newblock {\em ApJ}, 633:680--687.

\bibitem[{Berczik} et~al., 2006]{berczik:2006:embsbh}
{Berczik}, P., {Merritt}, D., {Spurzem}, R., and {Bischof}, H. (2006).
\newblock {Efficient Merger of Binary Supermassive Black Holes in
  Nonaxisymmetric Galaxies}.
\newblock {\em ApJ}, 642:L21--L24.

\bibitem[{Berti}, 2006]{berti:2006:lom}
{Berti}, E. (2006).
\newblock {LISA observations of massive black hole mergers: event rates and
  issues in waveform modelling}.
\newblock {\em Class.~Quantum~Grav.}, 23:785--+.

\bibitem[{Berti} et~al., 2005]{berti:2005:esb}
{Berti}, E., {Buonanno}, A., and {Will}, C.~M. (2005).
\newblock {Estimating spinning binary parameters and testing alternative
  theories of gravity with LISA}.
\newblock {\em Phys.~Rev.~D}, 71:084025--+.

\bibitem[{Berti} et~al., 2007]{berti:2007PhRvD..76j4044B}
{Berti}, E., {Cardoso}, J., {Cardoso}, V., and {Cavagli{\`a}}, M. (2007).
\newblock {Matched filtering and parameter estimation of ringdown waveforms}.
\newblock {\em Phys.~Rev.~D}, 76(10):104044--+.

\bibitem[Berti and Cardoso, 2006]{berti:2006:boson}
Berti, E. and Cardoso, V. (2006).
\newblock {Supermassive black holes or boson stars? Hair counting with
  gravitational wave detectors}.
\newblock {\em Int.J.Mod.Phys.}, D15:2209--2216.

\bibitem[{Berti} et~al., 2009]{berti:2009CQGra..26p3001B}
{Berti}, E., {Cardoso}, V., and {Starinets}, A.~O. (2009).
\newblock {TOPICAL REVIEW: Quasinormal modes of black holes and black branes}.
\newblock {\em Class.~Quantum~Grav.}, 26(16):163001.

\bibitem[{Berti} et~al., 2006]{berti:2006:gws}
{Berti}, E., {Cardoso}, V., and {Will}, C.~M. (2006).
\newblock {Gravitational-wave spectroscopy of massive black holes with the
  space interferometer LISA}.
\newblock {\em Phys.~Rev.~D}, 73:064030--+.

\bibitem[{Berti} et~al., 2011]{berti:2011arXiv1107.3528B}
{Berti}, E., {Gair}, J., and {Sesana}, A. (2011).
\newblock {Graviton mass bounds from space-based gravitational-wave
  observations of massive black hole populations}.
\newblock {\em ArXiv e-prints}.

\bibitem[{Berti} and {Volonteri}, 2008]{berti:2008:cbhse}
{Berti}, E. and {Volonteri}, M. (2008).
\newblock {Cosmological Black Hole Spin Evolution by Mergers and Accretion}.
\newblock {\em ApJ}, 684:822--828.

\bibitem[{Bertone} et~al., 2007]{bertone:2007:rgm}
{Bertone}, S., {De Lucia}, G., and {Thomas}, P.~A. (2007).
\newblock {The recycling of gas and metals in galaxy formation: predictions of
  a dynamical feedback model}.
\newblock {\em MNRAS}, 379:1143--1154.

\bibitem[{Bhattacharya} and {van den Heuvel}, 1991]{bhattacharya:1991:febmps}
{Bhattacharya}, D. and {van den Heuvel}, E.~P.~J. (1991).
\newblock {Formation and evolution of binary and millisecond radio pulsars}.
\newblock {\em Phys.~Rep.}, 203:1--124.

\bibitem[{Bildsten}, 1998]{bildsten:1998:grr}
{Bildsten}, L. (1998).
\newblock {Gravitational Radiation and Rotation of Accreting Neutron Stars}.
\newblock {\em ApJ}, 501:L89+.

\bibitem[{Bin{\'e}truy} et~al., 2012]{2012arXiv1201.0983B}
{Bin{\'e}truy}, P., {Boh{\'e}}, A., {Caprini}, C., and {Dufaux}, J.-F. (2012).
\newblock {Cosmological Backgrounds of Gravitational Waves and eLISA/NGO: Phase
  Transitions, Cosmic Strings and Other Sources}.
\newblock {\em ArXiv e-prints}.

\bibitem[{Bin{\'e}truy} et~al., 2010]{binetruy:2010PhRvD..82l6007B}
{Bin{\'e}truy}, P., {Boh{\'e}}, A., {Hertog}, T., and {Steer}, D.~A. (2010).
\newblock {Gravitational wave signatures from kink proliferation on cosmic
  (super-) strings}.
\newblock {\em Phys.~Rev.~D}, 82(12):126007--+.

\bibitem[Blanchet, 2006]{blanchet:2006:grp}
Blanchet, L. (2006).
\newblock Gravitational radiation from post-newtonian sources and inspiralling
  compact binaries.
\newblock {\em Living Reviews in Relativity}, 9(4).

\bibitem[{Blecha} et~al., 2011]{blecha:2011MNRAS.412.2154B}
{Blecha}, L., {Cox}, T.~J., {Loeb}, A., and {Hernquist}, L. (2011).
\newblock {Recoiling black holes in merging galaxies: relationship to active
  galactic nucleus lifetimes, starbursts and the M$_{BH}$-{$\sigma$}$_{*}$
  relation}.
\newblock {\em MNRAS}, 412:2154--2182.

\bibitem[{Blecha} and {Loeb}, 2008]{blecha:2008MNRAS.390.1311B}
{Blecha}, L. and {Loeb}, A. (2008).
\newblock {Effects of gravitational-wave recoil on the dynamics and growth of
  supermassive black holes}.
\newblock {\em MNRAS}, 390:1311--1325.

\bibitem[{Bogdanovi{\'c}} et~al., 2009]{bogdanoovic:2009ApJ...697..288B}
{Bogdanovi{\'c}}, T., {Eracleous}, M., and {Sigurdsson}, S. (2009).
\newblock {SDSS J092712.65+294344.0: Recoiling Black Hole or a Subparsec Binary
  Candidate?}
\newblock {\em ApJ}, 697:288--292.

\bibitem[{Bogdanovi{\'c}} et~al., 2008]{bogdanovic:2008ApJS..174..455B}
{Bogdanovi{\'c}}, T., {Smith}, B.~D., {Sigurdsson}, S., and {Eracleous}, M.
  (2008).
\newblock {Modeling of Emission Signatures of Massive Black Hole Binaries. I.
  Methods}.
\newblock {\em ApJS}, 174:455--480.

\bibitem[{Boh{\'e}}, 2011]{bohe:2011PhRvD..84f5016B}
{Boh{\'e}}, A. (2011).
\newblock {Gravitational power from cosmic string loops with many kinks}.
\newblock {\em Phys.~Rev.~D}, 84(6):065016--+.

\bibitem[{Boroson} and {Lauer}, 2009]{boroson:2009:subpcbh}
{Boroson}, T.~A. and {Lauer}, T.~R. (2009).
\newblock {A candidate sub-parsec supermassive binary black hole system}.
\newblock {\em Nature}, 458:53--55.

\bibitem[{Boylan-Kolchin} et~al., 2004]{boylankolchin:2004ApJ...613L..37B}
{Boylan-Kolchin}, M., {Ma}, C.-P., and {Quataert}, E. (2004).
\newblock {Core Formation in Galactic Nuclei due to Recoiling Black Holes}.
\newblock {\em ApJ}, 613:L37--L40.

\bibitem[{Boyle} and {Terlevich}, 1998]{boyle:1998MNRAS.293L..49B}
{Boyle}, B.~J. and {Terlevich}, R.~J. (1998).
\newblock {The cosmological evolution of the QSO luminosity density and of the
  star formation rate}.
\newblock {\em MNRAS}, 293:L49--L51.

\bibitem[{Bromm} et~al., 2002]{bromm:2002:ffs}
{Bromm}, V., {Coppi}, P.~S., and {Larson}, R.~B. (2002).
\newblock {The Formation of the First Stars. I. The Primordial Star-forming
  Cloud}.
\newblock {\em ApJ}, 564:23--51.

\bibitem[{Bromm} and {Loeb}, 2003]{bromm:2003:ffs}
{Bromm}, V. and {Loeb}, A. (2003).
\newblock {Formation of the First Supermassive Black Holes}.
\newblock {\em ApJ}, 596:34--46.

\bibitem[{Brown} et~al., 2009]{brown:2009:asd}
{Brown}, W.~R., {Geller}, M.~J., {Kenyon}, S.~J., and {Bromley}, B.~C. (2009).
\newblock {The Anisotropic Spatial Distribution of Hypervelocity Stars}.
\newblock {\em ApJ}, 690:L69--L71.

\bibitem[{Brown} et~al., 2011]{brown:2011ApJ...737L..23B}
{Brown}, W.~R., {Kilic}, M., {Hermes}, J.~J., {Allende Prieto}, C., {Kenyon},
  S.~J., and {Winget}, D.~E. (2011).
\newblock {A 12 Minute Orbital Period Detached White Dwarf Eclipsing Binary}.
\newblock {\em ApJ}, 737:L23+.

\bibitem[{Brustein} et~al., 1995]{brustein:1995:rgw}
{Brustein}, R., {Gasperini}, M., {Giovannini}, M., and {Veneziano}, G. (1995).
\newblock {Relic gravitational waves from string cosmology}.
\newblock {\em Physics Letters B}, 361:45--51.

\bibitem[{Bullock} et~al., 2001]{bullock:2001ApJ...555..240B}
{Bullock}, J.~S., {Dekel}, A., {Kolatt}, T.~S., {Kravtsov}, A.~V., {Klypin},
  A.~A., {Porciani}, C., and {Primack}, J.~R. (2001).
\newblock {A Universal Angular Momentum Profile for Galactic Halos}.
\newblock {\em ApJ}, 555:240--257.

\bibitem[{Buonanno}, 2003]{buonanno:2003:tasi}
{Buonanno}, A. (2003).
\newblock {TASI Lectures on Gravitational Waves from the Early Universe}.
\newblock {\em ArXiv e-prints}.

\bibitem[{Buonanno} et~al., 1997]{buonanno:1997:srg}
{Buonanno}, A., {Maggiore}, M., and {Ungarelli}, C. (1997).
\newblock {Spectrum of relic gravitational waves in string cosmology}.
\newblock {\em Phys.~Rev.~D}, 55:3330--3336.

\bibitem[{Callegari} et~al., 2011]{callegari:2011ApJ...729...85C}
{Callegari}, S., {Kazantzidis}, S., {Mayer}, L., {Colpi}, M., {Bellovary},
  J.~M., {Quinn}, T., and {Wadsley}, J. (2011).
\newblock {Growing Massive Black Hole Pairs in Minor Mergers of Disk Galaxies}.
\newblock {\em ApJ}, 729:85--+.

\bibitem[{Callegari} et~al., 2009]{callegari:2009ApJ...696L..89C}
{Callegari}, S., {Mayer}, L., {Kazantzidis}, S., {Colpi}, M., {Governato}, F.,
  {Quinn}, T., and {Wadsley}, J. (2009).
\newblock {Pairing of Supermassive Black Holes in Unequal-Mass Galaxy Mergers}.
\newblock {\em ApJ}, 696:L89--L92.

\bibitem[{Campanelli} et~al., 2006]{campanelli:2006:aeo}
{Campanelli}, M., {Lousto}, C.~O., {Marronetti}, P., and {Zlochower}, Y.
  (2006).
\newblock {Accurate Evolutions of Orbiting Black-Hole Binaries without
  Excision}.
\newblock {\em Phys.~Rev.~Lett.}, 96(11):111101--+.

\bibitem[{Campanelli} et~al., 2007]{campanelli:2007PhRvL..98w1102C}
{Campanelli}, M., {Lousto}, C.~O., {Zlochower}, Y., and {Merritt}, D. (2007).
\newblock {Maximum Gravitational Recoil}.
\newblock {\em Phys.~Rev.~Lett.}, 98(23):231102--+.

\bibitem[Caprini et~al., 2009]{Caprini:2009yp}
Caprini, C., Durrer, R., and Servant, G. (2009).
\newblock {The stochastic gravitational wave background from turbulence and
  magnetic fields generated by a first-order phase transition}.
\newblock {\em JCAP}, 0912:024.

\bibitem[Cardoso et~al., 2011]{Cardoso:2011xi}
Cardoso, V., Chakrabarti, S., Pani, P., Berti, E., and Gualtieri, L. (2011).
\newblock {Floating and sinking: The Imprint of massive scalars around rotating
  black holes}.
\newblock {\em Phys.Rev.Lett.}, 107:241101.

\bibitem[{Centrella} et~al., 2010]{centrella:2010ARNPS..60...75C}
{Centrella}, J., {Baker}, J.~G., {Kelly}, B.~J., and {van Meter}, J.~R. (2010).
\newblock {The Final Merger of Black-Hole Binaries}.
\newblock {\em Annual Review of Nuclear and Particle Science}, 60:75--100.

\bibitem[{Chakrabarty} et~al., 2003]{chkrabarty:2003:npms}
{Chakrabarty}, D., {Morgan}, E.~H., {Muno}, M.~P., {Galloway}, D.~K.,
  {Wijnands}, R., {van der Klis}, M., and {Markwardt}, C.~B. (2003).
\newblock {Nuclear-powered millisecond pulsars and the maximum spin frequency
  of neutron stars}.
\newblock {\em Nature}, 424:42--44.

\bibitem[{Chandrasekhar}, 1943]{chandrasekhar:1943:dfI}
{Chandrasekhar}, S. (1943).
\newblock {Dynamical Friction. I. General Considerations: the Coefficient of
  Dynamical Friction.}
\newblock {\em ApJ}, 97:255--+.

\bibitem[{Chandrasekhar} and {Detweiler}, 1975]{chandrasekhar:1975:qnm}
{Chandrasekhar}, S. and {Detweiler}, S. (1975).
\newblock {The quasi-normal modes of the Schwarzschild black hole}.
\newblock {\em Royal Soc. of London Proc. Series A}, 344:441--452.

\bibitem[{Chirenti} and {Rezzolla}, 2007]{chirenti:2007CQGra..24.4191C}
{Chirenti}, C.~B.~M.~H. and {Rezzolla}, L. (2007).
\newblock {How to tell a gravastar from a black hole}.
\newblock {\em Class.~Quantum~Grav.}, 24:4191--4206.

\bibitem[{Chongchitnan} and {Efstathiou}, 2006]{chongchitnan:2006:pdd}
{Chongchitnan}, S. and {Efstathiou}, G. (2006).
\newblock {Prospects for direct detection of primordial gravitational waves}.
\newblock {\em Phys.~Rev.~D}, 73(8):083511--+.

\bibitem[{Ciotti} et~al., 2010]{ciotti:2010ApJ...717..708C}
{Ciotti}, L., {Ostriker}, J.~P., and {Proga}, D. (2010).
\newblock {Feedback from Central Black Holes in Elliptical Galaxies. III.
  Models with Both Radiative and Mechanical Feedback}.
\newblock {\em ApJ}, 717:708--723.

\bibitem[{Civano} et~al., 2010]{civano:2010ApJ...717..209C}
{Civano}, F., {Elvis}, M., {Lanzuisi}, G., {Jahnke}, K., {Zamorani}, G.,
  {Blecha}, L., {Bongiorno}, A., {Brusa}, M., {Comastri}, A., {Hao}, H.,
  {Leauthaud}, A., {Loeb}, A., {Mainieri}, V., {Piconcelli}, E., {Salvato}, M.,
  {Scoville}, N., {Trump}, J., {Vignali}, C., {Aldcroft}, T., {Bolzonella}, M.,
  {Bressert}, E., {Finoguenov}, A., {Fruscione}, A., {Koekemoer}, A.~M.,
  {Cappelluti}, N., {Fiore}, F., {Giodini}, S., {Gilli}, R., {Impey}, C.~D.,
  {Lilly}, S.~J., {Lusso}, E., {Puccetti}, S., {Silverman}, J.~D., {Aussel},
  H., {Capak}, P., {Frayer}, D., {Le Floch}, E., {McCracken}, H.~J., {Sanders},
  D.~B., {Schiminovich}, D., and {Taniguchi}, Y. (2010).
\newblock {A Runaway Black Hole in COSMOS: Gravitational Wave or Slingshot
  Recoil?}
\newblock {\em ApJ}, 717:209--222.

\bibitem[{Clark} et~al., 2011]{clark:2011ApJ...727..110C}
{Clark}, P.~C., {Glover}, S.~C.~O., {Klessen}, R.~S., and {Bromm}, V. (2011).
\newblock {Gravitational Fragmentation in Turbulent Primordial Gas and the
  Initial Mass Function of Population III Stars}.
\newblock {\em ApJ}, 727:110--+.

\bibitem[{Collins} and {Hughes}, 2004]{collins:2004:tfm}
{Collins}, N.~A. and {Hughes}, S.~A. (2004).
\newblock {Towards a formalism for mapping the spacetimes of massive compact
  objects: Bumpy black holes and their orbits}.
\newblock {\em Phys.~Rev.~D}, 69(12):124022--+.

\bibitem[Colpi and Dotti, 2011]{colpi:2011:asl.2010.1210}
Colpi, M. and Dotti, M. (2011).
\newblock Massive binary black holes in the cosmic landscape.
\newblock {\em Advanced Science Letters}, 4(2):181--203.

\bibitem[{Colpi} et~al., 1999]{colpi:1999ApJ...525..720C}
{Colpi}, M., {Mayer}, L., and {Governato}, F. (1999).
\newblock {Dynamical Friction and the Evolution of Satellites in Virialized
  Halos: The Theory of Linear Response}.
\newblock {\em ApJ}, 525:720--733.

\bibitem[{Colpi} et~al., 1986]{colpi:1986PhRvL..57.2485C}
{Colpi}, M., {Shapiro}, S.~L., and {Wasserman}, I. (1986).
\newblock {Boson stars - Gravitational equilibria of self-interacting scalar
  fields}.
\newblock {\em Phys.~Rev.~Lett.}, 57:2485--2488.

\bibitem[{Copeland} et~al., 2004]{copeland:2004JHEP...06..013C}
{Copeland}, E.~J., {Myers}, R.~C., and {Polchinski}, J. (2004).
\newblock {Cosmic F- and D-strings}.
\newblock {\em Journal of High Energy Physics}, 6:13--+.

\bibitem[{Cornish} et~al., 2011]{cornish:2011PhRvD..84f2003C}
{Cornish}, N., {Sampson}, L., {Yunes}, N., and {Pretorius}, F. (2011).
\newblock {Gravitational wave tests of general relativity with the
  parameterized post-Einsteinian framework}.
\newblock {\em Phys.~Rev.~D}, 84(6):062003--+.

\bibitem[{Cornish}, 2011]{cornish:2011CQGra..28i4016C}
{Cornish}, N.~J. (2011).
\newblock {Detection strategies for extreme mass ratio inspirals}.
\newblock {\em Class.~Quantum~Grav.}, 28(9):094016--+.

\bibitem[{Cornish} and {Porter}, 2006]{cornish:2006:mes}
{Cornish}, N.~J. and {Porter}, E.~K. (2006).
\newblock {MCMC exploration of supermassive black hole binary inspirals}.
\newblock {\em Class.~Quantum~Grav.}, 23:761--+.

\bibitem[{Cox} and {Loeb}, 2008]{cox:2008MNRAS.386..461C}
{Cox}, T.~J. and {Loeb}, A. (2008).
\newblock {The collision between the Milky Way and Andromeda}.
\newblock {\em MNRAS}, 386:461--474.

\bibitem[{Croton} et~al., 2006]{croton:2006MNRAS.365...11C}
{Croton}, D.~J., {Springel}, V., {White}, S.~D.~M., {De Lucia}, G., {Frenk},
  C.~S., {Gao}, L., {Jenkins}, A., {Kauffmann}, G., {Navarro}, J.~F., and
  {Yoshida}, N. (2006).
\newblock {The many lives of active galactic nuclei: cooling flows, black holes
  and the luminosities and colours of galaxies}.
\newblock {\em MNRAS}, 365:11--28.

\bibitem[{Cuadra} et~al., 2009]{cuadra:2009MNRAS.393.1423C}
{Cuadra}, J., {Armitage}, P.~J., {Alexander}, R.~D., and {Begelman}, M.~C.
  (2009).
\newblock {Massive black hole binary mergers within subparsec scale gas discs}.
\newblock {\em MNRAS}, 393:1423--1432.

\bibitem[{Damour} et~al., 2011]{damour:2011PhRvD..83b4006D}
{Damour}, T., {Nagar}, A., and {Trias}, M. (2011).
\newblock {Accuracy and effectualness of closed-form, frequency-domain
  waveforms for nonspinning black hole binaries}.
\newblock {\em Phys.~Rev.~D}, 83(2):024006--+.

\bibitem[{Damour} and {Vilenkin}, 2005]{damour:2005:grc}
{Damour}, T. and {Vilenkin}, A. (2005).
\newblock {Gravitational radiation from cosmic (super)strings: Bursts,
  stochastic background, and observational windows}.
\newblock {\em Phys.~Rev.~D}, 71(6):063510--+.

\bibitem[{Dan} et~al., 2011]{dan:2011ApJ...737...89D}
{Dan}, M., {Rosswog}, S., {Guillochon}, J., and {Ramirez-Ruiz}, E. (2011).
\newblock {Prelude to A Double Degenerate Merger: The Onset of Mass Transfer
  and Its Impact on Gravitational Waves and Surface Detonations}.
\newblock {\em ApJ}, 737:89--+.

\bibitem[{Davies} et~al., 2011]{davies:2011ApJ...740L..42D}
{Davies}, M.~B., {Miller}, M.~C., and {Bellovary}, J.~M. (2011).
\newblock {Supermassive Black Hole Formation Via Gas Accretion in Nuclear
  Stellar Clusters}.
\newblock {\em ApJ}, 740:L42+.

\bibitem[{De Lucia} et~al., 2006]{delucia:2006:fheg}
{De Lucia}, G., {Springel}, V., {White}, S.~D.~M., {Croton}, D., and
  {Kauffmann}, G. (2006).
\newblock {The formation history of elliptical galaxies}.
\newblock {\em MNRAS}, 366:499--509.

\bibitem[{De Marco} et~al., 2011]{demarco:2011MNRAS.411.2277D}
{De Marco}, O., {Passy}, J.-C., {Moe}, M., {Herwig}, F., {Mac Low}, M.-M., and
  {Paxton}, B. (2011).
\newblock {On the {$\alpha$} formalism for the common envelope interaction}.
\newblock {\em MNRAS}, 411:2277--2292.

\bibitem[{De Propris} et~al., 2007]{depropris:2007:mgc}
{De Propris}, R., {Conselice}, C.~J., {Liske}, J., {Driver}, S.~P., {Patton},
  D.~R., {Graham}, A.~W., and {Allen}, P.~D. (2007).
\newblock {The Millennium Galaxy Catalogue: The Connection between Close Pairs
  and Asymmetry; Implications for the Galaxy Merger Rate}.
\newblock {\em ApJ}, 666:212--221.

\bibitem[{de Ravel} et~al., 2009]{deravel:2009:vimos}
{de Ravel}, L., {Le F{\`e}vre}, O., {Tresse}, L., {Bottini}, D., {Garilli}, B.,
  {Le Brun}, V., {Maccagni}, D., {Scaramella}, R., {Scodeggio}, M.,
  {Vettolani}, G., {Zanichelli}, A., {Adami}, C., {Arnouts}, S., {Bardelli},
  S., {Bolzonella}, M., {Cappi}, A., {Charlot}, S., {Ciliegi}, P., {Contini},
  T., {Foucaud}, S., {Franzetti}, P., {Gavignaud}, I., {Guzzo}, L., {Ilbert},
  O., {Iovino}, A., {Lamareille}, F., {McCracken}, H.~J., {Marano}, B.,
  {Marinoni}, C., {Mazure}, A., {Meneux}, B., {Merighi}, R., {Paltani}, S.,
  {Pell{\`o}}, R., {Pollo}, A., {Pozzetti}, L., {Radovich}, M., {Vergani}, D.,
  {Zamorani}, G., {Zucca}, E., {Bondi}, M., {Bongiorno}, A., {Brinchmann}, J.,
  {Cucciati}, O., {de La Torre}, S., {Gregorini}, L., {Memeo}, P.,
  {Perez-Montero}, E., {Mellier}, Y., {Merluzzi}, P., and {Temporin}, S.
  (2009).
\newblock {The VIMOS VLT Deep Survey. Evolution of the major merger rate since
  z \~{} 1 from spectroscopically confirmed galaxy pairs}.
\newblock {\em A\&A}, 498:379--397.

\bibitem[{Decarli} et~al., 2010]{decarli:2010ApJ...720L..93D}
{Decarli}, R., {Dotti}, M., {Montuori}, C., {Liimets}, T., and {Ederoclite}, A.
  (2010).
\newblock {The Peculiar Optical Spectrum of 4C+22.25: Imprint of a Massive
  Black Hole Binary?}
\newblock {\em ApJ}, 720:L93--L96.

\bibitem[{Detweiler}, 1980]{detweiler:1980PhRvD..22.2323D}
{Detweiler}, S. (1980).
\newblock {Klein-Gordon equation and rotating black holes}.
\newblock {\em Phys.~Rev.~D}, 22:2323--2326.

\bibitem[{Devecchi} et~al., 2009]{devecchi:2009MNRAS.394..633D}
{Devecchi}, B., {Rasia}, E., {Dotti}, M., {Volonteri}, M., and {Colpi}, M.
  (2009).
\newblock {Imprints of recoiling massive black holes on the hot gas of
  early-type galaxies}.
\newblock {\em MNRAS}, 394:633--640.

\bibitem[{Devecchi} and {Volonteri}, 2009]{devecchi:2009ApJ...694..302D}
{Devecchi}, B. and {Volonteri}, M. (2009).
\newblock {Formation of the First Nuclear Clusters and Massive Black Holes at
  High Redshift}.
\newblock {\em ApJ}, 694:302--313.

\bibitem[{Di Matteo} et~al., 2008]{dimatteo:2008:dcs}
{Di Matteo}, T., {Colberg}, J., {Springel}, V., {Hernquist}, L., and {Sijacki},
  D. (2008).
\newblock {Direct Cosmological Simulations of the Growth of Black Holes and
  Galaxies}.
\newblock {\em ApJ}, 676:33--53.

\bibitem[{Di Matteo} et~al., 2005]{dimatteo:2005:eiq}
{Di Matteo}, T., {Springel}, V., and {Hernquist}, L. (2005).
\newblock {Energy input from quasars regulates the growth and activity of black
  holes and their host galaxies}.
\newblock {\em Nature}, 433:604--607.

\bibitem[{Dijkstra} et~al., 2008]{dijkstra:2008MNRAS.391.1961D}
{Dijkstra}, M., {Haiman}, Z., {Mesinger}, A., and {Wyithe}, J.~S.~B. (2008).
\newblock {Fluctuations in the high-redshift Lyman-Werner background: close
  halo pairs as the origin of supermassive black holes}.
\newblock {\em MNRAS}, 391:1961--1972.

\bibitem[{Dotan} et~al., 2011]{dotan:2011MNRAS.tmp.1661D}
{Dotan}, C., {Rossi}, E.~M., and {Shaviv}, N.~J. (2011).
\newblock {A lower limit on the halo mass to form supermassive black holes}.
\newblock {\em MNRAS}, pages 1661--+.

\bibitem[{Dotti} et~al., 2006]{dotti:2006:lis}
{Dotti}, M., {Colpi}, M., and {Haardt}, F. (2006).
\newblock {Laser Interferometer Space Antenna double black holes: dynamics in
  gaseous nuclear discs}.
\newblock {\em mnras}, 367:103--112.

\bibitem[{Dotti} et~al., 2007]{dotti:2007:smbh}
{Dotti}, M., {Colpi}, M., {Haardt}, F., and {Mayer}, L. (2007).
\newblock {Supermassive black hole binaries in gaseous and stellar
  circumnuclear discs: orbital dynamics and gas accretion}.
\newblock {\em MNRAS}, 379:956--962.

\bibitem[{Dotti} et~al., 2009a]{dotti:2009MNRAS.398L..73D}
{Dotti}, M., {Montuori}, C., {Decarli}, R., {Volonteri}, M., {Colpi}, M., and
  {Haardt}, F. (2009a).
\newblock {SDSSJ092712.65+294344.0: a candidate massive black hole binary}.
\newblock {\em MNRAS}, 398:L73--L77.

\bibitem[{Dotti} et~al., 2009b]{dotti:2009MNRAS.396.1640D}
{Dotti}, M., {Ruszkowski}, M., {Paredi}, L., {Colpi}, M., {Volonteri}, M., and
  {Haardt}, F. (2009b).
\newblock {Dual black holes in merger remnants - I. Linking accretion to
  dynamics}.
\newblock {\em MNRAS}, 396:1640--1646.

\bibitem[{Dotti} et~al., 2010]{dotti:2010MNRAS.402..682D}
{Dotti}, M., {Volonteri}, M., {Perego}, A., {Colpi}, M., {Ruszkowski}, M., and
  {Haardt}, F. (2010).
\newblock {Dual black holes in merger remnants - II. Spin evolution and
  gravitational recoil}.
\newblock {\em MNRAS}, 402:682--690.

\bibitem[{Drasco} and {Hughes}, 2004]{drasco:2004:rbh}
{Drasco}, S. and {Hughes}, S.~A. (2004).
\newblock {Rotating black hole orbit functionals in the frequency domain}.
\newblock {\em Phys.~Rev.~D}, 69(4):044015--+.

\bibitem[{Drasco} and {Hughes}, 2006]{drasco:2006:gws}
{Drasco}, S. and {Hughes}, S.~A. (2006).
\newblock {Gravitational wave snapshots of generic extreme mass ratio
  inspirals}.
\newblock {\em Phys.~Rev.~D}, 73(2):024027--+.

\bibitem[{Dreyer} et~al., 2004]{dreyer:2004:bhs}
{Dreyer}, O., {Kelly}, B., {Krishnan}, B., {Finn}, L.~S., {Garrison}, D., and
  {Lopez-Aleman}, R. (2004).
\newblock {Black-hole spectroscopy: testing general relativity through
  gravitational-wave observations}.
\newblock {\em Class.~Quantum~Grav.}, 21:787--803.

\bibitem[{D'Souza} et~al., 2006]{dsouza:2006:dmt}
{D'Souza}, M.~C.~R., {Motl}, P.~M., {Tohline}, J.~E., and {Frank}, J. (2006).
\newblock {Numerical Simulations of the Onset and Stability of Dynamical Mass
  Transfer in Binaries}.
\newblock {\em ApJ}, 643:381--401.

\bibitem[Dufaux et~al., 2007]{Dufaux:2007pt}
Dufaux, J.~F., Bergman, A., Felder, G.~N., Kofman, L., and Uzan, J.-P. (2007).
\newblock {Theory and Numerics of Gravitational Waves from Preheating after
  Inflation}.
\newblock {\em Phys. Rev.}, D76:123517.

\bibitem[Dufaux et~al., 2009]{Dufaux:2008dn}
Dufaux, J.-F., Felder, G., Kofman, L., and Navros, O. (2009).
\newblock {Gravity Waves from Tachyonic Preheating after Hybrid Inflation}.
\newblock {\em JCAP}, 0903:001.

\bibitem[{Easther} and {Lim}, 2006]{easther:2006:sgw}
{Easther}, R. and {Lim}, E.~A. (2006).
\newblock {Stochastic gravitational wave production after inflation}.
\newblock {\em Journal of Cosmology and Astro-Particle Physics}, 4:10--+.

\bibitem[{Edlund} et~al., 2005]{edlund:2005:wdw}
{Edlund}, J.~A., {Tinto}, M., {Kr{\'o}lak}, A., and {Nelemans}, G. (2005).
\newblock {White-dwarf white-dwarf galactic background in the LISA data}.
\newblock {\em Phys.~Rev.~D}, 71:122003--+.

\bibitem[{Eisenhauer} et~al., 2005]{eisenhauer:2005:sinfoni}
{Eisenhauer}, F., {Genzel}, R., {Alexander}, T., {Abuter}, R., {Paumard}, T.,
  {Ott}, T., {Gilbert}, A., {Gillessen}, S., {Horrobin}, M., {Trippe}, S.,
  {Bonnet}, H., {Dumas}, C., {Hubin}, N., {Kaufer}, A., {Kissler-Patig}, M.,
  {Monnet}, G., {Str{\"o}bele}, S., {Szeifert}, T., {Eckart}, A.,
  {Sch{\"o}del}, R., and {Zucker}, S. (2005).
\newblock {SINFONI in the Galactic Center: Young Stars and Infrared Flares in
  the Central Light-Month}.
\newblock {\em ApJ}, 628:246--259.

\bibitem[{Enoki} et~al., 2004]{enoki:2004:gws}
{Enoki}, M., {Inoue}, K.~T., {Nagashima}, M., and {Sugiyama}, N. (2004).
\newblock {Gravitational Waves from Supermassive Black Hole Coalescence in a
  Hierarchical Galaxy Formation Model}.
\newblock {\em ApJ}, 615:19--28.

\bibitem[{Eracleous} et~al., 2011]{eracleous:2011arXiv1106.2952E}
{Eracleous}, M., {Boroson}, T.~A., {Halpern}, J.~P., and {Liu}, J. (2011).
\newblock {A Large Systematic Search for Recoiling and Close Supermassive
  Binary Black Holes}.
\newblock {\em ArXiv e-prints}.

\bibitem[{Escala} et~al., 2004]{escala:2004:rgm}
{Escala}, A., {Larson}, R.~B., {Coppi}, P.~S., and {Mardones}, D. (2004).
\newblock {The Role of Gas in the Merging of Massive Black Holes in Galactic
  Nuclei. I. Black Hole Merging in a Spherical Gas Cloud}.
\newblock {\em ApJ}, 607:765--777.

\bibitem[{Esposito-Far{\`e}se}, 2004]{esposito-farese:2004:stg}
{Esposito-Far{\`e}se}, G. (2004).
\newblock {Tests of Scalar-Tensor Gravity}.
\newblock In {C.~J.~A.~P.~Martins, P.~P.~Avelino, M.~S.~Costa, K.~Mack,
  M.~F.~Mota, \& M.~Parry}, editor, {\em Phi in the Sky: The Quest for
  Cosmological Scalar Fields}, volume 736 of {\em AIP~Conf.~Series}, pages
  35--52. AIP.

\bibitem[{Fabbiano} et~al., 2011]{fabbiano:2011Natur.477..431F}
{Fabbiano}, G., {Wang}, J., {Elvis}, M., and {Risaliti}, G. (2011).
\newblock {A close nuclear black-hole pair in the spiral galaxy NGC3393}.
\newblock {\em Nature}, 477:431--434.

\bibitem[{Fan} et~al., 2004]{fan:2004:sq_3}
{Fan}, X., {Hennawi}, J.~F., {Richards}, G.~T., {Strauss}, M.~A., {Schneider},
  D.~P., {Donley}, J.~L., {Young}, J.~E., {Annis}, J., {Lin}, H., {Lampeitl},
  H., {Lupton}, R.~H., {Gunn}, J.~E., {Knapp}, G.~R., {Brandt}, W.~N.,
  {Anderson}, S., {Bahcall}, N.~A., {Brinkmann}, J., {Brunner}, R.~J.,
  {Fukugita}, M., {Szalay}, A.~S., {Szokoly}, G.~P., and {York}, D.~G. (2004).
\newblock {A Survey of $z>5.7$ Quasars in the Sloan Digital Sky Survey. III.
  Discovery of Five Additional Quasars}.
\newblock {\em AJ}, 128:515--522.

\bibitem[{Fan} et~al., 2001]{fan:2001:sq_1}
{Fan}, X., {Narayanan}, V.~K., {Lupton}, R.~H., {Strauss}, M.~A., {Knapp},
  G.~R., {Becker}, R.~H., {White}, R.~L., {Pentericci}, L., {Leggett}, S.~K.,
  {Haiman}, Z., {Gunn}, J.~E., {Ivezi{\'c}}, {\v Z}., {Schneider}, D.~P.,
  {Anderson}, S.~F., {Brinkmann}, J., {Bahcall}, N.~A., {Connolly}, A.~J.,
  {Csabai}, I., {Doi}, M., {Fukugita}, M., {Geballe}, T., {Grebel}, E.~K.,
  {Harbeck}, D., {Hennessy}, G., {Lamb}, D.~Q., {Miknaitis}, G., {Munn}, J.~A.,
  {Nichol}, R., {Okamura}, S., {Pier}, J.~R., {Prada}, F., {Richards}, G.~T.,
  {Szalay}, A., and {York}, D.~G. (2001).
\newblock {A Survey of $z>5.8$ Quasars in the Sloan Digital Sky Survey. I.
  Discovery of Three New Quasars and the Spatial Density of Luminous Quasars at
  z\~{}6}.
\newblock {\em AJ}, 122:2833--2849.

\bibitem[{Fan} et~al., 2006a]{fan:2006AJ....132..117F}
{Fan}, X., {Strauss}, M.~A., {Becker}, R.~H., {White}, R.~L., {Gunn}, J.~E.,
  {Knapp}, G.~R., {Richards}, G.~T., {Schneider}, D.~P., {Brinkmann}, J., and
  {Fukugita}, M. (2006a).
\newblock {Constraining the Evolution of the Ionizing Background and the Epoch
  of Reionization with z\~{}6 Quasars. II. A Sample of 19 Quasars}.
\newblock {\em AJ}, 132:117--136.

\bibitem[{Fan} et~al., 2006b]{fan:2006:sq_4}
{Fan}, X., {Strauss}, M.~A., {Richards}, G.~T., {Hennawi}, J.~F., {Becker},
  R.~H., {White}, R.~L., {Diamond-Stanic}, A.~M., {Donley}, J.~L., {Jiang}, L.,
  {Kim}, J.~S., {Vestergaard}, M., {Young}, J.~E., {Gunn}, J.~E., {Lupton},
  R.~H., {Knapp}, G.~R., {Schneider}, D.~P., {Brandt}, W.~N., {Bahcall}, N.~A.,
  {Barentine}, J.~C., {Brinkmann}, J., {Brewington}, H.~J., {Fukugita}, M.,
  {Harvanek}, M., {Kleinman}, S.~J., {Krzesinski}, J., {Long}, D., {Neilsen},
  Jr., E.~H., {Nitta}, A., {Snedden}, S.~A., and {Voges}, W. (2006b).
\newblock {A Survey of $z>5.7$ Quasars in the Sloan Digital Sky Survey. IV.
  Discovery of Seven Additional Quasars}.
\newblock {\em AJ}, 131:1203--1209.

\bibitem[{Fan} et~al., 2003]{fan:2003:sq_2}
{Fan}, X., {Strauss}, M.~A., {Schneider}, D.~P., {Becker}, R.~H., {White},
  R.~L., {Haiman}, Z., {Gregg}, M., {Pentericci}, L., {Grebel}, E.~K.,
  {Narayanan}, V.~K., {Loh}, Y.-S., {Richards}, G.~T., {Gunn}, J.~E., {Lupton},
  R.~H., {Knapp}, G.~R., {Ivezi{\'c}}, {\v Z}., {Brandt}, W.~N., {Collinge},
  M., {Hao}, L., {Harbeck}, D., {Prada}, F., {Schaye}, J., {Strateva}, I.,
  {Zakamska}, N., {Anderson}, S., {Brinkmann}, J., {Bahcall}, N.~A., {Lamb},
  D.~Q., {Okamura}, S., {Szalay}, A., and {York}, D.~G. (2003).
\newblock {A Survey of $z>5.7$ Quasars in the Sloan Digital Sky Survey. II.
  Discovery of Three Additional Quasars at $z>6$}.
\newblock {\em AJ}, 125:1649--1659.

\bibitem[{Farmer} and {Phinney}, 2003]{farmer:2003:gwb}
{Farmer}, A.~J. and {Phinney}, E.~S. (2003).
\newblock {The gravitational wave background from cosmological compact
  binaries}.
\newblock {\em MNRAS}, 346:1197--1214.

\bibitem[{Ferrarese} et~al., 2006]{ferrarese:2006:smbh}
{Ferrarese}, L., {C{\^o}t{\'e}}, P., {Dalla Bont{\`a}}, E., {Peng}, E.~W.,
  {Merritt}, D., {Jord{\'a}n}, A., {Blakeslee}, J.~P., {Ha{\c s}egan}, M.,
  {Mei}, S., {Piatek}, S., {Tonry}, J.~L., and {West}, M.~J. (2006).
\newblock {A Fundamental Relation between Compact Stellar Nuclei, Supermassive
  Black Holes, and Their Host Galaxies}.
\newblock {\em ApJ}, 644:L21--L24.

\bibitem[{Ferrarese} and {Ford}, 2005]{ferrarese:2005:smbh}
{Ferrarese}, L. and {Ford}, H. (2005).
\newblock {Supermassive Black Holes in Galactic Nuclei: Past, Present and
  Future Research}.
\newblock {\em Space Science Reviews}, 116:523--624.

\bibitem[{Ferrarese} and {Merritt}, 2000]{ferrarese:2000:smbh}
{Ferrarese}, L. and {Merritt}, D. (2000).
\newblock {A Fundamental Relation between Supermassive Black Holes and Their
  Host Galaxies}.
\newblock {\em ApJ}, 539:L9--L12.

\bibitem[{Finn} and {Thorne}, 2000]{finn:2000:gwc}
{Finn}, L.~S. and {Thorne}, K.~S. (2000).
\newblock {Gravitational waves from a compact star in a circular, inspiral
  orbit, in the equatorial plane of a massive, spinning black hole, as observed
  by LISA}.
\newblock {\em Phys.~Rev.~D}, 6212:124021--+.

\bibitem[{Flanagan} and {Hughes}, 1998]{flanagan:1998:mgwa}
{Flanagan}, {\'E}.~{\'E}. and {Hughes}, S.~A. (1998).
\newblock {Measuring gravitational waves from binary black hole coalescences.
  I. Signal to noise for inspiral, merger, and ringdown}.
\newblock {\em Phys.~Rev.~D}, 57:4535--4565.

\bibitem[{Freitag} et~al., 2006a]{freitag:2006JPhCS..54..252F}
{Freitag}, M., {Amaro-Seoane}, P., and {Kalogera}, V. (2006a).
\newblock {Models of mass segregation at the Galactic Centre}.
\newblock {\em JPCS}, 54:252--258.

\bibitem[{Freitag} et~al., 2006b]{freitag:2006:srg}
{Freitag}, M., {Amaro-Seoane}, P., and {Kalogera}, V. (2006b).
\newblock {Stellar Remnants in Galactic Nuclei: Mass Segregation}.
\newblock {\em ApJ}, 649:91--117.

\bibitem[{Freitag} et~al., 2006c]{freitag:2006MNRAS.368..121F}
{Freitag}, M., {Rasio}, F.~A., and {Baumgardt}, H. (2006c).
\newblock {Runaway collisions in young star clusters - I. Methods and tests}.
\newblock {\em MNRAS}, 368:121--140.

\bibitem[{Fuller} and {Lai}, 2011]{fuller:2011arXiv1108.4910F}
{Fuller}, J. and {Lai}, D. (2011).
\newblock {Dynamical Tides in Compact White Dwarf Binaries: Tidal
  Synchronization and Dissipation}.
\newblock {\em ArXiv e-prints}.

\bibitem[{Gair} and {Yunes}, 2011]{gair:2011PhRvD..84f4016G}
{Gair}, J. and {Yunes}, N. (2011).
\newblock {Approximate waveforms for extreme-mass-ratio inspirals in modified
  gravity spacetimes}.
\newblock {\em Phys.~Rev.~D}, 84(6):064016--+.

\bibitem[{Gair}, 2009a]{gair:2009CQGra..26i4034G}
{Gair}, J.~R. (2009a).
\newblock {Probing black holes at low redshift using LISA EMRI observations}.
\newblock {\em Class.~Quantum~Grav.}, 26(9):094034--+.

\bibitem[{Gair}, 2009b]{gair:2008:pbh}
{Gair}, J.~R. (2009b).
\newblock {Probing black holes at low redshift using LISA EMRI observations}.
\newblock {\em Class.~Quantum~Grav.}, 26(9):094034--+.

\bibitem[{Gair} et~al., 2004]{gair:2004:ere}
{Gair}, J.~R., {Barack}, L., {Creighton}, T., {Cutler}, C., {Larson}, S.~L.,
  {Phinney}, E.~S., and {Vallisneri}, M. (2004).
\newblock {Event rate estimates for LISA extreme mass ratio capture sources}.
\newblock {\em Class.~Quantum~Grav.}, 21:S1595--S1606.

\bibitem[{Gair} et~al., 2008]{gair:2008:cmh}
{Gair}, J.~R., {Porter}, E., {Babak}, S., and {Barack}, L. (2008).
\newblock {A constrained Metropolis Hastings search for EMRIs in the Mock LISA
  Data Challenge 1B}.
\newblock {\em Class.~Quantum~Grav.}, 25(18):184030--+.

\bibitem[{Gair} et~al., 2010]{gair:2010:emri}
{Gair}, J.~R., {Tang}, C., and {Volonteri}, M. (2010).
\newblock {LISA extreme-mass-ratio inspiral events as probes of the black hole
  mass function}.
\newblock {\em Phys.~Rev.~D}, 81(10):104014--+.

\bibitem[{Galloway} et~al., 2002]{galloway:2002:hlamsp}
{Galloway}, D.~K., {Chakrabarty}, D., {Morgan}, E.~H., and {Remillard}, R.~A.
  (2002).
\newblock {Discovery of a High-Latitude Accreting Millisecond Pulsar in an
  Ultracompact Binary}.
\newblock {\em ApJ}, 576:L137--L140.

\bibitem[{Gammie} et~al., 2004]{gammie:2004:bhse}
{Gammie}, C.~F., {Shapiro}, S.~L., and {McKinney}, J.~C. (2004).
\newblock {Black Hole Spin Evolution}.
\newblock {\em ApJ}, 602:312--319.

\bibitem[Garcia-Bellido and Figueroa, 2007]{GarciaBellido:2007dg}
Garcia-Bellido, J. and Figueroa, D.~G. (2007).
\newblock {A stochastic background of gravitational waves from hybrid
  preheating}.
\newblock {\em Phys. Rev. Lett.}, 98:061302.

\bibitem[{Gebhardt} et~al., 2000]{gebhardt:2000:bhm}
{Gebhardt}, K., {Bender}, R., {Bower}, G., {Dressler}, A., {Faber}, S.~M.,
  {Filippenko}, A.~V., {Green}, R., {Grillmair}, C., {Ho}, L.~C., {Kormendy},
  J., {Lauer}, T.~R., {Magorrian}, J., {Pinkney}, J., {Richstone}, D., and
  {Tremaine}, S. (2000).
\newblock {A Relationship between Nuclear Black Hole Mass and Galaxy Velocity
  Dispersion}.
\newblock {\em ApJ}, 539:L13--L16.

\bibitem[{Genzel} et~al., 2010]{genzel:2010RvMP...82.3121G}
{Genzel}, R., {Eisenhauer}, F., and {Gillessen}, S. (2010).
\newblock {The Galactic Center massive black hole and nuclear star cluster}.
\newblock {\em Reviews of Modern Physics}, 82:3121--3195.

\bibitem[{Genzel} et~al., 2000]{genzel:2000MNRAS.317..348G}
{Genzel}, R., {Pichon}, C., {Eckart}, A., {Gerhard}, O.~E., and {Ott}, T.
  (2000).
\newblock {Stellar dynamics in the Galactic Centre: proper motions and
  anisotropy}.
\newblock {\em MNRAS}, 317:348--374.

\bibitem[{Genzel} et~al., 2006]{genzel:2006Natur.442..786G}
{Genzel}, R., {Tacconi}, L.~J., {Eisenhauer}, F., {F{\"o}rster Schreiber},
  N.~M., {Cimatti}, A., {Daddi}, E., {Bouch{\'e}}, N., {Davies}, R., {Lehnert},
  M.~D., {Lutz}, D., {Nesvadba}, N., {Verma}, A., {Abuter}, R., {Shapiro}, K.,
  {Sternberg}, A., {Renzini}, A., {Kong}, X., {Arimoto}, N., and {Mignoli}, M.
  (2006).
\newblock {The rapid formation of a large rotating disk galaxy three billion
  years after the Big Bang}.
\newblock {\em Nature}, 442:786--789.

\bibitem[{Ghez} et~al., 2003]{ghez:2003:fmsl}
{Ghez}, A.~M., {Duch{\^e}ne}, G., {Matthews}, K., {Hornstein}, S.~D., {Tanner},
  A., {Larkin}, J., {Morris}, M., {Becklin}, E.~E., {Salim}, S., {Kremenek},
  T., {Thompson}, D., {Soifer}, B.~T., {Neugebauer}, G., and {McLean}, I.
  (2003).
\newblock {The First Measurement of Spectral Lines in a Short-Period Star Bound
  to the Galaxy's Central Black Hole: A Paradox of Youth}.
\newblock {\em ApJ}, 586:L127--L131.

\bibitem[{Ghez} et~al., 2005]{ghez:2005:sogc}
{Ghez}, A.~M., {Salim}, S., {Hornstein}, S.~D., {Tanner}, A., {Lu}, J.~R.,
  {Morris}, M., {Becklin}, E.~E., and {Duch{\^e}ne}, G. (2005).
\newblock {Stellar Orbits around the Galactic Center Black Hole}.
\newblock {\em ApJ}, 620:744--757.

\bibitem[{Ghez} et~al., 2008]{ghez:2008:mdp}
{Ghez}, A.~M., {Salim}, S., {Weinberg}, N.~N., {Lu}, J.~R., {Do}, T., {Dunn},
  J.~K., {Matthews}, K., {Morris}, M.~R., {Yelda}, S., {Becklin}, E.~E.,
  {Kremenek}, T., {Milosavljevic}, M., and {Naiman}, J. (2008).
\newblock {Measuring Distance and Properties of the Milky Way's Central
  Supermassive Black Hole with Stellar Orbits}.
\newblock {\em ApJ}, 689:1044--1062.

\bibitem[{Gillessen} et~al., 2009]{gillessen:2009:mso}
{Gillessen}, S., {Eisenhauer}, F., {Trippe}, S., {Alexander}, T., {Genzel}, R.,
  {Martins}, F., and {Ott}, T. (2009).
\newblock {Monitoring Stellar Orbits Around the Massive Black Hole in the
  Galactic Center}.
\newblock {\em ApJ}, 692:1075--1109.

\bibitem[{Glampedakis} and {Babak}, 2006]{glampedakis:2006:msl}
{Glampedakis}, K. and {Babak}, S. (2006).
\newblock {Mapping spacetimes with LISA: inspiral of a test body in a
  'quasi-Kerr' field}.
\newblock {\em Class.~Quantum~Grav.}, 23:4167--4188.

\bibitem[{Gonz{\'a}lez} et~al., 2007]{gonzalez:2007PhRvL..98w1101G}
{Gonz{\'a}lez}, J.~A., {Hannam}, M., {Sperhake}, U., {Br{\"u}gmann}, B., and
  {Husa}, S. (2007).
\newblock {Supermassive Recoil Velocities for Binary Black-Hole Mergers with
  Antialigned Spins}.
\newblock {\em Phys.~Rev.~Lett.}, 98(23):231101--+.

\bibitem[{Gould} and {Rix}, 2000]{gould:2000ApJ...532L..29G}
{Gould}, A. and {Rix}, H.-W. (2000).
\newblock {Binary Black Hole Mergers from Planet-like Migrations}.
\newblock {\em ApJ}, 532:L29--L32.

\bibitem[{Graham} et~al., 2011]{graham:2011MNRAS.412.2211G}
{Graham}, A.~W., {Onken}, C.~A., {Athanassoula}, E., and {Combes}, F. (2011).
\newblock {An expanded M$_{bh}$-{$\sigma$} diagram, and a new calibration of
  active galactic nuclei masses}.
\newblock {\em MNRAS}, 412:2211--2228.

\bibitem[{Graham} and {Spitler}, 2009]{graham:2009MNRAS.397.2148G}
{Graham}, A.~W. and {Spitler}, L.~R. (2009).
\newblock {Quantifying the coexistence of massive black holes and dense nuclear
  star clusters}.
\newblock {\em MNRAS}, 397:2148--2162.

\bibitem[{Greene} and {Ho}, 2004]{greene:2004ApJ...610..722G}
{Greene}, J.~E. and {Ho}, L.~C. (2004).
\newblock {Active Galactic Nuclei with Candidate Intermediate-Mass Black
  Holes}.
\newblock {\em ApJ}, 610:722--736.

\bibitem[{Greene} et~al., 2008]{greene:2008ApJ...688..159G}
{Greene}, J.~E., {Ho}, L.~C., and {Barth}, A.~J. (2008).
\newblock {Black Holes in Pseudobulges and Spheroidals: A Change in the Black
  Hole-Bulge Scaling Relations at Low Mass}.
\newblock {\em ApJ}, 688:159--179.

\bibitem[{Gualandris} and {Merritt}, 2008]{gualandris:2008ApJ...678..780G}
{Gualandris}, A. and {Merritt}, D. (2008).
\newblock {Ejection of Supermassive Black Holes from Galaxy Cores}.
\newblock {\em ApJ}, 678:780--797.

\bibitem[{Guedes} et~al., 2009]{guedes:2009ApJ...702..890G}
{Guedes}, J., {Madau}, P., {Kuhlen}, M., {Diemand}, J., and {Zemp}, M. (2009).
\newblock {Simulations of Recoiling Massive Black Holes in the Via Lactea
  Halo}.
\newblock {\em ApJ}, 702:890--900.

\bibitem[{Guedes} et~al., 2011]{guedes:2011ApJ...729..125G}
{Guedes}, J., {Madau}, P., {Mayer}, L., and {Callegari}, S. (2011).
\newblock {Recoiling Massive Black Holes in Gas-rich Galaxy Mergers}.
\newblock {\em ApJ}, 729:125--+.

\bibitem[{G{\"u}ltekin} et~al., 2009]{gultekin:2009:ms}
{G{\"u}ltekin}, K., {Richstone}, D.~O., {Gebhardt}, K., {Lauer}, T.~R.,
  {Tremaine}, S., {Aller}, M.~C., {Bender}, R., {Dressler}, A., {Faber}, S.~M.,
  {Filippenko}, A.~V., {Green}, R., {Ho}, L.~C., {Kormendy}, J., {Magorrian},
  J., {Pinkney}, J., and {Siopis}, C. (2009).
\newblock {The M-{$\sigma$} and M-L Relations in Galactic Bulges, and
  Determinations of Their Intrinsic Scatter}.
\newblock {\em ApJ}, 698:198--221.

\bibitem[{Guo} et~al., 2011]{guo:2011:fds}
{Guo}, Q., {White}, S., {Boylan-Kolchin}, M., {De Lucia}, G., {Kauffmann}, G.,
  {Lemson}, G., {Li}, C., {Springel}, V., and {Weinmann}, S. (2011).
\newblock {From dwarf spheroidals to cD galaxies: simulating the galaxy
  population in a {$\Lambda$}CDM cosmology}.
\newblock {\em MNRAS}, 413:101--131.

\bibitem[{G{\"u}rkan} et~al., 2006]{guerkan:2006ApJ...640L..39G}
{G{\"u}rkan}, M.~A., {Fregeau}, J.~M., and {Rasio}, F.~A. (2006).
\newblock {Massive Black Hole Binaries from Collisional Runaways}.
\newblock {\em ApJ}, 640:L39--L42.

\bibitem[{G{\"u}rkan} et~al., 2004]{guerkan:2004ApJ...604..632G}
{G{\"u}rkan}, M.~A., {Freitag}, M., and {Rasio}, F.~A. (2004).
\newblock {Formation of Massive Black Holes in Dense Star Clusters. I. Mass
  Segregation and Core Collapse}.
\newblock {\em ApJ}, 604:632--652.

\bibitem[{G{\"u}rkan} and {Hopman}, 2007]{guerkan:2007:rr}
{G{\"u}rkan}, M.~A. and {Hopman}, C. (2007).
\newblock {Resonant relaxation near a massive black hole: the dependence on
  eccentricity}.
\newblock {\em MNRAS}, 379:1083--1088.

\bibitem[{Haehnelt}, 1994]{haehnelt:1994:lfg}
{Haehnelt}, M.~G. (1994).
\newblock {Low-Frequency Gravitational Waves from Supermassive Black-Holes}.
\newblock {\em MNRAS}, 269:199--+.

\bibitem[{Haehnelt} et~al., 1998]{haehnelt:1998MNRAS.300..817H}
{Haehnelt}, M.~G., {Natarajan}, P., and {Rees}, M.~J. (1998).
\newblock {High-redshift galaxies, their active nuclei and central black
  holes}.
\newblock {\em MNRAS}, 300:817--827.

\bibitem[{Haehnelt} and {Rees}, 1993]{haehnelt:1993MNRAS.263..168H}
{Haehnelt}, M.~G. and {Rees}, M.~J. (1993).
\newblock {The formation of nuclei in newly formed galaxies and the evolution
  of the quasar population}.
\newblock {\em MNRAS}, 263:168--178.

\bibitem[{Haiman} and {Loeb}, 1998]{haiman:1998ApJ...503..505H}
{Haiman}, Z. and {Loeb}, A. (1998).
\newblock {Observational Signatures of the First Quasars}.
\newblock {\em ApJ}, 503:505--+.

\bibitem[{Haiman} et~al., 1996]{haiman:1996ApJ...464..523H}
{Haiman}, Z., {Thoul}, A.~A., and {Loeb}, A. (1996).
\newblock {Cosmological Formation of Low-Mass Objects}.
\newblock {\em ApJ}, 464:523--+.

\bibitem[{H{\"a}ring} and {Rix}, 2004]{haering:2004:bhmbr}
{H{\"a}ring}, N. and {Rix}, H.-W. (2004).
\newblock {On the Black Hole Mass-Bulge Mass Relation}.
\newblock {\em ApJ}, 604:L89--L92.

\bibitem[{Hayasaki} et~al., 2008]{hayasaki:2008ApJ...682.1134H}
{Hayasaki}, K., {Mineshige}, S., and {Ho}, L.~C. (2008).
\newblock {A Supermassive Binary Black Hole with Triple Disks}.
\newblock {\em ApJ}, 682:1134--1140.

\bibitem[{Hayasaki} and {Okazaki}, 2009]{hayasaki:2009ApJ...691L...5H}
{Hayasaki}, K. and {Okazaki}, A.~T. (2009).
\newblock {A New Approach for Probing Circumbinary Disks}.
\newblock {\em ApJ}, 691:L5--L8.

\bibitem[{Heger} et~al., 2003]{heger:2003ApJ...591..288H}
{Heger}, A., {Fryer}, C.~L., {Woosley}, S.~E., {Langer}, N., and {Hartmann},
  D.~H. (2003).
\newblock {How Massive Single Stars End Their Life}.
\newblock {\em ApJ}, 591:288--300.

\bibitem[{Hogan}, 1986]{hogan:1986:grc}
{Hogan}, C.~J. (1986).
\newblock {Gravitational radiation from cosmological phase transitions}.
\newblock {\em MNRAS}, 218:629--636.

\bibitem[{Hogan}, 2000]{hogan:2000:gwm}
{Hogan}, C.~J. (2000).
\newblock {Gravitational Waves from Mesoscopic Dynamics of the Extra
  Dimensions}.
\newblock {\em Phys.~Rev.~Lett.}, 85:2044--2047.

\bibitem[{Hogan}, 2006]{hogan:2006:gws}
{Hogan}, C.~J. (2006).
\newblock {Gravitational Wave Sources from New Physics}.
\newblock In {\em Laser Interferometer Space Antenna: 6th International LISA
  Symposium}, volume 873 of {\em AIP Conf. Proc.}, pages 30--40. AIP.

\bibitem[{Hogan} and {Bender}, 2001]{hogan:2001:esg}
{Hogan}, C.~J. and {Bender}, P.~L. (2001).
\newblock {Estimating stochastic gravitational wave backgrounds with the Sagnac
  calibration}.
\newblock {\em Phys.~Rev.~D}, 64(6):062002--+.

\bibitem[{Holz} and {Hughes}, 2005]{holz:2005:ugw}
{Holz}, D.~E. and {Hughes}, S.~A. (2005).
\newblock {Using Gravitational-Wave Standard Sirens}.
\newblock {\em ApJ}, 629:15--22.

\bibitem[{Hopkins} et~al., 2006]{hopkins:2006:umm}
{Hopkins}, P.~F., {Hernquist}, L., {Cox}, T.~J., {Di Matteo}, T., {Robertson},
  B., and {Springel}, V. (2006).
\newblock {A Unified, Merger-driven Model of the Origin of Starbursts, Quasars,
  the Cosmic X-Ray Background, Supermassive Black Holes, and Galaxy Spheroids}.
\newblock {\em ApJS}, 163:1--49.

\bibitem[{Hopkins} et~al., 2008]{hopkins:2008ApJS..175..356H}
{Hopkins}, P.~F., {Hernquist}, L., {Cox}, T.~J., and {Kere{\v s}}, D. (2008).
\newblock {A Cosmological Framework for the Co-Evolution of Quasars,
  Supermassive Black Holes, and Elliptical Galaxies. I. Galaxy Mergers and
  Quasar Activity}.
\newblock {\em ApJS}, 175:356--389.

\bibitem[{Hopkins} et~al., 2009]{hopkins:2009MNRAS.398..303H}
{Hopkins}, P.~F., {Murray}, N., and {Thompson}, T.~A. (2009).
\newblock {The small scatter in BH-host correlations and the case for
  self-regulated BH growth}.
\newblock {\em MNRAS}, 398:303--311.

\bibitem[{Hopman}, 2009]{hopman:2009:emri}
{Hopman}, C. (2009).
\newblock {Extreme mass ratio inspiral rates: dependence on the massive black
  hole mass}.
\newblock {\em Class.~Quantum~Grav.}, 26(9):094028--+.

\bibitem[{Hopman} and {Alexander}, 2006a]{hopman:2006:rrn}
{Hopman}, C. and {Alexander}, T. (2006a).
\newblock {Resonant Relaxation near a Massive Black Hole: The Stellar
  Distribution and Gravitational Wave Sources}.
\newblock {\em ApJ}, 645:1152--1163.

\bibitem[{Hopman} and {Alexander}, 2006b]{hopman:2006:ems}
{Hopman}, C. and {Alexander}, T. (2006b).
\newblock {The Effect of Mass Segregation on Gravitational Wave Sources near
  Massive Black Holes}.
\newblock {\em ApJ}, 645:L133--L136.

\bibitem[Huber and Konstandin, 2008]{Huber:2008hg}
Huber, S.~J. and Konstandin, T. (2008).
\newblock {Gravitational Wave Production by Collisions: More Bubbles}.
\newblock {\em JCAP}, 0809:022.

\bibitem[{Huerta} and {Gair}, 2009]{huerta:2009PhRvD..79h4021H}
{Huerta}, E.~A. and {Gair}, J.~R. (2009).
\newblock {Influence of conservative corrections on parameter estimation for
  extreme-mass-ratio inspirals}.
\newblock {\em Phys.~Rev.~D}, 79(8):084021--+.

\bibitem[Hughes, 2006]{hughes:2006:abs}
Hughes, S.~A. (2006).
\newblock A brief survey of lisa sources and science.
\newblock In Merkowitz, S.~M. and Livas, J.~C., editors, {\em Laser
  Interferometer Space Antenna: 6th International LISA Symposium}, volume 873
  of {\em AIP~Conf.~Series}, pages 13--20. AIP.

\bibitem[{Hughes} and {Blandford}, 2003]{hughes:2003:bhm}
{Hughes}, S.~A. and {Blandford}, R.~D. (2003).
\newblock {Black Hole Mass and Spin Coevolution by Mergers}.
\newblock {\em ApJ}, 585:L101--L104.

\bibitem[{Huwyler} et~al., 2011]{huwyler:2011arXiv1108.1826H}
{Huwyler}, C., {Klein}, A., and {Jetzer}, P. (2011).
\newblock {Testing General Relativity with LISA including Spin Precession and
  Higher Harmonics in the Waveform}.
\newblock {\em ArXiv e-prints}.

\bibitem[{Ivanov} et~al., 1999]{ivanov:1999MNRAS.307...79I}
{Ivanov}, P.~B., {Papaloizou}, J.~C.~B., and {Polnarev}, A.~G. (1999).
\newblock {The evolution of a supermassive binary caused by an accretion disc}.
\newblock {\em MNRAS}, 307:79--90.

\bibitem[{Jiang} et~al., 2011a]{jiang:2011ApJ...737L..45J}
{Jiang}, Y.-F., {Greene}, J.~E., and {Ho}, L.~C. (2011a).
\newblock {Black Hole Mass and Bulge Luminosity for Low-mass Black Holes}.
\newblock {\em ApJ}, 737:L45+.

\bibitem[{Jiang} et~al., 2011b]{jiang:2011arXiv1107.4105J}
{Jiang}, Y.-F., {Greene}, J.~E., {Ho}, L.~C., {Xiao}, T., and {Barth}, A.~J.
  (2011b).
\newblock {The Host Galaxies of Low-mass Black Holes}.
\newblock {\em ArXiv e-prints}.

\bibitem[{Johansson} et~al., 2009]{johannson:2009ApJ...690..802J}
{Johansson}, P.~H., {Naab}, T., and {Burkert}, A. (2009).
\newblock {Equal- and Unequal-Mass Mergers of Disk and Elliptical Galaxies with
  Black Holes}.
\newblock {\em ApJ}, 690:802--821.

\bibitem[{Johnston} et~al., 2007]{johnston:2007PASA...24..174J}
{Johnston}, S., {Bailes}, M., {Bartel}, N., {Baugh}, C., {Bietenholz}, M.,
  {Blake}, C., {Braun}, R., {Brown}, J., {Chatterjee}, S., {Darling}, J.,
  {Deller}, A., {Dodson}, R., {Edwards}, P.~G., {Ekers}, R., {Ellingsen}, S.,
  {Feain}, I., {Gaensler}, B.~M., {Haverkorn}, M., {Hobbs}, G., {Hopkins}, A.,
  {Jackson}, C., {James}, C., {Joncas}, G., {Kaspi}, V., {Kilborn}, V.,
  {Koribalski}, B., {Kothes}, R., {Landecker}, T.~L., {Lenc}, E., {Lovell}, J.,
  {Macquart}, J.-P., {Manchester}, R., {Matthews}, D., {McClure-Griffiths},
  N.~M., {Norris}, R., {Pen}, U.-L., {Phillips}, C., {Power}, C., {Protheroe},
  R., {Sadler}, E., {Schmidt}, B., {Stairs}, I., {Staveley-Smith}, L., {Stil},
  J., {Taylor}, R., {Tingay}, S., {Tzioumis}, A., {Walker}, M., {Wall}, J., and
  {Wolleben}, M. (2007).
\newblock {Science with the Australian Square Kilometre Array Pathfinder}.
\newblock {\em PASA}, 24:174--188.

\bibitem[{Jonker} et~al., 2011]{jonker:2011ApJS..194...18J}
{Jonker}, P.~G., {Bassa}, C.~G., {Nelemans}, G., {Steeghs}, D., {Torres},
  M.~A.~P., {Maccarone}, T.~J., {Hynes}, R.~I., {Greiss}, S., {Clem}, J.,
  {Dieball}, A., {Mikles}, V.~J., {Britt}, C.~T., {Gossen}, L., {Collazzi},
  A.~C., {Wijnands}, R., {In't Zand}, J.~J.~M., {M{\'e}ndez}, M., {Rea}, N.,
  {Kuulkers}, E., {Ratti}, E.~M., {van Haaften}, L.~M., {Heinke}, C.,
  {{\"O}zel}, F., {Groot}, P.~J., and {Verbunt}, F. (2011).
\newblock {The Galactic Bulge Survey: Outline and X-ray Observations}.
\newblock {\em ApJS}, 194:18--+.

\bibitem[{Jonker} et~al., 2010]{jonker:2010MNRAS.407..645J}
{Jonker}, P.~G., {Torres}, M.~A.~P., {Fabian}, A.~C., {Heida}, M., {Miniutti},
  G., and {Pooley}, D. (2010).
\newblock {A bright off-nuclear X-ray source: a type IIn supernova, a bright
  ULX or a recoiling supermassive black hole in CXOJ122518.6+144545}.
\newblock {\em MNRAS}, 407:645--650.

\bibitem[{Kamaretsos} et~al., 2011]{kamaretsos:2011arXiv1107.0854K}
{Kamaretsos}, I., {Hannam}, M., {Husa}, S., and {Sathyaprakash}, B.~S. (2011).
\newblock {Black-hole hair loss: learning about binary progenitors from
  ringdown signals}.
\newblock {\em ArXiv e-prints}.

\bibitem[Kamionkowski et~al., 1994]{Kamionkowski:1993fg}
Kamionkowski, M., Kosowsky, A., and Turner, M.~S. (1994).
\newblock {Gravitational radiation from first order phase transitions}.
\newblock {\em Phys. Rev.}, D49:2837--2851.

\bibitem[{Kasliwal} et~al., 2010]{kasliwal:2010ApJ...723L..98K}
{Kasliwal}, M.~M., {Kulkarni}, S.~R., {Gal-Yam}, A., {Yaron}, O., {Quimby},
  R.~M., {Ofek}, E.~O., {Nugent}, P., {Poznanski}, D., {Jacobsen}, J.,
  {Sternberg}, A., {Arcavi}, I., {Howell}, D.~A., {Sullivan}, M., {Rich},
  D.~J., {Burke}, P.~F., {Brimacombe}, J., {Milisavljevic}, D., {Fesen}, R.,
  {Bildsten}, L., {Shen}, K., {Cenko}, S.~B., {Bloom}, J.~S., {Hsiao}, E.,
  {Law}, N.~M., {Gehrels}, N., {Immler}, S., {Dekany}, R., {Rahmer}, G.,
  {Hale}, D., {Smith}, R., {Zolkower}, J., {Velur}, V., {Walters}, R.,
  {Henning}, J., {Bui}, K., and {McKenna}, D. (2010).
\newblock {Rapidly Decaying Supernova 2010X: A Candidate ''.Ia'' Explosion}.
\newblock {\em ApJ}, 723:L98--L102.

\bibitem[{Kauffmann} and {Haehnelt}, 2000]{kauffmann:2000MNRAS.311..576K}
{Kauffmann}, G. and {Haehnelt}, M. (2000).
\newblock {A unified model for the evolution of galaxies and quasars}.
\newblock {\em MNRAS}, 311:576--588.

\bibitem[{Kesden} et~al., 2005]{kesden:2005:gws}
{Kesden}, M., {Gair}, J., and {Kamionkowski}, M. (2005).
\newblock {Gravitational-wave signature of an inspiral into a supermassive
  horizonless object}.
\newblock {\em Phys.~Rev.~D}, 71(4):044015--+.

\bibitem[{Kesden} et~al., 2010]{kesden:2010ApJ...715.1006K}
{Kesden}, M., {Sperhake}, U., and {Berti}, E. (2010).
\newblock {Relativistic Suppression of Black Hole Recoils}.
\newblock {\em ApJ}, 715:1006--1011.

\bibitem[{Khan} et~al., 2011]{khan:2011ApJ...732...89K}
{Khan}, F.~M., {Just}, A., and {Merritt}, D. (2011).
\newblock {Efficient Merger of Binary Supermassive Black Holes in Merging
  Galaxies}.
\newblock {\em ApJ}, 732:89--+.

\bibitem[{Khlebnikov} and {Tkachev}, 1997]{khlebnikov:1997:rgw}
{Khlebnikov}, S. and {Tkachev}, I. (1997).
\newblock {Relic gravitational waves produced after preheating}.
\newblock {\em Phys.~Rev.~D}, 56:653--660.

\bibitem[{King}, 2003]{king:2003ApJ...596L..27K}
{King}, A. (2003).
\newblock {Black Holes, Galaxy Formation, and the M$_{BH}$-{$\sigma$}
  Relation}.
\newblock {\em ApJ}, 596:L27--L29.

\bibitem[{King} et~al., 2005]{king:2005MNRAS.363...49K}
{King}, A.~R., {Lubow}, S.~H., {Ogilvie}, G.~I., and {Pringle}, J.~E. (2005).
\newblock {Aligning spinning black holes and accretion discs}.
\newblock {\em MNRAS}, 363:49--56.

\bibitem[{King} and {Pringle}, 2006]{king:2006MNRAS.373L..90K}
{King}, A.~R. and {Pringle}, J.~E. (2006).
\newblock {Growing supermassive black holes by chaotic accretion}.
\newblock {\em MNRAS}, 373:L90--L92.

\bibitem[{King} and {Pringle}, 2007]{king:2007MNRAS.377L..25K}
{King}, A.~R. and {Pringle}, J.~E. (2007).
\newblock {Fuelling active galactic nuclei}.
\newblock {\em MNRAS}, 377:L25--L28.

\bibitem[{Kocsis} et~al., 2011]{kocsis:2011PhRvD..84b4032K}
{Kocsis}, B., {Yunes}, N., and {Loeb}, A. (2011).
\newblock {Observable signatures of extreme mass-ratio inspiral black hole
  binaries embedded in thin accretion disks}.
\newblock {\em Phys.~Rev.~D}, 84(2):024032--+.

\bibitem[{Komossa} et~al., 2003]{komossa:2003ApJ...582L..15K}
{Komossa}, S., {Burwitz}, V., {Hasinger}, G., {Predehl}, P., {Kaastra}, J.~S.,
  and {Ikebe}, Y. (2003).
\newblock {Discovery of a Binary Active Galactic Nucleus in the Ultraluminous
  Infrared Galaxy NGC 6240 Using Chandra}.
\newblock {\em ApJ}, 582:L15--L19.

\bibitem[{Komossa} et~al., 2008]{komossa:2008ApJ...678L..81K}
{Komossa}, S., {Zhou}, H., and {Lu}, H. (2008).
\newblock {A Recoiling Supermassive Black Hole in the Quasar SDSS
  J092712.65+294344.0?}
\newblock {\em ApJ}, 678:L81--L84.

\bibitem[{Konstandin} et~al., 2010]{konstandin:2010PhRvD..82h3513K}
{Konstandin}, T., {Nardini}, G., and {Quiros}, M. (2010).
\newblock {Gravitational backreaction effects on the holographic phase
  transition}.
\newblock {\em Phys.~Rev.~D}, 82(8):083513--+.

\bibitem[{Konstantinidis} et~al., 2011]{konstatantinidis:2011arXiv1108.5175K}
{Konstantinidis}, S., {Amaro-Seoane}, P., and {Kokkotas}, K.~D. (2011).
\newblock {Kicking massive black holes off clusters: Intermediate-mass ratio
  inspirals}.
\newblock {\em ArXiv e-prints}.

\bibitem[{Kormendy} and {Bender}, 2009]{kormendy:2009ApJ...691L.142K}
{Kormendy}, J. and {Bender}, R. (2009).
\newblock {Correlations between Supermassive Black Holes, Velocity Dispersions,
  and Mass Deficits in Elliptical Galaxies with Cores}.
\newblock {\em ApJ}, 691:L142--L146.

\bibitem[{Koushiappas} et~al., 2004]{koushiappas:2004:mbhs}
{Koushiappas}, S.~M., {Bullock}, J.~S., and {Dekel}, A. (2004).
\newblock {Massive black hole seeds from low angular momentum material}.
\newblock {\em MNRAS}, 354:292--304.

\bibitem[{Koushiappas} and {Zentner}, 2006]{koushiappas:2006:tms}
{Koushiappas}, S.~M. and {Zentner}, A.~R. (2006).
\newblock {Testing Models of Supermassive Black Hole Seed Formation through
  Gravity Waves}.
\newblock {\em ApJ}, 639:7--22.

\bibitem[{Kramer} et~al., 2006]{kramer:2006:tgr}
{Kramer}, M., {Stairs}, I.~H., {Manchester}, R.~N., {McLaughlin}, M.~A.,
  {Lyne}, A.~G., {Ferdman}, R.~D., {Burgay}, M., {Lorimer}, D.~R., {Possenti},
  A., {D'Amico}, N., {Sarkissian}, J.~M., {Hobbs}, G.~B., {Reynolds}, J.~E.,
  {Freire}, P.~C.~C., and {Camilo}, F. (2006).
\newblock {Tests of General Relativity from Timing the Double Pulsar}.
\newblock {\em Science}, 314:97--102.

\bibitem[{Kramer} and {Wex}, 2009]{kramer:2009:tr}
{Kramer}, M. and {Wex}, N. (2009).
\newblock {TOPICAL REVIEW: The double pulsar system: a unique laboratory for
  gravity}.
\newblock {\em Class.~Quantum~Grav.}, 26(7):073001--+.

\bibitem[{Krolik}, 1999]{krolik:1999:agn}
{Krolik}, J.~H. (1999).
\newblock {\em {Active galactic nuclei: from the central black hole to the
  galactic environment}}.
\newblock Princeton University Press, Princeton, N.J.

\bibitem[{Kuo} et~al., 2011]{kuo:2011ApJ...727...20K}
{Kuo}, C.~Y., {Braatz}, J.~A., {Condon}, J.~J., {Impellizzeri}, C.~M.~V., {Lo},
  K.~Y., {Zaw}, I., {Schenker}, M., {Henkel}, C., {Reid}, M.~J., and {Greene},
  J.~E. (2011).
\newblock {The Megamaser Cosmology Project. III. Accurate Masses of Seven
  Supermassive Black Holes in Active Galaxies with Circumnuclear Megamaser
  Disks}.
\newblock {\em ApJ}, 727:20--+.

\bibitem[{Lacey} and {Cole}, 1993]{lacey:1993:mrhm}
{Lacey}, C. and {Cole}, S. (1993).
\newblock {Merger rates in hierarchical models of galaxy formation}.
\newblock {\em MNRAS}, 262:627--649.

\bibitem[{Lamastra} et~al., 2010]{lamastra:2010MNRAS.405...29L}
{Lamastra}, A., {Menci}, N., {Maiolino}, R., {Fiore}, F., and {Merloni}, A.
  (2010).
\newblock {The building up of the black hole-stellar mass relation}.
\newblock {\em MNRAS}, 405:29--40.

\bibitem[{Lauer} et~al., 2007]{lauer:2007:mnbh}
{Lauer}, T.~R., {Faber}, S.~M., {Richstone}, D., {Gebhardt}, K., {Tremaine},
  S., {Postman}, M., {Dressler}, A., {Aller}, M.~C., {Filippenko}, A.~V.,
  {Green}, R., {Ho}, L.~C., {Kormendy}, J., {Magorrian}, J., and {Pinkney}, J.
  (2007).
\newblock {The Masses of Nuclear Black Holes in Luminous Elliptical Galaxies
  and Implications for the Space Density of the Most Massive Black Holes}.
\newblock {\em ApJ}, 662:808--834.

\bibitem[{Leaver}, 1985]{leaver:1985RSPSA.402..285L}
{Leaver}, E.~W. (1985).
\newblock {An analytic representation for the quasi-normal modes of Kerr black
  holes}.
\newblock {\em Royal Society of London Proceedings Series A}, 402:285--298.

\bibitem[{Levan} et~al., 2006]{levan:2006:grb}
{Levan}, A.~J., {Wynn}, G.~A., {Chapman}, R., {Davies}, M.~B., {King}, A.~R.,
  {Priddey}, R.~S., and {Tanvir}, N.~R. (2006).
\newblock {Short gamma-ray bursts in old populations: magnetars from white
  dwarf-white dwarf mergers}.
\newblock {\em MNRAS}, 368:L1--L5.

\bibitem[{Levin}, 2007]{levin:2007:ssmbh}
{Levin}, Y. (2007).
\newblock {Starbursts near supermassive black holes: young stars in the
  Galactic Centre, and gravitational waves in LISA band}.
\newblock {\em MNRAS}, 374:515--524.

\bibitem[{Levitan} et~al., 2011]{levitan:2011ApJ...739...68L}
{Levitan}, D., {Fulton}, B.~J., {Groot}, P.~J., {Kulkarni}, S.~R., {Ofek},
  E.~O., {Prince}, T.~A., {Shporer}, A., {Bloom}, J.~S., {Cenko}, S.~B.,
  {Kasliwal}, M.~M., {Law}, N.~M., {Nugent}, P.~E., {Poznanski}, D., {Quimby},
  R.~M., {Horesh}, A., {Sesar}, B., and {Sternberg}, A. (2011).
\newblock {PTF1 J071912.13+485834.0: An Outbursting AM CVn System Discovered by
  a Synoptic Survey}.
\newblock {\em ApJ}, 739:68--+.

\bibitem[{Li} et~al., 2011]{li:2011arXiv1110.0530L}
{Li}, T.~G.~F., {Del Pozzo}, W., {Vitale}, S., {Van Den Broeck}, C., {Agathos},
  M., {Veitch}, J., {Grover}, K., {Sidery}, T., {Sturani}, R., and {Vecchio},
  A. (2011).
\newblock {Towards a generic test of the strong field dynamics of general
  relativity using compact binary coalescence}.
\newblock {\em ArXiv e-prints}.

\bibitem[{Lin} et~al., 2004]{lin:2004:deep2}
{Lin}, L., {Koo}, D.~C., {Willmer}, C.~N.~A., {Patton}, D.~R., {Conselice},
  C.~J., {Yan}, R., {Coil}, A.~L., {Cooper}, M.~C., {Davis}, M., {Faber},
  S.~M., {Gerke}, B.~F., {Guhathakurta}, P., and {Newman}, J.~A. (2004).
\newblock {The DEEP2 Galaxy Redshift Survey: Evolution of Close Galaxy Pairs
  and Major-Merger Rates up to z \~{} 1.2}.
\newblock {\em ApJ}, 617:L9--L12.

\bibitem[{Lin} et~al., 2008]{lin:2008:rew}
{Lin}, L., {Patton}, D.~R., {Koo}, D.~C., {Casteels}, K., {Conselice}, C.~J.,
  {Faber}, S.~M., {Lotz}, J., {Willmer}, C.~N.~A., {Hsieh}, B.~C., {Chiueh},
  T., {Newman}, J.~A., {Novak}, G.~S., {Weiner}, B.~J., and {Cooper}, M.~C.
  (2008).
\newblock {The Redshift Evolution of Wet, Dry, and Mixed Galaxy Mergers from
  Close Galaxy Pairs in the DEEP2 Galaxy Redshift Survey}.
\newblock {\em ApJ}, 681:232--243.

\bibitem[{Liu}, 2004]{liu:2004MNRAS.347.1357L}
{Liu}, F.~K. (2004).
\newblock {X-shaped radio galaxies as observational evidence for the
  interaction of supermassive binary black holes and accretion disc at parsec
  scale}.
\newblock {\em MNRAS}, 347:1357--1369.

\bibitem[{Lodato} and {Natarajan}, 2006]{lodato:2006:smbh}
{Lodato}, G. and {Natarajan}, P. (2006).
\newblock {Supermassive black hole formation during the assembly of
  pre-galactic discs}.
\newblock {\em MNRAS}, 371:1813--1823.

\bibitem[{Lodato} et~al., 2009]{lodato:2009MNRAS.398.1392L}
{Lodato}, G., {Nayakshin}, S., {King}, A.~R., and {Pringle}, J.~E. (2009).
\newblock {Black hole mergers: can gas discs solve the `final parsec' problem?}
\newblock {\em MNRAS}, 398:1392--1402.

\bibitem[{Loeb}, 2007]{loeb:2007PhRvL..99d1103L}
{Loeb}, A. (2007).
\newblock {Observable Signatures of a Black Hole Ejected by
  Gravitational-Radiation Recoil in a Galaxy Merger}.
\newblock {\em Phys.~Rev.~Lett.}, 99(4):041103--+.

\bibitem[{Loeb} and {Rasio}, 1994]{loeb:1994:cbg}
{Loeb}, A. and {Rasio}, F.~A. (1994).
\newblock {Collapse of primordial gas clouds and the formation of quasar black
  holes}.
\newblock {\em ApJ}, 432:52--61.

\bibitem[{Lorimer}, 2008]{lorimer:2008:bmp}
{Lorimer}, D.~R. (2008).
\newblock {Binary and Millisecond Pulsars}.
\newblock {\em Living Reviews in Relativity}, 11:8--+.

\bibitem[{Lousto} and {Zlochower}, 2011a]{lousto:2011arXiv1108.2009L}
{Lousto}, C.~O. and {Zlochower}, Y. (2011a).
\newblock {Hangup Kicks: Still Larger Recoils by Partial Spin/Orbit Alignment
  of Black-Hole Binaries}.
\newblock {\em ArXiv e-prints}.

\bibitem[{Lousto} and {Zlochower}, 2011b]{lousto:2011PhRvD..83b4003L}
{Lousto}, C.~O. and {Zlochower}, Y. (2011b).
\newblock {Modeling maximum astrophysical gravitational recoil velocities}.
\newblock {\em Phys.~Rev.~D}, 83(2):024003--+.

\bibitem[{LSST Science Collaborations} et~al., 2009]{lsst:2009arXiv0912.0201L}
{LSST Science Collaborations}, {Abell}, P.~A., {Allison}, J., {Anderson},
  S.~F., {Andrew}, J.~R., {Angel}, J.~R.~P., {Armus}, L., {Arnett}, D.,
  {Asztalos}, S.~J., {Axelrod}, T.~S., and et~al. (2009).
\newblock {LSST Science Book, Version 2.0}.
\newblock {\em ArXiv e-prints}.

\bibitem[{Lukes-Gerakopoulos} et~al.,
  2010]{lukes-gerakopoulos:2010PhRvD..81l4005L}
{Lukes-Gerakopoulos}, G., {Apostolatos}, T.~A., and {Contopoulos}, G. (2010).
\newblock {Observable signature of a background deviating from the Kerr
  metric}.
\newblock {\em Phys.~Rev.~D}, 81(12):124005--+.

\bibitem[{MacFadyen} and {Milosavljevi{\'c}},
  2008]{macfayden:2008ApJ...672...83M}
{MacFadyen}, A.~I. and {Milosavljevi{\'c}}, M. (2008).
\newblock {An Eccentric Circumbinary Accretion Disk and the Detection of Binary
  Massive Black Holes}.
\newblock {\em ApJ}, 672:83--93.

\bibitem[{MacLeod} and {Hogan}, 2008]{macleod:2008:phc}
{MacLeod}, C.~L. and {Hogan}, C.~J. (2008).
\newblock {Precision of Hubble constant derived using black hole binary
  absolute distances and statistical redshift information}.
\newblock {\em Phys.~Rev.~D}, 77(4):043512--+.

\bibitem[{Madau} and {Rees}, 2001]{madau:2001:mbhIII}
{Madau}, P. and {Rees}, M.~J. (2001).
\newblock {Massive Black Holes as Population III Remnants}.
\newblock {\em ApJ}, 551:L27--L30.

\bibitem[{Maggiore}, 2000]{maggiore:2000:gwe}
{Maggiore}, M. (2000).
\newblock {Gravitational wave experiments and early universe cosmology.}
\newblock {\em Phys.~Rep.}, 331:283--367.

\bibitem[{Magorrian} et~al., 1998]{magorrian:1998:dmdo}
{Magorrian}, J., {Tremaine}, S., {Richstone}, D., {Bender}, R., {Bower}, G.,
  {Dressler}, A., {Faber}, S.~M., {Gebhardt}, K., {Green}, R., {Grillmair}, C.,
  {Kormendy}, J., and {Lauer}, T. (1998).
\newblock {The Demography of Massive Dark Objects in Galaxy Centers}.
\newblock {\em AJ}, 115:2285--2305.

\bibitem[{Makino} and {Funato}, 2004]{makino:2004ApJ...602...93M}
{Makino}, J. and {Funato}, Y. (2004).
\newblock {Evolution of Massive Black Hole Binaries}.
\newblock {\em ApJ}, 602:93--102.

\bibitem[{Maoz}, 1998]{maoz:1998:dc}
{Maoz}, E. (1998).
\newblock {Dynamical Constraints on Alternatives to Supermassive Black Holes in
  Galactic Nuclei}.
\newblock {\em ApJ}, 494:L181+.

\bibitem[{Marconi} and {Hunt}, 2003]{marconi:2003ApJ...589L..21M}
{Marconi}, A. and {Hunt}, L.~K. (2003).
\newblock {The Relation between Black Hole Mass, Bulge Mass, and Near-Infrared
  Luminosity}.
\newblock {\em ApJ}, 589:L21--L24.

\bibitem[{Marconi} et~al., 2004]{marconi:2004:lsmbh}
{Marconi}, A., {Risaliti}, G., {Gilli}, R., {Hunt}, L.~K., {Maiolino}, R., and
  {Salvati}, M. (2004).
\newblock {Local supermassive black holes, relics of active galactic nuclei and
  the X-ray background}.
\newblock {\em MNRAS}, 351:169--185.

\bibitem[{Marronetti} et~al., 2008]{marronetti:2008PhRvD..77f4010M}
{Marronetti}, P., {Tichy}, W., {Br{\"u}gmann}, B., {Gonz{\'a}lez}, J., and
  {Sperhake}, U. (2008).
\newblock {High-spin binary black hole mergers}.
\newblock {\em Phys.~Rev.~D}, 77(6):064010--+.

\bibitem[{Marsh}, 2011]{marsh:2011CQGra..28i4019M}
{Marsh}, T.~R. (2011).
\newblock {Double white dwarfs and LISA}.
\newblock {\em Class.~Quantum~Grav.}, 28(9):094019--+.

\bibitem[{Marsh} et~al., 2004]{marsh:2004:mtwd}
{Marsh}, T.~R., {Nelemans}, G., and {Steeghs}, D. (2004).
\newblock {Mass transfer between double white dwarfs}.
\newblock {\em MNRAS}, 350:113--128.

\bibitem[{Marsh} and {Steeghs}, 2002]{marsh:2002MNRAS.331L...7M}
{Marsh}, T.~R. and {Steeghs}, D. (2002).
\newblock {V407 Vul: a direct impact accretor}.
\newblock {\em MNRAS}, 331:L7--L11.

\bibitem[{Mathur} and {Grupe}, 2005]{mathur:2005ApJ...633..688M}
{Mathur}, S. and {Grupe}, D. (2005).
\newblock {The Locus of Highly Accreting Active Galactic Nuclei on the
  M$_{BH}$-{$\sigma$} Plane: Selections, Limitations, and Implications}.
\newblock {\em ApJ}, 633:688--692.

\bibitem[{Mayer} et~al., 2010]{mayer:2010Natur.466.1082M}
{Mayer}, L., {Kazantzidis}, S., {Escala}, A., and {Callegari}, S. (2010).
\newblock {Direct formation of supermassive black holes via multi-scale gas
  inflows in galaxy mergers}.
\newblock {\em Nature}, 466:1082--1084.

\bibitem[{Mayer} et~al., 2007]{mayer:2007Sci...316.1874M}
{Mayer}, L., {Kazantzidis}, S., {Madau}, P., {Colpi}, M., {Quinn}, T., and
  {Wadsley}, J. (2007).
\newblock {Rapid Formation of Supermassive Black Hole Binaries in Galaxy
  Mergers with Gas}.
\newblock {\em Science}, 316:1874--.

\bibitem[{McWilliams}, 2010]{mcwilliams:2010PhRvL.104n1601M}
{McWilliams}, S.~T. (2010).
\newblock {Constraining the Braneworld with Gravitational Wave Observations}.
\newblock {\em Phys.~Rev.~Lett.}, 104(14):141601--+.

\bibitem[{Merloni}, 2004]{merloni:2004:ahg}
{Merloni}, A. (2004).
\newblock {The anti-hierarchical growth of supermassive black holes}.
\newblock {\em MNRAS}, 353:1035--1047.

\bibitem[{Merritt}, 2006]{merritt:2006ApJ...648..976M}
{Merritt}, D. (2006).
\newblock {Mass Deficits, Stalling Radii, and the Merger Histories of
  Elliptical Galaxies}.
\newblock {\em ApJ}, 648:976--986.

\bibitem[{Merritt} et~al., 2011]{merritt:2011PhRvD..84d4024M}
{Merritt}, D., {Alexander}, T., {Mikkola}, S., and {Will}, C.~M. (2011).
\newblock {Stellar dynamics of extreme-mass-ratio inspirals}.
\newblock {\em Phys.~Rev.~D}, 84(4):044024--+.

\bibitem[{Merritt} and {Ekers}, 2002]{merritt:2002Sci...297.1310M}
{Merritt}, D. and {Ekers}, R.~D. (2002).
\newblock {Tracing Black Hole Mergers Through Radio Lobe Morphology}.
\newblock {\em Science}, 297:1310--1313.

\bibitem[{Merritt} et~al., 2009a]{merritt:2009:eog}
{Merritt}, D., {Gualandris}, A., and {Mikkola}, S. (2009a).
\newblock {Explaining the Orbits of the Galactic Center S-Stars}.
\newblock {\em ApJ}, 693:L35--L38.

\bibitem[{Merritt} et~al., 2007]{merritt:2007ApJ...671...53M}
{Merritt}, D., {Mikkola}, S., and {Szell}, A. (2007).
\newblock {Long-Term Evolution of Massive Black Hole Binaries. III. Binary
  Evolution in Collisional Nuclei}.
\newblock {\em ApJ}, 671:53--72.

\bibitem[{Merritt} and {Milosavljevi{\'c}}, 2005]{merritt:2005:mbh}
{Merritt}, D. and {Milosavljevi{\'c}}, M. (2005).
\newblock {Massive Black Hole Binary Evolution}.
\newblock {\em Living Reviews in Relativity}, 8:8--+.

\bibitem[{Merritt} and {Poon}, 2004]{merritt:2004:clc}
{Merritt}, D. and {Poon}, M.~Y. (2004).
\newblock {Chaotic Loss Cones and Black Hole Fueling}.
\newblock {\em ApJ}, 606:788--798.

\bibitem[{Merritt} et~al., 2009b]{merritt:2009ApJ...699.1690M}
{Merritt}, D., {Schnittman}, J.~D., and {Komossa}, S. (2009b).
\newblock {Hypercompact Stellar Systems Around Recoiling Supermassive Black
  Holes}.
\newblock {\em ApJ}, 699:1690--1710.

\bibitem[{Mihos} and {Hernquist}, 1996]{mihos:1996ApJ...464..641M}
{Mihos}, J.~C. and {Hernquist}, L. (1996).
\newblock {Gasdynamics and Starbursts in Major Mergers}.
\newblock {\em ApJ}, 464:641--+.

\bibitem[{Miller}, 2009]{miller:2009CQGra..26i4031M}
{Miller}, M.~C. (2009).
\newblock {Intermediate-mass black holes as LISA sources}.
\newblock {\em Class.~Quantum~Grav.}, 26(9):094031--+.

\bibitem[{Miller} and {Colbert}, 2004]{miller:2004:imbh}
{Miller}, M.~C. and {Colbert}, E.~J.~M. (2004).
\newblock {Intermediate-Mass Black Holes}.
\newblock {\em International Journal of Modern Physics D}, 13:1--64.

\bibitem[{Miller} et~al., 2005]{miller:2005:bes}
{Miller}, M.~C., {Freitag}, M., {Hamilton}, D.~P., and {Lauburg}, V.~M. (2005).
\newblock {Binary Encounters with Supermassive Black Holes: Zero-Eccentricity
  LISA Events}.
\newblock {\em ApJ}, 631:L117--L120.

\bibitem[{Milosavljevi{\'c}} and {Merritt},
  2001]{milosavljevic:2001ApJ...563...34M}
{Milosavljevi{\'c}}, M. and {Merritt}, D. (2001).
\newblock {Formation of Galactic Nuclei}.
\newblock {\em ApJ}, 563:34--62.

\bibitem[{Milosavljevi{\'c}} and {Phinney}, 2005]{milosavljevic:2005:amb}
{Milosavljevi{\'c}}, M. and {Phinney}, E.~S. (2005).
\newblock {The Afterglow of Massive Black Hole Coalescence}.
\newblock {\em ApJ}, 622:L93--L96.

\bibitem[{Miralda-Escude}, 1998]{miralda-escude:1998ApJ...501...15M}
{Miralda-Escude}, J. (1998).
\newblock {Reionization of the Intergalactic Medium and the Damping Wing of the
  Gunn-Peterson Trough}.
\newblock {\em ApJ}, 501:15--+.

\bibitem[{Miralda-Escud{\'e}} and {Gould}, 2000]{miralda-escoude:2000:bhgc}
{Miralda-Escud{\'e}}, J. and {Gould}, A. (2000).
\newblock {A Cluster of Black Holes at the Galactic Center}.
\newblock {\em ApJ}, 545:847--853.

\bibitem[{Mirshekari} et~al., 2011]{mirshekari:2011arXiv1110.2720M}
{Mirshekari}, S., {Yunes}, N., and {Will}, C.~M. (2011).
\newblock {Constraining Generic Lorentz Violation and the Speed of the Graviton
  with Gravitational Waves}.
\newblock {\em ArXiv e-prints}.

\bibitem[{Mishra} et~al., 2010]{mishra:2010PhRvD..82f4010M}
{Mishra}, C.~K., {Arun}, K.~G., {Iyer}, B.~R., and {Sathyaprakash}, B.~S.
  (2010).
\newblock {Parametrized tests of post-Newtonian theory using Advanced LIGO and
  Einstein Telescope}.
\newblock {\em Phys.~Rev.~D}, 82(6):064010--+.

\bibitem[{Mo} et~al., 2010]{mo:2010gfe..book.....M}
{Mo}, H., {van den Bosch}, F.~C., and {White}, S. (2010).
\newblock {\em {Galaxy Formation and Evolution}}.
\newblock Cambridge University Press.

\bibitem[{Moderski} et~al., 1998]{moderski:1998MNRAS.301..142M}
{Moderski}, R., {Sikora}, M., and {Lasota}, J.-P. (1998).
\newblock {On the spin paradigm and the radio dichotomy of quasars}.
\newblock {\em MNRAS}, 301:142--148.

\bibitem[{Montero} et~al., 2011]{montero:2011arXiv1108.3090M}
{Montero}, P.~J., {Janka}, H.-T., and {Mueller}, E. (2011).
\newblock {Relativistic collapse and explosion of rotating supermassive stars
  with thermonuclear effects}.
\newblock {\em ArXiv e-prints}.

\bibitem[{Montuori} et~al., 2011]{montuori:2011MNRAS.412...26M}
{Montuori}, C., {Dotti}, M., {Colpi}, M., {Decarli}, R., and {Haardt}, F.
  (2011).
\newblock {Search for sub-parsec massive binary black holes through line
  diagnosis}.
\newblock {\em MNRAS}, 412:26--32.

\bibitem[{Mortlock} et~al., 2011]{mortlock:2011Natur.474..616M}
{Mortlock}, D.~J., {Warren}, S.~J., {Venemans}, B.~P., {Patel}, M., {Hewett},
  P.~C., {McMahon}, R.~G., {Simpson}, C., {Theuns}, T., {Gonz{\'a}les-Solares},
  E.~A., {Adamson}, A., {Dye}, S., {Hambly}, N.~C., {Hirst}, P., {Irwin},
  M.~J., {Kuiper}, E., {Lawrence}, A., and {R{\"o}ttgering}, H.~J.~A. (2011).
\newblock {A luminous quasar at a redshift of z = 7.085}.
\newblock {\em Nature}, 474:616--619.

\bibitem[{Mouawad} et~al., 2005]{mouawad:2005:wcgc}
{Mouawad}, N., {Eckart}, A., {Pfalzner}, S., {Sch{\"o}del}, R., {Moultaka}, J.,
  and {Spurzem}, R. (2005).
\newblock {Weighing the cusp at the Galactic Centre}.
\newblock {\em Astronomische Nachrichten}, 326:83--95.

\bibitem[{Nelemans} et~al., 2001]{nelemans:2001:ps2}
{Nelemans}, G., {Portegies Zwart}, S.~F., {Verbunt}, F., and {Yungelson}, L.~R.
  (2001).
\newblock {Population synthesis for double white dwarfs. II. Semi-detached
  systems: AM CVn stars}.
\newblock {\em A\&A}, 368:939--949.

\bibitem[{Nelemans} and {Tout}, 2005]{nelemans:2005MNRAS.356..753N}
{Nelemans}, G. and {Tout}, C.~A. (2005).
\newblock {Reconstructing the evolution of white dwarf binaries: further
  evidence for an alternative algorithm for the outcome of the common-envelope
  phase in close binaries}.
\newblock {\em MNRAS}, 356:753--764.

\bibitem[{Nelemans} et~al., 2004]{nelemans:2004:spc}
{Nelemans}, G., {Yungelson}, L.~R., and {Portegies Zwart}, S.~F. (2004).
\newblock {Short-period AM CVn systems as optical, X-ray and gravitational-wave
  sources}.
\newblock {\em MNRAS}, 349:181--192.

\bibitem[{O'Leary} and {Loeb}, 2009]{oleary:2009MNRAS.395..781O}
{O'Leary}, R.~M. and {Loeb}, A. (2009).
\newblock {Star clusters around recoiled black holes in the Milky Way halo}.
\newblock {\em MNRAS}, 395:781--786.

\bibitem[{Omukai} and {Palla}, 2001]{omukai:2001ApJ...561L..55O}
{Omukai}, K. and {Palla}, F. (2001).
\newblock {On the Formation of Massive Primordial Stars}.
\newblock {\em ApJ}, 561:L55--L58.

\bibitem[{Omukai} and {Palla}, 2003]{omukai:2003ApJ...589..677O}
{Omukai}, K. and {Palla}, F. (2003).
\newblock {Formation of the First Stars by Accretion}.
\newblock {\em ApJ}, 589:677--687.

\bibitem[{Orosz}, 2003]{orosz:2003IAUS..212..365O}
{Orosz}, J.~A. (2003).
\newblock {Inventory of black hole binaries}.
\newblock In {K.~van der Hucht, A.~Herrero, \& C.~Esteban}, editor, {\em A
  Massive Star Odyssey: From Main Sequence to Supernova}, volume 212 of {\em
  IAU Symposium}, pages 365--+.

\bibitem[{Ostriker}, 1999]{ostriker:1999ApJ...513..252O}
{Ostriker}, E.~C. (1999).
\newblock {Dynamical Friction in a Gaseous Medium}.
\newblock {\em ApJ}, 513:252--258.

\bibitem[{Paczynski}, 1976]{paczynski:1976:secbs}
{Paczynski}, B. (1976).
\newblock {Common Envelope Binaries}.
\newblock In {P.~Eggleton, S.~Mitton, \& J.~Whelan}, editor, {\em Structure and
  Evolution of Close Binary Systems}, volume~73 of {\em IAU Symposium}, pages
  75--+.

\bibitem[{Pakmor} et~al., 2010]{pakmor:2010:sltIa}
{Pakmor}, R., {Kromer}, M., {R{\"o}pke}, F.~K., {Sim}, S.~A., {Ruiter}, A.~J.,
  and {Hillebrandt}, W. (2010).
\newblock {Sub-luminous type Ia supernovae from the mergers of equal-mass white
  dwarfs with mass \~{}$0.9M_\text{solar}$}.
\newblock {\em Nature}, 463:61--64.

\bibitem[{Pan} et~al., 2011]{pan:2011arXiv1106.1021P}
{Pan}, Y., {Buonanno}, A., {Boyle}, M., {Buchman}, L.~T., {Kidder}, L.~E.,
  {Pfeiffer}, H.~P., and {Scheel}, M.~A. (2011).
\newblock {Inspiral-merger-ringdown multipolar waveforms of nonspinning
  black-hole binaries using the effective-one-body formalism}.
\newblock {\em ArXiv e-prints}.

\bibitem[{Pani} et~al., 2009]{pani:2009PhRvD..80l4047P}
{Pani}, P., {Berti}, E., {Cardoso}, V., {Chen}, Y., and {Norte}, R. (2009).
\newblock {Gravitational wave signatures of the absence of an event horizon:
  Nonradial oscillations of a thin-shell gravastar}.
\newblock {\em Phys.~Rev.~D}, 80(12):124047--+.

\bibitem[{Pani} and {Cardoso}, 2009]{pani:2009PhRvD..79h4031P}
{Pani}, P. and {Cardoso}, V. (2009).
\newblock {Are black holes in alternative theories serious astrophysical
  candidates? The case for Einstein-dilaton-Gauss-Bonnet black holes}.
\newblock {\em Phys.~Rev.~D}, 79(8):084031.

\bibitem[{Pani} et~al., 2011]{pani:2011PhRvD..83j4048P}
{Pani}, P., {Cardoso}, V., and {Gualtieri}, L. (2011).
\newblock {Gravitational waves from extreme mass-ratio inspirals in dynamical
  Chern-Simons gravity}.
\newblock {\em Phys.~Rev.~D}, 83(10):104048--+.

\bibitem[{Patruno} et~al., 2011]{patruno:2011arXiv1109.0536P}
{Patruno}, A., {Haskell}, B., and {D'Angelo}, C. (2011).
\newblock {Gravitational Waves and the Maximum Spin Frequency of Neutron
  Stars}.
\newblock {\em ArXiv e-prints}.

\bibitem[{Patton} et~al., 2002]{patton:2002:dcgp}
{Patton}, D.~R., {Pritchet}, C.~J., {Carlberg}, R.~G., {Marzke}, R.~O., {Yee},
  H.~K.~C., {Hall}, P.~B., {Lin}, H., {Morris}, S.~L., {Sawicki}, M.,
  {Shepherd}, C.~W., and {Wirth}, G.~D. (2002).
\newblock {Dynamically Close Galaxy Pairs and Merger Rate Evolution in the
  CNOC2 Redshift Survey}.
\newblock {\em ApJ}, 565:208--222.

\bibitem[{Peng}, 2007]{peng:2007ApJ...671.1098P}
{Peng}, C.~Y. (2007).
\newblock {How Mergers May Affect the Mass Scaling Relation between
  Gravitationally Bound Systems}.
\newblock {\em ApJ}, 671:1098--1107.

\bibitem[{Perego} et~al., 2009]{perego:2009MNRAS.399.2249P}
{Perego}, A., {Dotti}, M., {Colpi}, M., and {Volonteri}, M. (2009).
\newblock {Mass and spin co-evolution during the alignment of a black hole in a
  warped accretion disc}.
\newblock {\em MNRAS}, 399:2249--2263.

\bibitem[{Perets} and {Alexander}, 2008]{perets:2008:mpem}
{Perets}, H.~B. and {Alexander}, T. (2008).
\newblock {Massive Perturbers and the Efficient Merger of Binary Massive Black
  Holes}.
\newblock {\em ApJ}, 677:146--159.

\bibitem[{Perets} et~al., 2011]{perets:2011ApJ...730...89P}
{Perets}, H.~B., {Badenes}, C., {Arcavi}, I., {Simon}, J.~D., and {Gal-yam}, A.
  (2011).
\newblock {An Emerging Class of Bright, Fast-evolving Supernovae with Low-mass
  Ejecta}.
\newblock {\em ApJ}, 730:89--+.

\bibitem[{Perets} et~al., 2010]{perets:2010Natur.465..322P}
{Perets}, H.~B., {Gal-Yam}, A., {Mazzali}, P.~A., {Arnett}, D., {Kagan}, D.,
  {Filippenko}, A.~V., {Li}, W., {Arcavi}, I., {Cenko}, S.~B., {Fox}, D.~B.,
  {Leonard}, D.~C., {Moon}, D.-S., {Sand}, D.~J., {Soderberg}, A.~M.,
  {Anderson}, J.~P., {James}, P.~A., {Foley}, R.~J., {Ganeshalingam}, M.,
  {Ofek}, E.~O., {Bildsten}, L., {Nelemans}, G., {Shen}, K.~J., {Weinberg},
  N.~N., {Metzger}, B.~D., {Piro}, A.~L., {Quataert}, E., {Kiewe}, M., and
  {Poznanski}, D. (2010).
\newblock {A faint type of supernova from a white dwarf with a helium-rich
  companion}.
\newblock {\em Nature}, 465:322--325.

\bibitem[{Perlmutter} and {Riess}, 1999]{perlmutter:1999AIPC..478..129P}
{Perlmutter}, S. and {Riess}, A. (1999).
\newblock {Cosmological parameters from supernovae: Two groups' results agree}.
\newblock In {D.~O.~Caldwell}, editor, {\em COSMO-98}, volume 478 of {\em
  AIP~Conf.~Series}, pages 129--142.

\bibitem[{Peters}, 1964]{peters:1964PhRv..136.1224P}
{Peters}, P.~C. (1964).
\newblock {Gravitational Radiation and the Motion of Two Point Masses}.
\newblock {\em Physical Review}, 136:1224--1232.

\bibitem[{Peterson} et~al., 2005]{peterson:2005ApJ...632..799P}
{Peterson}, B.~M., {Bentz}, M.~C., {Desroches}, L.-B., {Filippenko}, A.~V.,
  {Ho}, L.~C., {Kaspi}, S., {Laor}, A., {Maoz}, D., {Moran}, E.~C., {Pogge},
  R.~W., and {Quillen}, A.~C. (2005).
\newblock {Multiwavelength Monitoring of the Dwarf Seyfert 1 Galaxy NGC 4395.
  I. A Reverberation-based Measurement of the Black Hole Mass}.
\newblock {\em ApJ}, 632:799--808.

\bibitem[{Petiteau} et~al., 2008]{petiteau:2008PhRvD..77b3002P}
{Petiteau}, A., {Auger}, G., {Halloin}, H., {Jeannin}, O., {Plagnol}, E.,
  {Pireaux}, S., {Regimbau}, T., and {Vinet}, J.-Y. (2008).
\newblock {LISACode: A scientific simulator of LISA}.
\newblock {\em Phys.~Rev.~D}, 77(2):023002.

\bibitem[{Petiteau} et~al., 2011]{petiteau:2011ApJ...732...82P}
{Petiteau}, A., {Babak}, S., and {Sesana}, A. (2011).
\newblock {Constraining the Dark Energy Equation of State Using LISA
  Observations of Spinning Massive Black Hole Binaries}.
\newblock {\em ApJ}, 732:82--+.

\bibitem[{Phinney}, 2009]{phinney:2009astro2010S.235P}
{Phinney}, E.~S. (2009).
\newblock {Finding and Using Electromagnetic Counterparts of Gravitational Wave
  Sources}.
\newblock In {\em astro2010: The Astronomy and Astrophysics Decadal Survey},
  volume 2010 of {\em Astronomy}, pages 235--+.

\bibitem[{Piro}, 2011]{piro:2011ApJ...740L..53P}
{Piro}, A.~L. (2011).
\newblock {Tidal Interactions in Merging White Dwarf Binaries}.
\newblock {\em ApJ}, 740:L53+.

\bibitem[{Popham} and {Gammie}, 1998]{popham:1998ApJ...504..419P}
{Popham}, R. and {Gammie}, C.~F. (1998).
\newblock {Advection-dominated Accretion Flows in the Kerr Metric. II. Steady
  State Global Solutions}.
\newblock {\em ApJ}, 504:419--+.

\bibitem[{Portegies Zwart} et~al., 2004]{portegieszwart:2004Natur.428..724P}
{Portegies Zwart}, S.~F., {Baumgardt}, H., {Hut}, P., {Makino}, J., and
  {McMillan}, S.~L.~W. (2004).
\newblock {Formation of massive black holes through runaway collisions in dense
  young star clusters}.
\newblock {\em Nature}, 428:724--726.

\bibitem[{Portegies Zwart} and {McMillan},
  2000]{portegieszwart:2000ApJ...528L..17P}
{Portegies Zwart}, S.~F. and {McMillan}, S.~L.~W. (2000).
\newblock {Black Hole Mergers in the Universe}.
\newblock {\em ApJ}, 528:L17--L20.

\bibitem[{Portegies Zwart} and {McMillan},
  2002]{portegieszwart:2002ApJ...576..899P}
{Portegies Zwart}, S.~F. and {McMillan}, S.~L.~W. (2002).
\newblock {The Runaway Growth of Intermediate-Mass Black Holes in Dense Star
  Clusters}.
\newblock {\em ApJ}, 576:899--907.

\bibitem[{Porter}, 2009]{porter:2009GWN.....1....4P}
{Porter}, E.~K. (2009).
\newblock {An Overview of LISA Data Analysis Algorithms}.
\newblock {\em GW Notes}, 1:4--+.

\bibitem[{Press} and {Schechter}, 1974]{press:1974:fgcg}
{Press}, W.~H. and {Schechter}, P. (1974).
\newblock {Formation of Galaxies and Clusters of Galaxies by Self-Similar
  Gravitational Condensation}.
\newblock {\em ApJ}, 187:425--438.

\bibitem[{Preto} and {Amaro-Seoane}, 2010]{preto:2010:ApJ...708L..42P}
{Preto}, M. and {Amaro-Seoane}, P. (2010).
\newblock {On Strong Mass Segregation Around a Massive Black Hole: Implications
  for Lower-Frequency Gravitational-Wave Astrophysics}.
\newblock {\em ApJ}, 708:L42--L46.

\bibitem[{Preto} et~al., 2011]{preto:2011ApJ...732L..26P}
{Preto}, M., {Berentzen}, I., {Berczik}, P., and {Spurzem}, R. (2011).
\newblock {Fast Coalescence of Massive Black Hole Binaries from Mergers of
  Galactic Nuclei: Implications for Low-frequency Gravitational-wave
  Astrophysics}.
\newblock {\em ApJ}, 732:L26+.

\bibitem[{Pretorius}, 2005]{pretorius:2005:ebs}
{Pretorius}, F. (2005).
\newblock {Evolution of Binary Black-Hole Spacetimes}.
\newblock {\em Phys.~Rev.~Lett.}, 95(12):121101--+.

\bibitem[{Quinlan}, 1996]{quinlan:1996:dembhb}
{Quinlan}, G.~D. (1996).
\newblock {The dynamical evolution of massive black hole binaries I. Hardening
  in a fixed stellar background}.
\newblock {\em NewA}, 1:35--56.

\bibitem[{Racine} et~al., 2007]{racine:2007:ndts}
{Racine}, {\'E}., {Phinney}, E.~S., and {Arras}, P. (2007).
\newblock {Non-dissipative tidal synchronization in accreting binary white
  dwarf systems}.
\newblock {\em MNRAS}, 380:381--398.

\bibitem[{Randall} and {Servant}, 2007]{randall:2006:gww}
{Randall}, L. and {Servant}, G. (2007).
\newblock {Gravitational Waves from Warped Spacetime}.
\newblock {\em Journal of High Energy Physics}, 5:54--+.

\bibitem[{Rau} et~al., 2010]{rau:2010ApJ...708..456R}
{Rau}, A., {Roelofs}, G.~H.~A., {Groot}, P.~J., {Marsh}, T.~R., {Nelemans}, G.,
  {Steeghs}, D., {Salvato}, M., and {Kasliwal}, M.~M. (2010).
\newblock {A Census of AM CVn Stars: Three New Candidates and One Confirmed
  48.3-Minute Binary}.
\newblock {\em ApJ}, 708:456--461.

\bibitem[{Rezzolla} et~al., 2008a]{rezzolla:2008PhRvD..78d4002R}
{Rezzolla}, L., {Barausse}, E., {Dorband}, E.~N., {Pollney}, D., {Reisswig},
  C., {Seiler}, J., and {Husa}, S. (2008a).
\newblock {Final spin from the coalescence of two black holes}.
\newblock {\em Phys.~Rev.~D}, 78(4):044002--+.

\bibitem[{Rezzolla} et~al., 2008b]{rezzolla:2008ApJ...674L..29R}
{Rezzolla}, L., {Diener}, P., {Dorband}, E.~N., {Pollney}, D., {Reisswig}, C.,
  {Schnetter}, E., and {Seiler}, J. (2008b).
\newblock {The Final Spin from the Coalescence of Aligned-Spin Black Hole
  Binaries}.
\newblock {\em ApJ}, 674:L29--L32.

\bibitem[{Rhook} and {Wyithe}, 2005]{rhook:2005:rer}
{Rhook}, K.~J. and {Wyithe}, J.~S.~B. (2005).
\newblock {Realistic event rates for detection of supermassive black hole
  coalescence by LISA}.
\newblock {\em MNRAS}, 361:1145--1152.

\bibitem[{Riess} et~al., 1998]{riess:1998ApJ...504..935R}
{Riess}, A.~G., {Nugent}, P., {Filippenko}, A.~V., {Kirshner}, R.~P., and
  {Perlmutter}, S. (1998).
\newblock {Snapshot Distances to Type IA Supernovae: All in ``One'' Night's
  Work}.
\newblock {\em ApJ}, 504:935--+.

\bibitem[{Ripamonti} et~al., 2002]{ripamonti:2002MNRAS.334..401R}
{Ripamonti}, E., {Haardt}, F., {Ferrara}, A., and {Colpi}, M. (2002).
\newblock {Radiation from the first forming stars}.
\newblock {\em MNRAS}, 334:401--418.

\bibitem[{Robertson} et~al., 2006]{robertson:2006ApJ...641...90R}
{Robertson}, B., {Hernquist}, L., {Cox}, T.~J., {Di Matteo}, T., {Hopkins},
  P.~F., {Martini}, P., and {Springel}, V. (2006).
\newblock {The Evolution of the M$_{BH}$-{$\sigma$} Relation}.
\newblock {\em ApJ}, 641:90--102.

\bibitem[{Rodriguez} et~al., 2006]{rodriguez:2006:csb}
{Rodriguez}, C., {Taylor}, G.~B., {Zavala}, R.~T., {Peck}, A.~B., {Pollack},
  L.~K., and {Romani}, R.~W. (2006).
\newblock {A Compact Supermassive Binary Black Hole System}.
\newblock {\em ApJ}, 646:49--60.

\bibitem[{Rodriguez} et~al., 2009]{rodirguez:2009:smbh}
{Rodriguez}, C., {Taylor}, G.~B., {Zavala}, R.~T., {Pihlstr{\"o}m}, Y.~M., and
  {Peck}, A.~B. (2009).
\newblock {H I Observations of the Supermassive Binary Black Hole System in
  0402+379}.
\newblock {\em ApJ}, 697:37--44.

\bibitem[{Roelofs} et~al., 2006]{roelofs:2006:kuh}
{Roelofs}, G.~H.~A., {Groot}, P.~J., {Nelemans}, G., {Marsh}, T.~R., and
  {Steeghs}, D. (2006).
\newblock {Kinematics of the ultracompact helium accretor AM Canum
  Venaticorum}.
\newblock {\em MNRAS}, 371:1231--1242.

\bibitem[{Roelofs} et~al., 2007]{roelofs:2007:sdss}
{Roelofs}, G.~H.~A., {Nelemans}, G., and {Groot}, P.~J. (2007).
\newblock {The population of AM CVn stars from the Sloan Digital Sky Survey}.
\newblock {\em MNRAS}, 382:685--692.

\bibitem[{Roelofs} et~al., 2010]{roelofs:2010:se}
{Roelofs}, G.~H.~A., {Rau}, A., {Marsh}, T.~R., {Steeghs}, D., {Groot}, P.~J.,
  and {Nelemans}, G. (2010).
\newblock {Spectroscopic Evidence for a 5.4 Minute Orbital Period in HM
  Cancri}.
\newblock {\em ApJ}, 711:L138--L142.

\bibitem[{Ruiter} et~al., 2009]{ruiter:2009:hwdb}
{Ruiter}, A.~J., {Belczynski}, K., {Benacquista}, M., and {Holley-Bockelmann},
  K. (2009).
\newblock {The Contribution of Halo White Dwarf Binaries to the Laser
  Interferometer Space Antenna Signal}.
\newblock {\em ApJ}, 693:383--387.

\bibitem[{Ruiter} et~al., 2010]{ruiter:2010ApJ...717.1006R}
{Ruiter}, A.~J., {Belczynski}, K., {Benacquista}, M., {Larson}, S.~L., and
  {Williams}, G. (2010).
\newblock {The LISA Gravitational Wave Foreground: A Study of Double White
  Dwarfs}.
\newblock {\em ApJ}, 717:1006--1021.

\bibitem[{Ryan}, 1995]{ryan:1995:gwi}
{Ryan}, F.~D. (1995).
\newblock {Gravitational waves from the inspiral of a compact object into a
  massive, axisymmetric body with arbitrary multipole moments}.
\newblock {\em Phys.~Rev.~D}, 52:5707--5718.

\bibitem[{Salpeter}, 1964]{Salpeter:1964}
{Salpeter}, E.~E. (1964).
\newblock {Accretion of Interstellar Matter by Massive Objects.}
\newblock {\em ApJ}, 140:796--800.

\bibitem[{Santamar{\'{\i}}a} et~al., 2010]{santamaria:2010PhRvD..82f4016S}
{Santamar{\'{\i}}a}, L., {Ohme}, F., {Ajith}, P., {Br{\"u}gmann}, B.,
  {Dorband}, N., {Hannam}, M., {Husa}, S., {M{\"o}sta}, P., {Pollney}, D.,
  {Reisswig}, C., {Robinson}, E.~L., {Seiler}, J., and {Krishnan}, B. (2010).
\newblock {Matching post-Newtonian and numerical relativity waveforms:
  Systematic errors and a new phenomenological model for nonprecessing black
  hole binaries}.
\newblock {\em Phys.~Rev.~D}, 82(6):064016--+.

\bibitem[{Schawinski} et~al., 2010]{schawinski:2010ApJ...714L.108S}
{Schawinski}, K., {Dowlin}, N., {Thomas}, D., {Urry}, C.~M., and {Edmondson},
  E. (2010).
\newblock {The Role of Mergers in Early-type Galaxy Evolution and Black Hole
  Growth}.
\newblock {\em ApJ}, 714:L108--L112.

\bibitem[{Schawinski} et~al., 2011]{schawinski:2011ApJ...727L..31S}
{Schawinski}, K., {Treister}, E., {Urry}, C.~M., {Cardamone}, C.~N., {Simmons},
  B., and {Yi}, S.~K. (2011).
\newblock {HST WFC3/IR Observations of Active Galactic Nucleus Host Galaxies at
  z \~{} 2: Supermassive Black Holes Grow in Disk Galaxies}.
\newblock {\em ApJ}, 727:L31+.

\bibitem[{Scheel} et~al., 2009]{scheel:2009PhRvD..79b4003S}
{Scheel}, M.~A., {Boyle}, M., {Chu}, T., {Kidder}, L.~E., {Matthews}, K.~D.,
  and {Pfeiffer}, H.~P. (2009).
\newblock {High-accuracy waveforms for binary black hole inspiral, merger, and
  ringdown}.
\newblock {\em Phys.~Rev.~D}, 79(2):024003--+.

\bibitem[{Scheuer} and {Feiler}, 1996]{scheuer:1996MNRAS.282..291S}
{Scheuer}, P.~A.~G. and {Feiler}, R. (1996).
\newblock {The realignment of a black hole misaligned with its accretion disc}.
\newblock {\em MNRAS}, 282:291--+.

\bibitem[{Schnittman}, 2007]{schnittman:2007ApJ...667L.133S}
{Schnittman}, J.~D. (2007).
\newblock {Retaining Black Holes with Very Large Recoil Velocities}.
\newblock {\em ApJ}, 667:L133--L136.

\bibitem[{Schnittman}, 2011]{schnittman:2011CQGra..28i4021S}
{Schnittman}, J.~D. (2011).
\newblock {Electromagnetic counterparts to black hole mergers}.
\newblock {\em Class.~Quantum~Grav.}, 28(9):094021--+.

\bibitem[{Schnittman} and {Buonanno}, 2007]{schnittman:2007ApJ...662L..63S}
{Schnittman}, J.~D. and {Buonanno}, A. (2007).
\newblock {The Distribution of Recoil Velocities from Merging Black Holes}.
\newblock {\em ApJ}, 662:L63--L66.

\bibitem[{Sch{\"o}del} et~al., 2003]{schoedel:2003:sdca}
{Sch{\"o}del}, R., {Ott}, T., {Genzel}, R., {Eckart}, A., {Mouawad}, N., and
  {Alexander}, T. (2003).
\newblock {Stellar Dynamics in the Central Arcsecond of Our Galaxy}.
\newblock {\em ApJ}, 596:1015--1034.

\bibitem[{Schutz}, 1986]{schutz:1986:hgw}
{Schutz}, B.~F. (1986).
\newblock {Determining the Hubble constant from gravitational wave
  observations}.
\newblock {\em Nature}, 323:310--+.

\bibitem[{Sesana}, 2010]{sesana:2010:scm}
{Sesana}, A. (2010).
\newblock {Self Consistent Model for the Evolution of Eccentric Massive Black
  Hole Binaries in Stellar Environments: Implications for Gravitational Wave
  Observations}.
\newblock {\em ApJ}, 719:851--864.

\bibitem[{Sesana} et~al., 2011]{sesana:2011:rmbh}
{Sesana}, A., {Gair}, J., {Berti}, E., and {Volonteri}, M. (2011).
\newblock {Reconstructing the massive black hole cosmic history through
  gravitational waves}.
\newblock {\em Phys.~Rev.~D}, 83(4):044036--+.

\bibitem[{Sesana} et~al., 2006]{sesana:2006:mbhb}
{Sesana}, A., {Haardt}, F., and {Madau}, P. (2006).
\newblock {Interaction of Massive Black Hole Binaries with Their Stellar
  Environment. I. Ejection of Hypervelocity Stars}.
\newblock {\em ApJ}, 651:392--400.

\bibitem[{Sesana} et~al., 2007a]{sesana:2007ApJ...660..546S}
{Sesana}, A., {Haardt}, F., and {Madau}, P. (2007a).
\newblock {Interaction of Massive Black Hole Binaries with Their Stellar
  Environment. II. Loss Cone Depletion and Binary Orbital Decay}.
\newblock {\em ApJ}, 660:546--555.

\bibitem[{Sesana} et~al., 2008a]{sesana:2008ApJ...686..432S}
{Sesana}, A., {Haardt}, F., and {Madau}, P. (2008a).
\newblock {Interaction of Massive Black Hole Binaries with Their Stellar
  Environment. III. Scattering of Bound Stars}.
\newblock {\em ApJ}, 686:432--447.

\bibitem[{Sesana} et~al., 2004]{sesana:2004:lfg}
{Sesana}, A., {Haardt}, F., {Madau}, P., and {Volonteri}, M. (2004).
\newblock {Low-Frequency Gravitational Radiation from Coalescing Massive Black
  Hole Binaries in Hierarchical Cosmologies}.
\newblock {\em ApJ}, 611:623--632.

\bibitem[{Sesana} et~al., 2005]{sesana:2005:gws}
{Sesana}, A., {Haardt}, F., {Madau}, P., and {Volonteri}, M. (2005).
\newblock {The Gravitational Wave Signal from Massive Black Hole Binaries and
  Its Contribution to the LISA Data Stream}.
\newblock {\em ApJ}, 623:23--30.

\bibitem[{Sesana} et~al., 2008b]{sesana:2008:sgwb}
{Sesana}, A., {Vecchio}, A., and {Colacino}, C.~N. (2008b).
\newblock {The stochastic gravitational-wave background from massive black hole
  binary systems: implications for observations with Pulsar Timing Arrays}.
\newblock {\em MNRAS}, 390:192--209.

\bibitem[{Sesana} et~al., 2009]{sesana:2009:gwrbb}
{Sesana}, A., {Vecchio}, A., and {Volonteri}, M. (2009).
\newblock {Gravitational waves from resolvable massive black hole binary
  systems and observations with Pulsar Timing Arrays}.
\newblock {\em MNRAS}, 394:2255--2265.

\bibitem[{Sesana} et~al., 2007b]{sesana:2007:imprint}
{Sesana}, A., {Volonteri}, M., and {Haardt}, F. (2007b).
\newblock {The imprint of massive black hole formation models on the LISA data
  stream}.
\newblock {\em MNRAS}, 377:1711--1716.

\bibitem[{Seth} et~al., 2008]{seth:2008ApJ...678..116S}
{Seth}, A., {Ag{\"u}eros}, M., {Lee}, D., and {Basu-Zych}, A. (2008).
\newblock {The Coincidence of Nuclear Star Clusters and Active Galactic
  Nuclei}.
\newblock {\em ApJ}, 678:116--130.

\bibitem[{Shakura} and {Sunyaev}, 1973]{shakura:1973A&A....24..337S}
{Shakura}, N.~I. and {Sunyaev}, R.~A. (1973).
\newblock {Black holes in binary systems. Observational appearance.}
\newblock {\em A\&A}, 24:337--355.

\bibitem[{Shapiro} and {Teukolsky}, 1979]{shapiro:1979:gc}
{Shapiro}, S.~L. and {Teukolsky}, S.~A. (1979).
\newblock {Gravitational collapse of supermassive stars to black holes -
  Numerical solution of the Einstein equations}.
\newblock {\em ApJ}, 234:L177--L181.

\bibitem[{Shapiro} and {Teukolsky}, 1986]{shapiro:1986bhwd.book.....S}
{Shapiro}, S.~L. and {Teukolsky}, S.~A. (1986).
\newblock {\em {Black Holes, White Dwarfs and Neutron Stars: The Physics of
  Compact Objects}}.
\newblock Wiley-VCH.

\bibitem[{Shen} and {Loeb}, 2010]{shen:2010ApJ...725..249S}
{Shen}, Y. and {Loeb}, A. (2010).
\newblock {Identifying Supermassive Black Hole Binaries with Broad Emission
  Line Diagnosis}.
\newblock {\em ApJ}, 725:249--260.

\bibitem[{Sheth} and {Tormen}, 1999]{sheth:1999:lsb}
{Sheth}, R.~K. and {Tormen}, G. (1999).
\newblock {Large-scale bias and the peak background split}.
\newblock {\em MNRAS}, 308:119--126.

\bibitem[{Shibata} and {Shapiro}, 2002]{shibata:2002ApJ...572L..39S}
{Shibata}, M. and {Shapiro}, S.~L. (2002).
\newblock {Collapse of a Rotating Supermassive Star to a Supermassive Black
  Hole: Fully Relativistic Simulations}.
\newblock {\em ApJ}, 572:L39--L43.

\bibitem[Shields et~al., 2009]{shields:0004-637X-707-2-936}
Shields, G.~A., Rosario, D.~J., Smith, K.~L., Bonning, E.~W., Salviander, S.,
  Kalirai, J.~S., Strickler, R., Ramirez-Ruiz, E., Dutton, A.~A., Treu, T., and
  Marshall, P.~J. (2009).
\newblock The quasar sdss j105041.35+345631.3: Black hole recoil or extreme
  double-peaked emitter?
\newblock {\em ApJ}, 707(2):936.

\bibitem[{Shlosman} et~al., 1989]{shlosman:1989Natur.338...45S}
{Shlosman}, I., {Frank}, J., and {Begelman}, M.~C. (1989).
\newblock {Bars within bars - A mechanism for fuelling active galactic nuclei}.
\newblock {\em Nature}, 338:45--47.

\bibitem[{Siemens} et~al., 2006]{siemens:2006:gwb}
{Siemens}, X., {Creighton}, J., {Maor}, I., {Majumder}, S.~R., {Cannon}, K.,
  and {Read}, J. (2006).
\newblock {Gravitational wave bursts from cosmic (super)strings: Quantitative
  analysis and constraints}.
\newblock {\em Phys.~Rev.~D}, 73(10):105001--+.

\bibitem[{Sigurdsson} and {Rees}, 1997]{sigurdsson:1997:csm}
{Sigurdsson}, S. and {Rees}, M.~J. (1997).
\newblock {Capture of stellar mass compact objects by massive black holes in
  galactic cusps}.
\newblock {\em MNRAS}, 284:318--326.

\bibitem[{Silk} and {Rees}, 1998]{silk:1998A&A...331L...1S}
{Silk}, J. and {Rees}, M.~J. (1998).
\newblock {Quasars and galaxy formation}.
\newblock {\em A\&A}, 331:L1--L4.

\bibitem[{Soltan}, 1982]{soltan:1982:moq}
{Soltan}, A. (1982).
\newblock {Masses of quasars}.
\newblock {\em MNRAS}, 200:115--122.

\bibitem[{Somerville} et~al., 2008]{somerville:2008MNRAS.391..481S}
{Somerville}, R.~S., {Hopkins}, P.~F., {Cox}, T.~J., {Robertson}, B.~E., and
  {Hernquist}, L. (2008).
\newblock {A semi-analytic model for the co-evolution of galaxies, black holes
  and active galactic nuclei}.
\newblock {\em MNRAS}, 391:481--506.

\bibitem[{Sopuerta}, 2010]{sopuerta:2010:emri}
{Sopuerta}, C.~F. (2010).
\newblock {A Roadmap to Fundamental Physics from LISA EMRI Observations}.
\newblock {\em GW Notes}, 4:3--47.

\bibitem[{Sopuerta} and {Yunes}, 2009]{sopuerta:2009PhRvD..80f4006S}
{Sopuerta}, C.~F. and {Yunes}, N. (2009).
\newblock {Extreme- and intermediate-mass ratio inspirals in dynamical
  Chern-Simons modified gravity}.
\newblock {\em Phys.~Rev.~D}, 80(6):064006--+.

\bibitem[{Springel} et~al., 2005]{springel:2005:sefc}
{Springel}, V., {White}, S.~D.~M., {Jenkins}, A., {Frenk}, C.~S., {Yoshida},
  N., {Gao}, L., {Navarro}, J., {Thacker}, R., {Croton}, D., {Helly}, J.,
  {Peacock}, J.~A., {Cole}, S., {Thomas}, P., {Couchman}, H., {Evrard}, A.,
  {Colberg}, J., and {Pearce}, F. (2005).
\newblock {Simulations of the formation, evolution and clustering of galaxies
  and quasars}.
\newblock {\em Nature}, 435:629--636.

\bibitem[{Stroeer} et~al., 2011]{stroeer:2011arXiv1109.4978S}
{Stroeer}, A., {Benacquista}, M., and {Ceballos}, F. (2011).
\newblock {Detecting Double Degenerate Progenitors of SNe Ia with LISA}.
\newblock {\em ArXiv e-prints}.

\bibitem[{Stroeer} and {Nelemans}, 2009]{stroeer:2009:stv}
{Stroeer}, A. and {Nelemans}, G. (2009).
\newblock {The influence of short-term variations in AM CVn systems on LISA
  measurements}.
\newblock {\em MNRAS}, 400:L24--L28.

\bibitem[{Stroeer} and {Vecchio}, 2006]{stroeer:2006:lvb}
{Stroeer}, A. and {Vecchio}, A. (2006).
\newblock {The LISA verification binaries}.
\newblock {\em Class.~Quantum~Grav.}, 23:809--+.

\bibitem[{Sullivan} et~al., 2011]{sullivan:2011ApJ...732..118S}
{Sullivan}, M., {Kasliwal}, M.~M., {Nugent}, P.~E., {Howell}, D.~A., {Thomas},
  R.~C., {Ofek}, E.~O., {Arcavi}, I., {Blake}, S., {Cooke}, J., {Gal-Yam}, A.,
  {Hook}, I.~M., {Mazzali}, P., {Podsiadlowski}, P., {Quimby}, R., {Bildsten},
  L., {Bloom}, J.~S., {Cenko}, S.~B., {Kulkarni}, S.~R., {Law}, N., and
  {Poznanski}, D. (2011).
\newblock {The Subluminous and Peculiar Type Ia Supernova PTF 09dav}.
\newblock {\em ApJ}, 732:118--+.

\bibitem[{Taam} and {Ricker}, 2010]{taam:2010NewAR..54...65T}
{Taam}, R.~E. and {Ricker}, P.~M. (2010).
\newblock {Common envelope evolution}.
\newblock {\em NewAR}, 54:65--71.

\bibitem[{Taam} and {Sandquist}, 2000]{taam:2000:cee}
{Taam}, R.~E. and {Sandquist}, E.~L. (2000).
\newblock {Common Envelope Evolution of Massive Binary Stars}.
\newblock {\em ARA\&A}, 38:113--141.

\bibitem[{Tegmark} et~al., 1997]{tegmark:1997ApJ...474....1T}
{Tegmark}, M., {Silk}, J., {Rees}, M.~J., {Blanchard}, A., {Abel}, T., and
  {Palla}, F. (1997).
\newblock {How Small Were the First Cosmological Objects?}
\newblock {\em ApJ}, 474:1--+.

\bibitem[{Teukolsky}, 1973]{teukolsky:1973:rbh}
{Teukolsky}, S.~A. (1973).
\newblock {Perturbations of a Rotating Black Hole. I. Fundamental Equations for
  Gravitational, Electromagnetic, and Neutrino-Field Perturbations}.
\newblock {\em ApJ}, 185:635--648.

\bibitem[{Thorne}, 1974]{thorne:1974:dabh}
{Thorne}, K.~S. (1974).
\newblock {Disk-Accretion onto a Black Hole. II. Evolution of the Hole}.
\newblock {\em ApJ}, 191:507--520.

\bibitem[{Tremaine} et~al., 2002]{tremaine:2002:slope}
{Tremaine}, S., {Gebhardt}, K., {Bender}, R., {Bower}, G., {Dressler}, A.,
  {Faber}, S.~M., {Filippenko}, A.~V., {Green}, R., {Grillmair}, C., {Ho},
  L.~C., {Kormendy}, J., {Lauer}, T.~R., {Magorrian}, J., {Pinkney}, J., and
  {Richstone}, D. (2002).
\newblock {The Slope of the Black Hole Mass versus Velocity Dispersion
  Correlation}.
\newblock {\em ApJ}, 574:740--753.

\bibitem[{Treu} et~al., 2007]{treu:2007ApJ...667..117T}
{Treu}, T., {Woo}, J.-H., {Malkan}, M.~A., and {Blandford}, R.~D. (2007).
\newblock {Cosmic Evolution of Black Holes and Spheroids. II. Scaling Relations
  at z=0.36}.
\newblock {\em ApJ}, 667:117--130.

\bibitem[{Tsalmantza} et~al., 2011]{tsalmantza:2011ApJ...738...20T}
{Tsalmantza}, P., {Decarli}, R., {Dotti}, M., and {Hogg}, D.~W. (2011).
\newblock {A Systematic Search for Massive Black Hole Binaries in the Sloan
  Digital Sky Survey Spectroscopic Sample}.
\newblock {\em ApJ}, 738:20--+.

\bibitem[{Tundo} et~al., 2007]{tundo:2007:bhmf}
{Tundo}, E., {Bernardi}, M., {Hyde}, J.~B., {Sheth}, R.~K., and {Pizzella}, A.
  (2007).
\newblock {On the Inconsistency between the Black Hole Mass Function Inferred
  from $M_\bullet$-$\sigma$ and $M_\bullet$-L Correlations}.
\newblock {\em ApJ}, 663:53--60.

\bibitem[{Valsecchi} et~al., 2011]{valsecchi:2011arXiv1105.4837V}
{Valsecchi}, F., {Farr}, W.~M., {Willems}, B., {Deloye}, C.~.~J., and
  {Kalogera}, V. (2011).
\newblock {Tidally-Induced Apsidal Precession in Double White Dwarfs: a new
  mass measurement tool with LISA}.
\newblock {\em ArXiv e-prints}.

\bibitem[{Valtonen} et~al., 2008]{valtonen:2008Natur.452..851V}
{Valtonen}, M.~J., {Lehto}, H.~J., {Nilsson}, K., {Heidt}, J., {Takalo}, L.~O.,
  {Sillanp{\"a}{\"a}}, A., {Villforth}, C., {Kidger}, M., {Poyner}, G.,
  {Pursimo}, T., {Zola}, S., {Wu}, J.-H., {Zhou}, X., {Sadakane}, K., {Drozdz},
  M., {Koziel}, D., {Marchev}, D., {Ogloza}, W., {Porowski}, C., {Siwak}, M.,
  {Stachowski}, G., {Winiarski}, M., {Hentunen}, V.-P., {Nissinen}, M.,
  {Liakos}, A., and {Dogru}, S. (2008).
\newblock {A massive binary black-hole system in OJ287 and a test of general
  relativity}.
\newblock {\em Nature}, 452:851--853.

\bibitem[{Valtonen} et~al., 2011]{valtonen:2011ApJ...729...33V}
{Valtonen}, M.~J., {Lehto}, H.~J., {Takalo}, L.~O., and {Sillanp{\"a}{\"a}}, A.
  (2011).
\newblock {Testing the 1995 Binary Black Hole Model of OJ287}.
\newblock {\em ApJ}, 729:33--+.

\bibitem[{Valtonen} et~al., 2010]{valtonen:2010ApJ...709..725V}
{Valtonen}, M.~J., {Mikkola}, S., {Merritt}, D., {Gopakumar}, A., {Lehto},
  H.~J., {Hyv{\"o}nen}, T., {Rampadarath}, H., {Saunders}, R., {Basta}, M., and
  {Hudec}, R. (2010).
\newblock {Measuring the Spin of the Primary Black Hole in OJ287}.
\newblock {\em ApJ}, 709:725--732.

\bibitem[{Valtonen} et~al., 2006]{valtonen:2006ApJ...643L...9V}
{Valtonen}, M.~J., {Nilsson}, K., {Sillanp{\"a}{\"a}}, A., {Takalo}, L.~O.,
  {Lehto}, H.~J., {Keel}, W.~C., {Haque}, S., {Cornwall}, D., and {Mattingly},
  A. (2006).
\newblock {The 2005 November Outburst in OJ 287 and the Binary Black Hole
  Model}.
\newblock {\em ApJ}, 643:L9--L12.

\bibitem[{van der Marel} et~al., 1994]{vandermarel:1994MNRAS.268..521V}
{van der Marel}, R.~P., {Rix}, H.~W., {Carter}, D., {Franx}, M., {White},
  S.~D.~M., and {de Zeeuw}, T. (1994).
\newblock {Velocity Profiles of Galaxies with Claimed Black-Holes - Part One -
  Observations of M31 M32 NGC3115 and NGC4594}.
\newblock {\em MNRAS}, 268:521--+.

\bibitem[{Vigeland} et~al., 2011]{vigeland:2011PhRvD..83j4027V}
{Vigeland}, S., {Yunes}, N., and {Stein}, L.~C. (2011).
\newblock {Bumpy black holes in alternative theories of gravity}.
\newblock {\em Phys.~Rev.~D}, 83(10):104027--+.

\bibitem[{Vigeland} and {Hughes}, 2010]{vigeland:2010PhRvD..81b4030V}
{Vigeland}, S.~J. and {Hughes}, S.~A. (2010).
\newblock {Spacetime and orbits of bumpy black holes}.
\newblock {\em Phys.~Rev.~D}, 81(2):024030--+.

\bibitem[{Volonteri}, 2010]{volonteri:2010A&ARv..18..279V}
{Volonteri}, M. (2010).
\newblock {Formation of supermassive black holes}.
\newblock {\em A\&A~Rev.}, 18:279--315.

\bibitem[{Volonteri} and {Begelman}, 2010]{volonteri:2010:qce}
{Volonteri}, M. and {Begelman}, M.~C. (2010).
\newblock {Quasi-stars and the cosmic evolution of massive black holes}.
\newblock {\em MNRAS}, 409:1022--1032.

\bibitem[{Volonteri} et~al., 2010]{volonteri:2010MNRAS.404.2143V}
{Volonteri}, M., {G{\"u}ltekin}, K., and {Dotti}, M. (2010).
\newblock {Gravitational recoil: effects on massive black hole occupation
  fraction over cosmic time}.
\newblock {\em MNRAS}, 404:2143--2150.

\bibitem[{Volonteri} et~al., 2003]{volonteri:2003:amh}
{Volonteri}, M., {Haardt}, F., and {Madau}, P. (2003).
\newblock {The Assembly and Merging History of Supermassive Black Holes in
  Hierarchical Models of Galaxy Formation}.
\newblock {\em ApJ}, 582:559--573.

\bibitem[{Volonteri} et~al., 2008]{volonteri:2008MNRAS.383.1079V}
{Volonteri}, M., {Lodato}, G., and {Natarajan}, P. (2008).
\newblock {The evolution of massive black hole seeds}.
\newblock {\em MNRAS}, 383:1079--1088.

\bibitem[{Volonteri} et~al., 2009]{volonteri:2009:subpc}
{Volonteri}, M., {Miller}, J.~M., and {Dotti}, M. (2009).
\newblock {Sub-Parsec Supermassive Binary Quasars: Expectations at $z< 1$}.
\newblock {\em ApJ}, 703:L86--L89.

\bibitem[{Volonteri} and {Natarajan}, 2009]{volonteri:2009MNRAS.400.1911V}
{Volonteri}, M. and {Natarajan}, P. (2009).
\newblock {Journey to the M$_{BH}$-{$\sigma$} relation: the fate of low-mass
  black holes in the Universe}.
\newblock {\em MNRAS}, 400:1911--1918.

\bibitem[{Volonteri} et~al., 2007]{volonteri:2007:bhs}
{Volonteri}, M., {Sikora}, M., and {Lasota}, J. (2007).
\newblock {Black Hole Spin and Galactic Morphology}.
\newblock {\em ApJ}, 667:704--713.

\bibitem[{Waldman} et~al., 2011]{waldman:2011ApJ...738...21W}
{Waldman}, R., {Sauer}, D., {Livne}, E., {Perets}, H., {Glasner}, A.,
  {Mazzali}, P., {Truran}, J.~W., and {Gal-Yam}, A. (2011).
\newblock {Helium Shell Detonations on Low-mass White Dwarfs as a Possible
  Explanation for SN 2005E}.
\newblock {\em ApJ}, 738:21--+.

\bibitem[{Wang} et~al., 2002]{wang:2002ApJ...572L..15W}
{Wang}, Y., {Holz}, D.~E., and {Munshi}, D. (2002).
\newblock {A Universal Probability Distribution Function for Weak-lensing
  Amplification}.
\newblock {\em ApJ}, 572:L15--L18.

\bibitem[{Watts} et~al., 2008]{watts:2008:gw}
{Watts}, A.~L., {Krishnan}, B., {Bildsten}, L., and {Schutz}, B.~F. (2008).
\newblock {Detecting gravitational wave emission from the known accreting
  neutron stars}.
\newblock {\em MNRAS}, 389:839--868.

\bibitem[{Webbink}, 1984]{webbink:1984:dwd}
{Webbink}, R.~F. (1984).
\newblock {Double white dwarfs as progenitors of R Coronae Borealis stars and
  Type I supernovae}.
\newblock {\em ApJ}, 277:355--360.

\bibitem[{Webbink}, 2010]{webbink:2010:wts}
{Webbink}, R.~F. (2010).
\newblock {Watch This Space: Observing Merging White Dwarfs}.
\newblock {\em AIP~Conf.~Series}, 1314:217--222.

\bibitem[{Wehner} and {Harris}, 2006]{wehner:2006ApJ...644L..17W}
{Wehner}, E.~H. and {Harris}, W.~E. (2006).
\newblock {From Supermassive Black Holes to Dwarf Elliptical Nuclei: A Mass
  Continuum}.
\newblock {\em ApJ}, 644:L17--L20.

\bibitem[{Weisberg} and {Taylor}, 2005]{weisberg:2005:rbp}
{Weisberg}, J.~M. and {Taylor}, J.~H. (2005).
\newblock {The Relativistic Binary Pulsar B1913+16: Thirty Years of
  Observations and Analysis}.
\newblock In {F.~A.~Rasio \& I.~H.~Stairs}, editor, {\em Binary Radio Pulsars},
  volume 328 of {\em Astronomical Society of the Pacific Conference Series},
  pages 25--+.

\bibitem[{White} and {Rees}, 1978]{white:1978MNRAS.183..341W}
{White}, S.~D.~M. and {Rees}, M.~J. (1978).
\newblock {Core condensation in heavy halos - A two-stage theory for galaxy
  formation and clustering}.
\newblock {\em MNRAS}, 183:341--358.

\bibitem[{Wijnands}, 2010]{wijnands:2010HiA....15..121W}
{Wijnands}, R. (2010).
\newblock {Accreting millisecond X-ray pulsars: recent developments}.
\newblock {\em Highlights of Astronomy}, 15:121--125.

\bibitem[{Will}, 2006]{will:2006:gre}
{Will}, C.~M. (2006).
\newblock {The Confrontation between General Relativity and Experiment}.
\newblock {\em Living Reviews in Relativity}, 9:3--+.

\bibitem[{Willems} et~al., 2010]{willems:2010ApJ...713..239W}
{Willems}, B., {Deloye}, C.~J., and {Kalogera}, V. (2010).
\newblock {Energy Dissipation Through Quasi-static Tides in White Dwarf
  Binaries}.
\newblock {\em ApJ}, 713:239--256.

\bibitem[{Witten}, 1984]{witten:1984:csp}
{Witten}, E. (1984).
\newblock {Cosmic separation of phases}.
\newblock {\em Phys.~Rev.~D}, 30:272--285.

\bibitem[{Woo} et~al., 2008]{woo:2008ApJ...681..925W}
{Woo}, J.-H., {Treu}, T., {Malkan}, M.~A., and {Blandford}, R.~D. (2008).
\newblock {Cosmic Evolution of Black Holes and Spheroids. III. The
  M$_{BH}$-{$\sigma$}$_{*}$ Relation in the Last Six Billion Years}.
\newblock {\em ApJ}, 681:925--930.

\bibitem[{Wyithe} and {Loeb}, 2002]{wyithe:2002ApJ...581..886W}
{Wyithe}, J.~S.~B. and {Loeb}, A. (2002).
\newblock {A Physical Model for the Luminosity Function of High-Redshift
  Quasars}.
\newblock {\em ApJ}, 581:886--894.

\bibitem[Wyithe and Loeb, 2003]{wyithe:2003:lfg}
Wyithe, J. S.~B. and Loeb, A. (2003).
\newblock Low-frequency gravitational waves from massive black hole binaries:
  Predictions for {LISA} and pulsar timing arrays.
\newblock {\em ApJ}, 590:691--706.

\bibitem[{Wyithe} and {Loeb}, 2003]{wyithe:2003ApJ...595..614W}
{Wyithe}, J.~S.~B. and {Loeb}, A. (2003).
\newblock {Self-regulated Growth of Supermassive Black Holes in Galaxies as the
  Origin of the Optical and X-Ray Luminosity Functions of Quasars}.
\newblock {\em ApJ}, 595:614--623.

\bibitem[{Xiao} et~al., 2011]{xiao:2011ApJ...739...28X}
{Xiao}, T., {Barth}, A.~J., {Greene}, J.~E., {Ho}, L.~C., {Bentz}, M.~C.,
  {Ludwig}, R.~R., and {Jiang}, Y. (2011).
\newblock {Exploring the Low-mass End of the M $_{BH}$-{$\sigma$}$_{*}$
  Relation with Active Galaxies}.
\newblock {\em ApJ}, 739:28--+.

\bibitem[{Xu} et~al., 2011]{xu:2011:cosmos}
{Xu}, C.~K., {Zhao}, Y., {Scoville}, N., {Capak}, P., {Drory}, N., and {Gao},
  Y. (2011).
\newblock {Galaxy Pairs in COSMOS -- Merger Rate Evolution Since z=1}.
\newblock {\em ArXiv e-prints}.

\bibitem[{Yagi} et~al., 2011]{yagi:2011PhRvD..83h4036Y}
{Yagi}, K., {Tanahashi}, N., and {Tanaka}, T. (2011).
\newblock {Probing the size of extra dimensions with gravitational wave
  astronomy}.
\newblock {\em Phys.~Rev.~D}, 83(8):084036--+.

\bibitem[{Yagi} and {Tanaka}, 2010]{yagi:2010PhRvD..81f4008Y}
{Yagi}, K. and {Tanaka}, T. (2010).
\newblock {Constraining alternative theories of gravity by gravitational waves
  from precessing eccentric compact binaries with LISA}.
\newblock {\em Phys.~Rev.~D}, 81(6):064008--+.

\bibitem[{Yoshida} et~al., 1994]{yoshida:1994PhRvD..50.6235Y}
{Yoshida}, S., {Eriguchi}, Y., and {Futamase}, T. (1994).
\newblock {Quasinormal modes of boson stars}.
\newblock {\em Phys.~Rev.~D}, 50:6235--6246.

\bibitem[{Yu} and {Tremaine}, 2002]{yu:2002:oc}
{Yu}, Q. and {Tremaine}, S. (2002).
\newblock {Observational constraints on growth of massive black holes}.
\newblock {\em MNRAS}, 335:965--976.

\bibitem[{Yu} and {Jeffery}, 2010]{yu:2010:gw}
{Yu}, S. and {Jeffery}, C.~S. (2010).
\newblock {The gravitational wave signal from diverse populations of double
  white dwarf binaries in the Galaxy}.
\newblock {\em A\&A}, 521:A85+.

\bibitem[{Yunes} et~al., 2011]{yunes:2011PhRvD..83d4030Y}
{Yunes}, N., {Coleman Miller}, M., and {Thornburg}, J. (2011).
\newblock {Effect of massive perturbers on extreme mass-ratio inspiral
  waveforms}.
\newblock {\em Phys.~Rev.~D}, 83(4):044030--+.

\bibitem[Yunes et~al., 2011]{Yunes:2011aaarXiv1112.3351}
Yunes, N., Pani, P., and Cardoso, V. (2011).
\newblock {Gravitational Waves from Extreme Mass-Ratio Inspirals as Probes of
  Scalar-Tensor Theories}.
\newblock {\em ArXiv e-prints}.

\bibitem[{Yunes} and {Pretorius}, 2009]{yunes:2009PhRvD..80l2003Y}
{Yunes}, N. and {Pretorius}, F. (2009).
\newblock {Fundamental theoretical bias in gravitational wave astrophysics and
  the parametrized post-Einsteinian framework}.
\newblock {\em Phys.~Rev.~D}, 80(12):122003--+.

\bibitem[{Yunes} et~al., 2010]{yunes:2010PhRvD..81f4018Y}
{Yunes}, N., {Pretorius}, F., and {Spergel}, D. (2010).
\newblock {Constraining the evolutionary history of Newton's constant with
  gravitational wave observations}.
\newblock {\em Phys.~Rev.~D}, 81(6):064018--+.

\bibitem[{Yungelson} et~al., 2002]{yungelson:2002:ucxb}
{Yungelson}, L.~R., {Nelemans}, G., and {van den Heuvel}, E.~P.~J. (2002).
\newblock {On the formation of neon-enriched donor stars in ultracompact X-ray
  binaries}.
\newblock {\em A\&A}, 388:546--551.

\bibitem[{Zel'dovich} and {Novikov}, 1964]{zeldovich:1964:rgw}
{Zel'dovich}, Y.~B. and {Novikov}, I.~D. (1964).
\newblock {The Radiation of Gravity Waves by Bodies Moving in the Field of a
  Collapsing Star}.
\newblock {\em Soviet Physics Doklady}, 9:246--+.

\bibitem[{Zinnecker} and {Yorke}, 2007]{zinnecker:2007ARA&A..45..481Z}
{Zinnecker}, H. and {Yorke}, H.~W. (2007).
\newblock {Toward Understanding Massive Star Formation}.
\newblock {\em ARA\&A}, 45:481--563.

\end{thebibliography}
\end{document}